\pdfoutput=1   %
\documentclass[letterpaper,twocolumn,10pt]{article}
\usepackage{usenix2019_v3.1}
\usepackage{graphicx}
\usepackage{booktabs}
\usepackage{amsmath}
\usepackage{amssymb}
\usepackage{pifont}
\usepackage{xspace}
\usepackage{multirow}
\usepackage{array}
\usepackage{tikz}
\usetikzlibrary{arrows.meta,positioning,calc,decorations.pathreplacing,patterns}
\usepackage[labelfont=bf,font=small,skip=4pt]{caption}
\usepackage{enumitem}
\setlist{nosep,leftmargin=*}
\usepackage{balance}
\usepackage{rotating}   %

\graphicspath{{figures/}}

\makeatletter
\newcommand{\R}[1]{\@ifundefined{bd@#1}{\textbf{\textcolor{red}{??#1}}\PackageWarning{berth}{undefined result key #1}}{\@nameuse{bd@#1}}}
\newcommand{\defR}[2]{\@namedef{bd@#1}{#2}}
\makeatother
\newcommand{\nPoints}{31}  %
\newcommand{\loadScaleFive}{67}  %
\newcommand{\loadScaleTen}{80}  %
\newcommand{\loadChurnPeak}{75}  %
\newcommand{\tabNotRun}{, including MPS's tenants beside BERT-large training and the GEMM job, where MPS did not start; MPS static partitions did not run that point (its mean covers the other 30 points).}  %
\newcommand{\nBaselines}{13}  %
\newcommand{\windowS}{90}  %
\newcommand{\warmupS}{5}  %
\newcommand{\feedMs}{1}  %
\newcommand{\hostDriver}{580.178.04}  %
\newcommand{\hostTorch}{2.8.0}  %
\newcommand{\hostCuda}{12.8}  %
\newcommand{\svcResNet}{5.3}  %
\newcommand{\svcRetina}{10.5}  %
\newcommand{\svcBert}{20.9}  %
\newcommand{\svcPercep}{11.8}  %
\newcommand{\thrAloneTrainBert}{1.62}  %
\newcommand{\thrAloneTrainResnet}{13.29}  %
\newcommand{\thrAloneGemm}{15.81}  %
\newcommand{\busyResNet}{68}  %
\newcommand{\tracePeak}{3.0}  %
\newcommand{\hostCpu}{AMD EPYC 7763}  %
\newcommand{\hostOs}{Ubuntu 22.04.5 LTS}  %
\newcommand{\kMaxTrainResnet}{1.7}  %
\newcommand{\kMaxTrainBert}{15.3}  %
\newcommand{\kTwTrainBert}{2.05}  %
\newcommand{\kMaxGemm}{62.8}  %
\newcommand{\tsPingAlone}{54}  %
\newcommand{\tsPreemptPfifty}{0.27}  %
\newcommand{\tsPreemptMax}{2.26}  %
\newcommand{\tsSwitchCost}{0.2}  %
\newcommand{\tsIdlePfifty}{0.69}  %
\newcommand{\tsIdleMax}{1.2}  %
\newcommand{\tsHogLong}{61}  %
\newcommand{\tsHogShort}{1.04}  %
\newcommand{\tsMediumMax}{34}  %
\newcommand{\tsLongMax}{43}  %
\newcommand{\tsRunAlone}{13.3}  %
\newcommand{\tsRunUninterrupted}{8.5}  %
\newcommand{\tsHogLoss}{1.2}  %
\newcommand{\tsNmin}{198}  %
\newcommand{\tsNmax}{351}  %
\newcommand{\hlRnAlone}{97.3}  %
\newcommand{\hlRnTs}{64.9}  %
\newcommand{\hlRnMpsprio}{0.0}  %
\newcommand{\hlRnTgs}{96.8}  %
\newcommand{\hlRnTgsThr}{6}  %
\newcommand{\aggBerthAtt}{95.8}  %
\newcommand{\aggBerthAttMin}{94.1}  %
\newcommand{\aggBerthThr}{48}  %
\newcommand{\aggBerthClose}{24}  %
\newcommand{\aggBerthNsbAttMin}{94.2}  %
\newcommand{\aggBerthNsbThr}{49}  %
\newcommand{\aggBerthNgAttMin}{91.9}  %
\newcommand{\aggFeedGrpAttMin}{82.9}  %
\newcommand{\aggFeedGrpThr}{52}  %
\newcommand{\aggLithosThr}{18}  %
\newcommand{\aggMpslimitsThr}{16}  %
\newcommand{\aggMpslimitsClose}{18}  %
\newcommand{\aggAloneAtt}{97.9}  %
\newcommand{\thrRatioRange}{2.7--3.5}  %
\newcommand{\nXsPoints}{27}  %
\newcommand{\zeroMps}{28}  %
\newcommand{\zeroMpsprio}{24}  %
\newcommand{\BerthScaleTenMin}{88.3}  %
\newcommand{\FeedGrpScaleTenMin}{59.3}  %
\newcommand{\BerthScaleFiveMin}{95.1}  %
\newcommand{\FeedGrpScaleFiveMin}{72.1}  %
\newcommand{\klBerthGemm}{95.1}  %
\newcommand{\klBerthGemmThr}{71}  %
\newcommand{\klBerthGemmPX}{1.1}  %
\newcommand{\klTsGemm}{67.0}  %
\newcommand{\klTgsGemm}{66.9}  %
\newcommand{\klFeedGrpGemm}{66.4}  %
\newcommand{\klFeedGrpGemmPX}{2.3}  %
\newcommand{\klBerthBertGemm}{99.6}  %
\newcommand{\klBerthBertGemmThr}{63}  %
\newcommand{\klBerthMixGemm}{96.4}  %
\newcommand{\klBerthMixGemmThr}{55}  %
\newcommand{\ghPingPnnDef}{2.25}  %
\newcommand{\ghPingPnnFive}{0.55}  %
\newcommand{\ghHogKernel}{61}  %
\newcommand{\shimRnPct}{5.9}  %
\newcommand{\shimTrainRnPct}{6.4}  %
\newcommand{\shimRnLaunches}{222}  %
\newcommand{\hundredGrpRnLo}{40}  %
\newcommand{\hundredGrpRnHi}{41}  %
\newcommand{\hundredGrpThr}{0.35}  %
\newcommand{\matGrpRnLo}{72}  %
\newcommand{\matGrpRnHi}{74}  %
\newcommand{\matGrpThr}{0.58}  %
\newcommand{\costMedian}{7}  %
\newcommand{\costScaleTen}{26}  %
\newcommand{\costTwobatch}{33}  %
\newcommand{\costScaleFive}{20}  %
\newcommand{\feasCornerAll}{62}  %
\newcommand{\frontierAll}{52.8}  %
\newcommand{\frontierAllWho}{Orion}  %
\newcommand{\scSliceRn}{95.2}  %
\newcommand{\scSliceRnGemm}{96.6}  %
\newcommand{\nScMulti}{12}  %
\newcommand{\scMultiSlice}{53.4}  %
\newcommand{\scMultiTierSlice}{60.3}  %
\newcommand{\scMultiBerth}{92.9}  %
\newcommand{\tierOnlyScaleFive}{95.7}  %
\newcommand{\tierOnlyScaleTen}{89.3}  %
\newcommand{\tierOnlyTightHalf}{75.3}  %
\newcommand{\graceBertSeventy}{94.9}  %
\newcommand{\graceBertSeventyNo}{67.8}  %
\newcommand{\graceMixSeventy}{94.6}  %
\newcommand{\graceMixSeventyNo}{90.3}  %
\newcommand{\fcFShortTight}{77.7}  %
\newcommand{\fcFShortMix}{95.1}  %
\newcommand{\fcLossLo}{23}  %
\newcommand{\fcLossHi}{69}  %
\newcommand{\fcLongCostTight}{10.1}  %
\newcommand{\rnSubGateOnly}{95.5}  %
\newcommand{\rnSubQbOnly}{95.4}  %
\newcommand{\llmBerthWorstMin}{99.8}  %
\newcommand{\llmBerthWorstMax}{100.0}  %
\newcommand{\llmBerthBeLo}{17}  %
\newcommand{\llmBerthBeHi}{38}  %
\newcommand{\llmBerthBeGemm}{46}  %
\newcommand{\llmBerthTpotLo}{15}  %
\newcommand{\llmBerthTpotHi}{22}  %
\newcommand{\llmBerthTtftHi}{126}  %
\newcommand{\llmBerthNsngWorstMin}{100.0}  %
\newcommand{\llmBerthNsngBeLo}{0}  %
\newcommand{\llmBerthNsngBeHi}{5}  %
\newcommand{\llmBerthNsngTpotLo}{10}  %
\newcommand{\llmBerthNsngTpotHi}{14}  %
\newcommand{\llmShareNames}{MPS limits, MPS static partitions, TPC quotas (LithOS-like) and MIG}  %
\newcommand{\llmShareWorstLo}{89.4}  %
\newcommand{\llmShareWorstHi}{100.0}  %
\newcommand{\llmShareBeLo}{11}  %
\newcommand{\llmShareBeHi}{54}  %
\newcommand{\llmMigSeventy}{0.0}  %
\newcommand{\llmLithosSeventy}{14.8}  %
\newcommand{\llmStockTpotLo}{35}  %
\newcommand{\llmStockTpotHi}{431}  %
\newcommand{\llmTgsBeMax}{1}  %
\newcommand{\llmStockWorstMax}{3.1}  %
\newcommand{\mxBerthRn}{79.6}  %
\newcommand{\mxBerthSvc}{91.5}  %
\newcommand{\mxBerthTpot}{27.7}  %
\newcommand{\mxBerthBe}{39}  %
\newcommand{\mxNsrRn}{51.5}  %
\newcommand{\mxNsrSvc}{11.9}  %
\newcommand{\mxNsngRn}{87.0}  %
\newcommand{\mxNsngSvc}{100.0}  %
\newcommand{\mxNsngBe}{22}  %
\newcommand{\mxBeCost}{6.5}  %
\newcommand{\mxSpatialLo}{98}  %
\newcommand{\mxSpatialBeLo}{9.5}  %
\newcommand{\mxSpatialBeHi}{9.9}  %
\newcommand{\mxThrRatio}{4}  %
\newcommand{\splitBerthBusyLo}{97}  %
\newcommand{\splitBerthBusyHi}{100}  %
\newcommand{\splitBerthLc}{38}  %
\newcommand{\splitBerthBe}{57}  %
\newcommand{\splitBerthCtxLo}{2}  %
\newcommand{\splitBerthCtxHi}{4}  %
\newcommand{\splitBerthBeThrLo}{54}  %
\newcommand{\splitBerthBeThrHi}{56}  %
\newcommand{\splitMigIdle}{31}  %
\newcommand{\splitTgsIdle}{55}  %
\newcommand{\splitTgsBe}{7}  %
\newcommand{\splitLimitsResident}{99.9}  %
\newcommand{\splitLimitsThrLo}{10}  %
\newcommand{\splitLimitsThrHi}{12}  %
\newcommand{\tjBerthWorst}{81.1}  %
\newcommand{\tjBerthTrain}{19.2}  %
\newcommand{\tjBerthGemm}{19.8}  %
\newcommand{\tjMpslimitsWorst}{87.9}  %
\newcommand{\tjMpslimitsThr}{10.7}  %
\newcommand{\tjLithosWorst}{89.0}  %
\newcommand{\tjLithosThr}{6.0}  %
\newcommand{\tjMigWorst}{90.7}  %
\newcommand{\tjMigThr}{0.0}  %
\newcommand{\tjTsWorst}{19.1}  %
\newcommand{\tjXsWorst}{44.8}  %
\newcommand{\tjlBerthTrain}{20.5}  %
\newcommand{\tjlBerthGemm}{20.3}  %
\newcommand{\tjlBerthTpot}{18.3}  %
\newcommand{\tjlNbaWorst}{99.8}  %
\newcommand{\tjlNbaTrain}{0.7}  %
\newcommand{\tjlNbaGemm}{45.1}  %
\newcommand{\tjlNbaTpot}{22.7}  %
\newcommand{\tjlMigTrain}{54.5}  %
\newcommand{\tjlLithosGemm}{53.6}  %
\newcommand{\tjlTsWorst}{1.6}  %
\newcommand{\tjlXsWorst}{2.4}  %
\newcommand{\nLlmPoints}{7}  %
\newcommand{\llmNotRunNote}{; --: not run}  %
\newcommand{\nOtherPoints}{21}  %
\newcommand{\nGhPoints}{30}  %
\newcommand{\ghBerthWorst}{94.2}  %
\newcommand{\ghBerthMean}{95.5}  %
\newcommand{\ghBerthThr}{51}  %
\newcommand{\ghXsThr}{59}  %
\newcommand{\ghBerthOnXsThr}{51}  %
\newcommand{\ghXsBehind}{6}  %
\newcommand{\ghTgsBehind}{0.2}  %
\newcommand{\ghLlmBerthThr}{51}  %
\newcommand{\ghLlmTsWorst}{91.3}  %
\newcommand{\ghLlmMpslimitsThr}{54}  %
\newcommand{\ghLlmMigThr}{36}  %
\newcommand{\sevenBBerthMetLo}{98.1}  %
\newcommand{\sevenBBerthMetHi}{100}  %
\newcommand{\sevenBBerthThrLo}{27}  %
\newcommand{\sevenBBerthThrHi}{52}  %
\newcommand{\sevenBTsMetLo}{1.9}  %
\newcommand{\sevenBTsMetHi}{46.3}  %
\newcommand{\sevenBMpslimitsMetLo}{7.8}  %
\newcommand{\sevenBMpslimitsMetHi}{98.2}  %
\newcommand{\sevenBMigMetLo}{99.9}  %
\newcommand{\sevenBMigMetHi}{100}  %
\newcommand{\sevenTwoNoArbMet}{93.8}  %
\newcommand{\sevenTwoArbMet}{100}  %
\newcommand{\sevenTwoNoArbTrain}{2}  %
\newcommand{\sevenTwoArbTrain}{22}  %
\newcommand{\ghXsThreeWorst}{92.8}  %
\newcommand{\ghXsThreeThr}{51}  %
\newcommand{\ghXsOneWorst}{89.0}  %
\newcommand{\ghXsOneThr}{60}  %
\newcommand{\ghBerthEightWorst}{95.4}  %
\newcommand{\ghBerthEightThr}{50}  %
\newcommand{\ghABerthWorst}{94.6}  %
\newcommand{\ghABerthAtt}{96.1}  %
\newcommand{\ghABerthThr}{49}  %
\newcommand{\ghLeadLo}{3}  %
\newcommand{\ghLeadHi}{16}  %
\newcommand{\ghThrVsMig}{1.6}  %
\newcommand{\ghThrVsLimits}{2.7}  %
\newcommand{\ghScaleTen}{99.1}  %
\newcommand{\ghScaleTenA}{88.3}  %
\newcommand{\ghPortPingDef}{2.00}  %
\newcommand{\ghPortPingFive}{0.33}  %
\newcommand{\ghPortHogCost}{0.5}  %
\newcommand{\tailBerthMed}{1.2}  %
\newcommand{\tailBerthMax}{5.0}  %
\newcommand{\tailTsMed}{7.7}  %
\newcommand{\tailXsMax}{17}  %
\newcommand{\tailShareMedLo}{1.5}  %
\newcommand{\tailShareMedHi}{1.7}  %
\newcommand{\tailShareMaxLo}{27}  %
\newcommand{\tailShareMaxHi}{2,052}  %
\newcommand{\scSliceTen}{0}  %
\newcommand{\xsThreeAMean}{67.6}  %
\newcommand{\xsOneAMean}{73.9}  %
\newcommand{\cgPairOff}{97.5}  %
\newcommand{\cgPairOn}{97.2}  %
\newcommand{\cgPairPlain}{100.0}  %
\newcommand{\cgMixOff}{95.3}  %
\newcommand{\cgMixOn}{94.6}  %
\newcommand{\cgMixPlain}{99.4}  %
\newcommand{\cgKernFwd}{249}  %
\newcommand{\cgKernBwd}{934}  %
\newcommand{\cgMsFwd}{193}  %
\newcommand{\cgMsBwd}{382}  %
\newcommand{\vlPair}{99.5}  %
\newcommand{\vlPairFixed}{100.0}  %
\newcommand{\vlPairDiff}{0.5}  %
\newcommand{\vlMix}{96.5}  %
\newcommand{\vlMixFixed}{99.4}  %
\newcommand{\vlMixDiff}{2.8}  %
\newcommand{\vlUnder}{2.9}  %
\newcommand{\smLcDramLlm}{38}  %
\newcommand{\smLcDramSeven}{47}  %
\newcommand{\cpuPairBerth}{57}  %
\newcommand{\cpuPairRef}{135}  %
\newcommand{\cpuBerthd}{0.9}  %
\newcommand{\cpuWatcher}{0.3}  %
\newcommand{\cpuLcShim}{1}  %
\newcommand{\sensTpotTenA}{17.9}  %
\newcommand{\sensBeTenA}{41}  %
\newcommand{\sensTpotThirtyA}{14.5}  %
\newcommand{\sensBeThirtyA}{34}  %
\newcommand{\sensTpotTenB}{19.6}  %
\newcommand{\sensBeTenB}{33}  %
\newcommand{\sensTpotThirtyB}{16.3}  %
\newcommand{\sensBeThirtyB}{27}  %
\newcommand{\sensCTpot}{0.3}  %
\defR{pair-resnet50-load30/berth-full/attmin}{96.7}
\defR{pair-resnet50-load30/berth-full/attmean}{96.7}
\defR{pair-resnet50-load30/berth-full/thr}{70}
\defR{pair-resnet50-load30/berth-full/n}{3}
\defR{pair-resnet50-load30/berth-full/att/resnet50}{96.7}
\defR{pair-resnet50-load30/berth-full/p99/resnet50}{18.4}
\defR{pair-resnet50-load30/berth-full/p99x/resnet50}{1.0}
\defR{pair-resnet50-load30/berth-full/thrabs/train_bert}{1.13}
\defR{pair-resnet50-load30/berth-full/thr/train_bert}{70}
\defR{pair-resnet50-load30/berth-nsr/attmin}{96.4}
\defR{pair-resnet50-load30/berth-nsr/attmean}{96.4}
\defR{pair-resnet50-load30/berth-nsr/thr}{70}
\defR{pair-resnet50-load30/berth-nsr/n}{3}
\defR{pair-resnet50-load30/berth-nsr/att/resnet50}{96.4}
\defR{pair-resnet50-load30/berth-nsr/p99/resnet50}{18.7}
\defR{pair-resnet50-load30/berth-nsr/p99x/resnet50}{1.0}
\defR{pair-resnet50-load30/berth-nsr/thrabs/train_bert}{1.13}
\defR{pair-resnet50-load30/berth-nsr/thr/train_bert}{70}
\defR{pair-resnet50-load30/berth-nba/attmin}{96.5}
\defR{pair-resnet50-load30/berth-nba/attmean}{96.5}
\defR{pair-resnet50-load30/berth-nba/thr}{69}
\defR{pair-resnet50-load30/berth-nba/n}{3}
\defR{pair-resnet50-load30/berth-nba/att/resnet50}{96.5}
\defR{pair-resnet50-load30/berth-nba/p99/resnet50}{19.0}
\defR{pair-resnet50-load30/berth-nba/p99x/resnet50}{1.0}
\defR{pair-resnet50-load30/berth-nba/thrabs/train_bert}{1.13}
\defR{pair-resnet50-load30/berth-nba/thr/train_bert}{69}
\defR{pair-resnet50-load30/berth-nsb/attmin}{96.9}
\defR{pair-resnet50-load30/berth-nsb/attmean}{96.9}
\defR{pair-resnet50-load30/berth-nsb/thr}{70}
\defR{pair-resnet50-load30/berth-nsb/n}{3}
\defR{pair-resnet50-load30/berth-nsb/att/resnet50}{96.9}
\defR{pair-resnet50-load30/berth-nsb/p99/resnet50}{18.3}
\defR{pair-resnet50-load30/berth-nsb/p99x/resnet50}{1.0}
\defR{pair-resnet50-load30/berth-nsb/thrabs/train_bert}{1.13}
\defR{pair-resnet50-load30/berth-nsb/thr/train_bert}{70}
\defR{pair-resnet50-load30/berth-ng/attmin}{95.6}
\defR{pair-resnet50-load30/berth-ng/attmean}{95.6}
\defR{pair-resnet50-load30/berth-ng/thr}{71}
\defR{pair-resnet50-load30/berth-ng/n}{3}
\defR{pair-resnet50-load30/berth-ng/att/resnet50}{95.6}
\defR{pair-resnet50-load30/berth-ng/p99/resnet50}{19.6}
\defR{pair-resnet50-load30/berth-ng/p99x/resnet50}{1.1}
\defR{pair-resnet50-load30/berth-ng/thrabs/train_bert}{1.15}
\defR{pair-resnet50-load30/berth-ng/thr/train_bert}{71}
\defR{pair-resnet50-load30/berth-nsng/attmin}{95.8}
\defR{pair-resnet50-load30/berth-nsng/attmean}{95.8}
\defR{pair-resnet50-load30/berth-nsng/thr}{71}
\defR{pair-resnet50-load30/berth-nsng/n}{3}
\defR{pair-resnet50-load30/berth-nsng/att/resnet50}{95.8}
\defR{pair-resnet50-load30/berth-nsng/p99/resnet50}{19.4}
\defR{pair-resnet50-load30/berth-nsng/p99x/resnet50}{1.1}
\defR{pair-resnet50-load30/berth-nsng/thrabs/train_bert}{1.15}
\defR{pair-resnet50-load30/berth-nsng/thr/train_bert}{71}
\defR{pair-resnet50-load30/berth-grouped/attmin}{95.0}
\defR{pair-resnet50-load30/berth-grouped/attmean}{95.0}
\defR{pair-resnet50-load30/berth-grouped/thr}{72}
\defR{pair-resnet50-load30/berth-grouped/n}{3}
\defR{pair-resnet50-load30/berth-grouped/att/resnet50}{95.0}
\defR{pair-resnet50-load30/berth-grouped/p99/resnet50}{21.2}
\defR{pair-resnet50-load30/berth-grouped/p99x/resnet50}{1.2}
\defR{pair-resnet50-load30/berth-grouped/thrabs/train_bert}{1.17}
\defR{pair-resnet50-load30/berth-grouped/thr/train_bert}{72}
\defR{pair-resnet50-load30/berth/attmin}{95.0}
\defR{pair-resnet50-load30/berth/attmean}{95.0}
\defR{pair-resnet50-load30/berth/thr}{71}
\defR{pair-resnet50-load30/berth/n}{3}
\defR{pair-resnet50-load30/berth/att/resnet50}{95.0}
\defR{pair-resnet50-load30/berth/p99/resnet50}{20.5}
\defR{pair-resnet50-load30/berth/p99x/resnet50}{1.1}
\defR{pair-resnet50-load30/berth/thrabs/train_bert}{1.16}
\defR{pair-resnet50-load30/berth/thr/train_bert}{71}
\defR{pair-resnet50-load30/timeslice/attmin}{64.9}
\defR{pair-resnet50-load30/timeslice/attmean}{64.9}
\defR{pair-resnet50-load30/timeslice/thr}{71}
\defR{pair-resnet50-load30/timeslice/n}{3}
\defR{pair-resnet50-load30/timeslice/att/resnet50}{64.9}
\defR{pair-resnet50-load30/timeslice/p99/resnet50}{46.3}
\defR{pair-resnet50-load30/timeslice/p99x/resnet50}{2.5}
\defR{pair-resnet50-load30/timeslice/thrabs/train_bert}{1.16}
\defR{pair-resnet50-load30/timeslice/thr/train_bert}{71}
\defR{pair-resnet50-load30/mps/attmin}{0.0}
\defR{pair-resnet50-load30/mps/attmean}{0.0}
\defR{pair-resnet50-load30/mps/thr}{93}
\defR{pair-resnet50-load30/mps/n}{3}
\defR{pair-resnet50-load30/mps/att/resnet50}{0.0}
\defR{pair-resnet50-load30/mps/p99/resnet50}{53787}
\defR{pair-resnet50-load30/mps/p99x/resnet50}{2953.3}
\defR{pair-resnet50-load30/mps/thrabs/train_bert}{1.51}
\defR{pair-resnet50-load30/mps/thr/train_bert}{93}
\defR{pair-resnet50-load30/mpsprio/attmin}{0.0}
\defR{pair-resnet50-load30/mpsprio/attmean}{0.0}
\defR{pair-resnet50-load30/mpsprio/thr}{83}
\defR{pair-resnet50-load30/mpsprio/n}{3}
\defR{pair-resnet50-load30/mpsprio/att/resnet50}{0.0}
\defR{pair-resnet50-load30/mpsprio/p99/resnet50}{12787}
\defR{pair-resnet50-load30/mpsprio/p99x/resnet50}{702.1}
\defR{pair-resnet50-load30/mpsprio/thrabs/train_bert}{1.35}
\defR{pair-resnet50-load30/mpsprio/thr/train_bert}{83}
\defR{pair-resnet50-load30/mpslimits/attmin}{98.5}
\defR{pair-resnet50-load30/mpslimits/attmean}{98.5}
\defR{pair-resnet50-load30/mpslimits/thr}{12}
\defR{pair-resnet50-load30/mpslimits/n}{3}
\defR{pair-resnet50-load30/mpslimits/att/resnet50}{98.5}
\defR{pair-resnet50-load30/mpslimits/p99/resnet50}{15.9}
\defR{pair-resnet50-load30/mpslimits/p99x/resnet50}{0.9}
\defR{pair-resnet50-load30/mpslimits/thrabs/train_bert}{0.19}
\defR{pair-resnet50-load30/mpslimits/thr/train_bert}{12}
\defR{pair-resnet50-load30/mpsparts/attmin}{98.0}
\defR{pair-resnet50-load30/mpsparts/attmean}{98.0}
\defR{pair-resnet50-load30/mpsparts/thr}{10}
\defR{pair-resnet50-load30/mpsparts/n}{2}
\defR{pair-resnet50-load30/mpsparts/att/resnet50}{98.0}
\defR{pair-resnet50-load30/mpsparts/p99/resnet50}{16.6}
\defR{pair-resnet50-load30/mpsparts/p99x/resnet50}{0.9}
\defR{pair-resnet50-load30/mpsparts/thrabs/train_bert}{0.16}
\defR{pair-resnet50-load30/mpsparts/thr/train_bert}{10}
\defR{pair-resnet50-load30/lithos/attmin}{98.0}
\defR{pair-resnet50-load30/lithos/attmean}{98.0}
\defR{pair-resnet50-load30/lithos/thr}{9}
\defR{pair-resnet50-load30/lithos/n}{3}
\defR{pair-resnet50-load30/lithos/att/resnet50}{98.0}
\defR{pair-resnet50-load30/lithos/p99/resnet50}{16.7}
\defR{pair-resnet50-load30/lithos/p99x/resnet50}{0.9}
\defR{pair-resnet50-load30/lithos/thrabs/train_bert}{0.14}
\defR{pair-resnet50-load30/lithos/thr/train_bert}{9}
\defR{pair-resnet50-load30/dysm/attmin}{0.0}
\defR{pair-resnet50-load30/dysm/attmean}{0.0}
\defR{pair-resnet50-load30/dysm/thr}{100}
\defR{pair-resnet50-load30/dysm/n}{3}
\defR{pair-resnet50-load30/dysm/att/resnet50}{0.0}
\defR{pair-resnet50-load30/dysm/thrabs/train_bert}{1.62}
\defR{pair-resnet50-load30/dysm/thr/train_bert}{100}
\defR{pair-resnet50-load30/xsched/attmin}{88.3}
\defR{pair-resnet50-load30/xsched/attmean}{88.3}
\defR{pair-resnet50-load30/xsched/thr}{70}
\defR{pair-resnet50-load30/xsched/n}{3}
\defR{pair-resnet50-load30/xsched/att/resnet50}{88.3}
\defR{pair-resnet50-load30/xsched/p99/resnet50}{28.4}
\defR{pair-resnet50-load30/xsched/p99x/resnet50}{1.6}
\defR{pair-resnet50-load30/xsched/thrabs/train_bert}{1.13}
\defR{pair-resnet50-load30/xsched/thr/train_bert}{70}
\defR{pair-resnet50-load30/multistream/attmin}{0.0}
\defR{pair-resnet50-load30/multistream/attmean}{0.0}
\defR{pair-resnet50-load30/multistream/thr}{84}
\defR{pair-resnet50-load30/multistream/n}{3}
\defR{pair-resnet50-load30/multistream/att/resnet50}{0.0}
\defR{pair-resnet50-load30/multistream/p99/resnet50}{17740}
\defR{pair-resnet50-load30/multistream/p99x/resnet50}{974.1}
\defR{pair-resnet50-load30/multistream/thrabs/train_bert}{1.37}
\defR{pair-resnet50-load30/multistream/thr/train_bert}{84}
\defR{pair-resnet50-load30/tgs/attmin}{96.8}
\defR{pair-resnet50-load30/tgs/attmean}{96.8}
\defR{pair-resnet50-load30/tgs/thr}{6}
\defR{pair-resnet50-load30/tgs/n}{3}
\defR{pair-resnet50-load30/tgs/att/resnet50}{96.8}
\defR{pair-resnet50-load30/tgs/p99/resnet50}{18.5}
\defR{pair-resnet50-load30/tgs/p99x/resnet50}{1.0}
\defR{pair-resnet50-load30/tgs/thrabs/train_bert}{0.09}
\defR{pair-resnet50-load30/tgs/thr/train_bert}{6}
\defR{pair-resnet50-load30/mig/attmin}{95.3}
\defR{pair-resnet50-load30/mig/attmean}{95.3}
\defR{pair-resnet50-load30/mig/thr}{45}
\defR{pair-resnet50-load30/mig/n}{3}
\defR{pair-resnet50-load30/mig/att/resnet50}{95.3}
\defR{pair-resnet50-load30/mig/p99/resnet50}{20.1}
\defR{pair-resnet50-load30/mig/p99x/resnet50}{1.1}
\defR{pair-resnet50-load30/mig/thrabs/train_bert}{0.73}
\defR{pair-resnet50-load30/mig/thr/train_bert}{45}
\defR{pair-resnet50-load30/tally/attmin}{0.0}
\defR{pair-resnet50-load30/tally/attmean}{0.0}
\defR{pair-resnet50-load30/tally/thr}{1}
\defR{pair-resnet50-load30/tally/n}{3}
\defR{pair-resnet50-load30/tally/att/resnet50}{0.0}
\defR{pair-resnet50-load30/tally/p99/resnet50}{49175}
\defR{pair-resnet50-load30/tally/p99x/resnet50}{2700.1}
\defR{pair-resnet50-load30/tally/thrabs/train_bert}{0.01}
\defR{pair-resnet50-load30/tally/thr/train_bert}{1}
\defR{pair-resnet50-load30/orion/attmin}{93.2}
\defR{pair-resnet50-load30/orion/attmean}{93.2}
\defR{pair-resnet50-load30/orion/thr}{76}
\defR{pair-resnet50-load30/orion/n}{3}
\defR{pair-resnet50-load30/orion/att/resnet50}{93.2}
\defR{pair-resnet50-load30/orion/p99/resnet50}{31.7}
\defR{pair-resnet50-load30/orion/p99x/resnet50}{1.7}
\defR{pair-resnet50-load30/orion/thrabs/train_bert}{1.23}
\defR{pair-resnet50-load30/orion/thr/train_bert}{76}
\defR{pair-resnet50-load50/berth-full/attmin}{87.0}
\defR{pair-resnet50-load50/berth-full/attmean}{87.0}
\defR{pair-resnet50-load50/berth-full/thr}{49}
\defR{pair-resnet50-load50/berth-full/n}{3}
\defR{pair-resnet50-load50/berth-full/att/resnet50}{87.0}
\defR{pair-resnet50-load50/berth-full/p99/resnet50}{28.2}
\defR{pair-resnet50-load50/berth-full/p99x/resnet50}{1.1}
\defR{pair-resnet50-load50/berth-full/thrabs/train_bert}{0.80}
\defR{pair-resnet50-load50/berth-full/thr/train_bert}{49}
\defR{pair-resnet50-load50/berth-nsr/attmin}{87.5}
\defR{pair-resnet50-load50/berth-nsr/attmean}{87.5}
\defR{pair-resnet50-load50/berth-nsr/thr}{49}
\defR{pair-resnet50-load50/berth-nsr/n}{3}
\defR{pair-resnet50-load50/berth-nsr/att/resnet50}{87.5}
\defR{pair-resnet50-load50/berth-nsr/p99/resnet50}{27.6}
\defR{pair-resnet50-load50/berth-nsr/p99x/resnet50}{1.1}
\defR{pair-resnet50-load50/berth-nsr/thrabs/train_bert}{0.80}
\defR{pair-resnet50-load50/berth-nsr/thr/train_bert}{49}
\defR{pair-resnet50-load50/berth-nba/attmin}{88.2}
\defR{pair-resnet50-load50/berth-nba/attmean}{88.2}
\defR{pair-resnet50-load50/berth-nba/thr}{50}
\defR{pair-resnet50-load50/berth-nba/n}{3}
\defR{pair-resnet50-load50/berth-nba/att/resnet50}{88.2}
\defR{pair-resnet50-load50/berth-nba/p99/resnet50}{27.3}
\defR{pair-resnet50-load50/berth-nba/p99x/resnet50}{1.1}
\defR{pair-resnet50-load50/berth-nba/thrabs/train_bert}{0.80}
\defR{pair-resnet50-load50/berth-nba/thr/train_bert}{50}
\defR{pair-resnet50-load50/berth-nsb/attmin}{88.8}
\defR{pair-resnet50-load50/berth-nsb/attmean}{88.8}
\defR{pair-resnet50-load50/berth-nsb/thr}{50}
\defR{pair-resnet50-load50/berth-nsb/n}{3}
\defR{pair-resnet50-load50/berth-nsb/att/resnet50}{88.8}
\defR{pair-resnet50-load50/berth-nsb/p99/resnet50}{27.0}
\defR{pair-resnet50-load50/berth-nsb/p99x/resnet50}{1.0}
\defR{pair-resnet50-load50/berth-nsb/thrabs/train_bert}{0.81}
\defR{pair-resnet50-load50/berth-nsb/thr/train_bert}{50}
\defR{pair-resnet50-load50/berth-ng/attmin}{84.7}
\defR{pair-resnet50-load50/berth-ng/attmean}{84.7}
\defR{pair-resnet50-load50/berth-ng/thr}{52}
\defR{pair-resnet50-load50/berth-ng/n}{3}
\defR{pair-resnet50-load50/berth-ng/att/resnet50}{84.7}
\defR{pair-resnet50-load50/berth-ng/p99/resnet50}{29.9}
\defR{pair-resnet50-load50/berth-ng/p99x/resnet50}{1.2}
\defR{pair-resnet50-load50/berth-ng/thrabs/train_bert}{0.84}
\defR{pair-resnet50-load50/berth-ng/thr/train_bert}{52}
\defR{pair-resnet50-load50/berth-nsng/attmin}{85.6}
\defR{pair-resnet50-load50/berth-nsng/attmean}{85.6}
\defR{pair-resnet50-load50/berth-nsng/thr}{52}
\defR{pair-resnet50-load50/berth-nsng/n}{3}
\defR{pair-resnet50-load50/berth-nsng/att/resnet50}{85.6}
\defR{pair-resnet50-load50/berth-nsng/p99/resnet50}{28.9}
\defR{pair-resnet50-load50/berth-nsng/p99x/resnet50}{1.1}
\defR{pair-resnet50-load50/berth-nsng/thrabs/train_bert}{0.84}
\defR{pair-resnet50-load50/berth-nsng/thr/train_bert}{52}
\defR{pair-resnet50-load50/berth-grouped/attmin}{85.2}
\defR{pair-resnet50-load50/berth-grouped/attmean}{85.2}
\defR{pair-resnet50-load50/berth-grouped/thr}{52}
\defR{pair-resnet50-load50/berth-grouped/n}{3}
\defR{pair-resnet50-load50/berth-grouped/att/resnet50}{85.2}
\defR{pair-resnet50-load50/berth-grouped/p99/resnet50}{31.2}
\defR{pair-resnet50-load50/berth-grouped/p99x/resnet50}{1.2}
\defR{pair-resnet50-load50/berth-grouped/thrabs/train_bert}{0.84}
\defR{pair-resnet50-load50/berth-grouped/thr/train_bert}{52}
\defR{pair-resnet50-load50/berth/attmin}{84.6}
\defR{pair-resnet50-load50/berth/attmean}{84.6}
\defR{pair-resnet50-load50/berth/thr}{52}
\defR{pair-resnet50-load50/berth/n}{3}
\defR{pair-resnet50-load50/berth/att/resnet50}{84.6}
\defR{pair-resnet50-load50/berth/p99/resnet50}{30.8}
\defR{pair-resnet50-load50/berth/p99x/resnet50}{1.2}
\defR{pair-resnet50-load50/berth/thrabs/train_bert}{0.84}
\defR{pair-resnet50-load50/berth/thr/train_bert}{52}
\defR{pair-resnet50-load50/timeslice/attmin}{20.5}
\defR{pair-resnet50-load50/timeslice/attmean}{20.5}
\defR{pair-resnet50-load50/timeslice/thr}{52}
\defR{pair-resnet50-load50/timeslice/n}{3}
\defR{pair-resnet50-load50/timeslice/att/resnet50}{20.5}
\defR{pair-resnet50-load50/timeslice/p99/resnet50}{182}
\defR{pair-resnet50-load50/timeslice/p99x/resnet50}{7.0}
\defR{pair-resnet50-load50/timeslice/thrabs/train_bert}{0.85}
\defR{pair-resnet50-load50/timeslice/thr/train_bert}{52}
\defR{pair-resnet50-load50/mps/attmin}{0.0}
\defR{pair-resnet50-load50/mps/attmean}{0.0}
\defR{pair-resnet50-load50/mps/thr}{93}
\defR{pair-resnet50-load50/mps/n}{3}
\defR{pair-resnet50-load50/mps/att/resnet50}{0.0}
\defR{pair-resnet50-load50/mps/p99/resnet50}{70292}
\defR{pair-resnet50-load50/mps/p99x/resnet50}{2725.7}
\defR{pair-resnet50-load50/mps/thrabs/train_bert}{1.51}
\defR{pair-resnet50-load50/mps/thr/train_bert}{93}
\defR{pair-resnet50-load50/mpsprio/attmin}{0.0}
\defR{pair-resnet50-load50/mpsprio/attmean}{0.0}
\defR{pair-resnet50-load50/mpsprio/thr}{83}
\defR{pair-resnet50-load50/mpsprio/n}{3}
\defR{pair-resnet50-load50/mpsprio/att/resnet50}{0.0}
\defR{pair-resnet50-load50/mpsprio/p99/resnet50}{45637}
\defR{pair-resnet50-load50/mpsprio/p99x/resnet50}{1769.7}
\defR{pair-resnet50-load50/mpsprio/thrabs/train_bert}{1.35}
\defR{pair-resnet50-load50/mpsprio/thr/train_bert}{83}
\defR{pair-resnet50-load50/mpslimits/attmin}{90.6}
\defR{pair-resnet50-load50/mpslimits/attmean}{90.6}
\defR{pair-resnet50-load50/mpslimits/thr}{12}
\defR{pair-resnet50-load50/mpslimits/n}{3}
\defR{pair-resnet50-load50/mpslimits/att/resnet50}{90.6}
\defR{pair-resnet50-load50/mpslimits/p99/resnet50}{25.5}
\defR{pair-resnet50-load50/mpslimits/p99x/resnet50}{1.0}
\defR{pair-resnet50-load50/mpslimits/thrabs/train_bert}{0.19}
\defR{pair-resnet50-load50/mpslimits/thr/train_bert}{12}
\defR{pair-resnet50-load50/mpsparts/attmin}{91.5}
\defR{pair-resnet50-load50/mpsparts/attmean}{91.5}
\defR{pair-resnet50-load50/mpsparts/thr}{9}
\defR{pair-resnet50-load50/mpsparts/n}{2}
\defR{pair-resnet50-load50/mpsparts/att/resnet50}{91.5}
\defR{pair-resnet50-load50/mpsparts/p99/resnet50}{24.6}
\defR{pair-resnet50-load50/mpsparts/p99x/resnet50}{1.0}
\defR{pair-resnet50-load50/mpsparts/thrabs/train_bert}{0.14}
\defR{pair-resnet50-load50/mpsparts/thr/train_bert}{9}
\defR{pair-resnet50-load50/lithos/attmin}{90.5}
\defR{pair-resnet50-load50/lithos/attmean}{90.5}
\defR{pair-resnet50-load50/lithos/thr}{9}
\defR{pair-resnet50-load50/lithos/n}{3}
\defR{pair-resnet50-load50/lithos/att/resnet50}{90.5}
\defR{pair-resnet50-load50/lithos/p99/resnet50}{25.4}
\defR{pair-resnet50-load50/lithos/p99x/resnet50}{1.0}
\defR{pair-resnet50-load50/lithos/thrabs/train_bert}{0.14}
\defR{pair-resnet50-load50/lithos/thr/train_bert}{9}
\defR{pair-resnet50-load50/dysm/attmin}{0.0}
\defR{pair-resnet50-load50/dysm/attmean}{0.0}
\defR{pair-resnet50-load50/dysm/thr}{100}
\defR{pair-resnet50-load50/dysm/n}{3}
\defR{pair-resnet50-load50/dysm/att/resnet50}{0.0}
\defR{pair-resnet50-load50/dysm/thrabs/train_bert}{1.62}
\defR{pair-resnet50-load50/dysm/thr/train_bert}{100}
\defR{pair-resnet50-load50/xsched/attmin}{67.7}
\defR{pair-resnet50-load50/xsched/attmean}{67.7}
\defR{pair-resnet50-load50/xsched/thr}{51}
\defR{pair-resnet50-load50/xsched/n}{3}
\defR{pair-resnet50-load50/xsched/att/resnet50}{67.7}
\defR{pair-resnet50-load50/xsched/p99/resnet50}{55.5}
\defR{pair-resnet50-load50/xsched/p99x/resnet50}{2.2}
\defR{pair-resnet50-load50/xsched/thrabs/train_bert}{0.83}
\defR{pair-resnet50-load50/xsched/thr/train_bert}{51}
\defR{pair-resnet50-load50/multistream/attmin}{0.0}
\defR{pair-resnet50-load50/multistream/attmean}{0.0}
\defR{pair-resnet50-load50/multistream/thr}{84}
\defR{pair-resnet50-load50/multistream/n}{3}
\defR{pair-resnet50-load50/multistream/att/resnet50}{0.0}
\defR{pair-resnet50-load50/multistream/p99/resnet50}{48493}
\defR{pair-resnet50-load50/multistream/p99x/resnet50}{1880.4}
\defR{pair-resnet50-load50/multistream/thrabs/train_bert}{1.37}
\defR{pair-resnet50-load50/multistream/thr/train_bert}{84}
\defR{pair-resnet50-load50/tgs/attmin}{66.6}
\defR{pair-resnet50-load50/tgs/attmean}{66.6}
\defR{pair-resnet50-load50/tgs/thr}{21}
\defR{pair-resnet50-load50/tgs/n}{3}
\defR{pair-resnet50-load50/tgs/att/resnet50}{66.6}
\defR{pair-resnet50-load50/tgs/p99/resnet50}{85.3}
\defR{pair-resnet50-load50/tgs/p99x/resnet50}{3.3}
\defR{pair-resnet50-load50/tgs/thrabs/train_bert}{0.35}
\defR{pair-resnet50-load50/tgs/thr/train_bert}{21}
\defR{pair-resnet50-load50/mig/attmin}{91.2}
\defR{pair-resnet50-load50/mig/attmean}{91.2}
\defR{pair-resnet50-load50/mig/thr}{0}
\defR{pair-resnet50-load50/mig/n}{3}
\defR{pair-resnet50-load50/mig/att/resnet50}{91.2}
\defR{pair-resnet50-load50/mig/p99/resnet50}{24.8}
\defR{pair-resnet50-load50/mig/p99x/resnet50}{1.0}
\defR{pair-resnet50-load50/mig/thrabs/train_bert}{0.00}
\defR{pair-resnet50-load50/mig/thr/train_bert}{0}
\defR{pair-resnet50-load50/tally/attmin}{0.0}
\defR{pair-resnet50-load50/tally/attmean}{0.0}
\defR{pair-resnet50-load50/tally/thr}{1}
\defR{pair-resnet50-load50/tally/n}{3}
\defR{pair-resnet50-load50/tally/att/resnet50}{0.0}
\defR{pair-resnet50-load50/tally/p99/resnet50}{81170}
\defR{pair-resnet50-load50/tally/p99x/resnet50}{3147.5}
\defR{pair-resnet50-load50/tally/thrabs/train_bert}{0.01}
\defR{pair-resnet50-load50/tally/thr/train_bert}{1}
\defR{pair-resnet50-load50/orion/attmin}{87.2}
\defR{pair-resnet50-load50/orion/attmean}{87.2}
\defR{pair-resnet50-load50/orion/thr}{61}
\defR{pair-resnet50-load50/orion/n}{3}
\defR{pair-resnet50-load50/orion/att/resnet50}{87.2}
\defR{pair-resnet50-load50/orion/p99/resnet50}{39.2}
\defR{pair-resnet50-load50/orion/p99x/resnet50}{1.5}
\defR{pair-resnet50-load50/orion/thrabs/train_bert}{0.99}
\defR{pair-resnet50-load50/orion/thr/train_bert}{61}
\defR{pair-resnet50-load70/berth-full/attmin}{66.9}
\defR{pair-resnet50-load70/berth-full/attmean}{66.9}
\defR{pair-resnet50-load70/berth-full/thr}{28}
\defR{pair-resnet50-load70/berth-full/n}{3}
\defR{pair-resnet50-load70/berth-full/att/resnet50}{66.9}
\defR{pair-resnet50-load70/berth-full/p99/resnet50}{51.9}
\defR{pair-resnet50-load70/berth-full/p99x/resnet50}{1.1}
\defR{pair-resnet50-load70/berth-full/thrabs/train_bert}{0.46}
\defR{pair-resnet50-load70/berth-full/thr/train_bert}{28}
\defR{pair-resnet50-load70/berth-nsr/attmin}{67.0}
\defR{pair-resnet50-load70/berth-nsr/attmean}{67.0}
\defR{pair-resnet50-load70/berth-nsr/thr}{28}
\defR{pair-resnet50-load70/berth-nsr/n}{3}
\defR{pair-resnet50-load70/berth-nsr/att/resnet50}{67.0}
\defR{pair-resnet50-load70/berth-nsr/p99/resnet50}{52.0}
\defR{pair-resnet50-load70/berth-nsr/p99x/resnet50}{1.1}
\defR{pair-resnet50-load70/berth-nsr/thrabs/train_bert}{0.46}
\defR{pair-resnet50-load70/berth-nsr/thr/train_bert}{28}
\defR{pair-resnet50-load70/berth-nba/attmin}{66.4}
\defR{pair-resnet50-load70/berth-nba/attmean}{66.4}
\defR{pair-resnet50-load70/berth-nba/thr}{28}
\defR{pair-resnet50-load70/berth-nba/n}{3}
\defR{pair-resnet50-load70/berth-nba/att/resnet50}{66.4}
\defR{pair-resnet50-load70/berth-nba/p99/resnet50}{65.2}
\defR{pair-resnet50-load70/berth-nba/p99x/resnet50}{1.4}
\defR{pair-resnet50-load70/berth-nba/thrabs/train_bert}{0.46}
\defR{pair-resnet50-load70/berth-nba/thr/train_bert}{28}
\defR{pair-resnet50-load70/berth-nsb/attmin}{68.9}
\defR{pair-resnet50-load70/berth-nsb/attmean}{68.9}
\defR{pair-resnet50-load70/berth-nsb/thr}{29}
\defR{pair-resnet50-load70/berth-nsb/n}{3}
\defR{pair-resnet50-load70/berth-nsb/att/resnet50}{68.9}
\defR{pair-resnet50-load70/berth-nsb/p99/resnet50}{48.2}
\defR{pair-resnet50-load70/berth-nsb/p99x/resnet50}{1.1}
\defR{pair-resnet50-load70/berth-nsb/thrabs/train_bert}{0.48}
\defR{pair-resnet50-load70/berth-nsb/thr/train_bert}{29}
\defR{pair-resnet50-load70/berth-ng/attmin}{63.5}
\defR{pair-resnet50-load70/berth-ng/attmean}{63.5}
\defR{pair-resnet50-load70/berth-ng/thr}{32}
\defR{pair-resnet50-load70/berth-ng/n}{3}
\defR{pair-resnet50-load70/berth-ng/att/resnet50}{63.5}
\defR{pair-resnet50-load70/berth-ng/p99/resnet50}{55.0}
\defR{pair-resnet50-load70/berth-ng/p99x/resnet50}{1.2}
\defR{pair-resnet50-load70/berth-ng/thrabs/train_bert}{0.53}
\defR{pair-resnet50-load70/berth-ng/thr/train_bert}{32}
\defR{pair-resnet50-load70/berth-nsng/attmin}{63.8}
\defR{pair-resnet50-load70/berth-nsng/attmean}{63.8}
\defR{pair-resnet50-load70/berth-nsng/thr}{33}
\defR{pair-resnet50-load70/berth-nsng/n}{3}
\defR{pair-resnet50-load70/berth-nsng/att/resnet50}{63.8}
\defR{pair-resnet50-load70/berth-nsng/p99/resnet50}{53.8}
\defR{pair-resnet50-load70/berth-nsng/p99x/resnet50}{1.2}
\defR{pair-resnet50-load70/berth-nsng/thrabs/train_bert}{0.53}
\defR{pair-resnet50-load70/berth-nsng/thr/train_bert}{33}
\defR{pair-resnet50-load70/berth-grouped/attmin}{65.8}
\defR{pair-resnet50-load70/berth-grouped/attmean}{65.8}
\defR{pair-resnet50-load70/berth-grouped/thr}{34}
\defR{pair-resnet50-load70/berth-grouped/n}{3}
\defR{pair-resnet50-load70/berth-grouped/att/resnet50}{65.8}
\defR{pair-resnet50-load70/berth-grouped/p99/resnet50}{56.6}
\defR{pair-resnet50-load70/berth-grouped/p99x/resnet50}{1.2}
\defR{pair-resnet50-load70/berth-grouped/thrabs/train_bert}{0.55}
\defR{pair-resnet50-load70/berth-grouped/thr/train_bert}{34}
\defR{pair-resnet50-load70/berth/attmin}{65.1}
\defR{pair-resnet50-load70/berth/attmean}{65.1}
\defR{pair-resnet50-load70/berth/thr}{34}
\defR{pair-resnet50-load70/berth/n}{3}
\defR{pair-resnet50-load70/berth/att/resnet50}{65.1}
\defR{pair-resnet50-load70/berth/p99/resnet50}{58.4}
\defR{pair-resnet50-load70/berth/p99x/resnet50}{1.3}
\defR{pair-resnet50-load70/berth/thrabs/train_bert}{0.54}
\defR{pair-resnet50-load70/berth/thr/train_bert}{34}
\defR{pair-resnet50-load70/timeslice/attmin}{0.0}
\defR{pair-resnet50-load70/timeslice/attmean}{0.0}
\defR{pair-resnet50-load70/timeslice/thr}{46}
\defR{pair-resnet50-load70/timeslice/n}{3}
\defR{pair-resnet50-load70/timeslice/att/resnet50}{0.0}
\defR{pair-resnet50-load70/timeslice/p99/resnet50}{18109}
\defR{pair-resnet50-load70/timeslice/p99x/resnet50}{396.5}
\defR{pair-resnet50-load70/timeslice/thrabs/train_bert}{0.75}
\defR{pair-resnet50-load70/timeslice/thr/train_bert}{46}
\defR{pair-resnet50-load70/mps/attmin}{0.0}
\defR{pair-resnet50-load70/mps/attmean}{0.0}
\defR{pair-resnet50-load70/mps/thr}{93}
\defR{pair-resnet50-load70/mps/n}{3}
\defR{pair-resnet50-load70/mps/att/resnet50}{0.0}
\defR{pair-resnet50-load70/mps/p99/resnet50}{77524}
\defR{pair-resnet50-load70/mps/p99x/resnet50}{1697.2}
\defR{pair-resnet50-load70/mps/thrabs/train_bert}{1.51}
\defR{pair-resnet50-load70/mps/thr/train_bert}{93}
\defR{pair-resnet50-load70/mpsprio/attmin}{0.0}
\defR{pair-resnet50-load70/mpsprio/attmean}{0.0}
\defR{pair-resnet50-load70/mpsprio/thr}{83}
\defR{pair-resnet50-load70/mpsprio/n}{3}
\defR{pair-resnet50-load70/mpsprio/att/resnet50}{0.0}
\defR{pair-resnet50-load70/mpsprio/p99/resnet50}{59820}
\defR{pair-resnet50-load70/mpsprio/p99x/resnet50}{1309.6}
\defR{pair-resnet50-load70/mpsprio/thrabs/train_bert}{1.35}
\defR{pair-resnet50-load70/mpsprio/thr/train_bert}{83}
\defR{pair-resnet50-load70/mpslimits/attmin}{71.8}
\defR{pair-resnet50-load70/mpslimits/attmean}{71.8}
\defR{pair-resnet50-load70/mpslimits/thr}{11}
\defR{pair-resnet50-load70/mpslimits/n}{3}
\defR{pair-resnet50-load70/mpslimits/att/resnet50}{71.8}
\defR{pair-resnet50-load70/mpslimits/p99/resnet50}{48.5}
\defR{pair-resnet50-load70/mpslimits/p99x/resnet50}{1.1}
\defR{pair-resnet50-load70/mpslimits/thrabs/train_bert}{0.19}
\defR{pair-resnet50-load70/mpslimits/thr/train_bert}{11}
\defR{pair-resnet50-load70/mpsparts/attmin}{66.9}
\defR{pair-resnet50-load70/mpsparts/attmean}{66.9}
\defR{pair-resnet50-load70/mpsparts/thr}{9}
\defR{pair-resnet50-load70/mpsparts/n}{2}
\defR{pair-resnet50-load70/mpsparts/att/resnet50}{66.9}
\defR{pair-resnet50-load70/mpsparts/p99/resnet50}{58.4}
\defR{pair-resnet50-load70/mpsparts/p99x/resnet50}{1.3}
\defR{pair-resnet50-load70/mpsparts/thrabs/train_bert}{0.14}
\defR{pair-resnet50-load70/mpsparts/thr/train_bert}{9}
\defR{pair-resnet50-load70/lithos/attmin}{69.0}
\defR{pair-resnet50-load70/lithos/attmean}{69.0}
\defR{pair-resnet50-load70/lithos/thr}{8}
\defR{pair-resnet50-load70/lithos/n}{3}
\defR{pair-resnet50-load70/lithos/att/resnet50}{69.0}
\defR{pair-resnet50-load70/lithos/p99/resnet50}{51.8}
\defR{pair-resnet50-load70/lithos/p99x/resnet50}{1.1}
\defR{pair-resnet50-load70/lithos/thrabs/train_bert}{0.13}
\defR{pair-resnet50-load70/lithos/thr/train_bert}{8}
\defR{pair-resnet50-load70/dysm/attmin}{0.0}
\defR{pair-resnet50-load70/dysm/attmean}{0.0}
\defR{pair-resnet50-load70/dysm/thr}{100}
\defR{pair-resnet50-load70/dysm/n}{3}
\defR{pair-resnet50-load70/dysm/att/resnet50}{0.0}
\defR{pair-resnet50-load70/dysm/thrabs/train_bert}{1.62}
\defR{pair-resnet50-load70/dysm/thr/train_bert}{100}
\defR{pair-resnet50-load70/xsched/attmin}{25.6}
\defR{pair-resnet50-load70/xsched/attmean}{25.6}
\defR{pair-resnet50-load70/xsched/thr}{32}
\defR{pair-resnet50-load70/xsched/n}{3}
\defR{pair-resnet50-load70/xsched/att/resnet50}{25.6}
\defR{pair-resnet50-load70/xsched/p99/resnet50}{164}
\defR{pair-resnet50-load70/xsched/p99x/resnet50}{3.6}
\defR{pair-resnet50-load70/xsched/thrabs/train_bert}{0.53}
\defR{pair-resnet50-load70/xsched/thr/train_bert}{32}
\defR{pair-resnet50-load70/multistream/attmin}{0.0}
\defR{pair-resnet50-load70/multistream/attmean}{0.0}
\defR{pair-resnet50-load70/multistream/thr}{84}
\defR{pair-resnet50-load70/multistream/n}{3}
\defR{pair-resnet50-load70/multistream/att/resnet50}{0.0}
\defR{pair-resnet50-load70/multistream/p99/resnet50}{61808}
\defR{pair-resnet50-load70/multistream/p99x/resnet50}{1353.1}
\defR{pair-resnet50-load70/multistream/thrabs/train_bert}{1.37}
\defR{pair-resnet50-load70/multistream/thr/train_bert}{84}
\defR{pair-resnet50-load70/tgs/attmin}{61.0}
\defR{pair-resnet50-load70/tgs/attmean}{61.0}
\defR{pair-resnet50-load70/tgs/thr}{5}
\defR{pair-resnet50-load70/tgs/n}{3}
\defR{pair-resnet50-load70/tgs/att/resnet50}{61.0}
\defR{pair-resnet50-load70/tgs/p99/resnet50}{468}
\defR{pair-resnet50-load70/tgs/p99x/resnet50}{10.2}
\defR{pair-resnet50-load70/tgs/thrabs/train_bert}{0.09}
\defR{pair-resnet50-load70/tgs/thr/train_bert}{5}
\defR{pair-resnet50-load70/mig/attmin}{67.7}
\defR{pair-resnet50-load70/mig/attmean}{67.7}
\defR{pair-resnet50-load70/mig/thr}{0}
\defR{pair-resnet50-load70/mig/n}{3}
\defR{pair-resnet50-load70/mig/att/resnet50}{67.7}
\defR{pair-resnet50-load70/mig/p99/resnet50}{53.9}
\defR{pair-resnet50-load70/mig/p99x/resnet50}{1.2}
\defR{pair-resnet50-load70/mig/thrabs/train_bert}{0.00}
\defR{pair-resnet50-load70/mig/thr/train_bert}{0}
\defR{pair-resnet50-load70/tally/attmin}{0.0}
\defR{pair-resnet50-load70/tally/attmean}{0.0}
\defR{pair-resnet50-load70/tally/thr}{1}
\defR{pair-resnet50-load70/tally/n}{3}
\defR{pair-resnet50-load70/tally/att/resnet50}{0.0}
\defR{pair-resnet50-load70/tally/p99/resnet50}{94117}
\defR{pair-resnet50-load70/tally/p99x/resnet50}{2060.4}
\defR{pair-resnet50-load70/tally/thrabs/train_bert}{0.02}
\defR{pair-resnet50-load70/tally/thr/train_bert}{1}
\defR{pair-resnet50-load70/orion/attmin}{72.8}
\defR{pair-resnet50-load70/orion/attmean}{72.8}
\defR{pair-resnet50-load70/orion/thr}{45}
\defR{pair-resnet50-load70/orion/n}{3}
\defR{pair-resnet50-load70/orion/att/resnet50}{72.8}
\defR{pair-resnet50-load70/orion/p99/resnet50}{51.2}
\defR{pair-resnet50-load70/orion/p99x/resnet50}{1.1}
\defR{pair-resnet50-load70/orion/thrabs/train_bert}{0.72}
\defR{pair-resnet50-load70/orion/thr/train_bert}{45}
\defR{pair-retinanet-load30/berth-full/attmin}{100.0}
\defR{pair-retinanet-load30/berth-full/attmean}{100.0}
\defR{pair-retinanet-load30/berth-full/thr}{65}
\defR{pair-retinanet-load30/berth-full/n}{3}
\defR{pair-retinanet-load30/berth-full/att/retinanet}{100.0}
\defR{pair-retinanet-load30/berth-full/p99/retinanet}{38.8}
\defR{pair-retinanet-load30/berth-full/p99x/retinanet}{1.2}
\defR{pair-retinanet-load30/berth-full/thrabs/train_bert}{1.06}
\defR{pair-retinanet-load30/berth-full/thr/train_bert}{65}
\defR{pair-retinanet-load30/berth-nsr/attmin}{100.0}
\defR{pair-retinanet-load30/berth-nsr/attmean}{100.0}
\defR{pair-retinanet-load30/berth-nsr/thr}{65}
\defR{pair-retinanet-load30/berth-nsr/n}{3}
\defR{pair-retinanet-load30/berth-nsr/att/retinanet}{100.0}
\defR{pair-retinanet-load30/berth-nsr/p99/retinanet}{39.4}
\defR{pair-retinanet-load30/berth-nsr/p99x/retinanet}{1.2}
\defR{pair-retinanet-load30/berth-nsr/thrabs/train_bert}{1.05}
\defR{pair-retinanet-load30/berth-nsr/thr/train_bert}{65}
\defR{pair-retinanet-load30/berth-nba/attmin}{100.0}
\defR{pair-retinanet-load30/berth-nba/attmean}{100.0}
\defR{pair-retinanet-load30/berth-nba/thr}{65}
\defR{pair-retinanet-load30/berth-nba/n}{3}
\defR{pair-retinanet-load30/berth-nba/att/retinanet}{100.0}
\defR{pair-retinanet-load30/berth-nba/p99/retinanet}{38.3}
\defR{pair-retinanet-load30/berth-nba/p99x/retinanet}{1.2}
\defR{pair-retinanet-load30/berth-nba/thrabs/train_bert}{1.05}
\defR{pair-retinanet-load30/berth-nba/thr/train_bert}{65}
\defR{pair-retinanet-load30/berth-nsb/attmin}{100.0}
\defR{pair-retinanet-load30/berth-nsb/attmean}{100.0}
\defR{pair-retinanet-load30/berth-nsb/thr}{65}
\defR{pair-retinanet-load30/berth-nsb/n}{3}
\defR{pair-retinanet-load30/berth-nsb/att/retinanet}{100.0}
\defR{pair-retinanet-load30/berth-nsb/p99/retinanet}{39.3}
\defR{pair-retinanet-load30/berth-nsb/p99x/retinanet}{1.2}
\defR{pair-retinanet-load30/berth-nsb/thrabs/train_bert}{1.05}
\defR{pair-retinanet-load30/berth-nsb/thr/train_bert}{65}
\defR{pair-retinanet-load30/berth-ng/attmin}{100.0}
\defR{pair-retinanet-load30/berth-ng/attmean}{100.0}
\defR{pair-retinanet-load30/berth-ng/thr}{65}
\defR{pair-retinanet-load30/berth-ng/n}{3}
\defR{pair-retinanet-load30/berth-ng/att/retinanet}{100.0}
\defR{pair-retinanet-load30/berth-ng/p99/retinanet}{46.2}
\defR{pair-retinanet-load30/berth-ng/p99x/retinanet}{1.4}
\defR{pair-retinanet-load30/berth-ng/thrabs/train_bert}{1.06}
\defR{pair-retinanet-load30/berth-ng/thr/train_bert}{65}
\defR{pair-retinanet-load30/berth-nsng/attmin}{100.0}
\defR{pair-retinanet-load30/berth-nsng/attmean}{100.0}
\defR{pair-retinanet-load30/berth-nsng/thr}{65}
\defR{pair-retinanet-load30/berth-nsng/n}{3}
\defR{pair-retinanet-load30/berth-nsng/att/retinanet}{100.0}
\defR{pair-retinanet-load30/berth-nsng/p99/retinanet}{45.0}
\defR{pair-retinanet-load30/berth-nsng/p99x/retinanet}{1.4}
\defR{pair-retinanet-load30/berth-nsng/thrabs/train_bert}{1.05}
\defR{pair-retinanet-load30/berth-nsng/thr/train_bert}{65}
\defR{pair-retinanet-load30/berth-grouped/attmin}{100.0}
\defR{pair-retinanet-load30/berth-grouped/attmean}{100.0}
\defR{pair-retinanet-load30/berth-grouped/thr}{67}
\defR{pair-retinanet-load30/berth-grouped/n}{3}
\defR{pair-retinanet-load30/berth-grouped/att/retinanet}{100.0}
\defR{pair-retinanet-load30/berth-grouped/p99/retinanet}{41.1}
\defR{pair-retinanet-load30/berth-grouped/p99x/retinanet}{1.3}
\defR{pair-retinanet-load30/berth-grouped/thrabs/train_bert}{1.09}
\defR{pair-retinanet-load30/berth-grouped/thr/train_bert}{67}
\defR{pair-retinanet-load30/berth/attmin}{100.0}
\defR{pair-retinanet-load30/berth/attmean}{100.0}
\defR{pair-retinanet-load30/berth/thr}{68}
\defR{pair-retinanet-load30/berth/n}{3}
\defR{pair-retinanet-load30/berth/att/retinanet}{100.0}
\defR{pair-retinanet-load30/berth/p99/retinanet}{41.3}
\defR{pair-retinanet-load30/berth/p99x/retinanet}{1.3}
\defR{pair-retinanet-load30/berth/thrabs/train_bert}{1.10}
\defR{pair-retinanet-load30/berth/thr/train_bert}{68}
\defR{pair-retinanet-load30/timeslice/attmin}{81.0}
\defR{pair-retinanet-load30/timeslice/attmean}{81.0}
\defR{pair-retinanet-load30/timeslice/thr}{61}
\defR{pair-retinanet-load30/timeslice/n}{3}
\defR{pair-retinanet-load30/timeslice/att/retinanet}{81.0}
\defR{pair-retinanet-load30/timeslice/p99/retinanet}{295}
\defR{pair-retinanet-load30/timeslice/p99x/retinanet}{9.3}
\defR{pair-retinanet-load30/timeslice/thrabs/train_bert}{0.99}
\defR{pair-retinanet-load30/timeslice/thr/train_bert}{61}
\defR{pair-retinanet-load30/mps/attmin}{0.0}
\defR{pair-retinanet-load30/mps/attmean}{0.0}
\defR{pair-retinanet-load30/mps/thr}{92}
\defR{pair-retinanet-load30/mps/n}{3}
\defR{pair-retinanet-load30/mps/att/retinanet}{0.0}
\defR{pair-retinanet-load30/mps/p99/retinanet}{60230}
\defR{pair-retinanet-load30/mps/p99x/retinanet}{1888.0}
\defR{pair-retinanet-load30/mps/thrabs/train_bert}{1.49}
\defR{pair-retinanet-load30/mps/thr/train_bert}{92}
\defR{pair-retinanet-load30/mpsprio/attmin}{0.0}
\defR{pair-retinanet-load30/mpsprio/attmean}{0.0}
\defR{pair-retinanet-load30/mpsprio/thr}{83}
\defR{pair-retinanet-load30/mpsprio/n}{3}
\defR{pair-retinanet-load30/mpsprio/att/retinanet}{0.0}
\defR{pair-retinanet-load30/mpsprio/p99/retinanet}{29924}
\defR{pair-retinanet-load30/mpsprio/p99x/retinanet}{938.0}
\defR{pair-retinanet-load30/mpsprio/thrabs/train_bert}{1.34}
\defR{pair-retinanet-load30/mpsprio/thr/train_bert}{83}
\defR{pair-retinanet-load30/mpslimits/attmin}{100.0}
\defR{pair-retinanet-load30/mpslimits/attmean}{100.0}
\defR{pair-retinanet-load30/mpslimits/thr}{46}
\defR{pair-retinanet-load30/mpslimits/n}{3}
\defR{pair-retinanet-load30/mpslimits/att/retinanet}{100.0}
\defR{pair-retinanet-load30/mpslimits/p99/retinanet}{60.3}
\defR{pair-retinanet-load30/mpslimits/p99x/retinanet}{1.9}
\defR{pair-retinanet-load30/mpslimits/thrabs/train_bert}{0.74}
\defR{pair-retinanet-load30/mpslimits/thr/train_bert}{46}
\defR{pair-retinanet-load30/mpsparts/attmin}{100.0}
\defR{pair-retinanet-load30/mpsparts/attmean}{100.0}
\defR{pair-retinanet-load30/mpsparts/thr}{46}
\defR{pair-retinanet-load30/mpsparts/n}{2}
\defR{pair-retinanet-load30/mpsparts/att/retinanet}{100.0}
\defR{pair-retinanet-load30/mpsparts/p99/retinanet}{54.8}
\defR{pair-retinanet-load30/mpsparts/p99x/retinanet}{1.7}
\defR{pair-retinanet-load30/mpsparts/thrabs/train_bert}{0.74}
\defR{pair-retinanet-load30/mpsparts/thr/train_bert}{46}
\defR{pair-retinanet-load30/lithos/attmin}{100.0}
\defR{pair-retinanet-load30/lithos/attmean}{100.0}
\defR{pair-retinanet-load30/lithos/thr}{46}
\defR{pair-retinanet-load30/lithos/n}{3}
\defR{pair-retinanet-load30/lithos/att/retinanet}{100.0}
\defR{pair-retinanet-load30/lithos/p99/retinanet}{58.0}
\defR{pair-retinanet-load30/lithos/p99x/retinanet}{1.8}
\defR{pair-retinanet-load30/lithos/thrabs/train_bert}{0.74}
\defR{pair-retinanet-load30/lithos/thr/train_bert}{46}
\defR{pair-retinanet-load30/dysm/attmin}{53.5}
\defR{pair-retinanet-load30/dysm/attmean}{53.5}
\defR{pair-retinanet-load30/dysm/thr}{73}
\defR{pair-retinanet-load30/dysm/n}{3}
\defR{pair-retinanet-load30/dysm/att/retinanet}{53.5}
\defR{pair-retinanet-load30/dysm/p99/retinanet}{569}
\defR{pair-retinanet-load30/dysm/p99x/retinanet}{17.8}
\defR{pair-retinanet-load30/dysm/thrabs/train_bert}{1.19}
\defR{pair-retinanet-load30/dysm/thr/train_bert}{73}
\defR{pair-retinanet-load30/xsched/attmin}{100.0}
\defR{pair-retinanet-load30/xsched/attmean}{100.0}
\defR{pair-retinanet-load30/xsched/thr}{66}
\defR{pair-retinanet-load30/xsched/n}{3}
\defR{pair-retinanet-load30/xsched/att/retinanet}{100.0}
\defR{pair-retinanet-load30/xsched/p99/retinanet}{47.5}
\defR{pair-retinanet-load30/xsched/p99x/retinanet}{1.5}
\defR{pair-retinanet-load30/xsched/thrabs/train_bert}{1.07}
\defR{pair-retinanet-load30/xsched/thr/train_bert}{66}
\defR{pair-retinanet-load30/multistream/attmin}{0.0}
\defR{pair-retinanet-load30/multistream/attmean}{0.0}
\defR{pair-retinanet-load30/multistream/thr}{84}
\defR{pair-retinanet-load30/multistream/n}{3}
\defR{pair-retinanet-load30/multistream/att/retinanet}{0.0}
\defR{pair-retinanet-load30/multistream/p99/retinanet}{32816}
\defR{pair-retinanet-load30/multistream/p99x/retinanet}{1028.7}
\defR{pair-retinanet-load30/multistream/thrabs/train_bert}{1.36}
\defR{pair-retinanet-load30/multistream/thr/train_bert}{84}
\defR{pair-retinanet-load30/tgs/attmin}{100.0}
\defR{pair-retinanet-load30/tgs/attmean}{100.0}
\defR{pair-retinanet-load30/tgs/thr}{3}
\defR{pair-retinanet-load30/tgs/n}{3}
\defR{pair-retinanet-load30/tgs/att/retinanet}{100.0}
\defR{pair-retinanet-load30/tgs/p99/retinanet}{34.6}
\defR{pair-retinanet-load30/tgs/p99x/retinanet}{1.1}
\defR{pair-retinanet-load30/tgs/thrabs/train_bert}{0.05}
\defR{pair-retinanet-load30/tgs/thr/train_bert}{3}
\defR{pair-retinanet-load30/mig/attmin}{99.1}
\defR{pair-retinanet-load30/mig/attmean}{99.1}
\defR{pair-retinanet-load30/mig/thr}{55}
\defR{pair-retinanet-load30/mig/n}{3}
\defR{pair-retinanet-load30/mig/att/retinanet}{99.1}
\defR{pair-retinanet-load30/mig/p99/retinanet}{98.4}
\defR{pair-retinanet-load30/mig/p99x/retinanet}{3.1}
\defR{pair-retinanet-load30/mig/thrabs/train_bert}{0.90}
\defR{pair-retinanet-load30/mig/thr/train_bert}{55}
\defR{pair-retinanet-load30/tally/attmin}{72.6}
\defR{pair-retinanet-load30/tally/attmean}{72.6}
\defR{pair-retinanet-load30/tally/thr}{14}
\defR{pair-retinanet-load30/tally/n}{3}
\defR{pair-retinanet-load30/tally/att/retinanet}{72.6}
\defR{pair-retinanet-load30/tally/p99/retinanet}{364}
\defR{pair-retinanet-load30/tally/p99x/retinanet}{11.4}
\defR{pair-retinanet-load30/tally/thrabs/train_bert}{0.23}
\defR{pair-retinanet-load30/tally/thr/train_bert}{14}
\defR{pair-retinanet-load30/orion/attmin}{100.0}
\defR{pair-retinanet-load30/orion/attmean}{100.0}
\defR{pair-retinanet-load30/orion/thr}{71}
\defR{pair-retinanet-load30/orion/n}{3}
\defR{pair-retinanet-load30/orion/att/retinanet}{100.0}
\defR{pair-retinanet-load30/orion/p99/retinanet}{43.1}
\defR{pair-retinanet-load30/orion/p99x/retinanet}{1.4}
\defR{pair-retinanet-load30/orion/thrabs/train_bert}{1.14}
\defR{pair-retinanet-load30/orion/thr/train_bert}{71}
\defR{pair-retinanet-load50/berth-full/attmin}{100.0}
\defR{pair-retinanet-load50/berth-full/attmean}{100.0}
\defR{pair-retinanet-load50/berth-full/thr}{44}
\defR{pair-retinanet-load50/berth-full/n}{3}
\defR{pair-retinanet-load50/berth-full/att/retinanet}{100.0}
\defR{pair-retinanet-load50/berth-full/p99/retinanet}{58.6}
\defR{pair-retinanet-load50/berth-full/p99x/retinanet}{1.3}
\defR{pair-retinanet-load50/berth-full/thrabs/train_bert}{0.72}
\defR{pair-retinanet-load50/berth-full/thr/train_bert}{44}
\defR{pair-retinanet-load50/berth-nsr/attmin}{100.0}
\defR{pair-retinanet-load50/berth-nsr/attmean}{100.0}
\defR{pair-retinanet-load50/berth-nsr/thr}{44}
\defR{pair-retinanet-load50/berth-nsr/n}{3}
\defR{pair-retinanet-load50/berth-nsr/att/retinanet}{100.0}
\defR{pair-retinanet-load50/berth-nsr/p99/retinanet}{56.2}
\defR{pair-retinanet-load50/berth-nsr/p99x/retinanet}{1.2}
\defR{pair-retinanet-load50/berth-nsr/thrabs/train_bert}{0.72}
\defR{pair-retinanet-load50/berth-nsr/thr/train_bert}{44}
\defR{pair-retinanet-load50/berth-nba/attmin}{100.0}
\defR{pair-retinanet-load50/berth-nba/attmean}{100.0}
\defR{pair-retinanet-load50/berth-nba/thr}{44}
\defR{pair-retinanet-load50/berth-nba/n}{3}
\defR{pair-retinanet-load50/berth-nba/att/retinanet}{100.0}
\defR{pair-retinanet-load50/berth-nba/p99/retinanet}{55.9}
\defR{pair-retinanet-load50/berth-nba/p99x/retinanet}{1.2}
\defR{pair-retinanet-load50/berth-nba/thrabs/train_bert}{0.72}
\defR{pair-retinanet-load50/berth-nba/thr/train_bert}{44}
\defR{pair-retinanet-load50/berth-nsb/attmin}{100.0}
\defR{pair-retinanet-load50/berth-nsb/attmean}{100.0}
\defR{pair-retinanet-load50/berth-nsb/thr}{44}
\defR{pair-retinanet-load50/berth-nsb/n}{3}
\defR{pair-retinanet-load50/berth-nsb/att/retinanet}{100.0}
\defR{pair-retinanet-load50/berth-nsb/p99/retinanet}{58.6}
\defR{pair-retinanet-load50/berth-nsb/p99x/retinanet}{1.3}
\defR{pair-retinanet-load50/berth-nsb/thrabs/train_bert}{0.72}
\defR{pair-retinanet-load50/berth-nsb/thr/train_bert}{44}
\defR{pair-retinanet-load50/berth-ng/attmin}{99.9}
\defR{pair-retinanet-load50/berth-ng/attmean}{99.9}
\defR{pair-retinanet-load50/berth-ng/thr}{44}
\defR{pair-retinanet-load50/berth-ng/n}{3}
\defR{pair-retinanet-load50/berth-ng/att/retinanet}{99.9}
\defR{pair-retinanet-load50/berth-ng/p99/retinanet}{72.4}
\defR{pair-retinanet-load50/berth-ng/p99x/retinanet}{1.6}
\defR{pair-retinanet-load50/berth-ng/thrabs/train_bert}{0.71}
\defR{pair-retinanet-load50/berth-ng/thr/train_bert}{44}
\defR{pair-retinanet-load50/berth-nsng/attmin}{99.8}
\defR{pair-retinanet-load50/berth-nsng/attmean}{99.8}
\defR{pair-retinanet-load50/berth-nsng/thr}{44}
\defR{pair-retinanet-load50/berth-nsng/n}{3}
\defR{pair-retinanet-load50/berth-nsng/att/retinanet}{99.8}
\defR{pair-retinanet-load50/berth-nsng/p99/retinanet}{71.3}
\defR{pair-retinanet-load50/berth-nsng/p99x/retinanet}{1.5}
\defR{pair-retinanet-load50/berth-nsng/thrabs/train_bert}{0.71}
\defR{pair-retinanet-load50/berth-nsng/thr/train_bert}{44}
\defR{pair-retinanet-load50/berth-grouped/attmin}{99.8}
\defR{pair-retinanet-load50/berth-grouped/attmean}{99.8}
\defR{pair-retinanet-load50/berth-grouped/thr}{46}
\defR{pair-retinanet-load50/berth-grouped/n}{3}
\defR{pair-retinanet-load50/berth-grouped/att/retinanet}{99.8}
\defR{pair-retinanet-load50/berth-grouped/p99/retinanet}{65.8}
\defR{pair-retinanet-load50/berth-grouped/p99x/retinanet}{1.4}
\defR{pair-retinanet-load50/berth-grouped/thrabs/train_bert}{0.75}
\defR{pair-retinanet-load50/berth-grouped/thr/train_bert}{46}
\defR{pair-retinanet-load50/berth/attmin}{99.8}
\defR{pair-retinanet-load50/berth/attmean}{99.8}
\defR{pair-retinanet-load50/berth/thr}{45}
\defR{pair-retinanet-load50/berth/n}{3}
\defR{pair-retinanet-load50/berth/att/retinanet}{99.8}
\defR{pair-retinanet-load50/berth/p99/retinanet}{68.0}
\defR{pair-retinanet-load50/berth/p99x/retinanet}{1.5}
\defR{pair-retinanet-load50/berth/thrabs/train_bert}{0.73}
\defR{pair-retinanet-load50/berth/thr/train_bert}{45}
\defR{pair-retinanet-load50/timeslice/attmin}{0.0}
\defR{pair-retinanet-load50/timeslice/attmean}{0.0}
\defR{pair-retinanet-load50/timeslice/thr}{48}
\defR{pair-retinanet-load50/timeslice/n}{3}
\defR{pair-retinanet-load50/timeslice/att/retinanet}{0.0}
\defR{pair-retinanet-load50/timeslice/p99/retinanet}{18804}
\defR{pair-retinanet-load50/timeslice/p99x/retinanet}{404.3}
\defR{pair-retinanet-load50/timeslice/thrabs/train_bert}{0.77}
\defR{pair-retinanet-load50/timeslice/thr/train_bert}{48}
\defR{pair-retinanet-load50/mps/attmin}{0.0}
\defR{pair-retinanet-load50/mps/attmean}{0.0}
\defR{pair-retinanet-load50/mps/thr}{92}
\defR{pair-retinanet-load50/mps/n}{3}
\defR{pair-retinanet-load50/mps/att/retinanet}{0.0}
\defR{pair-retinanet-load50/mps/p99/retinanet}{74076}
\defR{pair-retinanet-load50/mps/p99x/retinanet}{1592.6}
\defR{pair-retinanet-load50/mps/thrabs/train_bert}{1.49}
\defR{pair-retinanet-load50/mps/thr/train_bert}{92}
\defR{pair-retinanet-load50/mpsprio/attmin}{0.0}
\defR{pair-retinanet-load50/mpsprio/attmean}{0.0}
\defR{pair-retinanet-load50/mpsprio/thr}{83}
\defR{pair-retinanet-load50/mpsprio/n}{3}
\defR{pair-retinanet-load50/mpsprio/att/retinanet}{0.0}
\defR{pair-retinanet-load50/mpsprio/p99/retinanet}{55602}
\defR{pair-retinanet-load50/mpsprio/p99x/retinanet}{1195.4}
\defR{pair-retinanet-load50/mpsprio/thrabs/train_bert}{1.34}
\defR{pair-retinanet-load50/mpsprio/thr/train_bert}{83}
\defR{pair-retinanet-load50/mpslimits/attmin}{99.9}
\defR{pair-retinanet-load50/mpslimits/attmean}{99.9}
\defR{pair-retinanet-load50/mpslimits/thr}{18}
\defR{pair-retinanet-load50/mpslimits/n}{3}
\defR{pair-retinanet-load50/mpslimits/att/retinanet}{99.9}
\defR{pair-retinanet-load50/mpslimits/p99/retinanet}{57.5}
\defR{pair-retinanet-load50/mpslimits/p99x/retinanet}{1.2}
\defR{pair-retinanet-load50/mpslimits/thrabs/train_bert}{0.29}
\defR{pair-retinanet-load50/mpslimits/thr/train_bert}{18}
\defR{pair-retinanet-load50/mpsparts/attmin}{99.9}
\defR{pair-retinanet-load50/mpsparts/attmean}{99.9}
\defR{pair-retinanet-load50/mpsparts/thr}{18}
\defR{pair-retinanet-load50/mpsparts/n}{2}
\defR{pair-retinanet-load50/mpsparts/att/retinanet}{99.9}
\defR{pair-retinanet-load50/mpsparts/p99/retinanet}{55.9}
\defR{pair-retinanet-load50/mpsparts/p99x/retinanet}{1.2}
\defR{pair-retinanet-load50/mpsparts/thrabs/train_bert}{0.29}
\defR{pair-retinanet-load50/mpsparts/thr/train_bert}{18}
\defR{pair-retinanet-load50/lithos/attmin}{99.9}
\defR{pair-retinanet-load50/lithos/attmean}{99.9}
\defR{pair-retinanet-load50/lithos/thr}{17}
\defR{pair-retinanet-load50/lithos/n}{3}
\defR{pair-retinanet-load50/lithos/att/retinanet}{99.9}
\defR{pair-retinanet-load50/lithos/p99/retinanet}{60.4}
\defR{pair-retinanet-load50/lithos/p99x/retinanet}{1.3}
\defR{pair-retinanet-load50/lithos/thrabs/train_bert}{0.28}
\defR{pair-retinanet-load50/lithos/thr/train_bert}{17}
\defR{pair-retinanet-load50/dysm/attmin}{51.1}
\defR{pair-retinanet-load50/dysm/attmean}{51.1}
\defR{pair-retinanet-load50/dysm/thr}{54}
\defR{pair-retinanet-load50/dysm/n}{3}
\defR{pair-retinanet-load50/dysm/att/retinanet}{51.1}
\defR{pair-retinanet-load50/dysm/p99/retinanet}{669}
\defR{pair-retinanet-load50/dysm/p99x/retinanet}{14.4}
\defR{pair-retinanet-load50/dysm/thrabs/train_bert}{0.88}
\defR{pair-retinanet-load50/dysm/thr/train_bert}{54}
\defR{pair-retinanet-load50/xsched/attmin}{99.7}
\defR{pair-retinanet-load50/xsched/attmean}{99.7}
\defR{pair-retinanet-load50/xsched/thr}{45}
\defR{pair-retinanet-load50/xsched/n}{3}
\defR{pair-retinanet-load50/xsched/att/retinanet}{99.7}
\defR{pair-retinanet-load50/xsched/p99/retinanet}{77.3}
\defR{pair-retinanet-load50/xsched/p99x/retinanet}{1.7}
\defR{pair-retinanet-load50/xsched/thrabs/train_bert}{0.74}
\defR{pair-retinanet-load50/xsched/thr/train_bert}{45}
\defR{pair-retinanet-load50/multistream/attmin}{0.0}
\defR{pair-retinanet-load50/multistream/attmean}{0.0}
\defR{pair-retinanet-load50/multistream/thr}{84}
\defR{pair-retinanet-load50/multistream/n}{3}
\defR{pair-retinanet-load50/multistream/att/retinanet}{0.0}
\defR{pair-retinanet-load50/multistream/p99/retinanet}{57599}
\defR{pair-retinanet-load50/multistream/p99x/retinanet}{1238.3}
\defR{pair-retinanet-load50/multistream/thrabs/train_bert}{1.36}
\defR{pair-retinanet-load50/multistream/thr/train_bert}{84}
\defR{pair-retinanet-load50/tgs/attmin}{98.9}
\defR{pair-retinanet-load50/tgs/attmean}{98.9}
\defR{pair-retinanet-load50/tgs/thr}{5}
\defR{pair-retinanet-load50/tgs/n}{3}
\defR{pair-retinanet-load50/tgs/att/retinanet}{98.9}
\defR{pair-retinanet-load50/tgs/p99/retinanet}{116}
\defR{pair-retinanet-load50/tgs/p99x/retinanet}{2.5}
\defR{pair-retinanet-load50/tgs/thrabs/train_bert}{0.08}
\defR{pair-retinanet-load50/tgs/thr/train_bert}{5}
\defR{pair-retinanet-load50/mig/attmin}{100.0}
\defR{pair-retinanet-load50/mig/attmean}{100.0}
\defR{pair-retinanet-load50/mig/thr}{0}
\defR{pair-retinanet-load50/mig/n}{3}
\defR{pair-retinanet-load50/mig/att/retinanet}{100.0}
\defR{pair-retinanet-load50/mig/p99/retinanet}{49.6}
\defR{pair-retinanet-load50/mig/p99x/retinanet}{1.1}
\defR{pair-retinanet-load50/mig/thrabs/train_bert}{0.00}
\defR{pair-retinanet-load50/mig/thr/train_bert}{0}
\defR{pair-retinanet-load50/tally/attmin}{0.0}
\defR{pair-retinanet-load50/tally/attmean}{0.0}
\defR{pair-retinanet-load50/tally/thr}{0}
\defR{pair-retinanet-load50/tally/n}{3}
\defR{pair-retinanet-load50/tally/att/retinanet}{0.0}
\defR{pair-retinanet-load50/tally/p99/retinanet}{30302}
\defR{pair-retinanet-load50/tally/p99x/retinanet}{651.5}
\defR{pair-retinanet-load50/tally/thrabs/train_bert}{0.00}
\defR{pair-retinanet-load50/tally/thr/train_bert}{0}
\defR{pair-retinanet-load50/orion/attmin}{100.0}
\defR{pair-retinanet-load50/orion/attmean}{100.0}
\defR{pair-retinanet-load50/orion/thr}{52}
\defR{pair-retinanet-load50/orion/n}{3}
\defR{pair-retinanet-load50/orion/att/retinanet}{100.0}
\defR{pair-retinanet-load50/orion/p99/retinanet}{56.0}
\defR{pair-retinanet-load50/orion/p99x/retinanet}{1.2}
\defR{pair-retinanet-load50/orion/thrabs/train_bert}{0.84}
\defR{pair-retinanet-load50/orion/thr/train_bert}{52}
\defR{pair-retinanet-load70/berth-full/attmin}{99.3}
\defR{pair-retinanet-load70/berth-full/attmean}{99.3}
\defR{pair-retinanet-load70/berth-full/thr}{26}
\defR{pair-retinanet-load70/berth-full/n}{3}
\defR{pair-retinanet-load70/berth-full/att/retinanet}{99.3}
\defR{pair-retinanet-load70/berth-full/p99/retinanet}{102}
\defR{pair-retinanet-load70/berth-full/p99x/retinanet}{1.3}
\defR{pair-retinanet-load70/berth-full/thrabs/train_bert}{0.42}
\defR{pair-retinanet-load70/berth-full/thr/train_bert}{26}
\defR{pair-retinanet-load70/berth-nsr/attmin}{99.2}
\defR{pair-retinanet-load70/berth-nsr/attmean}{99.2}
\defR{pair-retinanet-load70/berth-nsr/thr}{26}
\defR{pair-retinanet-load70/berth-nsr/n}{3}
\defR{pair-retinanet-load70/berth-nsr/att/retinanet}{99.2}
\defR{pair-retinanet-load70/berth-nsr/p99/retinanet}{101}
\defR{pair-retinanet-load70/berth-nsr/p99x/retinanet}{1.3}
\defR{pair-retinanet-load70/berth-nsr/thrabs/train_bert}{0.41}
\defR{pair-retinanet-load70/berth-nsr/thr/train_bert}{26}
\defR{pair-retinanet-load70/berth-nba/attmin}{99.1}
\defR{pair-retinanet-load70/berth-nba/attmean}{99.1}
\defR{pair-retinanet-load70/berth-nba/thr}{26}
\defR{pair-retinanet-load70/berth-nba/n}{3}
\defR{pair-retinanet-load70/berth-nba/att/retinanet}{99.1}
\defR{pair-retinanet-load70/berth-nba/p99/retinanet}{99.5}
\defR{pair-retinanet-load70/berth-nba/p99x/retinanet}{1.3}
\defR{pair-retinanet-load70/berth-nba/thrabs/train_bert}{0.42}
\defR{pair-retinanet-load70/berth-nba/thr/train_bert}{26}
\defR{pair-retinanet-load70/berth-nsb/attmin}{99.1}
\defR{pair-retinanet-load70/berth-nsb/attmean}{99.1}
\defR{pair-retinanet-load70/berth-nsb/thr}{26}
\defR{pair-retinanet-load70/berth-nsb/n}{3}
\defR{pair-retinanet-load70/berth-nsb/att/retinanet}{99.1}
\defR{pair-retinanet-load70/berth-nsb/p99/retinanet}{104}
\defR{pair-retinanet-load70/berth-nsb/p99x/retinanet}{1.3}
\defR{pair-retinanet-load70/berth-nsb/thrabs/train_bert}{0.41}
\defR{pair-retinanet-load70/berth-nsb/thr/train_bert}{26}
\defR{pair-retinanet-load70/berth-ng/attmin}{96.8}
\defR{pair-retinanet-load70/berth-ng/attmean}{96.8}
\defR{pair-retinanet-load70/berth-ng/thr}{24}
\defR{pair-retinanet-load70/berth-ng/n}{3}
\defR{pair-retinanet-load70/berth-ng/att/retinanet}{96.8}
\defR{pair-retinanet-load70/berth-ng/p99/retinanet}{149}
\defR{pair-retinanet-load70/berth-ng/p99x/retinanet}{1.9}
\defR{pair-retinanet-load70/berth-ng/thrabs/train_bert}{0.39}
\defR{pair-retinanet-load70/berth-ng/thr/train_bert}{24}
\defR{pair-retinanet-load70/berth-nsng/attmin}{97.3}
\defR{pair-retinanet-load70/berth-nsng/attmean}{97.3}
\defR{pair-retinanet-load70/berth-nsng/thr}{24}
\defR{pair-retinanet-load70/berth-nsng/n}{3}
\defR{pair-retinanet-load70/berth-nsng/att/retinanet}{97.3}
\defR{pair-retinanet-load70/berth-nsng/p99/retinanet}{132}
\defR{pair-retinanet-load70/berth-nsng/p99x/retinanet}{1.7}
\defR{pair-retinanet-load70/berth-nsng/thrabs/train_bert}{0.39}
\defR{pair-retinanet-load70/berth-nsng/thr/train_bert}{24}
\defR{pair-retinanet-load70/berth-grouped/attmin}{98.2}
\defR{pair-retinanet-load70/berth-grouped/attmean}{98.2}
\defR{pair-retinanet-load70/berth-grouped/thr}{28}
\defR{pair-retinanet-load70/berth-grouped/n}{3}
\defR{pair-retinanet-load70/berth-grouped/att/retinanet}{98.2}
\defR{pair-retinanet-load70/berth-grouped/p99/retinanet}{124}
\defR{pair-retinanet-load70/berth-grouped/p99x/retinanet}{1.6}
\defR{pair-retinanet-load70/berth-grouped/thrabs/train_bert}{0.45}
\defR{pair-retinanet-load70/berth-grouped/thr/train_bert}{28}
\defR{pair-retinanet-load70/berth/attmin}{98.1}
\defR{pair-retinanet-load70/berth/attmean}{98.1}
\defR{pair-retinanet-load70/berth/thr}{28}
\defR{pair-retinanet-load70/berth/n}{3}
\defR{pair-retinanet-load70/berth/att/retinanet}{98.1}
\defR{pair-retinanet-load70/berth/p99/retinanet}{119}
\defR{pair-retinanet-load70/berth/p99x/retinanet}{1.5}
\defR{pair-retinanet-load70/berth/thrabs/train_bert}{0.45}
\defR{pair-retinanet-load70/berth/thr/train_bert}{28}
\defR{pair-retinanet-load70/timeslice/attmin}{0.0}
\defR{pair-retinanet-load70/timeslice/attmean}{0.0}
\defR{pair-retinanet-load70/timeslice/thr}{48}
\defR{pair-retinanet-load70/timeslice/n}{3}
\defR{pair-retinanet-load70/timeslice/att/retinanet}{0.0}
\defR{pair-retinanet-load70/timeslice/p99/retinanet}{40707}
\defR{pair-retinanet-load70/timeslice/p99x/retinanet}{512.3}
\defR{pair-retinanet-load70/timeslice/thrabs/train_bert}{0.79}
\defR{pair-retinanet-load70/timeslice/thr/train_bert}{48}
\defR{pair-retinanet-load70/mps/attmin}{0.0}
\defR{pair-retinanet-load70/mps/attmean}{0.0}
\defR{pair-retinanet-load70/mps/thr}{92}
\defR{pair-retinanet-load70/mps/n}{3}
\defR{pair-retinanet-load70/mps/att/retinanet}{0.0}
\defR{pair-retinanet-load70/mps/p99/retinanet}{80305}
\defR{pair-retinanet-load70/mps/p99x/retinanet}{1010.7}
\defR{pair-retinanet-load70/mps/thrabs/train_bert}{1.50}
\defR{pair-retinanet-load70/mps/thr/train_bert}{92}
\defR{pair-retinanet-load70/mpsprio/attmin}{0.0}
\defR{pair-retinanet-load70/mpsprio/attmean}{0.0}
\defR{pair-retinanet-load70/mpsprio/thr}{83}
\defR{pair-retinanet-load70/mpsprio/n}{3}
\defR{pair-retinanet-load70/mpsprio/att/retinanet}{0.0}
\defR{pair-retinanet-load70/mpsprio/p99/retinanet}{66914}
\defR{pair-retinanet-load70/mpsprio/p99x/retinanet}{842.2}
\defR{pair-retinanet-load70/mpsprio/thrabs/train_bert}{1.34}
\defR{pair-retinanet-load70/mpsprio/thr/train_bert}{83}
\defR{pair-retinanet-load70/mpslimits/attmin}{97.8}
\defR{pair-retinanet-load70/mpslimits/attmean}{97.8}
\defR{pair-retinanet-load70/mpslimits/thr}{11}
\defR{pair-retinanet-load70/mpslimits/n}{3}
\defR{pair-retinanet-load70/mpslimits/att/retinanet}{97.8}
\defR{pair-retinanet-load70/mpslimits/p99/retinanet}{128}
\defR{pair-retinanet-load70/mpslimits/p99x/retinanet}{1.6}
\defR{pair-retinanet-load70/mpslimits/thrabs/train_bert}{0.18}
\defR{pair-retinanet-load70/mpslimits/thr/train_bert}{11}
\defR{pair-retinanet-load70/mpsparts/attmin}{98.3}
\defR{pair-retinanet-load70/mpsparts/attmean}{98.3}
\defR{pair-retinanet-load70/mpsparts/thr}{9}
\defR{pair-retinanet-load70/mpsparts/n}{2}
\defR{pair-retinanet-load70/mpsparts/att/retinanet}{98.3}
\defR{pair-retinanet-load70/mpsparts/p99/retinanet}{123}
\defR{pair-retinanet-load70/mpsparts/p99x/retinanet}{1.5}
\defR{pair-retinanet-load70/mpsparts/thrabs/train_bert}{0.14}
\defR{pair-retinanet-load70/mpsparts/thr/train_bert}{9}
\defR{pair-retinanet-load70/lithos/attmin}{98.8}
\defR{pair-retinanet-load70/lithos/attmean}{98.8}
\defR{pair-retinanet-load70/lithos/thr}{8}
\defR{pair-retinanet-load70/lithos/n}{3}
\defR{pair-retinanet-load70/lithos/att/retinanet}{98.8}
\defR{pair-retinanet-load70/lithos/p99/retinanet}{112}
\defR{pair-retinanet-load70/lithos/p99x/retinanet}{1.4}
\defR{pair-retinanet-load70/lithos/thrabs/train_bert}{0.13}
\defR{pair-retinanet-load70/lithos/thr/train_bert}{8}
\defR{pair-retinanet-load70/dysm/attmin}{0.0}
\defR{pair-retinanet-load70/dysm/attmean}{0.0}
\defR{pair-retinanet-load70/dysm/thr}{100}
\defR{pair-retinanet-load70/dysm/n}{3}
\defR{pair-retinanet-load70/dysm/att/retinanet}{0.0}
\defR{pair-retinanet-load70/dysm/thrabs/train_bert}{1.62}
\defR{pair-retinanet-load70/dysm/thr/train_bert}{100}
\defR{pair-retinanet-load70/xsched/attmin}{94.5}
\defR{pair-retinanet-load70/xsched/attmean}{94.5}
\defR{pair-retinanet-load70/xsched/thr}{25}
\defR{pair-retinanet-load70/xsched/n}{3}
\defR{pair-retinanet-load70/xsched/att/retinanet}{94.5}
\defR{pair-retinanet-load70/xsched/p99/retinanet}{171}
\defR{pair-retinanet-load70/xsched/p99x/retinanet}{2.2}
\defR{pair-retinanet-load70/xsched/thrabs/train_bert}{0.41}
\defR{pair-retinanet-load70/xsched/thr/train_bert}{25}
\defR{pair-retinanet-load70/multistream/attmin}{0.0}
\defR{pair-retinanet-load70/multistream/attmean}{0.0}
\defR{pair-retinanet-load70/multistream/thr}{84}
\defR{pair-retinanet-load70/multistream/n}{3}
\defR{pair-retinanet-load70/multistream/att/retinanet}{0.0}
\defR{pair-retinanet-load70/multistream/p99/retinanet}{68333}
\defR{pair-retinanet-load70/multistream/p99x/retinanet}{860.0}
\defR{pair-retinanet-load70/multistream/thrabs/train_bert}{1.36}
\defR{pair-retinanet-load70/multistream/thr/train_bert}{84}
\defR{pair-retinanet-load70/tgs/attmin}{91.6}
\defR{pair-retinanet-load70/tgs/attmean}{91.6}
\defR{pair-retinanet-load70/tgs/thr}{3}
\defR{pair-retinanet-load70/tgs/n}{3}
\defR{pair-retinanet-load70/tgs/att/retinanet}{91.6}
\defR{pair-retinanet-load70/tgs/p99/retinanet}{464}
\defR{pair-retinanet-load70/tgs/p99x/retinanet}{5.8}
\defR{pair-retinanet-load70/tgs/thrabs/train_bert}{0.06}
\defR{pair-retinanet-load70/tgs/thr/train_bert}{3}
\defR{pair-retinanet-load70/mig/attmin}{99.2}
\defR{pair-retinanet-load70/mig/attmean}{99.2}
\defR{pair-retinanet-load70/mig/thr}{0}
\defR{pair-retinanet-load70/mig/n}{3}
\defR{pair-retinanet-load70/mig/att/retinanet}{99.2}
\defR{pair-retinanet-load70/mig/p99/retinanet}{96.6}
\defR{pair-retinanet-load70/mig/p99x/retinanet}{1.2}
\defR{pair-retinanet-load70/mig/thrabs/train_bert}{0.00}
\defR{pair-retinanet-load70/mig/thr/train_bert}{0}
\defR{pair-retinanet-load70/tally/attmin}{0.0}
\defR{pair-retinanet-load70/tally/attmean}{0.0}
\defR{pair-retinanet-load70/tally/thr}{0}
\defR{pair-retinanet-load70/tally/n}{3}
\defR{pair-retinanet-load70/tally/att/retinanet}{0.0}
\defR{pair-retinanet-load70/tally/p99/retinanet}{57739}
\defR{pair-retinanet-load70/tally/p99x/retinanet}{726.7}
\defR{pair-retinanet-load70/tally/thrabs/train_bert}{0.00}
\defR{pair-retinanet-load70/tally/thr/train_bert}{0}
\defR{pair-retinanet-load70/orion/attmin}{98.9}
\defR{pair-retinanet-load70/orion/attmean}{98.9}
\defR{pair-retinanet-load70/orion/thr}{33}
\defR{pair-retinanet-load70/orion/n}{3}
\defR{pair-retinanet-load70/orion/att/retinanet}{98.9}
\defR{pair-retinanet-load70/orion/p99/retinanet}{102}
\defR{pair-retinanet-load70/orion/p99x/retinanet}{1.3}
\defR{pair-retinanet-load70/orion/thrabs/train_bert}{0.53}
\defR{pair-retinanet-load70/orion/thr/train_bert}{33}
\defR{pair-bertlarge-load30/berth-full/attmin}{100.0}
\defR{pair-bertlarge-load30/berth-full/attmean}{100.0}
\defR{pair-bertlarge-load30/berth-full/thr}{66}
\defR{pair-bertlarge-load30/berth-full/n}{3}
\defR{pair-bertlarge-load30/berth-full/att/bertlarge}{100.0}
\defR{pair-bertlarge-load30/berth-full/p99/bertlarge}{74.2}
\defR{pair-bertlarge-load30/berth-full/p99x/bertlarge}{1.3}
\defR{pair-bertlarge-load30/berth-full/thrabs/train_bert}{1.07}
\defR{pair-bertlarge-load30/berth-full/thr/train_bert}{66}
\defR{pair-bertlarge-load30/berth-nsr/attmin}{99.9}
\defR{pair-bertlarge-load30/berth-nsr/attmean}{99.9}
\defR{pair-bertlarge-load30/berth-nsr/thr}{66}
\defR{pair-bertlarge-load30/berth-nsr/n}{3}
\defR{pair-bertlarge-load30/berth-nsr/att/bertlarge}{99.9}
\defR{pair-bertlarge-load30/berth-nsr/p99/bertlarge}{77.4}
\defR{pair-bertlarge-load30/berth-nsr/p99x/bertlarge}{1.3}
\defR{pair-bertlarge-load30/berth-nsr/thrabs/train_bert}{1.07}
\defR{pair-bertlarge-load30/berth-nsr/thr/train_bert}{66}
\defR{pair-bertlarge-load30/berth-nba/attmin}{100.0}
\defR{pair-bertlarge-load30/berth-nba/attmean}{100.0}
\defR{pair-bertlarge-load30/berth-nba/thr}{66}
\defR{pair-bertlarge-load30/berth-nba/n}{3}
\defR{pair-bertlarge-load30/berth-nba/att/bertlarge}{100.0}
\defR{pair-bertlarge-load30/berth-nba/p99/bertlarge}{72.1}
\defR{pair-bertlarge-load30/berth-nba/p99x/bertlarge}{1.2}
\defR{pair-bertlarge-load30/berth-nba/thrabs/train_bert}{1.07}
\defR{pair-bertlarge-load30/berth-nba/thr/train_bert}{66}
\defR{pair-bertlarge-load30/berth-nsb/attmin}{100.0}
\defR{pair-bertlarge-load30/berth-nsb/attmean}{100.0}
\defR{pair-bertlarge-load30/berth-nsb/thr}{66}
\defR{pair-bertlarge-load30/berth-nsb/n}{3}
\defR{pair-bertlarge-load30/berth-nsb/att/bertlarge}{100.0}
\defR{pair-bertlarge-load30/berth-nsb/p99/bertlarge}{76.0}
\defR{pair-bertlarge-load30/berth-nsb/p99x/bertlarge}{1.3}
\defR{pair-bertlarge-load30/berth-nsb/thrabs/train_bert}{1.07}
\defR{pair-bertlarge-load30/berth-nsb/thr/train_bert}{66}
\defR{pair-bertlarge-load30/berth-ng/attmin}{100.0}
\defR{pair-bertlarge-load30/berth-ng/attmean}{100.0}
\defR{pair-bertlarge-load30/berth-ng/thr}{66}
\defR{pair-bertlarge-load30/berth-ng/n}{3}
\defR{pair-bertlarge-load30/berth-ng/att/bertlarge}{100.0}
\defR{pair-bertlarge-load30/berth-ng/p99/bertlarge}{95.2}
\defR{pair-bertlarge-load30/berth-ng/p99x/bertlarge}{1.6}
\defR{pair-bertlarge-load30/berth-ng/thrabs/train_bert}{1.07}
\defR{pair-bertlarge-load30/berth-ng/thr/train_bert}{66}
\defR{pair-bertlarge-load30/berth-nsng/attmin}{100.0}
\defR{pair-bertlarge-load30/berth-nsng/attmean}{100.0}
\defR{pair-bertlarge-load30/berth-nsng/thr}{65}
\defR{pair-bertlarge-load30/berth-nsng/n}{3}
\defR{pair-bertlarge-load30/berth-nsng/att/bertlarge}{100.0}
\defR{pair-bertlarge-load30/berth-nsng/p99/bertlarge}{96.5}
\defR{pair-bertlarge-load30/berth-nsng/p99x/bertlarge}{1.7}
\defR{pair-bertlarge-load30/berth-nsng/thrabs/train_bert}{1.06}
\defR{pair-bertlarge-load30/berth-nsng/thr/train_bert}{65}
\defR{pair-bertlarge-load30/berth-grouped/attmin}{99.9}
\defR{pair-bertlarge-load30/berth-grouped/attmean}{99.9}
\defR{pair-bertlarge-load30/berth-grouped/thr}{67}
\defR{pair-bertlarge-load30/berth-grouped/n}{3}
\defR{pair-bertlarge-load30/berth-grouped/att/bertlarge}{99.9}
\defR{pair-bertlarge-load30/berth-grouped/p99/bertlarge}{99.4}
\defR{pair-bertlarge-load30/berth-grouped/p99x/bertlarge}{1.7}
\defR{pair-bertlarge-load30/berth-grouped/thrabs/train_bert}{1.09}
\defR{pair-bertlarge-load30/berth-grouped/thr/train_bert}{67}
\defR{pair-bertlarge-load30/berth/attmin}{99.9}
\defR{pair-bertlarge-load30/berth/attmean}{99.9}
\defR{pair-bertlarge-load30/berth/thr}{67}
\defR{pair-bertlarge-load30/berth/n}{3}
\defR{pair-bertlarge-load30/berth/att/bertlarge}{99.9}
\defR{pair-bertlarge-load30/berth/p99/bertlarge}{96.0}
\defR{pair-bertlarge-load30/berth/p99x/bertlarge}{1.6}
\defR{pair-bertlarge-load30/berth/thrabs/train_bert}{1.09}
\defR{pair-bertlarge-load30/berth/thr/train_bert}{67}
\defR{pair-bertlarge-load30/timeslice/attmin}{48.4}
\defR{pair-bertlarge-load30/timeslice/attmean}{48.4}
\defR{pair-bertlarge-load30/timeslice/thr}{61}
\defR{pair-bertlarge-load30/timeslice/n}{3}
\defR{pair-bertlarge-load30/timeslice/att/bertlarge}{48.4}
\defR{pair-bertlarge-load30/timeslice/p99/bertlarge}{902}
\defR{pair-bertlarge-load30/timeslice/p99x/bertlarge}{15.4}
\defR{pair-bertlarge-load30/timeslice/thrabs/train_bert}{0.98}
\defR{pair-bertlarge-load30/timeslice/thr/train_bert}{61}
\defR{pair-bertlarge-load30/mps/attmin}{0.0}
\defR{pair-bertlarge-load30/mps/attmean}{0.0}
\defR{pair-bertlarge-load30/mps/thr}{89}
\defR{pair-bertlarge-load30/mps/n}{3}
\defR{pair-bertlarge-load30/mps/att/bertlarge}{0.0}
\defR{pair-bertlarge-load30/mps/p99/bertlarge}{48991}
\defR{pair-bertlarge-load30/mps/p99x/bertlarge}{838.0}
\defR{pair-bertlarge-load30/mps/thrabs/train_bert}{1.45}
\defR{pair-bertlarge-load30/mps/thr/train_bert}{89}
\defR{pair-bertlarge-load30/mpsprio/attmin}{4.7}
\defR{pair-bertlarge-load30/mpsprio/attmean}{4.7}
\defR{pair-bertlarge-load30/mpsprio/thr}{75}
\defR{pair-bertlarge-load30/mpsprio/n}{3}
\defR{pair-bertlarge-load30/mpsprio/att/bertlarge}{4.7}
\defR{pair-bertlarge-load30/mpsprio/p99/bertlarge}{3116}
\defR{pair-bertlarge-load30/mpsprio/p99x/bertlarge}{53.3}
\defR{pair-bertlarge-load30/mpsprio/thrabs/train_bert}{1.21}
\defR{pair-bertlarge-load30/mpsprio/thr/train_bert}{75}
\defR{pair-bertlarge-load30/mpslimits/attmin}{99.9}
\defR{pair-bertlarge-load30/mpslimits/attmean}{99.9}
\defR{pair-bertlarge-load30/mpslimits/thr}{21}
\defR{pair-bertlarge-load30/mpslimits/n}{3}
\defR{pair-bertlarge-load30/mpslimits/att/bertlarge}{99.9}
\defR{pair-bertlarge-load30/mpslimits/p99/bertlarge}{99.3}
\defR{pair-bertlarge-load30/mpslimits/p99x/bertlarge}{1.7}
\defR{pair-bertlarge-load30/mpslimits/thrabs/train_bert}{0.33}
\defR{pair-bertlarge-load30/mpslimits/thr/train_bert}{21}
\defR{pair-bertlarge-load30/mpsparts/attmin}{100.0}
\defR{pair-bertlarge-load30/mpsparts/attmean}{100.0}
\defR{pair-bertlarge-load30/mpsparts/thr}{22}
\defR{pair-bertlarge-load30/mpsparts/n}{2}
\defR{pair-bertlarge-load30/mpsparts/att/bertlarge}{100.0}
\defR{pair-bertlarge-load30/mpsparts/p99/bertlarge}{95.5}
\defR{pair-bertlarge-load30/mpsparts/p99x/bertlarge}{1.6}
\defR{pair-bertlarge-load30/mpsparts/thrabs/train_bert}{0.36}
\defR{pair-bertlarge-load30/mpsparts/thr/train_bert}{22}
\defR{pair-bertlarge-load30/lithos/attmin}{99.4}
\defR{pair-bertlarge-load30/lithos/attmean}{99.4}
\defR{pair-bertlarge-load30/lithos/thr}{21}
\defR{pair-bertlarge-load30/lithos/n}{3}
\defR{pair-bertlarge-load30/lithos/att/bertlarge}{99.4}
\defR{pair-bertlarge-load30/lithos/p99/bertlarge}{120}
\defR{pair-bertlarge-load30/lithos/p99x/bertlarge}{2.1}
\defR{pair-bertlarge-load30/lithos/thrabs/train_bert}{0.33}
\defR{pair-bertlarge-load30/lithos/thr/train_bert}{21}
\defR{pair-bertlarge-load30/dysm/attmin}{67.9}
\defR{pair-bertlarge-load30/dysm/attmean}{67.9}
\defR{pair-bertlarge-load30/dysm/thr}{61}
\defR{pair-bertlarge-load30/dysm/n}{3}
\defR{pair-bertlarge-load30/dysm/att/bertlarge}{67.9}
\defR{pair-bertlarge-load30/dysm/p99/bertlarge}{483}
\defR{pair-bertlarge-load30/dysm/p99x/bertlarge}{8.3}
\defR{pair-bertlarge-load30/dysm/thrabs/train_bert}{0.99}
\defR{pair-bertlarge-load30/dysm/thr/train_bert}{61}
\defR{pair-bertlarge-load30/xsched/attmin}{99.9}
\defR{pair-bertlarge-load30/xsched/attmean}{99.9}
\defR{pair-bertlarge-load30/xsched/thr}{65}
\defR{pair-bertlarge-load30/xsched/n}{3}
\defR{pair-bertlarge-load30/xsched/att/bertlarge}{99.9}
\defR{pair-bertlarge-load30/xsched/p99/bertlarge}{97.3}
\defR{pair-bertlarge-load30/xsched/p99x/bertlarge}{1.7}
\defR{pair-bertlarge-load30/xsched/thrabs/train_bert}{1.06}
\defR{pair-bertlarge-load30/xsched/thr/train_bert}{65}
\defR{pair-bertlarge-load30/multistream/attmin}{5.7}
\defR{pair-bertlarge-load30/multistream/attmean}{5.7}
\defR{pair-bertlarge-load30/multistream/thr}{75}
\defR{pair-bertlarge-load30/multistream/n}{3}
\defR{pair-bertlarge-load30/multistream/att/bertlarge}{5.7}
\defR{pair-bertlarge-load30/multistream/p99/bertlarge}{2962}
\defR{pair-bertlarge-load30/multistream/p99x/bertlarge}{50.7}
\defR{pair-bertlarge-load30/multistream/thrabs/train_bert}{1.22}
\defR{pair-bertlarge-load30/multistream/thr/train_bert}{75}
\defR{pair-bertlarge-load30/tgs/attmin}{100.0}
\defR{pair-bertlarge-load30/tgs/attmean}{100.0}
\defR{pair-bertlarge-load30/tgs/thr}{1}
\defR{pair-bertlarge-load30/tgs/n}{3}
\defR{pair-bertlarge-load30/tgs/att/bertlarge}{100.0}
\defR{pair-bertlarge-load30/tgs/p99/bertlarge}{61.9}
\defR{pair-bertlarge-load30/tgs/p99x/bertlarge}{1.1}
\defR{pair-bertlarge-load30/tgs/thrabs/train_bert}{0.02}
\defR{pair-bertlarge-load30/tgs/thr/train_bert}{1}
\defR{pair-bertlarge-load30/mig/attmin}{97.6}
\defR{pair-bertlarge-load30/mig/attmean}{97.6}
\defR{pair-bertlarge-load30/mig/thr}{44}
\defR{pair-bertlarge-load30/mig/n}{3}
\defR{pair-bertlarge-load30/mig/att/bertlarge}{97.6}
\defR{pair-bertlarge-load30/mig/p99/bertlarge}{169}
\defR{pair-bertlarge-load30/mig/p99x/bertlarge}{2.9}
\defR{pair-bertlarge-load30/mig/thrabs/train_bert}{0.71}
\defR{pair-bertlarge-load30/mig/thr/train_bert}{44}
\defR{pair-bertlarge-load30/tally/attmin}{72.4}
\defR{pair-bertlarge-load30/tally/attmean}{72.4}
\defR{pair-bertlarge-load30/tally/thr}{44}
\defR{pair-bertlarge-load30/tally/n}{3}
\defR{pair-bertlarge-load30/tally/att/bertlarge}{72.4}
\defR{pair-bertlarge-load30/tally/p99/bertlarge}{403}
\defR{pair-bertlarge-load30/tally/p99x/bertlarge}{6.9}
\defR{pair-bertlarge-load30/tally/thrabs/train_bert}{0.71}
\defR{pair-bertlarge-load30/tally/thr/train_bert}{44}
\defR{pair-bertlarge-load30/orion/attmin}{100.0}
\defR{pair-bertlarge-load30/orion/attmean}{100.0}
\defR{pair-bertlarge-load30/orion/thr}{70}
\defR{pair-bertlarge-load30/orion/n}{3}
\defR{pair-bertlarge-load30/orion/att/bertlarge}{100.0}
\defR{pair-bertlarge-load30/orion/p99/bertlarge}{70.1}
\defR{pair-bertlarge-load30/orion/p99x/bertlarge}{1.2}
\defR{pair-bertlarge-load30/orion/thrabs/train_bert}{1.13}
\defR{pair-bertlarge-load30/orion/thr/train_bert}{70}
\defR{pair-bertlarge-load50/berth-full/attmin}{99.6}
\defR{pair-bertlarge-load50/berth-full/attmean}{99.6}
\defR{pair-bertlarge-load50/berth-full/thr}{46}
\defR{pair-bertlarge-load50/berth-full/n}{3}
\defR{pair-bertlarge-load50/berth-full/att/bertlarge}{99.6}
\defR{pair-bertlarge-load50/berth-full/p99/bertlarge}{113}
\defR{pair-bertlarge-load50/berth-full/p99x/bertlarge}{1.2}
\defR{pair-bertlarge-load50/berth-full/thrabs/train_bert}{0.74}
\defR{pair-bertlarge-load50/berth-full/thr/train_bert}{46}
\defR{pair-bertlarge-load50/berth-nsr/attmin}{99.7}
\defR{pair-bertlarge-load50/berth-nsr/attmean}{99.7}
\defR{pair-bertlarge-load50/berth-nsr/thr}{46}
\defR{pair-bertlarge-load50/berth-nsr/n}{3}
\defR{pair-bertlarge-load50/berth-nsr/att/bertlarge}{99.7}
\defR{pair-bertlarge-load50/berth-nsr/p99/bertlarge}{112}
\defR{pair-bertlarge-load50/berth-nsr/p99x/bertlarge}{1.2}
\defR{pair-bertlarge-load50/berth-nsr/thrabs/train_bert}{0.74}
\defR{pair-bertlarge-load50/berth-nsr/thr/train_bert}{46}
\defR{pair-bertlarge-load50/berth-nba/attmin}{99.7}
\defR{pair-bertlarge-load50/berth-nba/attmean}{99.7}
\defR{pair-bertlarge-load50/berth-nba/thr}{45}
\defR{pair-bertlarge-load50/berth-nba/n}{3}
\defR{pair-bertlarge-load50/berth-nba/att/bertlarge}{99.7}
\defR{pair-bertlarge-load50/berth-nba/p99/bertlarge}{111}
\defR{pair-bertlarge-load50/berth-nba/p99x/bertlarge}{1.2}
\defR{pair-bertlarge-load50/berth-nba/thrabs/train_bert}{0.74}
\defR{pair-bertlarge-load50/berth-nba/thr/train_bert}{45}
\defR{pair-bertlarge-load50/berth-nsb/attmin}{99.5}
\defR{pair-bertlarge-load50/berth-nsb/attmean}{99.5}
\defR{pair-bertlarge-load50/berth-nsb/thr}{46}
\defR{pair-bertlarge-load50/berth-nsb/n}{3}
\defR{pair-bertlarge-load50/berth-nsb/att/bertlarge}{99.5}
\defR{pair-bertlarge-load50/berth-nsb/p99/bertlarge}{114}
\defR{pair-bertlarge-load50/berth-nsb/p99x/bertlarge}{1.2}
\defR{pair-bertlarge-load50/berth-nsb/thrabs/train_bert}{0.74}
\defR{pair-bertlarge-load50/berth-nsb/thr/train_bert}{46}
\defR{pair-bertlarge-load50/berth-ng/attmin}{96.3}
\defR{pair-bertlarge-load50/berth-ng/attmean}{96.3}
\defR{pair-bertlarge-load50/berth-ng/thr}{43}
\defR{pair-bertlarge-load50/berth-ng/n}{3}
\defR{pair-bertlarge-load50/berth-ng/att/bertlarge}{96.3}
\defR{pair-bertlarge-load50/berth-ng/p99/bertlarge}{192}
\defR{pair-bertlarge-load50/berth-ng/p99x/bertlarge}{2.1}
\defR{pair-bertlarge-load50/berth-ng/thrabs/train_bert}{0.70}
\defR{pair-bertlarge-load50/berth-ng/thr/train_bert}{43}
\defR{pair-bertlarge-load50/berth-nsng/attmin}{97.1}
\defR{pair-bertlarge-load50/berth-nsng/attmean}{97.1}
\defR{pair-bertlarge-load50/berth-nsng/thr}{44}
\defR{pair-bertlarge-load50/berth-nsng/n}{3}
\defR{pair-bertlarge-load50/berth-nsng/att/bertlarge}{97.1}
\defR{pair-bertlarge-load50/berth-nsng/p99/bertlarge}{169}
\defR{pair-bertlarge-load50/berth-nsng/p99x/bertlarge}{1.8}
\defR{pair-bertlarge-load50/berth-nsng/thrabs/train_bert}{0.71}
\defR{pair-bertlarge-load50/berth-nsng/thr/train_bert}{44}
\defR{pair-bertlarge-load50/berth-grouped/attmin}{97.0}
\defR{pair-bertlarge-load50/berth-grouped/attmean}{97.0}
\defR{pair-bertlarge-load50/berth-grouped/thr}{47}
\defR{pair-bertlarge-load50/berth-grouped/n}{3}
\defR{pair-bertlarge-load50/berth-grouped/att/bertlarge}{97.0}
\defR{pair-bertlarge-load50/berth-grouped/p99/bertlarge}{187}
\defR{pair-bertlarge-load50/berth-grouped/p99x/bertlarge}{2.0}
\defR{pair-bertlarge-load50/berth-grouped/thrabs/train_bert}{0.76}
\defR{pair-bertlarge-load50/berth-grouped/thr/train_bert}{47}
\defR{pair-bertlarge-load50/berth/attmin}{97.0}
\defR{pair-bertlarge-load50/berth/attmean}{97.0}
\defR{pair-bertlarge-load50/berth/thr}{47}
\defR{pair-bertlarge-load50/berth/n}{3}
\defR{pair-bertlarge-load50/berth/att/bertlarge}{97.0}
\defR{pair-bertlarge-load50/berth/p99/bertlarge}{184}
\defR{pair-bertlarge-load50/berth/p99x/bertlarge}{2.0}
\defR{pair-bertlarge-load50/berth/thrabs/train_bert}{0.76}
\defR{pair-bertlarge-load50/berth/thr/train_bert}{47}
\defR{pair-bertlarge-load50/timeslice/attmin}{0.0}
\defR{pair-bertlarge-load50/timeslice/attmean}{0.0}
\defR{pair-bertlarge-load50/timeslice/thr}{51}
\defR{pair-bertlarge-load50/timeslice/n}{3}
\defR{pair-bertlarge-load50/timeslice/att/bertlarge}{0.0}
\defR{pair-bertlarge-load50/timeslice/p99/bertlarge}{24406}
\defR{pair-bertlarge-load50/timeslice/p99x/bertlarge}{265.3}
\defR{pair-bertlarge-load50/timeslice/thrabs/train_bert}{0.83}
\defR{pair-bertlarge-load50/timeslice/thr/train_bert}{51}
\defR{pair-bertlarge-load50/mps/attmin}{0.0}
\defR{pair-bertlarge-load50/mps/attmean}{0.0}
\defR{pair-bertlarge-load50/mps/thr}{89}
\defR{pair-bertlarge-load50/mps/n}{3}
\defR{pair-bertlarge-load50/mps/att/bertlarge}{0.0}
\defR{pair-bertlarge-load50/mps/p99/bertlarge}{67312}
\defR{pair-bertlarge-load50/mps/p99x/bertlarge}{731.6}
\defR{pair-bertlarge-load50/mps/thrabs/train_bert}{1.45}
\defR{pair-bertlarge-load50/mps/thr/train_bert}{89}
\defR{pair-bertlarge-load50/mpsprio/attmin}{0.0}
\defR{pair-bertlarge-load50/mpsprio/attmean}{0.0}
\defR{pair-bertlarge-load50/mpsprio/thr}{75}
\defR{pair-bertlarge-load50/mpsprio/n}{3}
\defR{pair-bertlarge-load50/mpsprio/att/bertlarge}{0.0}
\defR{pair-bertlarge-load50/mpsprio/p99/bertlarge}{38233}
\defR{pair-bertlarge-load50/mpsprio/p99x/bertlarge}{415.5}
\defR{pair-bertlarge-load50/mpsprio/thrabs/train_bert}{1.21}
\defR{pair-bertlarge-load50/mpsprio/thr/train_bert}{75}
\defR{pair-bertlarge-load50/mpslimits/attmin}{99.6}
\defR{pair-bertlarge-load50/mpslimits/attmean}{99.6}
\defR{pair-bertlarge-load50/mpslimits/thr}{12}
\defR{pair-bertlarge-load50/mpslimits/n}{3}
\defR{pair-bertlarge-load50/mpslimits/att/bertlarge}{99.6}
\defR{pair-bertlarge-load50/mpslimits/p99/bertlarge}{108}
\defR{pair-bertlarge-load50/mpslimits/p99x/bertlarge}{1.2}
\defR{pair-bertlarge-load50/mpslimits/thrabs/train_bert}{0.19}
\defR{pair-bertlarge-load50/mpslimits/thr/train_bert}{12}
\defR{pair-bertlarge-load50/mpsparts/attmin}{99.8}
\defR{pair-bertlarge-load50/mpsparts/attmean}{99.8}
\defR{pair-bertlarge-load50/mpsparts/thr}{10}
\defR{pair-bertlarge-load50/mpsparts/n}{2}
\defR{pair-bertlarge-load50/mpsparts/att/bertlarge}{99.8}
\defR{pair-bertlarge-load50/mpsparts/p99/bertlarge}{99.5}
\defR{pair-bertlarge-load50/mpsparts/p99x/bertlarge}{1.1}
\defR{pair-bertlarge-load50/mpsparts/thrabs/train_bert}{0.16}
\defR{pair-bertlarge-load50/mpsparts/thr/train_bert}{10}
\defR{pair-bertlarge-load50/lithos/attmin}{99.8}
\defR{pair-bertlarge-load50/lithos/attmean}{99.8}
\defR{pair-bertlarge-load50/lithos/thr}{9}
\defR{pair-bertlarge-load50/lithos/n}{3}
\defR{pair-bertlarge-load50/lithos/att/bertlarge}{99.8}
\defR{pair-bertlarge-load50/lithos/p99/bertlarge}{102}
\defR{pair-bertlarge-load50/lithos/p99x/bertlarge}{1.1}
\defR{pair-bertlarge-load50/lithos/thrabs/train_bert}{0.14}
\defR{pair-bertlarge-load50/lithos/thr/train_bert}{9}
\defR{pair-bertlarge-load50/dysm/attmin}{0.0}
\defR{pair-bertlarge-load50/dysm/attmean}{0.0}
\defR{pair-bertlarge-load50/dysm/thr}{99}
\defR{pair-bertlarge-load50/dysm/n}{3}
\defR{pair-bertlarge-load50/dysm/att/bertlarge}{0.0}
\defR{pair-bertlarge-load50/dysm/thrabs/train_bert}{1.61}
\defR{pair-bertlarge-load50/dysm/thr/train_bert}{99}
\defR{pair-bertlarge-load50/xsched/attmin}{96.6}
\defR{pair-bertlarge-load50/xsched/attmean}{96.6}
\defR{pair-bertlarge-load50/xsched/thr}{45}
\defR{pair-bertlarge-load50/xsched/n}{3}
\defR{pair-bertlarge-load50/xsched/att/bertlarge}{96.6}
\defR{pair-bertlarge-load50/xsched/p99/bertlarge}{193}
\defR{pair-bertlarge-load50/xsched/p99x/bertlarge}{2.1}
\defR{pair-bertlarge-load50/xsched/thrabs/train_bert}{0.73}
\defR{pair-bertlarge-load50/xsched/thr/train_bert}{45}
\defR{pair-bertlarge-load50/multistream/attmin}{0.0}
\defR{pair-bertlarge-load50/multistream/attmean}{0.0}
\defR{pair-bertlarge-load50/multistream/thr}{75}
\defR{pair-bertlarge-load50/multistream/n}{3}
\defR{pair-bertlarge-load50/multistream/att/bertlarge}{0.0}
\defR{pair-bertlarge-load50/multistream/p99/bertlarge}{37849}
\defR{pair-bertlarge-load50/multistream/p99x/bertlarge}{411.3}
\defR{pair-bertlarge-load50/multistream/thrabs/train_bert}{1.21}
\defR{pair-bertlarge-load50/multistream/thr/train_bert}{75}
\defR{pair-bertlarge-load50/tgs/attmin}{99.9}
\defR{pair-bertlarge-load50/tgs/attmean}{99.9}
\defR{pair-bertlarge-load50/tgs/thr}{1}
\defR{pair-bertlarge-load50/tgs/n}{3}
\defR{pair-bertlarge-load50/tgs/att/bertlarge}{99.9}
\defR{pair-bertlarge-load50/tgs/p99/bertlarge}{95.5}
\defR{pair-bertlarge-load50/tgs/p99x/bertlarge}{1.0}
\defR{pair-bertlarge-load50/tgs/thrabs/train_bert}{0.02}
\defR{pair-bertlarge-load50/tgs/thr/train_bert}{1}
\defR{pair-bertlarge-load50/mig/attmin}{99.9}
\defR{pair-bertlarge-load50/mig/attmean}{99.9}
\defR{pair-bertlarge-load50/mig/thr}{0}
\defR{pair-bertlarge-load50/mig/n}{3}
\defR{pair-bertlarge-load50/mig/att/bertlarge}{99.9}
\defR{pair-bertlarge-load50/mig/p99/bertlarge}{94.5}
\defR{pair-bertlarge-load50/mig/p99x/bertlarge}{1.0}
\defR{pair-bertlarge-load50/mig/thrabs/train_bert}{0.00}
\defR{pair-bertlarge-load50/mig/thr/train_bert}{0}
\defR{pair-bertlarge-load50/tally/attmin}{0.0}
\defR{pair-bertlarge-load50/tally/attmean}{0.0}
\defR{pair-bertlarge-load50/tally/thr}{34}
\defR{pair-bertlarge-load50/tally/n}{3}
\defR{pair-bertlarge-load50/tally/att/bertlarge}{0.0}
\defR{pair-bertlarge-load50/tally/p99/bertlarge}{28274}
\defR{pair-bertlarge-load50/tally/p99x/bertlarge}{307.3}
\defR{pair-bertlarge-load50/tally/thrabs/train_bert}{0.55}
\defR{pair-bertlarge-load50/tally/thr/train_bert}{34}
\defR{pair-bertlarge-load50/orion/attmin}{99.9}
\defR{pair-bertlarge-load50/orion/attmean}{99.9}
\defR{pair-bertlarge-load50/orion/thr}{50}
\defR{pair-bertlarge-load50/orion/n}{3}
\defR{pair-bertlarge-load50/orion/att/bertlarge}{99.9}
\defR{pair-bertlarge-load50/orion/p99/bertlarge}{103}
\defR{pair-bertlarge-load50/orion/p99x/bertlarge}{1.1}
\defR{pair-bertlarge-load50/orion/thrabs/train_bert}{0.82}
\defR{pair-bertlarge-load50/orion/thr/train_bert}{50}
\defR{pair-bertlarge-load70/berth-full/attmin}{94.7}
\defR{pair-bertlarge-load70/berth-full/attmean}{94.7}
\defR{pair-bertlarge-load70/berth-full/thr}{26}
\defR{pair-bertlarge-load70/berth-full/n}{3}
\defR{pair-bertlarge-load70/berth-full/att/bertlarge}{94.7}
\defR{pair-bertlarge-load70/berth-full/p99/bertlarge}{221}
\defR{pair-bertlarge-load70/berth-full/p99x/bertlarge}{1.2}
\defR{pair-bertlarge-load70/berth-full/thrabs/train_bert}{0.43}
\defR{pair-bertlarge-load70/berth-full/thr/train_bert}{26}
\defR{pair-bertlarge-load70/berth-nsr/attmin}{93.6}
\defR{pair-bertlarge-load70/berth-nsr/attmean}{93.6}
\defR{pair-bertlarge-load70/berth-nsr/thr}{26}
\defR{pair-bertlarge-load70/berth-nsr/n}{3}
\defR{pair-bertlarge-load70/berth-nsr/att/bertlarge}{93.6}
\defR{pair-bertlarge-load70/berth-nsr/p99/bertlarge}{224}
\defR{pair-bertlarge-load70/berth-nsr/p99x/bertlarge}{1.2}
\defR{pair-bertlarge-load70/berth-nsr/thrabs/train_bert}{0.43}
\defR{pair-bertlarge-load70/berth-nsr/thr/train_bert}{26}
\defR{pair-bertlarge-load70/berth-nba/attmin}{94.8}
\defR{pair-bertlarge-load70/berth-nba/attmean}{94.8}
\defR{pair-bertlarge-load70/berth-nba/thr}{26}
\defR{pair-bertlarge-load70/berth-nba/n}{3}
\defR{pair-bertlarge-load70/berth-nba/att/bertlarge}{94.8}
\defR{pair-bertlarge-load70/berth-nba/p99/bertlarge}{213}
\defR{pair-bertlarge-load70/berth-nba/p99x/bertlarge}{1.2}
\defR{pair-bertlarge-load70/berth-nba/thrabs/train_bert}{0.43}
\defR{pair-bertlarge-load70/berth-nba/thr/train_bert}{26}
\defR{pair-bertlarge-load70/berth-nsb/attmin}{94.9}
\defR{pair-bertlarge-load70/berth-nsb/attmean}{94.9}
\defR{pair-bertlarge-load70/berth-nsb/thr}{26}
\defR{pair-bertlarge-load70/berth-nsb/n}{3}
\defR{pair-bertlarge-load70/berth-nsb/att/bertlarge}{94.9}
\defR{pair-bertlarge-load70/berth-nsb/p99/bertlarge}{211}
\defR{pair-bertlarge-load70/berth-nsb/p99x/bertlarge}{1.1}
\defR{pair-bertlarge-load70/berth-nsb/thrabs/train_bert}{0.43}
\defR{pair-bertlarge-load70/berth-nsb/thr/train_bert}{26}
\defR{pair-bertlarge-load70/berth-ng/attmin}{67.8}
\defR{pair-bertlarge-load70/berth-ng/attmean}{67.8}
\defR{pair-bertlarge-load70/berth-ng/thr}{22}
\defR{pair-bertlarge-load70/berth-ng/n}{3}
\defR{pair-bertlarge-load70/berth-ng/att/bertlarge}{67.8}
\defR{pair-bertlarge-load70/berth-ng/p99/bertlarge}{553}
\defR{pair-bertlarge-load70/berth-ng/p99x/bertlarge}{3.0}
\defR{pair-bertlarge-load70/berth-ng/thrabs/train_bert}{0.36}
\defR{pair-bertlarge-load70/berth-ng/thr/train_bert}{22}
\defR{pair-bertlarge-load70/berth-nsng/attmin}{68.7}
\defR{pair-bertlarge-load70/berth-nsng/attmean}{68.7}
\defR{pair-bertlarge-load70/berth-nsng/thr}{22}
\defR{pair-bertlarge-load70/berth-nsng/n}{3}
\defR{pair-bertlarge-load70/berth-nsng/att/bertlarge}{68.7}
\defR{pair-bertlarge-load70/berth-nsng/p99/bertlarge}{546}
\defR{pair-bertlarge-load70/berth-nsng/p99x/bertlarge}{3.0}
\defR{pair-bertlarge-load70/berth-nsng/thrabs/train_bert}{0.36}
\defR{pair-bertlarge-load70/berth-nsng/thr/train_bert}{22}
\defR{pair-bertlarge-load70/berth-grouped/attmin}{68.1}
\defR{pair-bertlarge-load70/berth-grouped/attmean}{68.1}
\defR{pair-bertlarge-load70/berth-grouped/thr}{26}
\defR{pair-bertlarge-load70/berth-grouped/n}{3}
\defR{pair-bertlarge-load70/berth-grouped/att/bertlarge}{68.1}
\defR{pair-bertlarge-load70/berth-grouped/p99/bertlarge}{553}
\defR{pair-bertlarge-load70/berth-grouped/p99x/bertlarge}{3.0}
\defR{pair-bertlarge-load70/berth-grouped/thrabs/train_bert}{0.42}
\defR{pair-bertlarge-load70/berth-grouped/thr/train_bert}{26}
\defR{pair-bertlarge-load70/berth/attmin}{67.9}
\defR{pair-bertlarge-load70/berth/attmean}{67.9}
\defR{pair-bertlarge-load70/berth/thr}{26}
\defR{pair-bertlarge-load70/berth/n}{3}
\defR{pair-bertlarge-load70/berth/att/bertlarge}{67.9}
\defR{pair-bertlarge-load70/berth/p99/bertlarge}{574}
\defR{pair-bertlarge-load70/berth/p99x/bertlarge}{3.1}
\defR{pair-bertlarge-load70/berth/thrabs/train_bert}{0.42}
\defR{pair-bertlarge-load70/berth/thr/train_bert}{26}
\defR{pair-bertlarge-load70/timeslice/attmin}{0.0}
\defR{pair-bertlarge-load70/timeslice/attmean}{0.0}
\defR{pair-bertlarge-load70/timeslice/thr}{51}
\defR{pair-bertlarge-load70/timeslice/n}{3}
\defR{pair-bertlarge-load70/timeslice/att/bertlarge}{0.0}
\defR{pair-bertlarge-load70/timeslice/p99/bertlarge}{44553}
\defR{pair-bertlarge-load70/timeslice/p99x/bertlarge}{242.1}
\defR{pair-bertlarge-load70/timeslice/thrabs/train_bert}{0.83}
\defR{pair-bertlarge-load70/timeslice/thr/train_bert}{51}
\defR{pair-bertlarge-load70/mps/attmin}{0.0}
\defR{pair-bertlarge-load70/mps/attmean}{0.0}
\defR{pair-bertlarge-load70/mps/thr}{89}
\defR{pair-bertlarge-load70/mps/n}{3}
\defR{pair-bertlarge-load70/mps/att/bertlarge}{0.0}
\defR{pair-bertlarge-load70/mps/p99/bertlarge}{75309}
\defR{pair-bertlarge-load70/mps/p99x/bertlarge}{409.3}
\defR{pair-bertlarge-load70/mps/thrabs/train_bert}{1.45}
\defR{pair-bertlarge-load70/mps/thr/train_bert}{89}
\defR{pair-bertlarge-load70/mpsprio/attmin}{0.0}
\defR{pair-bertlarge-load70/mpsprio/attmean}{0.0}
\defR{pair-bertlarge-load70/mpsprio/thr}{75}
\defR{pair-bertlarge-load70/mpsprio/n}{3}
\defR{pair-bertlarge-load70/mpsprio/att/bertlarge}{0.0}
\defR{pair-bertlarge-load70/mpsprio/p99/bertlarge}{54483}
\defR{pair-bertlarge-load70/mpsprio/p99x/bertlarge}{296.1}
\defR{pair-bertlarge-load70/mpsprio/thrabs/train_bert}{1.21}
\defR{pair-bertlarge-load70/mpsprio/thr/train_bert}{75}
\defR{pair-bertlarge-load70/mpslimits/attmin}{93.1}
\defR{pair-bertlarge-load70/mpslimits/attmean}{93.1}
\defR{pair-bertlarge-load70/mpslimits/thr}{12}
\defR{pair-bertlarge-load70/mpslimits/n}{3}
\defR{pair-bertlarge-load70/mpslimits/att/bertlarge}{93.1}
\defR{pair-bertlarge-load70/mpslimits/p99/bertlarge}{244}
\defR{pair-bertlarge-load70/mpslimits/p99x/bertlarge}{1.3}
\defR{pair-bertlarge-load70/mpslimits/thrabs/train_bert}{0.19}
\defR{pair-bertlarge-load70/mpslimits/thr/train_bert}{12}
\defR{pair-bertlarge-load70/mpsparts/attmin}{95.3}
\defR{pair-bertlarge-load70/mpsparts/attmean}{95.3}
\defR{pair-bertlarge-load70/mpsparts/thr}{9}
\defR{pair-bertlarge-load70/mpsparts/n}{2}
\defR{pair-bertlarge-load70/mpsparts/att/bertlarge}{95.3}
\defR{pair-bertlarge-load70/mpsparts/p99/bertlarge}{248}
\defR{pair-bertlarge-load70/mpsparts/p99x/bertlarge}{1.3}
\defR{pair-bertlarge-load70/mpsparts/thrabs/train_bert}{0.14}
\defR{pair-bertlarge-load70/mpsparts/thr/train_bert}{9}
\defR{pair-bertlarge-load70/lithos/attmin}{95.0}
\defR{pair-bertlarge-load70/lithos/attmean}{95.0}
\defR{pair-bertlarge-load70/lithos/thr}{9}
\defR{pair-bertlarge-load70/lithos/n}{3}
\defR{pair-bertlarge-load70/lithos/att/bertlarge}{95.0}
\defR{pair-bertlarge-load70/lithos/p99/bertlarge}{218}
\defR{pair-bertlarge-load70/lithos/p99x/bertlarge}{1.2}
\defR{pair-bertlarge-load70/lithos/thrabs/train_bert}{0.14}
\defR{pair-bertlarge-load70/lithos/thr/train_bert}{9}
\defR{pair-bertlarge-load70/dysm/attmin}{0.0}
\defR{pair-bertlarge-load70/dysm/attmean}{0.0}
\defR{pair-bertlarge-load70/dysm/thr}{99}
\defR{pair-bertlarge-load70/dysm/n}{3}
\defR{pair-bertlarge-load70/dysm/att/bertlarge}{0.0}
\defR{pair-bertlarge-load70/dysm/thrabs/train_bert}{1.61}
\defR{pair-bertlarge-load70/dysm/thr/train_bert}{99}
\defR{pair-bertlarge-load70/xsched/attmin}{71.4}
\defR{pair-bertlarge-load70/xsched/attmean}{71.4}
\defR{pair-bertlarge-load70/xsched/thr}{25}
\defR{pair-bertlarge-load70/xsched/n}{3}
\defR{pair-bertlarge-load70/xsched/att/bertlarge}{71.4}
\defR{pair-bertlarge-load70/xsched/p99/bertlarge}{528}
\defR{pair-bertlarge-load70/xsched/p99x/bertlarge}{2.9}
\defR{pair-bertlarge-load70/xsched/thrabs/train_bert}{0.40}
\defR{pair-bertlarge-load70/xsched/thr/train_bert}{25}
\defR{pair-bertlarge-load70/multistream/attmin}{0.0}
\defR{pair-bertlarge-load70/multistream/attmean}{0.0}
\defR{pair-bertlarge-load70/multistream/thr}{75}
\defR{pair-bertlarge-load70/multistream/n}{3}
\defR{pair-bertlarge-load70/multistream/att/bertlarge}{0.0}
\defR{pair-bertlarge-load70/multistream/p99/bertlarge}{54470}
\defR{pair-bertlarge-load70/multistream/p99x/bertlarge}{296.0}
\defR{pair-bertlarge-load70/multistream/thrabs/train_bert}{1.21}
\defR{pair-bertlarge-load70/multistream/thr/train_bert}{75}
\defR{pair-bertlarge-load70/tgs/attmin}{97.5}
\defR{pair-bertlarge-load70/tgs/attmean}{97.5}
\defR{pair-bertlarge-load70/tgs/thr}{1}
\defR{pair-bertlarge-load70/tgs/n}{3}
\defR{pair-bertlarge-load70/tgs/att/bertlarge}{97.5}
\defR{pair-bertlarge-load70/tgs/p99/bertlarge}{154}
\defR{pair-bertlarge-load70/tgs/p99x/bertlarge}{0.8}
\defR{pair-bertlarge-load70/tgs/thrabs/train_bert}{0.02}
\defR{pair-bertlarge-load70/tgs/thr/train_bert}{1}
\defR{pair-bertlarge-load70/mig/attmin}{96.8}
\defR{pair-bertlarge-load70/mig/attmean}{96.8}
\defR{pair-bertlarge-load70/mig/thr}{0}
\defR{pair-bertlarge-load70/mig/n}{3}
\defR{pair-bertlarge-load70/mig/att/bertlarge}{96.8}
\defR{pair-bertlarge-load70/mig/p99/bertlarge}{196}
\defR{pair-bertlarge-load70/mig/p99x/bertlarge}{1.1}
\defR{pair-bertlarge-load70/mig/thrabs/train_bert}{0.00}
\defR{pair-bertlarge-load70/mig/thr/train_bert}{0}
\defR{pair-bertlarge-load70/tally/attmin}{0.0}
\defR{pair-bertlarge-load70/tally/attmean}{0.0}
\defR{pair-bertlarge-load70/tally/thr}{26}
\defR{pair-bertlarge-load70/tally/n}{3}
\defR{pair-bertlarge-load70/tally/att/bertlarge}{0.0}
\defR{pair-bertlarge-load70/tally/p99/bertlarge}{39357}
\defR{pair-bertlarge-load70/tally/p99x/bertlarge}{213.9}
\defR{pair-bertlarge-load70/tally/thrabs/train_bert}{0.43}
\defR{pair-bertlarge-load70/tally/thr/train_bert}{26}
\defR{pair-bertlarge-load70/orion/attmin}{95.4}
\defR{pair-bertlarge-load70/orion/attmean}{95.4}
\defR{pair-bertlarge-load70/orion/thr}{31}
\defR{pair-bertlarge-load70/orion/n}{3}
\defR{pair-bertlarge-load70/orion/att/bertlarge}{95.4}
\defR{pair-bertlarge-load70/orion/p99/bertlarge}{182}
\defR{pair-bertlarge-load70/orion/p99x/bertlarge}{1.0}
\defR{pair-bertlarge-load70/orion/thrabs/train_bert}{0.51}
\defR{pair-bertlarge-load70/orion/thr/train_bert}{31}
\defR{pair-perception-30fps/berth-full/attmin}{100.0}
\defR{pair-perception-30fps/berth-full/attmean}{100.0}
\defR{pair-perception-30fps/berth-full/thr}{55}
\defR{pair-perception-30fps/berth-full/n}{3}
\defR{pair-perception-30fps/berth-full/att/perception}{100.0}
\defR{pair-perception-30fps/berth-full/p99/perception}{19.7}
\defR{pair-perception-30fps/berth-full/p99x/perception}{1.4}
\defR{pair-perception-30fps/berth-full/thrabs/train_bert}{0.90}
\defR{pair-perception-30fps/berth-full/thr/train_bert}{55}
\defR{pair-perception-30fps/berth-nsr/attmin}{100.0}
\defR{pair-perception-30fps/berth-nsr/attmean}{100.0}
\defR{pair-perception-30fps/berth-nsr/thr}{55}
\defR{pair-perception-30fps/berth-nsr/n}{3}
\defR{pair-perception-30fps/berth-nsr/att/perception}{100.0}
\defR{pair-perception-30fps/berth-nsr/p99/perception}{19.8}
\defR{pair-perception-30fps/berth-nsr/p99x/perception}{1.4}
\defR{pair-perception-30fps/berth-nsr/thrabs/train_bert}{0.89}
\defR{pair-perception-30fps/berth-nsr/thr/train_bert}{55}
\defR{pair-perception-30fps/berth-nba/attmin}{100.0}
\defR{pair-perception-30fps/berth-nba/attmean}{100.0}
\defR{pair-perception-30fps/berth-nba/thr}{55}
\defR{pair-perception-30fps/berth-nba/n}{3}
\defR{pair-perception-30fps/berth-nba/att/perception}{100.0}
\defR{pair-perception-30fps/berth-nba/p99/perception}{19.7}
\defR{pair-perception-30fps/berth-nba/p99x/perception}{1.4}
\defR{pair-perception-30fps/berth-nba/thrabs/train_bert}{0.90}
\defR{pair-perception-30fps/berth-nba/thr/train_bert}{55}
\defR{pair-perception-30fps/berth-nsb/attmin}{100.0}
\defR{pair-perception-30fps/berth-nsb/attmean}{100.0}
\defR{pair-perception-30fps/berth-nsb/thr}{56}
\defR{pair-perception-30fps/berth-nsb/n}{3}
\defR{pair-perception-30fps/berth-nsb/att/perception}{100.0}
\defR{pair-perception-30fps/berth-nsb/p99/perception}{19.4}
\defR{pair-perception-30fps/berth-nsb/p99x/perception}{1.4}
\defR{pair-perception-30fps/berth-nsb/thrabs/train_bert}{0.90}
\defR{pair-perception-30fps/berth-nsb/thr/train_bert}{56}
\defR{pair-perception-30fps/berth-ng/attmin}{100.0}
\defR{pair-perception-30fps/berth-ng/attmean}{100.0}
\defR{pair-perception-30fps/berth-ng/thr}{56}
\defR{pair-perception-30fps/berth-ng/n}{3}
\defR{pair-perception-30fps/berth-ng/att/perception}{100.0}
\defR{pair-perception-30fps/berth-ng/p99/perception}{19.9}
\defR{pair-perception-30fps/berth-ng/p99x/perception}{1.5}
\defR{pair-perception-30fps/berth-ng/thrabs/train_bert}{0.91}
\defR{pair-perception-30fps/berth-ng/thr/train_bert}{56}
\defR{pair-perception-30fps/berth-nsng/attmin}{100.0}
\defR{pair-perception-30fps/berth-nsng/attmean}{100.0}
\defR{pair-perception-30fps/berth-nsng/thr}{56}
\defR{pair-perception-30fps/berth-nsng/n}{3}
\defR{pair-perception-30fps/berth-nsng/att/perception}{100.0}
\defR{pair-perception-30fps/berth-nsng/p99/perception}{19.9}
\defR{pair-perception-30fps/berth-nsng/p99x/perception}{1.4}
\defR{pair-perception-30fps/berth-nsng/thrabs/train_bert}{0.91}
\defR{pair-perception-30fps/berth-nsng/thr/train_bert}{56}
\defR{pair-perception-30fps/berth-grouped/attmin}{100.0}
\defR{pair-perception-30fps/berth-grouped/attmean}{100.0}
\defR{pair-perception-30fps/berth-grouped/thr}{61}
\defR{pair-perception-30fps/berth-grouped/n}{3}
\defR{pair-perception-30fps/berth-grouped/att/perception}{100.0}
\defR{pair-perception-30fps/berth-grouped/p99/perception}{27.6}
\defR{pair-perception-30fps/berth-grouped/p99x/perception}{2.0}
\defR{pair-perception-30fps/berth-grouped/thrabs/train_bert}{1.00}
\defR{pair-perception-30fps/berth-grouped/thr/train_bert}{61}
\defR{pair-perception-30fps/berth/attmin}{100.0}
\defR{pair-perception-30fps/berth/attmean}{100.0}
\defR{pair-perception-30fps/berth/thr}{61}
\defR{pair-perception-30fps/berth/n}{3}
\defR{pair-perception-30fps/berth/att/perception}{100.0}
\defR{pair-perception-30fps/berth/p99/perception}{27.6}
\defR{pair-perception-30fps/berth/p99x/perception}{2.0}
\defR{pair-perception-30fps/berth/thrabs/train_bert}{1.00}
\defR{pair-perception-30fps/berth/thr/train_bert}{61}
\defR{pair-perception-30fps/timeslice/attmin}{100.0}
\defR{pair-perception-30fps/timeslice/attmean}{100.0}
\defR{pair-perception-30fps/timeslice/thr}{53}
\defR{pair-perception-30fps/timeslice/n}{3}
\defR{pair-perception-30fps/timeslice/att/perception}{100.0}
\defR{pair-perception-30fps/timeslice/p99/perception}{30.9}
\defR{pair-perception-30fps/timeslice/p99x/perception}{2.3}
\defR{pair-perception-30fps/timeslice/thrabs/train_bert}{0.86}
\defR{pair-perception-30fps/timeslice/thr/train_bert}{53}
\defR{pair-perception-30fps/mps/attmin}{0.0}
\defR{pair-perception-30fps/mps/attmean}{0.0}
\defR{pair-perception-30fps/mps/thr}{84}
\defR{pair-perception-30fps/mps/n}{3}
\defR{pair-perception-30fps/mps/att/perception}{0.0}
\defR{pair-perception-30fps/mps/p99/perception}{43850}
\defR{pair-perception-30fps/mps/p99x/perception}{3200.3}
\defR{pair-perception-30fps/mps/thrabs/train_bert}{1.36}
\defR{pair-perception-30fps/mps/thr/train_bert}{84}
\defR{pair-perception-30fps/mpsprio/attmin}{0.0}
\defR{pair-perception-30fps/mpsprio/attmean}{0.0}
\defR{pair-perception-30fps/mpsprio/thr}{69}
\defR{pair-perception-30fps/mpsprio/n}{3}
\defR{pair-perception-30fps/mpsprio/att/perception}{0.0}
\defR{pair-perception-30fps/mpsprio/p99/perception}{1803}
\defR{pair-perception-30fps/mpsprio/p99x/perception}{131.6}
\defR{pair-perception-30fps/mpsprio/thrabs/train_bert}{1.13}
\defR{pair-perception-30fps/mpsprio/thr/train_bert}{69}
\defR{pair-perception-30fps/mpslimits/attmin}{100.0}
\defR{pair-perception-30fps/mpslimits/attmean}{100.0}
\defR{pair-perception-30fps/mpslimits/thr}{47}
\defR{pair-perception-30fps/mpslimits/n}{3}
\defR{pair-perception-30fps/mpslimits/att/perception}{100.0}
\defR{pair-perception-30fps/mpslimits/p99/perception}{23.7}
\defR{pair-perception-30fps/mpslimits/p99x/perception}{1.7}
\defR{pair-perception-30fps/mpslimits/thrabs/train_bert}{0.76}
\defR{pair-perception-30fps/mpslimits/thr/train_bert}{47}
\defR{pair-perception-30fps/mpsparts/attmin}{100.0}
\defR{pair-perception-30fps/mpsparts/attmean}{100.0}
\defR{pair-perception-30fps/mpsparts/thr}{49}
\defR{pair-perception-30fps/mpsparts/n}{2}
\defR{pair-perception-30fps/mpsparts/att/perception}{100.0}
\defR{pair-perception-30fps/mpsparts/p99/perception}{25.1}
\defR{pair-perception-30fps/mpsparts/p99x/perception}{1.8}
\defR{pair-perception-30fps/mpsparts/thrabs/train_bert}{0.80}
\defR{pair-perception-30fps/mpsparts/thr/train_bert}{49}
\defR{pair-perception-30fps/lithos/attmin}{100.0}
\defR{pair-perception-30fps/lithos/attmean}{100.0}
\defR{pair-perception-30fps/lithos/thr}{49}
\defR{pair-perception-30fps/lithos/n}{3}
\defR{pair-perception-30fps/lithos/att/perception}{100.0}
\defR{pair-perception-30fps/lithos/p99/perception}{27.7}
\defR{pair-perception-30fps/lithos/p99x/perception}{2.0}
\defR{pair-perception-30fps/lithos/thrabs/train_bert}{0.80}
\defR{pair-perception-30fps/lithos/thr/train_bert}{49}
\defR{pair-perception-30fps/dysm/attmin}{98.7}
\defR{pair-perception-30fps/dysm/attmean}{98.7}
\defR{pair-perception-30fps/dysm/thr}{61}
\defR{pair-perception-30fps/dysm/n}{3}
\defR{pair-perception-30fps/dysm/att/perception}{98.7}
\defR{pair-perception-30fps/dysm/p99/perception}{34.5}
\defR{pair-perception-30fps/dysm/p99x/perception}{2.5}
\defR{pair-perception-30fps/dysm/thrabs/train_bert}{0.98}
\defR{pair-perception-30fps/dysm/thr/train_bert}{61}
\defR{pair-perception-30fps/xsched/attmin}{100.0}
\defR{pair-perception-30fps/xsched/attmean}{100.0}
\defR{pair-perception-30fps/xsched/thr}{60}
\defR{pair-perception-30fps/xsched/n}{3}
\defR{pair-perception-30fps/xsched/att/perception}{100.0}
\defR{pair-perception-30fps/xsched/p99/perception}{30.4}
\defR{pair-perception-30fps/xsched/p99x/perception}{2.2}
\defR{pair-perception-30fps/xsched/thrabs/train_bert}{0.97}
\defR{pair-perception-30fps/xsched/thr/train_bert}{60}
\defR{pair-perception-30fps/multistream/attmin}{0.0}
\defR{pair-perception-30fps/multistream/attmean}{0.0}
\defR{pair-perception-30fps/multistream/thr}{70}
\defR{pair-perception-30fps/multistream/n}{3}
\defR{pair-perception-30fps/multistream/att/perception}{0.0}
\defR{pair-perception-30fps/multistream/p99/perception}{3928}
\defR{pair-perception-30fps/multistream/p99x/perception}{286.7}
\defR{pair-perception-30fps/multistream/thrabs/train_bert}{1.14}
\defR{pair-perception-30fps/multistream/thr/train_bert}{70}
\defR{pair-perception-30fps/tgs/attmin}{100.0}
\defR{pair-perception-30fps/tgs/attmean}{100.0}
\defR{pair-perception-30fps/tgs/thr}{18}
\defR{pair-perception-30fps/tgs/n}{3}
\defR{pair-perception-30fps/tgs/att/perception}{100.0}
\defR{pair-perception-30fps/tgs/p99/perception}{20.8}
\defR{pair-perception-30fps/tgs/p99x/perception}{1.5}
\defR{pair-perception-30fps/tgs/thrabs/train_bert}{0.29}
\defR{pair-perception-30fps/tgs/thr/train_bert}{18}
\defR{pair-perception-30fps/mig/attmin}{100.0}
\defR{pair-perception-30fps/mig/attmean}{100.0}
\defR{pair-perception-30fps/mig/thr}{55}
\defR{pair-perception-30fps/mig/n}{3}
\defR{pair-perception-30fps/mig/att/perception}{100.0}
\defR{pair-perception-30fps/mig/p99/perception}{24.5}
\defR{pair-perception-30fps/mig/p99x/perception}{1.8}
\defR{pair-perception-30fps/mig/thrabs/train_bert}{0.89}
\defR{pair-perception-30fps/mig/thr/train_bert}{55}
\defR{pair-perception-30fps/tally/attmin}{95.9}
\defR{pair-perception-30fps/tally/attmean}{95.9}
\defR{pair-perception-30fps/tally/thr}{6}
\defR{pair-perception-30fps/tally/n}{3}
\defR{pair-perception-30fps/tally/att/perception}{95.9}
\defR{pair-perception-30fps/tally/p99/perception}{42.2}
\defR{pair-perception-30fps/tally/p99x/perception}{3.1}
\defR{pair-perception-30fps/tally/thrabs/train_bert}{0.10}
\defR{pair-perception-30fps/tally/thr/train_bert}{6}
\defR{pair-perception-30fps/orion/attmin}{98.1}
\defR{pair-perception-30fps/orion/attmean}{98.1}
\defR{pair-perception-30fps/orion/thr}{64}
\defR{pair-perception-30fps/orion/n}{3}
\defR{pair-perception-30fps/orion/att/perception}{98.1}
\defR{pair-perception-30fps/orion/p99/perception}{37.8}
\defR{pair-perception-30fps/orion/p99x/perception}{2.8}
\defR{pair-perception-30fps/orion/thrabs/train_bert}{1.04}
\defR{pair-perception-30fps/orion/thr/train_bert}{64}
\defR{mix-load40/berth-full/attmin}{99.4}
\defR{mix-load40/berth-full/attmean}{99.8}
\defR{mix-load40/berth-full/thr}{57}
\defR{mix-load40/berth-full/n}{3}
\defR{mix-load40/berth-full/att/resnet50}{99.4}
\defR{mix-load40/berth-full/att/retinanet}{100.0}
\defR{mix-load40/berth-full/att/bertlarge}{99.9}
\defR{mix-load40/berth-full/p99/resnet50}{14.3}
\defR{mix-load40/berth-full/p99/retinanet}{38.1}
\defR{mix-load40/berth-full/p99/bertlarge}{91.9}
\defR{mix-load40/berth-full/p99x/resnet50}{1.1}
\defR{mix-load40/berth-full/p99x/retinanet}{1.1}
\defR{mix-load40/berth-full/p99x/bertlarge}{1.3}
\defR{mix-load40/berth-full/thrabs/train_bert}{0.92}
\defR{mix-load40/berth-full/thr/train_bert}{57}
\defR{mix-load40/berth-nsr/attmin}{99.3}
\defR{mix-load40/berth-nsr/attmean}{99.7}
\defR{mix-load40/berth-nsr/thr}{56}
\defR{mix-load40/berth-nsr/n}{3}
\defR{mix-load40/berth-nsr/att/resnet50}{99.3}
\defR{mix-load40/berth-nsr/att/retinanet}{100.0}
\defR{mix-load40/berth-nsr/att/bertlarge}{99.9}
\defR{mix-load40/berth-nsr/p99/resnet50}{14.5}
\defR{mix-load40/berth-nsr/p99/retinanet}{38.3}
\defR{mix-load40/berth-nsr/p99/bertlarge}{93.3}
\defR{mix-load40/berth-nsr/p99x/resnet50}{1.1}
\defR{mix-load40/berth-nsr/p99x/retinanet}{1.1}
\defR{mix-load40/berth-nsr/p99x/bertlarge}{1.3}
\defR{mix-load40/berth-nsr/thrabs/train_bert}{0.91}
\defR{mix-load40/berth-nsr/thr/train_bert}{56}
\defR{mix-load40/berth-nba/attmin}{99.4}
\defR{mix-load40/berth-nba/attmean}{99.8}
\defR{mix-load40/berth-nba/thr}{57}
\defR{mix-load40/berth-nba/n}{3}
\defR{mix-load40/berth-nba/att/resnet50}{99.4}
\defR{mix-load40/berth-nba/att/retinanet}{100.0}
\defR{mix-load40/berth-nba/att/bertlarge}{99.9}
\defR{mix-load40/berth-nba/p99/resnet50}{14.0}
\defR{mix-load40/berth-nba/p99/retinanet}{39.8}
\defR{mix-load40/berth-nba/p99/bertlarge}{93.5}
\defR{mix-load40/berth-nba/p99x/resnet50}{1.0}
\defR{mix-load40/berth-nba/p99x/retinanet}{1.1}
\defR{mix-load40/berth-nba/p99x/bertlarge}{1.3}
\defR{mix-load40/berth-nba/thrabs/train_bert}{0.92}
\defR{mix-load40/berth-nba/thr/train_bert}{57}
\defR{mix-load40/berth-nsb/attmin}{99.3}
\defR{mix-load40/berth-nsb/attmean}{99.7}
\defR{mix-load40/berth-nsb/thr}{56}
\defR{mix-load40/berth-nsb/n}{3}
\defR{mix-load40/berth-nsb/att/resnet50}{99.3}
\defR{mix-load40/berth-nsb/att/retinanet}{100.0}
\defR{mix-load40/berth-nsb/att/bertlarge}{99.7}
\defR{mix-load40/berth-nsb/p99/resnet50}{14.4}
\defR{mix-load40/berth-nsb/p99/retinanet}{39.8}
\defR{mix-load40/berth-nsb/p99/bertlarge}{98.2}
\defR{mix-load40/berth-nsb/p99x/resnet50}{1.1}
\defR{mix-load40/berth-nsb/p99x/retinanet}{1.1}
\defR{mix-load40/berth-nsb/p99x/bertlarge}{1.4}
\defR{mix-load40/berth-nsb/thrabs/train_bert}{0.91}
\defR{mix-load40/berth-nsb/thr/train_bert}{56}
\defR{mix-load40/berth-ng/attmin}{99.0}
\defR{mix-load40/berth-ng/attmean}{99.5}
\defR{mix-load40/berth-ng/thr}{57}
\defR{mix-load40/berth-ng/n}{3}
\defR{mix-load40/berth-ng/att/resnet50}{99.0}
\defR{mix-load40/berth-ng/att/retinanet}{100.0}
\defR{mix-load40/berth-ng/att/bertlarge}{99.5}
\defR{mix-load40/berth-ng/p99/resnet50}{14.9}
\defR{mix-load40/berth-ng/p99/retinanet}{42.4}
\defR{mix-load40/berth-ng/p99/bertlarge}{109}
\defR{mix-load40/berth-ng/p99x/resnet50}{1.1}
\defR{mix-load40/berth-ng/p99x/retinanet}{1.2}
\defR{mix-load40/berth-ng/p99x/bertlarge}{1.5}
\defR{mix-load40/berth-ng/thrabs/train_bert}{0.92}
\defR{mix-load40/berth-ng/thr/train_bert}{57}
\defR{mix-load40/berth-nsng/attmin}{99.2}
\defR{mix-load40/berth-nsng/attmean}{99.6}
\defR{mix-load40/berth-nsng/thr}{57}
\defR{mix-load40/berth-nsng/n}{3}
\defR{mix-load40/berth-nsng/att/resnet50}{99.2}
\defR{mix-load40/berth-nsng/att/retinanet}{100.0}
\defR{mix-load40/berth-nsng/att/bertlarge}{99.6}
\defR{mix-load40/berth-nsng/p99/resnet50}{14.7}
\defR{mix-load40/berth-nsng/p99/retinanet}{41.0}
\defR{mix-load40/berth-nsng/p99/bertlarge}{99.6}
\defR{mix-load40/berth-nsng/p99x/resnet50}{1.1}
\defR{mix-load40/berth-nsng/p99x/retinanet}{1.2}
\defR{mix-load40/berth-nsng/p99x/bertlarge}{1.4}
\defR{mix-load40/berth-nsng/thrabs/train_bert}{0.92}
\defR{mix-load40/berth-nsng/thr/train_bert}{57}
\defR{mix-load40/berth-grouped/attmin}{94.7}
\defR{mix-load40/berth-grouped/attmean}{98.2}
\defR{mix-load40/berth-grouped/thr}{61}
\defR{mix-load40/berth-grouped/n}{3}
\defR{mix-load40/berth-grouped/att/resnet50}{94.7}
\defR{mix-load40/berth-grouped/att/retinanet}{100.0}
\defR{mix-load40/berth-grouped/att/bertlarge}{99.9}
\defR{mix-load40/berth-grouped/p99/resnet50}{22.0}
\defR{mix-load40/berth-grouped/p99/retinanet}{48.3}
\defR{mix-load40/berth-grouped/p99/bertlarge}{71.7}
\defR{mix-load40/berth-grouped/p99x/resnet50}{1.6}
\defR{mix-load40/berth-grouped/p99x/retinanet}{1.4}
\defR{mix-load40/berth-grouped/p99x/bertlarge}{1.0}
\defR{mix-load40/berth-grouped/thrabs/train_bert}{0.98}
\defR{mix-load40/berth-grouped/thr/train_bert}{61}
\defR{mix-load40/berth/attmin}{94.2}
\defR{mix-load40/berth/attmean}{98.0}
\defR{mix-load40/berth/thr}{58}
\defR{mix-load40/berth/n}{3}
\defR{mix-load40/berth/att/resnet50}{94.2}
\defR{mix-load40/berth/att/retinanet}{100.0}
\defR{mix-load40/berth/att/bertlarge}{99.9}
\defR{mix-load40/berth/p99/resnet50}{22.8}
\defR{mix-load40/berth/p99/retinanet}{51.5}
\defR{mix-load40/berth/p99/bertlarge}{94.7}
\defR{mix-load40/berth/p99x/resnet50}{1.7}
\defR{mix-load40/berth/p99x/retinanet}{1.5}
\defR{mix-load40/berth/p99x/bertlarge}{1.3}
\defR{mix-load40/berth/thrabs/train_bert}{0.94}
\defR{mix-load40/berth/thr/train_bert}{58}
\defR{mix-load40/timeslice/attmin}{62.7}
\defR{mix-load40/timeslice/attmean}{77.7}
\defR{mix-load40/timeslice/thr}{54}
\defR{mix-load40/timeslice/n}{3}
\defR{mix-load40/timeslice/att/resnet50}{62.7}
\defR{mix-load40/timeslice/att/retinanet}{96.2}
\defR{mix-load40/timeslice/att/bertlarge}{74.3}
\defR{mix-load40/timeslice/p99/resnet50}{43.3}
\defR{mix-load40/timeslice/p99/retinanet}{135}
\defR{mix-load40/timeslice/p99/bertlarge}{296}
\defR{mix-load40/timeslice/p99x/resnet50}{3.2}
\defR{mix-load40/timeslice/p99x/retinanet}{3.9}
\defR{mix-load40/timeslice/p99x/bertlarge}{4.1}
\defR{mix-load40/timeslice/thrabs/train_bert}{0.88}
\defR{mix-load40/timeslice/thr/train_bert}{54}
\defR{mix-load40/mps/attmin}{0.0}
\defR{mix-load40/mps/attmean}{0.0}
\defR{mix-load40/mps/thr}{80}
\defR{mix-load40/mps/n}{3}
\defR{mix-load40/mps/att/resnet50}{0.0}
\defR{mix-load40/mps/att/retinanet}{0.0}
\defR{mix-load40/mps/att/bertlarge}{0.1}
\defR{mix-load40/mps/p99/resnet50}{31302}
\defR{mix-load40/mps/p99/retinanet}{42441}
\defR{mix-load40/mps/p99/bertlarge}{6706}
\defR{mix-load40/mps/p99x/resnet50}{2336.5}
\defR{mix-load40/mps/p99x/retinanet}{1220.4}
\defR{mix-load40/mps/p99x/bertlarge}{92.7}
\defR{mix-load40/mps/thrabs/train_bert}{1.30}
\defR{mix-load40/mps/thr/train_bert}{80}
\defR{mix-load40/mpsprio/attmin}{0.1}
\defR{mix-load40/mpsprio/attmean}{43.0}
\defR{mix-load40/mpsprio/thr}{70}
\defR{mix-load40/mpsprio/n}{3}
\defR{mix-load40/mpsprio/att/resnet50}{0.1}
\defR{mix-load40/mpsprio/att/retinanet}{51.7}
\defR{mix-load40/mpsprio/att/bertlarge}{77.4}
\defR{mix-load40/mpsprio/p99/resnet50}{142}
\defR{mix-load40/mpsprio/p99/retinanet}{387}
\defR{mix-load40/mpsprio/p99/bertlarge}{265}
\defR{mix-load40/mpsprio/p99x/resnet50}{10.6}
\defR{mix-load40/mpsprio/p99x/retinanet}{11.1}
\defR{mix-load40/mpsprio/p99x/bertlarge}{3.7}
\defR{mix-load40/mpsprio/thrabs/train_bert}{1.14}
\defR{mix-load40/mpsprio/thr/train_bert}{70}
\defR{mix-load40/mpslimits/attmin}{88.8}
\defR{mix-load40/mpslimits/attmean}{94.8}
\defR{mix-load40/mpslimits/thr}{11}
\defR{mix-load40/mpslimits/n}{3}
\defR{mix-load40/mpslimits/att/resnet50}{99.1}
\defR{mix-load40/mpslimits/att/retinanet}{96.4}
\defR{mix-load40/mpslimits/att/bertlarge}{88.8}
\defR{mix-load40/mpslimits/p99/resnet50}{14.7}
\defR{mix-load40/mpslimits/p99/retinanet}{136}
\defR{mix-load40/mpslimits/p99/bertlarge}{214}
\defR{mix-load40/mpslimits/p99x/resnet50}{1.1}
\defR{mix-load40/mpslimits/p99x/retinanet}{3.9}
\defR{mix-load40/mpslimits/p99x/bertlarge}{3.0}
\defR{mix-load40/mpslimits/thrabs/train_bert}{0.19}
\defR{mix-load40/mpslimits/thr/train_bert}{11}
\defR{mix-load40/mpsparts/attmin}{87.6}
\defR{mix-load40/mpsparts/attmean}{93.4}
\defR{mix-load40/mpsparts/thr}{18}
\defR{mix-load40/mpsparts/n}{2}
\defR{mix-load40/mpsparts/att/resnet50}{98.9}
\defR{mix-load40/mpsparts/att/retinanet}{93.7}
\defR{mix-load40/mpsparts/att/bertlarge}{87.6}
\defR{mix-load40/mpsparts/p99/resnet50}{15.3}
\defR{mix-load40/mpsparts/p99/retinanet}{156}
\defR{mix-load40/mpsparts/p99/bertlarge}{214}
\defR{mix-load40/mpsparts/p99x/resnet50}{1.1}
\defR{mix-load40/mpsparts/p99x/retinanet}{4.5}
\defR{mix-load40/mpsparts/p99x/bertlarge}{3.0}
\defR{mix-load40/mpsparts/thrabs/train_bert}{0.30}
\defR{mix-load40/mpsparts/thr/train_bert}{18}
\defR{mix-load40/lithos/attmin}{87.5}
\defR{mix-load40/lithos/attmean}{93.9}
\defR{mix-load40/lithos/thr}{14}
\defR{mix-load40/lithos/n}{3}
\defR{mix-load40/lithos/att/resnet50}{98.9}
\defR{mix-load40/lithos/att/retinanet}{95.4}
\defR{mix-load40/lithos/att/bertlarge}{87.5}
\defR{mix-load40/lithos/p99/resnet50}{15.1}
\defR{mix-load40/lithos/p99/retinanet}{145}
\defR{mix-load40/lithos/p99/bertlarge}{220}
\defR{mix-load40/lithos/p99x/resnet50}{1.1}
\defR{mix-load40/lithos/p99x/retinanet}{4.2}
\defR{mix-load40/lithos/p99x/bertlarge}{3.0}
\defR{mix-load40/lithos/thrabs/train_bert}{0.22}
\defR{mix-load40/lithos/thr/train_bert}{14}
\defR{mix-load40/dysm/attmin}{0.0}
\defR{mix-load40/dysm/attmean}{24.5}
\defR{mix-load40/dysm/thr}{85}
\defR{mix-load40/dysm/n}{3}
\defR{mix-load40/dysm/att/resnet50}{0.0}
\defR{mix-load40/dysm/att/retinanet}{73.6}
\defR{mix-load40/dysm/att/bertlarge}{0.0}
\defR{mix-load40/dysm/p99/retinanet}{283}
\defR{mix-load40/dysm/p99x/retinanet}{8.1}
\defR{mix-load40/dysm/thrabs/train_bert}{1.38}
\defR{mix-load40/dysm/thr/train_bert}{85}
\defR{mix-load40/xsched/attmin}{88.0}
\defR{mix-load40/xsched/attmean}{95.9}
\defR{mix-load40/xsched/thr}{56}
\defR{mix-load40/xsched/n}{3}
\defR{mix-load40/xsched/att/resnet50}{88.0}
\defR{mix-load40/xsched/att/retinanet}{100.0}
\defR{mix-load40/xsched/att/bertlarge}{99.8}
\defR{mix-load40/xsched/p99/resnet50}{27.9}
\defR{mix-load40/xsched/p99/retinanet}{57.9}
\defR{mix-load40/xsched/p99/bertlarge}{105}
\defR{mix-load40/xsched/p99x/resnet50}{2.1}
\defR{mix-load40/xsched/p99x/retinanet}{1.7}
\defR{mix-load40/xsched/p99x/bertlarge}{1.5}
\defR{mix-load40/xsched/thrabs/train_bert}{0.91}
\defR{mix-load40/xsched/thr/train_bert}{56}
\defR{mix-load40/multistream/attmin}{0.0}
\defR{mix-load40/multistream/attmean}{43.5}
\defR{mix-load40/multistream/thr}{71}
\defR{mix-load40/multistream/n}{3}
\defR{mix-load40/multistream/att/resnet50}{0.0}
\defR{mix-load40/multistream/att/retinanet}{51.9}
\defR{mix-load40/multistream/att/bertlarge}{78.7}
\defR{mix-load40/multistream/p99/resnet50}{157}
\defR{mix-load40/multistream/p99/retinanet}{391}
\defR{mix-load40/multistream/p99/bertlarge}{255}
\defR{mix-load40/multistream/p99x/resnet50}{11.7}
\defR{mix-load40/multistream/p99x/retinanet}{11.2}
\defR{mix-load40/multistream/p99x/bertlarge}{3.5}
\defR{mix-load40/multistream/thrabs/train_bert}{1.15}
\defR{mix-load40/multistream/thr/train_bert}{71}
\defR{mix-load40/tgs/attmin}{96.1}
\defR{mix-load40/tgs/attmean}{98.6}
\defR{mix-load40/tgs/thr}{4}
\defR{mix-load40/tgs/n}{3}
\defR{mix-load40/tgs/att/resnet50}{96.1}
\defR{mix-load40/tgs/att/retinanet}{100.0}
\defR{mix-load40/tgs/att/bertlarge}{99.8}
\defR{mix-load40/tgs/p99/resnet50}{20.6}
\defR{mix-load40/tgs/p99/retinanet}{42.2}
\defR{mix-load40/tgs/p99/bertlarge}{88.4}
\defR{mix-load40/tgs/p99x/resnet50}{1.5}
\defR{mix-load40/tgs/p99x/retinanet}{1.2}
\defR{mix-load40/tgs/p99x/bertlarge}{1.2}
\defR{mix-load40/tgs/thrabs/train_bert}{0.06}
\defR{mix-load40/tgs/thr/train_bert}{4}
\defR{mix-load40/mig/attmin}{90.6}
\defR{mix-load40/mig/attmean}{95.8}
\defR{mix-load40/mig/thr}{0}
\defR{mix-load40/mig/n}{3}
\defR{mix-load40/mig/att/resnet50}{98.5}
\defR{mix-load40/mig/att/retinanet}{98.4}
\defR{mix-load40/mig/att/bertlarge}{90.6}
\defR{mix-load40/mig/p99/resnet50}{16.5}
\defR{mix-load40/mig/p99/retinanet}{109}
\defR{mix-load40/mig/p99/bertlarge}{207}
\defR{mix-load40/mig/p99x/resnet50}{1.2}
\defR{mix-load40/mig/p99x/retinanet}{3.1}
\defR{mix-load40/mig/p99x/bertlarge}{2.9}
\defR{mix-load40/mig/thrabs/train_bert}{0.00}
\defR{mix-load40/mig/thr/train_bert}{0}
\defR{mix-load40/tally/attmin}{0.0}
\defR{mix-load40/tally/attmean}{32.9}
\defR{mix-load40/tally/thr}{26}
\defR{mix-load40/tally/n}{3}
\defR{mix-load40/tally/att/resnet50}{0.0}
\defR{mix-load40/tally/att/retinanet}{4.2}
\defR{mix-load40/tally/att/bertlarge}{94.5}
\defR{mix-load40/tally/p99/resnet50}{94201}
\defR{mix-load40/tally/p99/retinanet}{2652}
\defR{mix-load40/tally/p99/bertlarge}{175}
\defR{mix-load40/tally/p99x/resnet50}{7031.3}
\defR{mix-load40/tally/p99x/retinanet}{76.3}
\defR{mix-load40/tally/p99x/bertlarge}{2.4}
\defR{mix-load40/tally/thrabs/train_bert}{0.42}
\defR{mix-load40/tally/thr/train_bert}{26}
\defR{mix-load40/orion/attmin}{0.0}
\defR{mix-load40/orion/attmean}{34.6}
\defR{mix-load40/orion/thr}{73}
\defR{mix-load40/orion/n}{3}
\defR{mix-load40/orion/att/resnet50}{94.5}
\defR{mix-load40/orion/att/retinanet}{0.0}
\defR{mix-load40/orion/att/bertlarge}{9.2}
\defR{mix-load40/orion/p99/resnet50}{29.0}
\defR{mix-load40/orion/p99/retinanet}{70258}
\defR{mix-load40/orion/p99/bertlarge}{1603}
\defR{mix-load40/orion/p99x/resnet50}{2.2}
\defR{mix-load40/orion/p99x/retinanet}{2020.2}
\defR{mix-load40/orion/p99x/bertlarge}{22.2}
\defR{mix-load40/orion/thrabs/train_bert}{1.18}
\defR{mix-load40/orion/thr/train_bert}{73}
\defR{mix-load70/berth-full/attmin}{93.8}
\defR{mix-load70/berth-full/attmean}{97.2}
\defR{mix-load70/berth-full/thr}{30}
\defR{mix-load70/berth-full/n}{3}
\defR{mix-load70/berth-full/att/resnet50}{97.9}
\defR{mix-load70/berth-full/att/retinanet}{100.0}
\defR{mix-load70/berth-full/att/bertlarge}{93.8}
\defR{mix-load70/berth-full/p99/resnet50}{16.8}
\defR{mix-load70/berth-full/p99/retinanet}{63.1}
\defR{mix-load70/berth-full/p99/bertlarge}{187}
\defR{mix-load70/berth-full/p99x/resnet50}{1.0}
\defR{mix-load70/berth-full/p99x/retinanet}{1.8}
\defR{mix-load70/berth-full/p99x/bertlarge}{2.8}
\defR{mix-load70/berth-full/thrabs/train_bert}{0.48}
\defR{mix-load70/berth-full/thr/train_bert}{30}
\defR{mix-load70/berth-nsr/attmin}{93.3}
\defR{mix-load70/berth-nsr/attmean}{97.1}
\defR{mix-load70/berth-nsr/thr}{29}
\defR{mix-load70/berth-nsr/n}{3}
\defR{mix-load70/berth-nsr/att/resnet50}{98.0}
\defR{mix-load70/berth-nsr/att/retinanet}{100.0}
\defR{mix-load70/berth-nsr/att/bertlarge}{93.3}
\defR{mix-load70/berth-nsr/p99/resnet50}{16.9}
\defR{mix-load70/berth-nsr/p99/retinanet}{60.9}
\defR{mix-load70/berth-nsr/p99/bertlarge}{201}
\defR{mix-load70/berth-nsr/p99x/resnet50}{1.0}
\defR{mix-load70/berth-nsr/p99x/retinanet}{1.7}
\defR{mix-load70/berth-nsr/p99x/bertlarge}{3.1}
\defR{mix-load70/berth-nsr/thrabs/train_bert}{0.47}
\defR{mix-load70/berth-nsr/thr/train_bert}{29}
\defR{mix-load70/berth-nba/attmin}{94.6}
\defR{mix-load70/berth-nba/attmean}{97.6}
\defR{mix-load70/berth-nba/thr}{29}
\defR{mix-load70/berth-nba/n}{3}
\defR{mix-load70/berth-nba/att/resnet50}{98.1}
\defR{mix-load70/berth-nba/att/retinanet}{100.0}
\defR{mix-load70/berth-nba/att/bertlarge}{94.6}
\defR{mix-load70/berth-nba/p99/resnet50}{16.5}
\defR{mix-load70/berth-nba/p99/retinanet}{60.9}
\defR{mix-load70/berth-nba/p99/bertlarge}{188}
\defR{mix-load70/berth-nba/p99x/resnet50}{1.0}
\defR{mix-load70/berth-nba/p99x/retinanet}{1.7}
\defR{mix-load70/berth-nba/p99x/bertlarge}{2.9}
\defR{mix-load70/berth-nba/thrabs/train_bert}{0.47}
\defR{mix-load70/berth-nba/thr/train_bert}{29}
\defR{mix-load70/berth-nsb/attmin}{94.6}
\defR{mix-load70/berth-nsb/attmean}{97.6}
\defR{mix-load70/berth-nsb/thr}{30}
\defR{mix-load70/berth-nsb/n}{3}
\defR{mix-load70/berth-nsb/att/resnet50}{98.2}
\defR{mix-load70/berth-nsb/att/retinanet}{100.0}
\defR{mix-load70/berth-nsb/att/bertlarge}{94.6}
\defR{mix-load70/berth-nsb/p99/resnet50}{16.3}
\defR{mix-load70/berth-nsb/p99/retinanet}{60.2}
\defR{mix-load70/berth-nsb/p99/bertlarge}{187}
\defR{mix-load70/berth-nsb/p99x/resnet50}{1.0}
\defR{mix-load70/berth-nsb/p99x/retinanet}{1.7}
\defR{mix-load70/berth-nsb/p99x/bertlarge}{2.8}
\defR{mix-load70/berth-nsb/thrabs/train_bert}{0.48}
\defR{mix-load70/berth-nsb/thr/train_bert}{30}
\defR{mix-load70/berth-ng/attmin}{90.3}
\defR{mix-load70/berth-ng/attmean}{96.0}
\defR{mix-load70/berth-ng/thr}{29}
\defR{mix-load70/berth-ng/n}{3}
\defR{mix-load70/berth-ng/att/resnet50}{97.6}
\defR{mix-load70/berth-ng/att/retinanet}{100.0}
\defR{mix-load70/berth-ng/att/bertlarge}{90.3}
\defR{mix-load70/berth-ng/p99/resnet50}{17.2}
\defR{mix-load70/berth-ng/p99/retinanet}{64.6}
\defR{mix-load70/berth-ng/p99/bertlarge}{225}
\defR{mix-load70/berth-ng/p99x/resnet50}{1.0}
\defR{mix-load70/berth-ng/p99x/retinanet}{1.8}
\defR{mix-load70/berth-ng/p99x/bertlarge}{3.4}
\defR{mix-load70/berth-ng/thrabs/train_bert}{0.48}
\defR{mix-load70/berth-ng/thr/train_bert}{29}
\defR{mix-load70/berth-nsng/attmin}{90.3}
\defR{mix-load70/berth-nsng/attmean}{96.0}
\defR{mix-load70/berth-nsng/thr}{29}
\defR{mix-load70/berth-nsng/n}{3}
\defR{mix-load70/berth-nsng/att/resnet50}{97.8}
\defR{mix-load70/berth-nsng/att/retinanet}{100.0}
\defR{mix-load70/berth-nsng/att/bertlarge}{90.3}
\defR{mix-load70/berth-nsng/p99/resnet50}{17.1}
\defR{mix-load70/berth-nsng/p99/retinanet}{64.0}
\defR{mix-load70/berth-nsng/p99/bertlarge}{231}
\defR{mix-load70/berth-nsng/p99x/resnet50}{1.0}
\defR{mix-load70/berth-nsng/p99x/retinanet}{1.8}
\defR{mix-load70/berth-nsng/p99x/bertlarge}{3.5}
\defR{mix-load70/berth-nsng/thrabs/train_bert}{0.48}
\defR{mix-load70/berth-nsng/thr/train_bert}{29}
\defR{mix-load70/berth-grouped/attmin}{86.1}
\defR{mix-load70/berth-grouped/attmean}{95.2}
\defR{mix-load70/berth-grouped/thr}{37}
\defR{mix-load70/berth-grouped/n}{3}
\defR{mix-load70/berth-grouped/att/resnet50}{86.1}
\defR{mix-load70/berth-grouped/att/retinanet}{99.8}
\defR{mix-load70/berth-grouped/att/bertlarge}{99.8}
\defR{mix-load70/berth-grouped/p99/resnet50}{31.3}
\defR{mix-load70/berth-grouped/p99/retinanet}{75.0}
\defR{mix-load70/berth-grouped/p99/bertlarge}{110}
\defR{mix-load70/berth-grouped/p99x/resnet50}{1.8}
\defR{mix-load70/berth-grouped/p99x/retinanet}{2.1}
\defR{mix-load70/berth-grouped/p99x/bertlarge}{1.7}
\defR{mix-load70/berth-grouped/thrabs/train_bert}{0.60}
\defR{mix-load70/berth-grouped/thr/train_bert}{37}
\defR{mix-load70/berth/attmin}{79.4}
\defR{mix-load70/berth/attmean}{90.5}
\defR{mix-load70/berth/thr}{29}
\defR{mix-load70/berth/n}{3}
\defR{mix-load70/berth/att/resnet50}{79.4}
\defR{mix-load70/berth/att/retinanet}{99.0}
\defR{mix-load70/berth/att/bertlarge}{93.1}
\defR{mix-load70/berth/p99/resnet50}{38.7}
\defR{mix-load70/berth/p99/retinanet}{104}
\defR{mix-load70/berth/p99/bertlarge}{207}
\defR{mix-load70/berth/p99x/resnet50}{2.3}
\defR{mix-load70/berth/p99x/retinanet}{3.0}
\defR{mix-load70/berth/p99x/bertlarge}{3.1}
\defR{mix-load70/berth/thrabs/train_bert}{0.48}
\defR{mix-load70/berth/thr/train_bert}{29}
\defR{mix-load70/timeslice/attmin}{0.0}
\defR{mix-load70/timeslice/attmean}{15.2}
\defR{mix-load70/timeslice/thr}{29}
\defR{mix-load70/timeslice/n}{3}
\defR{mix-load70/timeslice/att/resnet50}{19.9}
\defR{mix-load70/timeslice/att/retinanet}{25.6}
\defR{mix-load70/timeslice/att/bertlarge}{0.0}
\defR{mix-load70/timeslice/p99/resnet50}{163}
\defR{mix-load70/timeslice/p99/retinanet}{874}
\defR{mix-load70/timeslice/p99/bertlarge}{12219}
\defR{mix-load70/timeslice/p99x/resnet50}{9.5}
\defR{mix-load70/timeslice/p99x/retinanet}{24.9}
\defR{mix-load70/timeslice/p99x/bertlarge}{185.8}
\defR{mix-load70/timeslice/thrabs/train_bert}{0.47}
\defR{mix-load70/timeslice/thr/train_bert}{29}
\defR{mix-load70/mps/attmin}{0.0}
\defR{mix-load70/mps/attmean}{0.0}
\defR{mix-load70/mps/thr}{80}
\defR{mix-load70/mps/n}{3}
\defR{mix-load70/mps/att/resnet50}{0.0}
\defR{mix-load70/mps/att/retinanet}{0.0}
\defR{mix-load70/mps/att/bertlarge}{0.0}
\defR{mix-load70/mps/p99/resnet50}{58854}
\defR{mix-load70/mps/p99/retinanet}{65268}
\defR{mix-load70/mps/p99/bertlarge}{42680}
\defR{mix-load70/mps/p99x/resnet50}{3427.8}
\defR{mix-load70/mps/p99x/retinanet}{1860.9}
\defR{mix-load70/mps/p99x/bertlarge}{649.1}
\defR{mix-load70/mps/thrabs/train_bert}{1.30}
\defR{mix-load70/mps/thr/train_bert}{80}
\defR{mix-load70/mpsprio/attmin}{0.0}
\defR{mix-load70/mpsprio/attmean}{10.9}
\defR{mix-load70/mpsprio/thr}{55}
\defR{mix-load70/mpsprio/n}{3}
\defR{mix-load70/mpsprio/att/resnet50}{0.0}
\defR{mix-load70/mpsprio/att/retinanet}{0.0}
\defR{mix-load70/mpsprio/att/bertlarge}{32.6}
\defR{mix-load70/mpsprio/p99/resnet50}{10088}
\defR{mix-load70/mpsprio/p99/retinanet}{23565}
\defR{mix-load70/mpsprio/p99/bertlarge}{661}
\defR{mix-load70/mpsprio/p99x/resnet50}{587.6}
\defR{mix-load70/mpsprio/p99x/retinanet}{671.9}
\defR{mix-load70/mpsprio/p99x/bertlarge}{10.0}
\defR{mix-load70/mpsprio/thrabs/train_bert}{0.90}
\defR{mix-load70/mpsprio/thr/train_bert}{55}
\defR{mix-load70/mpslimits/attmin}{69.6}
\defR{mix-load70/mpslimits/attmean}{79.9}
\defR{mix-load70/mpslimits/thr}{12}
\defR{mix-load70/mpslimits/n}{3}
\defR{mix-load70/mpslimits/att/resnet50}{93.2}
\defR{mix-load70/mpslimits/att/retinanet}{77.0}
\defR{mix-load70/mpslimits/att/bertlarge}{69.6}
\defR{mix-load70/mpslimits/p99/resnet50}{22.4}
\defR{mix-load70/mpslimits/p99/retinanet}{243}
\defR{mix-load70/mpslimits/p99/bertlarge}{348}
\defR{mix-load70/mpslimits/p99x/resnet50}{1.3}
\defR{mix-load70/mpslimits/p99x/retinanet}{6.9}
\defR{mix-load70/mpslimits/p99x/bertlarge}{5.3}
\defR{mix-load70/mpslimits/thrabs/train_bert}{0.19}
\defR{mix-load70/mpslimits/thr/train_bert}{12}
\defR{mix-load70/mpsparts/attmin}{68.6}
\defR{mix-load70/mpsparts/attmean}{83.8}
\defR{mix-load70/mpsparts/thr}{9}
\defR{mix-load70/mpsparts/n}{2}
\defR{mix-load70/mpsparts/att/resnet50}{93.3}
\defR{mix-load70/mpsparts/att/retinanet}{89.6}
\defR{mix-load70/mpsparts/att/bertlarge}{68.6}
\defR{mix-load70/mpsparts/p99/resnet50}{22.7}
\defR{mix-load70/mpsparts/p99/retinanet}{182}
\defR{mix-load70/mpsparts/p99/bertlarge}{347}
\defR{mix-load70/mpsparts/p99x/resnet50}{1.3}
\defR{mix-load70/mpsparts/p99x/retinanet}{5.2}
\defR{mix-load70/mpsparts/p99x/bertlarge}{5.3}
\defR{mix-load70/mpsparts/thrabs/train_bert}{0.14}
\defR{mix-load70/mpsparts/thr/train_bert}{9}
\defR{mix-load70/lithos/attmin}{70.8}
\defR{mix-load70/lithos/attmean}{81.1}
\defR{mix-load70/lithos/thr}{11}
\defR{mix-load70/lithos/n}{3}
\defR{mix-load70/lithos/att/resnet50}{91.8}
\defR{mix-load70/lithos/att/retinanet}{80.7}
\defR{mix-load70/lithos/att/bertlarge}{70.8}
\defR{mix-load70/lithos/p99/resnet50}{23.6}
\defR{mix-load70/lithos/p99/retinanet}{219}
\defR{mix-load70/lithos/p99/bertlarge}{343}
\defR{mix-load70/lithos/p99x/resnet50}{1.4}
\defR{mix-load70/lithos/p99x/retinanet}{6.2}
\defR{mix-load70/lithos/p99x/bertlarge}{5.2}
\defR{mix-load70/lithos/thrabs/train_bert}{0.18}
\defR{mix-load70/lithos/thr/train_bert}{11}
\defR{mix-load70/dysm/attmin}{0.0}
\defR{mix-load70/dysm/attmean}{19.7}
\defR{mix-load70/dysm/thr}{79}
\defR{mix-load70/dysm/n}{3}
\defR{mix-load70/dysm/att/resnet50}{0.0}
\defR{mix-load70/dysm/att/retinanet}{59.0}
\defR{mix-load70/dysm/att/bertlarge}{0.0}
\defR{mix-load70/dysm/p99/retinanet}{554}
\defR{mix-load70/dysm/p99x/retinanet}{15.8}
\defR{mix-load70/dysm/thrabs/train_bert}{1.27}
\defR{mix-load70/dysm/thr/train_bert}{79}
\defR{mix-load70/xsched/attmin}{64.9}
\defR{mix-load70/xsched/attmean}{84.5}
\defR{mix-load70/xsched/thr}{27}
\defR{mix-load70/xsched/n}{3}
\defR{mix-load70/xsched/att/resnet50}{64.9}
\defR{mix-load70/xsched/att/retinanet}{97.0}
\defR{mix-load70/xsched/att/bertlarge}{91.6}
\defR{mix-load70/xsched/p99/resnet50}{54.0}
\defR{mix-load70/xsched/p99/retinanet}{134}
\defR{mix-load70/xsched/p99/bertlarge}{226}
\defR{mix-load70/xsched/p99x/resnet50}{3.1}
\defR{mix-load70/xsched/p99x/retinanet}{3.8}
\defR{mix-load70/xsched/p99x/bertlarge}{3.4}
\defR{mix-load70/xsched/thrabs/train_bert}{0.43}
\defR{mix-load70/xsched/thr/train_bert}{27}
\defR{mix-load70/multistream/attmin}{0.0}
\defR{mix-load70/multistream/attmean}{11.6}
\defR{mix-load70/multistream/thr}{56}
\defR{mix-load70/multistream/n}{3}
\defR{mix-load70/multistream/att/resnet50}{0.0}
\defR{mix-load70/multistream/att/retinanet}{0.0}
\defR{mix-load70/multistream/att/bertlarge}{34.7}
\defR{mix-load70/multistream/p99/resnet50}{11799}
\defR{mix-load70/multistream/p99/retinanet}{21492}
\defR{mix-load70/multistream/p99/bertlarge}{629}
\defR{mix-load70/multistream/p99x/resnet50}{687.2}
\defR{mix-load70/multistream/p99x/retinanet}{612.8}
\defR{mix-load70/multistream/p99x/bertlarge}{9.6}
\defR{mix-load70/multistream/thrabs/train_bert}{0.90}
\defR{mix-load70/multistream/thr/train_bert}{56}
\defR{mix-load70/tgs/attmin}{81.1}
\defR{mix-load70/tgs/attmean}{90.0}
\defR{mix-load70/tgs/thr}{3}
\defR{mix-load70/tgs/n}{3}
\defR{mix-load70/tgs/att/resnet50}{81.1}
\defR{mix-load70/tgs/att/retinanet}{97.8}
\defR{mix-load70/tgs/att/bertlarge}{91.0}
\defR{mix-load70/tgs/p99/resnet50}{45.5}
\defR{mix-load70/tgs/p99/retinanet}{135}
\defR{mix-load70/tgs/p99/bertlarge}{346}
\defR{mix-load70/tgs/p99x/resnet50}{2.7}
\defR{mix-load70/tgs/p99x/retinanet}{3.8}
\defR{mix-load70/tgs/p99x/bertlarge}{5.3}
\defR{mix-load70/tgs/thrabs/train_bert}{0.06}
\defR{mix-load70/tgs/thr/train_bert}{3}
\defR{mix-load70/mig/attmin}{60.0}
\defR{mix-load70/mig/attmean}{79.3}
\defR{mix-load70/mig/thr}{0}
\defR{mix-load70/mig/n}{3}
\defR{mix-load70/mig/att/resnet50}{94.1}
\defR{mix-load70/mig/att/retinanet}{84.0}
\defR{mix-load70/mig/att/bertlarge}{60.0}
\defR{mix-load70/mig/p99/resnet50}{21.9}
\defR{mix-load70/mig/p99/retinanet}{200}
\defR{mix-load70/mig/p99/bertlarge}{397}
\defR{mix-load70/mig/p99x/resnet50}{1.3}
\defR{mix-load70/mig/p99x/retinanet}{5.7}
\defR{mix-load70/mig/p99x/bertlarge}{6.0}
\defR{mix-load70/mig/thrabs/train_bert}{0.00}
\defR{mix-load70/mig/thr/train_bert}{0}
\defR{mix-load70/tally/attmin}{0.0}
\defR{mix-load70/tally/attmean}{27.1}
\defR{mix-load70/tally/thr}{14}
\defR{mix-load70/tally/n}{3}
\defR{mix-load70/tally/att/resnet50}{0.0}
\defR{mix-load70/tally/att/retinanet}{0.0}
\defR{mix-load70/tally/att/bertlarge}{81.4}
\defR{mix-load70/tally/p99/resnet50}{94130}
\defR{mix-load70/tally/p99/retinanet}{59936}
\defR{mix-load70/tally/p99/bertlarge}{293}
\defR{mix-load70/tally/p99x/resnet50}{5482.4}
\defR{mix-load70/tally/p99x/retinanet}{1708.9}
\defR{mix-load70/tally/p99x/bertlarge}{4.5}
\defR{mix-load70/tally/thrabs/train_bert}{0.23}
\defR{mix-load70/tally/thr/train_bert}{14}
\defR{mix-load70/orion/attmin}{0.0}
\defR{mix-load70/orion/attmean}{30.3}
\defR{mix-load70/orion/thr}{61}
\defR{mix-load70/orion/n}{3}
\defR{mix-load70/orion/att/resnet50}{90.8}
\defR{mix-load70/orion/att/retinanet}{0.0}
\defR{mix-load70/orion/att/bertlarge}{0.2}
\defR{mix-load70/orion/p99/resnet50}{30.7}
\defR{mix-load70/orion/p99/retinanet}{90270}
\defR{mix-load70/orion/p99/bertlarge}{15760}
\defR{mix-load70/orion/p99x/resnet50}{1.8}
\defR{mix-load70/orion/p99x/retinanet}{2573.7}
\defR{mix-load70/orion/p99x/bertlarge}{239.7}
\defR{mix-load70/orion/thrabs/train_bert}{0.99}
\defR{mix-load70/orion/thr/train_bert}{61}
\defR{step-load40/berth-full/attmin}{98.6}
\defR{step-load40/berth-full/attmean}{99.4}
\defR{step-load40/berth-full/thr}{53}
\defR{step-load40/berth-full/n}{3}
\defR{step-load40/berth-full/att/resnet50}{98.6}
\defR{step-load40/berth-full/att/retinanet}{100.0}
\defR{step-load40/berth-full/att/bertlarge}{99.7}
\defR{step-load40/berth-full/p99/resnet50}{15.7}
\defR{step-load40/berth-full/p99/retinanet}{50.1}
\defR{step-load40/berth-full/p99/bertlarge}{104}
\defR{step-load40/berth-full/p99x/resnet50}{1.0}
\defR{step-load40/berth-full/p99x/retinanet}{1.4}
\defR{step-load40/berth-full/p99x/bertlarge}{1.4}
\defR{step-load40/berth-full/thrabs/train_bert}{0.85}
\defR{step-load40/berth-full/thr/train_bert}{53}
\defR{step-load40/berth-nsr/attmin}{98.6}
\defR{step-load40/berth-nsr/attmean}{99.5}
\defR{step-load40/berth-nsr/thr}{53}
\defR{step-load40/berth-nsr/n}{3}
\defR{step-load40/berth-nsr/att/resnet50}{98.6}
\defR{step-load40/berth-nsr/att/retinanet}{100.0}
\defR{step-load40/berth-nsr/att/bertlarge}{99.8}
\defR{step-load40/berth-nsr/p99/resnet50}{15.7}
\defR{step-load40/berth-nsr/p99/retinanet}{46.2}
\defR{step-load40/berth-nsr/p99/bertlarge}{105}
\defR{step-load40/berth-nsr/p99x/resnet50}{1.0}
\defR{step-load40/berth-nsr/p99x/retinanet}{1.3}
\defR{step-load40/berth-nsr/p99x/bertlarge}{1.5}
\defR{step-load40/berth-nsr/thrabs/train_bert}{0.85}
\defR{step-load40/berth-nsr/thr/train_bert}{53}
\defR{step-load40/berth-nba/attmin}{98.5}
\defR{step-load40/berth-nba/attmean}{99.4}
\defR{step-load40/berth-nba/thr}{53}
\defR{step-load40/berth-nba/n}{3}
\defR{step-load40/berth-nba/att/resnet50}{98.5}
\defR{step-load40/berth-nba/att/retinanet}{100.0}
\defR{step-load40/berth-nba/att/bertlarge}{99.7}
\defR{step-load40/berth-nba/p99/resnet50}{15.9}
\defR{step-load40/berth-nba/p99/retinanet}{47.6}
\defR{step-load40/berth-nba/p99/bertlarge}{105}
\defR{step-load40/berth-nba/p99x/resnet50}{1.0}
\defR{step-load40/berth-nba/p99x/retinanet}{1.4}
\defR{step-load40/berth-nba/p99x/bertlarge}{1.5}
\defR{step-load40/berth-nba/thrabs/train_bert}{0.86}
\defR{step-load40/berth-nba/thr/train_bert}{53}
\defR{step-load40/berth-nsb/attmin}{98.7}
\defR{step-load40/berth-nsb/attmean}{99.5}
\defR{step-load40/berth-nsb/thr}{53}
\defR{step-load40/berth-nsb/n}{3}
\defR{step-load40/berth-nsb/att/resnet50}{98.7}
\defR{step-load40/berth-nsb/att/retinanet}{100.0}
\defR{step-load40/berth-nsb/att/bertlarge}{99.7}
\defR{step-load40/berth-nsb/p99/resnet50}{15.5}
\defR{step-load40/berth-nsb/p99/retinanet}{46.3}
\defR{step-load40/berth-nsb/p99/bertlarge}{107}
\defR{step-load40/berth-nsb/p99x/resnet50}{1.0}
\defR{step-load40/berth-nsb/p99x/retinanet}{1.3}
\defR{step-load40/berth-nsb/p99x/bertlarge}{1.5}
\defR{step-load40/berth-nsb/thrabs/train_bert}{0.86}
\defR{step-load40/berth-nsb/thr/train_bert}{53}
\defR{step-load40/berth-ng/attmin}{98.1}
\defR{step-load40/berth-ng/attmean}{99.1}
\defR{step-load40/berth-ng/thr}{53}
\defR{step-load40/berth-ng/n}{3}
\defR{step-load40/berth-ng/att/resnet50}{98.1}
\defR{step-load40/berth-ng/att/retinanet}{100.0}
\defR{step-load40/berth-ng/att/bertlarge}{99.3}
\defR{step-load40/berth-ng/p99/resnet50}{16.5}
\defR{step-load40/berth-ng/p99/retinanet}{48.6}
\defR{step-load40/berth-ng/p99/bertlarge}{123}
\defR{step-load40/berth-ng/p99x/resnet50}{1.0}
\defR{step-load40/berth-ng/p99x/retinanet}{1.4}
\defR{step-load40/berth-ng/p99x/bertlarge}{1.7}
\defR{step-load40/berth-ng/thrabs/train_bert}{0.86}
\defR{step-load40/berth-ng/thr/train_bert}{53}
\defR{step-load40/berth-nsng/attmin}{98.2}
\defR{step-load40/berth-nsng/attmean}{99.2}
\defR{step-load40/berth-nsng/thr}{53}
\defR{step-load40/berth-nsng/n}{3}
\defR{step-load40/berth-nsng/att/resnet50}{98.2}
\defR{step-load40/berth-nsng/att/retinanet}{100.0}
\defR{step-load40/berth-nsng/att/bertlarge}{99.3}
\defR{step-load40/berth-nsng/p99/resnet50}{16.4}
\defR{step-load40/berth-nsng/p99/retinanet}{48.9}
\defR{step-load40/berth-nsng/p99/bertlarge}{117}
\defR{step-load40/berth-nsng/p99x/resnet50}{1.0}
\defR{step-load40/berth-nsng/p99x/retinanet}{1.4}
\defR{step-load40/berth-nsng/p99x/bertlarge}{1.6}
\defR{step-load40/berth-nsng/thrabs/train_bert}{0.86}
\defR{step-load40/berth-nsng/thr/train_bert}{53}
\defR{step-load40/berth-grouped/attmin}{92.1}
\defR{step-load40/berth-grouped/attmean}{97.4}
\defR{step-load40/berth-grouped/thr}{58}
\defR{step-load40/berth-grouped/n}{3}
\defR{step-load40/berth-grouped/att/resnet50}{92.1}
\defR{step-load40/berth-grouped/att/retinanet}{100.0}
\defR{step-load40/berth-grouped/att/bertlarge}{100.0}
\defR{step-load40/berth-grouped/p99/resnet50}{28.2}
\defR{step-load40/berth-grouped/p99/retinanet}{44.1}
\defR{step-load40/berth-grouped/p99/bertlarge}{71.5}
\defR{step-load40/berth-grouped/p99x/resnet50}{1.7}
\defR{step-load40/berth-grouped/p99x/retinanet}{1.3}
\defR{step-load40/berth-grouped/p99x/bertlarge}{1.0}
\defR{step-load40/berth-grouped/thrabs/train_bert}{0.93}
\defR{step-load40/berth-grouped/thr/train_bert}{58}
\defR{step-load40/berth/attmin}{90.4}
\defR{step-load40/berth/attmean}{96.7}
\defR{step-load40/berth/thr}{54}
\defR{step-load40/berth/n}{3}
\defR{step-load40/berth/att/resnet50}{90.4}
\defR{step-load40/berth/att/retinanet}{100.0}
\defR{step-load40/berth/att/bertlarge}{99.6}
\defR{step-load40/berth/p99/resnet50}{29.8}
\defR{step-load40/berth/p99/retinanet}{55.5}
\defR{step-load40/berth/p99/bertlarge}{112}
\defR{step-load40/berth/p99x/resnet50}{1.8}
\defR{step-load40/berth/p99x/retinanet}{1.6}
\defR{step-load40/berth/p99x/bertlarge}{1.5}
\defR{step-load40/berth/thrabs/train_bert}{0.88}
\defR{step-load40/berth/thr/train_bert}{54}
\defR{step-load40/timeslice/attmin}{47.8}
\defR{step-load40/timeslice/attmean}{69.9}
\defR{step-load40/timeslice/thr}{50}
\defR{step-load40/timeslice/n}{3}
\defR{step-load40/timeslice/att/resnet50}{47.8}
\defR{step-load40/timeslice/att/retinanet}{94.2}
\defR{step-load40/timeslice/att/bertlarge}{67.6}
\defR{step-load40/timeslice/p99/resnet50}{104}
\defR{step-load40/timeslice/p99/retinanet}{151}
\defR{step-load40/timeslice/p99/bertlarge}{357}
\defR{step-load40/timeslice/p99x/resnet50}{6.3}
\defR{step-load40/timeslice/p99x/retinanet}{4.3}
\defR{step-load40/timeslice/p99x/bertlarge}{4.9}
\defR{step-load40/timeslice/thrabs/train_bert}{0.81}
\defR{step-load40/timeslice/thr/train_bert}{50}
\defR{step-load40/mps/attmin}{0.0}
\defR{step-load40/mps/attmean}{0.0}
\defR{step-load40/mps/thr}{80}
\defR{step-load40/mps/n}{3}
\defR{step-load40/mps/att/resnet50}{0.0}
\defR{step-load40/mps/att/retinanet}{0.0}
\defR{step-load40/mps/att/bertlarge}{0.1}
\defR{step-load40/mps/p99/resnet50}{46164}
\defR{step-load40/mps/p99/retinanet}{42713}
\defR{step-load40/mps/p99/bertlarge}{6776}
\defR{step-load40/mps/p99x/resnet50}{2826.5}
\defR{step-load40/mps/p99x/retinanet}{1228.2}
\defR{step-load40/mps/p99x/bertlarge}{93.7}
\defR{step-load40/mps/thrabs/train_bert}{1.30}
\defR{step-load40/mps/thr/train_bert}{80}
\defR{step-load40/mpsprio/attmin}{0.1}
\defR{step-load40/mpsprio/attmean}{42.1}
\defR{step-load40/mpsprio/thr}{68}
\defR{step-load40/mpsprio/n}{3}
\defR{step-load40/mpsprio/att/resnet50}{0.1}
\defR{step-load40/mpsprio/att/retinanet}{49.3}
\defR{step-load40/mpsprio/att/bertlarge}{77.0}
\defR{step-load40/mpsprio/p99/resnet50}{5277}
\defR{step-load40/mpsprio/p99/retinanet}{403}
\defR{step-load40/mpsprio/p99/bertlarge}{261}
\defR{step-load40/mpsprio/p99x/resnet50}{323.1}
\defR{step-load40/mpsprio/p99x/retinanet}{11.6}
\defR{step-load40/mpsprio/p99x/bertlarge}{3.6}
\defR{step-load40/mpsprio/thrabs/train_bert}{1.10}
\defR{step-load40/mpsprio/thr/train_bert}{68}
\defR{step-load40/mpslimits/attmin}{88.4}
\defR{step-load40/mpslimits/attmean}{94.0}
\defR{step-load40/mpslimits/thr}{11}
\defR{step-load40/mpslimits/n}{3}
\defR{step-load40/mpslimits/att/resnet50}{97.1}
\defR{step-load40/mpslimits/att/retinanet}{96.3}
\defR{step-load40/mpslimits/att/bertlarge}{88.4}
\defR{step-load40/mpslimits/p99/resnet50}{18.3}
\defR{step-load40/mpslimits/p99/retinanet}{135}
\defR{step-load40/mpslimits/p99/bertlarge}{218}
\defR{step-load40/mpslimits/p99x/resnet50}{1.1}
\defR{step-load40/mpslimits/p99x/retinanet}{3.9}
\defR{step-load40/mpslimits/p99x/bertlarge}{3.0}
\defR{step-load40/mpslimits/thrabs/train_bert}{0.19}
\defR{step-load40/mpslimits/thr/train_bert}{11}
\defR{step-load40/mpsparts/attmin}{87.6}
\defR{step-load40/mpsparts/attmean}{92.4}
\defR{step-load40/mpsparts/thr}{18}
\defR{step-load40/mpsparts/n}{2}
\defR{step-load40/mpsparts/att/resnet50}{96.8}
\defR{step-load40/mpsparts/att/retinanet}{92.9}
\defR{step-load40/mpsparts/att/bertlarge}{87.6}
\defR{step-load40/mpsparts/p99/resnet50}{19.0}
\defR{step-load40/mpsparts/p99/retinanet}{157}
\defR{step-load40/mpsparts/p99/bertlarge}{219}
\defR{step-load40/mpsparts/p99x/resnet50}{1.2}
\defR{step-load40/mpsparts/p99x/retinanet}{4.5}
\defR{step-load40/mpsparts/p99x/bertlarge}{3.0}
\defR{step-load40/mpsparts/thrabs/train_bert}{0.30}
\defR{step-load40/mpsparts/thr/train_bert}{18}
\defR{step-load40/lithos/attmin}{86.4}
\defR{step-load40/lithos/attmean}{92.9}
\defR{step-load40/lithos/thr}{14}
\defR{step-load40/lithos/n}{3}
\defR{step-load40/lithos/att/resnet50}{96.9}
\defR{step-load40/lithos/att/retinanet}{95.5}
\defR{step-load40/lithos/att/bertlarge}{86.4}
\defR{step-load40/lithos/p99/resnet50}{18.7}
\defR{step-load40/lithos/p99/retinanet}{143}
\defR{step-load40/lithos/p99/bertlarge}{230}
\defR{step-load40/lithos/p99x/resnet50}{1.1}
\defR{step-load40/lithos/p99x/retinanet}{4.1}
\defR{step-load40/lithos/p99x/bertlarge}{3.2}
\defR{step-load40/lithos/thrabs/train_bert}{0.22}
\defR{step-load40/lithos/thr/train_bert}{14}
\defR{step-load40/dysm/attmin}{0.0}
\defR{step-load40/dysm/attmean}{24.1}
\defR{step-load40/dysm/thr}{85}
\defR{step-load40/dysm/n}{3}
\defR{step-load40/dysm/att/resnet50}{0.0}
\defR{step-load40/dysm/att/retinanet}{72.4}
\defR{step-load40/dysm/att/bertlarge}{0.0}
\defR{step-load40/dysm/p99/retinanet}{326}
\defR{step-load40/dysm/p99x/retinanet}{9.4}
\defR{step-load40/dysm/thrabs/train_bert}{1.38}
\defR{step-load40/dysm/thr/train_bert}{85}
\defR{step-load40/xsched/attmin}{81.5}
\defR{step-load40/xsched/attmean}{93.5}
\defR{step-load40/xsched/thr}{53}
\defR{step-load40/xsched/n}{3}
\defR{step-load40/xsched/att/resnet50}{81.5}
\defR{step-load40/xsched/att/retinanet}{100.0}
\defR{step-load40/xsched/att/bertlarge}{99.2}
\defR{step-load40/xsched/p99/resnet50}{38.6}
\defR{step-load40/xsched/p99/retinanet}{66.0}
\defR{step-load40/xsched/p99/bertlarge}{122}
\defR{step-load40/xsched/p99x/resnet50}{2.4}
\defR{step-load40/xsched/p99x/retinanet}{1.9}
\defR{step-load40/xsched/p99x/bertlarge}{1.7}
\defR{step-load40/xsched/thrabs/train_bert}{0.85}
\defR{step-load40/xsched/thr/train_bert}{53}
\defR{step-load40/multistream/attmin}{0.0}
\defR{step-load40/multistream/attmean}{42.7}
\defR{step-load40/multistream/thr}{68}
\defR{step-load40/multistream/n}{3}
\defR{step-load40/multistream/att/resnet50}{0.0}
\defR{step-load40/multistream/att/retinanet}{50.6}
\defR{step-load40/multistream/att/bertlarge}{77.4}
\defR{step-load40/multistream/p99/resnet50}{6165}
\defR{step-load40/multistream/p99/retinanet}{402}
\defR{step-load40/multistream/p99/bertlarge}{267}
\defR{step-load40/multistream/p99x/resnet50}{377.4}
\defR{step-load40/multistream/p99x/retinanet}{11.6}
\defR{step-load40/multistream/p99x/bertlarge}{3.7}
\defR{step-load40/multistream/thrabs/train_bert}{1.11}
\defR{step-load40/multistream/thr/train_bert}{68}
\defR{step-load40/tgs/attmin}{93.5}
\defR{step-load40/tgs/attmean}{97.8}
\defR{step-load40/tgs/thr}{2}
\defR{step-load40/tgs/n}{3}
\defR{step-load40/tgs/att/resnet50}{93.5}
\defR{step-load40/tgs/att/retinanet}{100.0}
\defR{step-load40/tgs/att/bertlarge}{99.9}
\defR{step-load40/tgs/p99/resnet50}{25.7}
\defR{step-load40/tgs/p99/retinanet}{46.6}
\defR{step-load40/tgs/p99/bertlarge}{90.6}
\defR{step-load40/tgs/p99x/resnet50}{1.6}
\defR{step-load40/tgs/p99x/retinanet}{1.3}
\defR{step-load40/tgs/p99x/bertlarge}{1.3}
\defR{step-load40/tgs/thrabs/train_bert}{0.03}
\defR{step-load40/tgs/thr/train_bert}{2}
\defR{step-load40/mig/attmin}{90.6}
\defR{step-load40/mig/attmean}{94.8}
\defR{step-load40/mig/thr}{0}
\defR{step-load40/mig/n}{3}
\defR{step-load40/mig/att/resnet50}{95.5}
\defR{step-load40/mig/att/retinanet}{98.4}
\defR{step-load40/mig/att/bertlarge}{90.6}
\defR{step-load40/mig/p99/resnet50}{20.3}
\defR{step-load40/mig/p99/retinanet}{109}
\defR{step-load40/mig/p99/bertlarge}{207}
\defR{step-load40/mig/p99x/resnet50}{1.2}
\defR{step-load40/mig/p99x/retinanet}{3.1}
\defR{step-load40/mig/p99x/bertlarge}{2.9}
\defR{step-load40/mig/thrabs/train_bert}{0.00}
\defR{step-load40/mig/thr/train_bert}{0}
\defR{step-load40/tally/attmin}{0.0}
\defR{step-load40/tally/attmean}{33.9}
\defR{step-load40/tally/thr}{26}
\defR{step-load40/tally/n}{3}
\defR{step-load40/tally/att/resnet50}{0.0}
\defR{step-load40/tally/att/retinanet}{5.5}
\defR{step-load40/tally/att/bertlarge}{96.1}
\defR{step-load40/tally/p99/resnet50}{93897}
\defR{step-load40/tally/p99/retinanet}{2610}
\defR{step-load40/tally/p99/bertlarge}{156}
\defR{step-load40/tally/p99x/resnet50}{5749.0}
\defR{step-load40/tally/p99x/retinanet}{75.0}
\defR{step-load40/tally/p99x/bertlarge}{2.2}
\defR{step-load40/tally/thrabs/train_bert}{0.42}
\defR{step-load40/tally/thr/train_bert}{26}
\defR{step-load40/orion/attmin}{0.0}
\defR{step-load40/orion/attmean}{37.0}
\defR{step-load40/orion/thr}{58}
\defR{step-load40/orion/n}{3}
\defR{step-load40/orion/att/resnet50}{92.2}
\defR{step-load40/orion/att/retinanet}{0.0}
\defR{step-load40/orion/att/bertlarge}{18.8}
\defR{step-load40/orion/p99/resnet50}{29.4}
\defR{step-load40/orion/p99/retinanet}{50953}
\defR{step-load40/orion/p99/bertlarge}{1187}
\defR{step-load40/orion/p99x/resnet50}{1.8}
\defR{step-load40/orion/p99x/retinanet}{1465.1}
\defR{step-load40/orion/p99x/bertlarge}{16.4}
\defR{step-load40/orion/thrabs/train_bert}{0.93}
\defR{step-load40/orion/thr/train_bert}{58}
\defR{tight-x0.50/berth-full/attmin}{77.5}
\defR{tight-x0.50/berth-full/attmean}{90.9}
\defR{tight-x0.50/berth-full/thr}{57}
\defR{tight-x0.50/berth-full/n}{3}
\defR{tight-x0.50/berth-full/att/resnet50}{77.5}
\defR{tight-x0.50/berth-full/att/retinanet}{99.8}
\defR{tight-x0.50/berth-full/att/bertlarge}{95.4}
\defR{tight-x0.50/berth-full/p99/resnet50}{14.3}
\defR{tight-x0.50/berth-full/p99/retinanet}{40.1}
\defR{tight-x0.50/berth-full/p99/bertlarge}{81.7}
\defR{tight-x0.50/berth-full/p99x/resnet50}{1.1}
\defR{tight-x0.50/berth-full/p99x/retinanet}{1.2}
\defR{tight-x0.50/berth-full/p99x/bertlarge}{1.1}
\defR{tight-x0.50/berth-full/thrabs/train_bert}{0.92}
\defR{tight-x0.50/berth-full/thr/train_bert}{57}
\defR{tight-x0.50/berth-nsr/attmin}{77.0}
\defR{tight-x0.50/berth-nsr/attmean}{90.5}
\defR{tight-x0.50/berth-nsr/thr}{56}
\defR{tight-x0.50/berth-nsr/n}{3}
\defR{tight-x0.50/berth-nsr/att/resnet50}{77.0}
\defR{tight-x0.50/berth-nsr/att/retinanet}{99.8}
\defR{tight-x0.50/berth-nsr/att/bertlarge}{94.6}
\defR{tight-x0.50/berth-nsr/p99/resnet50}{14.6}
\defR{tight-x0.50/berth-nsr/p99/retinanet}{41.0}
\defR{tight-x0.50/berth-nsr/p99/bertlarge}{87.4}
\defR{tight-x0.50/berth-nsr/p99x/resnet50}{1.1}
\defR{tight-x0.50/berth-nsr/p99x/retinanet}{1.2}
\defR{tight-x0.50/berth-nsr/p99x/bertlarge}{1.2}
\defR{tight-x0.50/berth-nsr/thrabs/train_bert}{0.91}
\defR{tight-x0.50/berth-nsr/thr/train_bert}{56}
\defR{tight-x0.50/berth-nba/attmin}{77.6}
\defR{tight-x0.50/berth-nba/attmean}{90.8}
\defR{tight-x0.50/berth-nba/thr}{57}
\defR{tight-x0.50/berth-nba/n}{3}
\defR{tight-x0.50/berth-nba/att/resnet50}{77.6}
\defR{tight-x0.50/berth-nba/att/retinanet}{99.8}
\defR{tight-x0.50/berth-nba/att/bertlarge}{95.1}
\defR{tight-x0.50/berth-nba/p99/resnet50}{14.1}
\defR{tight-x0.50/berth-nba/p99/retinanet}{42.1}
\defR{tight-x0.50/berth-nba/p99/bertlarge}{85.1}
\defR{tight-x0.50/berth-nba/p99x/resnet50}{1.1}
\defR{tight-x0.50/berth-nba/p99x/retinanet}{1.2}
\defR{tight-x0.50/berth-nba/p99x/bertlarge}{1.2}
\defR{tight-x0.50/berth-nba/thrabs/train_bert}{0.92}
\defR{tight-x0.50/berth-nba/thr/train_bert}{57}
\defR{tight-x0.50/berth-nsb/attmin}{76.9}
\defR{tight-x0.50/berth-nsb/attmean}{90.3}
\defR{tight-x0.50/berth-nsb/thr}{57}
\defR{tight-x0.50/berth-nsb/n}{3}
\defR{tight-x0.50/berth-nsb/att/resnet50}{76.9}
\defR{tight-x0.50/berth-nsb/att/retinanet}{99.5}
\defR{tight-x0.50/berth-nsb/att/bertlarge}{94.6}
\defR{tight-x0.50/berth-nsb/p99/resnet50}{14.4}
\defR{tight-x0.50/berth-nsb/p99/retinanet}{43.5}
\defR{tight-x0.50/berth-nsb/p99/bertlarge}{81.0}
\defR{tight-x0.50/berth-nsb/p99x/resnet50}{1.1}
\defR{tight-x0.50/berth-nsb/p99x/retinanet}{1.2}
\defR{tight-x0.50/berth-nsb/p99x/bertlarge}{1.1}
\defR{tight-x0.50/berth-nsb/thrabs/train_bert}{0.92}
\defR{tight-x0.50/berth-nsb/thr/train_bert}{57}
\defR{tight-x0.50/berth-ng/attmin}{74.4}
\defR{tight-x0.50/berth-ng/attmean}{88.9}
\defR{tight-x0.50/berth-ng/thr}{57}
\defR{tight-x0.50/berth-ng/n}{3}
\defR{tight-x0.50/berth-ng/att/resnet50}{74.4}
\defR{tight-x0.50/berth-ng/att/retinanet}{99.8}
\defR{tight-x0.50/berth-ng/att/bertlarge}{92.6}
\defR{tight-x0.50/berth-ng/p99/resnet50}{14.9}
\defR{tight-x0.50/berth-ng/p99/retinanet}{43.0}
\defR{tight-x0.50/berth-ng/p99/bertlarge}{93.0}
\defR{tight-x0.50/berth-ng/p99x/resnet50}{1.1}
\defR{tight-x0.50/berth-ng/p99x/retinanet}{1.2}
\defR{tight-x0.50/berth-ng/p99x/bertlarge}{1.3}
\defR{tight-x0.50/berth-ng/thrabs/train_bert}{0.93}
\defR{tight-x0.50/berth-ng/thr/train_bert}{57}
\defR{tight-x0.50/berth-nsng/attmin}{74.8}
\defR{tight-x0.50/berth-nsng/attmean}{89.0}
\defR{tight-x0.50/berth-nsng/thr}{57}
\defR{tight-x0.50/berth-nsng/n}{3}
\defR{tight-x0.50/berth-nsng/att/resnet50}{74.8}
\defR{tight-x0.50/berth-nsng/att/retinanet}{99.5}
\defR{tight-x0.50/berth-nsng/att/bertlarge}{92.8}
\defR{tight-x0.50/berth-nsng/p99/resnet50}{15.0}
\defR{tight-x0.50/berth-nsng/p99/retinanet}{42.3}
\defR{tight-x0.50/berth-nsng/p99/bertlarge}{92.5}
\defR{tight-x0.50/berth-nsng/p99x/resnet50}{1.1}
\defR{tight-x0.50/berth-nsng/p99x/retinanet}{1.2}
\defR{tight-x0.50/berth-nsng/p99x/bertlarge}{1.3}
\defR{tight-x0.50/berth-nsng/thrabs/train_bert}{0.93}
\defR{tight-x0.50/berth-nsng/thr/train_bert}{57}
\defR{tight-x0.50/berth-grouped/attmin}{59.0}
\defR{tight-x0.50/berth-grouped/attmean}{85.5}
\defR{tight-x0.50/berth-grouped/thr}{61}
\defR{tight-x0.50/berth-grouped/n}{3}
\defR{tight-x0.50/berth-grouped/att/resnet50}{59.0}
\defR{tight-x0.50/berth-grouped/att/retinanet}{99.6}
\defR{tight-x0.50/berth-grouped/att/bertlarge}{97.8}
\defR{tight-x0.50/berth-grouped/p99/resnet50}{22.4}
\defR{tight-x0.50/berth-grouped/p99/retinanet}{44.2}
\defR{tight-x0.50/berth-grouped/p99/bertlarge}{69.8}
\defR{tight-x0.50/berth-grouped/p99x/resnet50}{1.7}
\defR{tight-x0.50/berth-grouped/p99x/retinanet}{1.3}
\defR{tight-x0.50/berth-grouped/p99x/bertlarge}{1.0}
\defR{tight-x0.50/berth-grouped/thrabs/train_bert}{0.99}
\defR{tight-x0.50/berth-grouped/thr/train_bert}{61}
\defR{tight-x0.50/berth/attmin}{52.9}
\defR{tight-x0.50/berth/attmean}{81.5}
\defR{tight-x0.50/berth/thr}{58}
\defR{tight-x0.50/berth/n}{3}
\defR{tight-x0.50/berth/att/resnet50}{52.9}
\defR{tight-x0.50/berth/att/retinanet}{98.7}
\defR{tight-x0.50/berth/att/bertlarge}{92.8}
\defR{tight-x0.50/berth/p99/resnet50}{23.0}
\defR{tight-x0.50/berth/p99/retinanet}{52.0}
\defR{tight-x0.50/berth/p99/bertlarge}{98.7}
\defR{tight-x0.50/berth/p99x/resnet50}{1.7}
\defR{tight-x0.50/berth/p99x/retinanet}{1.5}
\defR{tight-x0.50/berth/p99x/bertlarge}{1.4}
\defR{tight-x0.50/berth/thrabs/train_bert}{0.94}
\defR{tight-x0.50/berth/thr/train_bert}{58}
\defR{tight-x0.50/timeslice/attmin}{1.4}
\defR{tight-x0.50/timeslice/attmean}{29.2}
\defR{tight-x0.50/timeslice/thr}{55}
\defR{tight-x0.50/timeslice/n}{3}
\defR{tight-x0.50/timeslice/att/resnet50}{1.4}
\defR{tight-x0.50/timeslice/att/retinanet}{72.4}
\defR{tight-x0.50/timeslice/att/bertlarge}{13.6}
\defR{tight-x0.50/timeslice/p99/resnet50}{41.7}
\defR{tight-x0.50/timeslice/p99/retinanet}{130}
\defR{tight-x0.50/timeslice/p99/bertlarge}{295}
\defR{tight-x0.50/timeslice/p99x/resnet50}{3.1}
\defR{tight-x0.50/timeslice/p99x/retinanet}{3.7}
\defR{tight-x0.50/timeslice/p99x/bertlarge}{4.1}
\defR{tight-x0.50/timeslice/thrabs/train_bert}{0.89}
\defR{tight-x0.50/timeslice/thr/train_bert}{55}
\defR{tight-x0.50/mps/attmin}{0.0}
\defR{tight-x0.50/mps/attmean}{0.0}
\defR{tight-x0.50/mps/thr}{80}
\defR{tight-x0.50/mps/n}{3}
\defR{tight-x0.50/mps/att/resnet50}{0.0}
\defR{tight-x0.50/mps/att/retinanet}{0.0}
\defR{tight-x0.50/mps/att/bertlarge}{0.0}
\defR{tight-x0.50/mps/p99/resnet50}{31525}
\defR{tight-x0.50/mps/p99/retinanet}{42472}
\defR{tight-x0.50/mps/p99/bertlarge}{6773}
\defR{tight-x0.50/mps/p99x/resnet50}{2353.1}
\defR{tight-x0.50/mps/p99x/retinanet}{1221.3}
\defR{tight-x0.50/mps/p99x/bertlarge}{93.6}
\defR{tight-x0.50/mps/thrabs/train_bert}{1.30}
\defR{tight-x0.50/mps/thr/train_bert}{80}
\defR{tight-x0.50/mpsprio/attmin}{0.0}
\defR{tight-x0.50/mpsprio/attmean}{2.5}
\defR{tight-x0.50/mpsprio/thr}{70}
\defR{tight-x0.50/mpsprio/n}{3}
\defR{tight-x0.50/mpsprio/att/resnet50}{0.0}
\defR{tight-x0.50/mpsprio/att/retinanet}{5.4}
\defR{tight-x0.50/mpsprio/att/bertlarge}{2.1}
\defR{tight-x0.50/mpsprio/p99/resnet50}{141}
\defR{tight-x0.50/mpsprio/p99/retinanet}{420}
\defR{tight-x0.50/mpsprio/p99/bertlarge}{262}
\defR{tight-x0.50/mpsprio/p99x/resnet50}{10.5}
\defR{tight-x0.50/mpsprio/p99x/retinanet}{12.1}
\defR{tight-x0.50/mpsprio/p99x/bertlarge}{3.6}
\defR{tight-x0.50/mpsprio/thrabs/train_bert}{1.13}
\defR{tight-x0.50/mpsprio/thr/train_bert}{70}
\defR{tight-x0.50/mpslimits/attmin}{22.2}
\defR{tight-x0.50/mpslimits/attmean}{59.2}
\defR{tight-x0.50/mpslimits/thr}{12}
\defR{tight-x0.50/mpslimits/n}{3}
\defR{tight-x0.50/mpslimits/att/resnet50}{22.2}
\defR{tight-x0.50/mpslimits/att/retinanet}{86.6}
\defR{tight-x0.50/mpslimits/att/bertlarge}{69.0}
\defR{tight-x0.50/mpslimits/p99/resnet50}{18.9}
\defR{tight-x0.50/mpslimits/p99/retinanet}{89.4}
\defR{tight-x0.50/mpslimits/p99/bertlarge}{170}
\defR{tight-x0.50/mpslimits/p99x/resnet50}{1.4}
\defR{tight-x0.50/mpslimits/p99x/retinanet}{2.6}
\defR{tight-x0.50/mpslimits/p99x/bertlarge}{2.3}
\defR{tight-x0.50/mpslimits/thrabs/train_bert}{0.19}
\defR{tight-x0.50/mpslimits/thr/train_bert}{12}
\defR{tight-x0.50/mpsparts/attmin}{20.9}
\defR{tight-x0.50/mpsparts/attmean}{59.7}
\defR{tight-x0.50/mpsparts/thr}{14}
\defR{tight-x0.50/mpsparts/n}{2}
\defR{tight-x0.50/mpsparts/att/resnet50}{20.9}
\defR{tight-x0.50/mpsparts/att/retinanet}{86.5}
\defR{tight-x0.50/mpsparts/att/bertlarge}{71.9}
\defR{tight-x0.50/mpsparts/p99/resnet50}{18.5}
\defR{tight-x0.50/mpsparts/p99/retinanet}{91.6}
\defR{tight-x0.50/mpsparts/p99/bertlarge}{161}
\defR{tight-x0.50/mpsparts/p99x/resnet50}{1.4}
\defR{tight-x0.50/mpsparts/p99x/retinanet}{2.6}
\defR{tight-x0.50/mpsparts/p99x/bertlarge}{2.2}
\defR{tight-x0.50/mpsparts/thrabs/train_bert}{0.22}
\defR{tight-x0.50/mpsparts/thr/train_bert}{14}
\defR{tight-x0.50/lithos/attmin}{23.6}
\defR{tight-x0.50/lithos/attmean}{60.3}
\defR{tight-x0.50/lithos/thr}{11}
\defR{tight-x0.50/lithos/n}{3}
\defR{tight-x0.50/lithos/att/resnet50}{23.6}
\defR{tight-x0.50/lithos/att/retinanet}{87.4}
\defR{tight-x0.50/lithos/att/bertlarge}{69.8}
\defR{tight-x0.50/lithos/p99/resnet50}{19.9}
\defR{tight-x0.50/lithos/p99/retinanet}{87.0}
\defR{tight-x0.50/lithos/p99/bertlarge}{164}
\defR{tight-x0.50/lithos/p99x/resnet50}{1.5}
\defR{tight-x0.50/lithos/p99x/retinanet}{2.5}
\defR{tight-x0.50/lithos/p99x/bertlarge}{2.3}
\defR{tight-x0.50/lithos/thrabs/train_bert}{0.19}
\defR{tight-x0.50/lithos/thr/train_bert}{11}
\defR{tight-x0.50/dysm/attmin}{0.0}
\defR{tight-x0.50/dysm/attmean}{28.2}
\defR{tight-x0.50/dysm/thr}{77}
\defR{tight-x0.50/dysm/n}{3}
\defR{tight-x0.50/dysm/att/resnet50}{0.0}
\defR{tight-x0.50/dysm/att/retinanet}{84.6}
\defR{tight-x0.50/dysm/att/bertlarge}{0.0}
\defR{tight-x0.50/dysm/p99/retinanet}{101}
\defR{tight-x0.50/dysm/p99x/retinanet}{2.9}
\defR{tight-x0.50/dysm/thrabs/train_bert}{1.25}
\defR{tight-x0.50/dysm/thr/train_bert}{77}
\defR{tight-x0.50/xsched/attmin}{39.2}
\defR{tight-x0.50/xsched/attmean}{75.9}
\defR{tight-x0.50/xsched/thr}{57}
\defR{tight-x0.50/xsched/n}{3}
\defR{tight-x0.50/xsched/att/resnet50}{39.2}
\defR{tight-x0.50/xsched/att/retinanet}{97.7}
\defR{tight-x0.50/xsched/att/bertlarge}{90.7}
\defR{tight-x0.50/xsched/p99/resnet50}{29.1}
\defR{tight-x0.50/xsched/p99/retinanet}{59.7}
\defR{tight-x0.50/xsched/p99/bertlarge}{100}
\defR{tight-x0.50/xsched/p99x/resnet50}{2.2}
\defR{tight-x0.50/xsched/p99x/retinanet}{1.7}
\defR{tight-x0.50/xsched/p99x/bertlarge}{1.4}
\defR{tight-x0.50/xsched/thrabs/train_bert}{0.92}
\defR{tight-x0.50/xsched/thr/train_bert}{57}
\defR{tight-x0.50/multistream/attmin}{0.0}
\defR{tight-x0.50/multistream/attmean}{4.4}
\defR{tight-x0.50/multistream/thr}{71}
\defR{tight-x0.50/multistream/n}{3}
\defR{tight-x0.50/multistream/att/resnet50}{0.0}
\defR{tight-x0.50/multistream/att/retinanet}{11.3}
\defR{tight-x0.50/multistream/att/bertlarge}{1.8}
\defR{tight-x0.50/multistream/p99/resnet50}{154}
\defR{tight-x0.50/multistream/p99/retinanet}{360}
\defR{tight-x0.50/multistream/p99/bertlarge}{258}
\defR{tight-x0.50/multistream/p99x/resnet50}{11.5}
\defR{tight-x0.50/multistream/p99x/retinanet}{10.3}
\defR{tight-x0.50/multistream/p99x/bertlarge}{3.6}
\defR{tight-x0.50/multistream/thrabs/train_bert}{1.15}
\defR{tight-x0.50/multistream/thr/train_bert}{71}
\defR{tight-x0.50/tgs/attmin}{65.9}
\defR{tight-x0.50/tgs/attmean}{87.5}
\defR{tight-x0.50/tgs/thr}{2}
\defR{tight-x0.50/tgs/n}{3}
\defR{tight-x0.50/tgs/att/resnet50}{65.9}
\defR{tight-x0.50/tgs/att/retinanet}{99.8}
\defR{tight-x0.50/tgs/att/bertlarge}{96.7}
\defR{tight-x0.50/tgs/p99/resnet50}{19.8}
\defR{tight-x0.50/tgs/p99/retinanet}{40.7}
\defR{tight-x0.50/tgs/p99/bertlarge}{81.6}
\defR{tight-x0.50/tgs/p99x/resnet50}{1.5}
\defR{tight-x0.50/tgs/p99x/retinanet}{1.2}
\defR{tight-x0.50/tgs/p99x/bertlarge}{1.1}
\defR{tight-x0.50/tgs/thrabs/train_bert}{0.03}
\defR{tight-x0.50/tgs/thr/train_bert}{2}
\defR{tight-x0.50/mig/attmin}{60.7}
\defR{tight-x0.50/mig/attmean}{74.1}
\defR{tight-x0.50/mig/thr}{0}
\defR{tight-x0.50/mig/n}{3}
\defR{tight-x0.50/mig/att/resnet50}{82.8}
\defR{tight-x0.50/mig/att/retinanet}{78.8}
\defR{tight-x0.50/mig/att/bertlarge}{60.7}
\defR{tight-x0.50/mig/p99/resnet50}{16.5}
\defR{tight-x0.50/mig/p99/retinanet}{109}
\defR{tight-x0.50/mig/p99/bertlarge}{207}
\defR{tight-x0.50/mig/p99x/resnet50}{1.2}
\defR{tight-x0.50/mig/p99x/retinanet}{3.1}
\defR{tight-x0.50/mig/p99x/bertlarge}{2.9}
\defR{tight-x0.50/mig/thrabs/train_bert}{0.00}
\defR{tight-x0.50/mig/thr/train_bert}{0}
\defR{tight-x0.50/tally/attmin}{0.0}
\defR{tight-x0.50/tally/attmean}{24.1}
\defR{tight-x0.50/tally/thr}{26}
\defR{tight-x0.50/tally/n}{3}
\defR{tight-x0.50/tally/att/resnet50}{0.0}
\defR{tight-x0.50/tally/att/retinanet}{0.0}
\defR{tight-x0.50/tally/att/bertlarge}{72.2}
\defR{tight-x0.50/tally/p99/resnet50}{94201}
\defR{tight-x0.50/tally/p99/retinanet}{3576}
\defR{tight-x0.50/tally/p99/bertlarge}{158}
\defR{tight-x0.50/tally/p99x/resnet50}{7031.3}
\defR{tight-x0.50/tally/p99x/retinanet}{102.8}
\defR{tight-x0.50/tally/p99x/bertlarge}{2.2}
\defR{tight-x0.50/tally/thrabs/train_bert}{0.42}
\defR{tight-x0.50/tally/thr/train_bert}{26}
\defR{tight-x0.50/orion/attmin}{0.0}
\defR{tight-x0.50/orion/attmean}{23.0}
\defR{tight-x0.50/orion/thr}{73}
\defR{tight-x0.50/orion/n}{3}
\defR{tight-x0.50/orion/att/resnet50}{69.0}
\defR{tight-x0.50/orion/att/retinanet}{0.0}
\defR{tight-x0.50/orion/att/bertlarge}{0.0}
\defR{tight-x0.50/orion/p99/resnet50}{27.9}
\defR{tight-x0.50/orion/p99/retinanet}{70199}
\defR{tight-x0.50/orion/p99/bertlarge}{1725}
\defR{tight-x0.50/orion/p99x/resnet50}{2.1}
\defR{tight-x0.50/orion/p99x/retinanet}{2018.5}
\defR{tight-x0.50/orion/p99x/bertlarge}{23.8}
\defR{tight-x0.50/orion/thrabs/train_bert}{1.18}
\defR{tight-x0.50/orion/thr/train_bert}{73}
\defR{tight-x0.75/berth-full/attmin}{95.7}
\defR{tight-x0.75/berth-full/attmean}{98.4}
\defR{tight-x0.75/berth-full/thr}{57}
\defR{tight-x0.75/berth-full/n}{3}
\defR{tight-x0.75/berth-full/att/resnet50}{95.7}
\defR{tight-x0.75/berth-full/att/retinanet}{100.0}
\defR{tight-x0.75/berth-full/att/bertlarge}{99.5}
\defR{tight-x0.75/berth-full/p99/resnet50}{14.4}
\defR{tight-x0.75/berth-full/p99/retinanet}{37.8}
\defR{tight-x0.75/berth-full/p99/bertlarge}{89.9}
\defR{tight-x0.75/berth-full/p99x/resnet50}{1.1}
\defR{tight-x0.75/berth-full/p99x/retinanet}{1.1}
\defR{tight-x0.75/berth-full/p99x/bertlarge}{1.2}
\defR{tight-x0.75/berth-full/thrabs/train_bert}{0.92}
\defR{tight-x0.75/berth-full/thr/train_bert}{57}
\defR{tight-x0.75/berth-nsr/attmin}{95.7}
\defR{tight-x0.75/berth-nsr/attmean}{98.4}
\defR{tight-x0.75/berth-nsr/thr}{56}
\defR{tight-x0.75/berth-nsr/n}{3}
\defR{tight-x0.75/berth-nsr/att/resnet50}{95.7}
\defR{tight-x0.75/berth-nsr/att/retinanet}{100.0}
\defR{tight-x0.75/berth-nsr/att/bertlarge}{99.4}
\defR{tight-x0.75/berth-nsr/p99/resnet50}{14.1}
\defR{tight-x0.75/berth-nsr/p99/retinanet}{38.2}
\defR{tight-x0.75/berth-nsr/p99/bertlarge}{89.6}
\defR{tight-x0.75/berth-nsr/p99x/resnet50}{1.1}
\defR{tight-x0.75/berth-nsr/p99x/retinanet}{1.1}
\defR{tight-x0.75/berth-nsr/p99x/bertlarge}{1.2}
\defR{tight-x0.75/berth-nsr/thrabs/train_bert}{0.91}
\defR{tight-x0.75/berth-nsr/thr/train_bert}{56}
\defR{tight-x0.75/berth-nba/attmin}{95.9}
\defR{tight-x0.75/berth-nba/attmean}{98.4}
\defR{tight-x0.75/berth-nba/thr}{57}
\defR{tight-x0.75/berth-nba/n}{3}
\defR{tight-x0.75/berth-nba/att/resnet50}{95.9}
\defR{tight-x0.75/berth-nba/att/retinanet}{100.0}
\defR{tight-x0.75/berth-nba/att/bertlarge}{99.4}
\defR{tight-x0.75/berth-nba/p99/resnet50}{14.1}
\defR{tight-x0.75/berth-nba/p99/retinanet}{39.5}
\defR{tight-x0.75/berth-nba/p99/bertlarge}{90.9}
\defR{tight-x0.75/berth-nba/p99x/resnet50}{1.1}
\defR{tight-x0.75/berth-nba/p99x/retinanet}{1.1}
\defR{tight-x0.75/berth-nba/p99x/bertlarge}{1.3}
\defR{tight-x0.75/berth-nba/thrabs/train_bert}{0.92}
\defR{tight-x0.75/berth-nba/thr/train_bert}{57}
\defR{tight-x0.75/berth-nsb/attmin}{95.6}
\defR{tight-x0.75/berth-nsb/attmean}{98.4}
\defR{tight-x0.75/berth-nsb/thr}{57}
\defR{tight-x0.75/berth-nsb/n}{3}
\defR{tight-x0.75/berth-nsb/att/resnet50}{95.6}
\defR{tight-x0.75/berth-nsb/att/retinanet}{100.0}
\defR{tight-x0.75/berth-nsb/att/bertlarge}{99.5}
\defR{tight-x0.75/berth-nsb/p99/resnet50}{13.9}
\defR{tight-x0.75/berth-nsb/p99/retinanet}{41.3}
\defR{tight-x0.75/berth-nsb/p99/bertlarge}{87.3}
\defR{tight-x0.75/berth-nsb/p99x/resnet50}{1.0}
\defR{tight-x0.75/berth-nsb/p99x/retinanet}{1.2}
\defR{tight-x0.75/berth-nsb/p99x/bertlarge}{1.2}
\defR{tight-x0.75/berth-nsb/thrabs/train_bert}{0.92}
\defR{tight-x0.75/berth-nsb/thr/train_bert}{57}
\defR{tight-x0.75/berth-ng/attmin}{94.6}
\defR{tight-x0.75/berth-ng/attmean}{97.7}
\defR{tight-x0.75/berth-ng/thr}{57}
\defR{tight-x0.75/berth-ng/n}{3}
\defR{tight-x0.75/berth-ng/att/resnet50}{94.6}
\defR{tight-x0.75/berth-ng/att/retinanet}{99.9}
\defR{tight-x0.75/berth-ng/att/bertlarge}{98.6}
\defR{tight-x0.75/berth-ng/p99/resnet50}{14.8}
\defR{tight-x0.75/berth-ng/p99/retinanet}{42.7}
\defR{tight-x0.75/berth-ng/p99/bertlarge}{102}
\defR{tight-x0.75/berth-ng/p99x/resnet50}{1.1}
\defR{tight-x0.75/berth-ng/p99x/retinanet}{1.2}
\defR{tight-x0.75/berth-ng/p99x/bertlarge}{1.4}
\defR{tight-x0.75/berth-ng/thrabs/train_bert}{0.93}
\defR{tight-x0.75/berth-ng/thr/train_bert}{57}
\defR{tight-x0.75/berth-nsng/attmin}{94.8}
\defR{tight-x0.75/berth-nsng/attmean}{97.8}
\defR{tight-x0.75/berth-nsng/thr}{57}
\defR{tight-x0.75/berth-nsng/n}{3}
\defR{tight-x0.75/berth-nsng/att/resnet50}{94.8}
\defR{tight-x0.75/berth-nsng/att/retinanet}{100.0}
\defR{tight-x0.75/berth-nsng/att/bertlarge}{98.7}
\defR{tight-x0.75/berth-nsng/p99/resnet50}{14.5}
\defR{tight-x0.75/berth-nsng/p99/retinanet}{42.0}
\defR{tight-x0.75/berth-nsng/p99/bertlarge}{101}
\defR{tight-x0.75/berth-nsng/p99x/resnet50}{1.1}
\defR{tight-x0.75/berth-nsng/p99x/retinanet}{1.2}
\defR{tight-x0.75/berth-nsng/p99x/bertlarge}{1.4}
\defR{tight-x0.75/berth-nsng/thrabs/train_bert}{0.93}
\defR{tight-x0.75/berth-nsng/thr/train_bert}{57}
\defR{tight-x0.75/berth-grouped/attmin}{87.5}
\defR{tight-x0.75/berth-grouped/attmean}{95.8}
\defR{tight-x0.75/berth-grouped/thr}{61}
\defR{tight-x0.75/berth-grouped/n}{3}
\defR{tight-x0.75/berth-grouped/att/resnet50}{87.5}
\defR{tight-x0.75/berth-grouped/att/retinanet}{100.0}
\defR{tight-x0.75/berth-grouped/att/bertlarge}{99.9}
\defR{tight-x0.75/berth-grouped/p99/resnet50}{22.6}
\defR{tight-x0.75/berth-grouped/p99/retinanet}{43.8}
\defR{tight-x0.75/berth-grouped/p99/bertlarge}{66.6}
\defR{tight-x0.75/berth-grouped/p99x/resnet50}{1.7}
\defR{tight-x0.75/berth-grouped/p99x/retinanet}{1.3}
\defR{tight-x0.75/berth-grouped/p99x/bertlarge}{0.9}
\defR{tight-x0.75/berth-grouped/thrabs/train_bert}{0.99}
\defR{tight-x0.75/berth-grouped/thr/train_bert}{61}
\defR{tight-x0.75/berth/attmin}{85.0}
\defR{tight-x0.75/berth/attmean}{94.6}
\defR{tight-x0.75/berth/thr}{58}
\defR{tight-x0.75/berth/n}{3}
\defR{tight-x0.75/berth/att/resnet50}{85.0}
\defR{tight-x0.75/berth/att/retinanet}{99.9}
\defR{tight-x0.75/berth/att/bertlarge}{99.0}
\defR{tight-x0.75/berth/p99/resnet50}{23.6}
\defR{tight-x0.75/berth/p99/retinanet}{52.2}
\defR{tight-x0.75/berth/p99/bertlarge}{96.8}
\defR{tight-x0.75/berth/p99x/resnet50}{1.8}
\defR{tight-x0.75/berth/p99x/retinanet}{1.5}
\defR{tight-x0.75/berth/p99x/bertlarge}{1.3}
\defR{tight-x0.75/berth/thrabs/train_bert}{0.94}
\defR{tight-x0.75/berth/thr/train_bert}{58}
\defR{tight-x0.75/timeslice/attmin}{35.7}
\defR{tight-x0.75/timeslice/attmean}{61.5}
\defR{tight-x0.75/timeslice/thr}{55}
\defR{tight-x0.75/timeslice/n}{3}
\defR{tight-x0.75/timeslice/att/resnet50}{35.7}
\defR{tight-x0.75/timeslice/att/retinanet}{89.9}
\defR{tight-x0.75/timeslice/att/bertlarge}{58.8}
\defR{tight-x0.75/timeslice/p99/resnet50}{41.5}
\defR{tight-x0.75/timeslice/p99/retinanet}{132}
\defR{tight-x0.75/timeslice/p99/bertlarge}{292}
\defR{tight-x0.75/timeslice/p99x/resnet50}{3.1}
\defR{tight-x0.75/timeslice/p99x/retinanet}{3.8}
\defR{tight-x0.75/timeslice/p99x/bertlarge}{4.0}
\defR{tight-x0.75/timeslice/thrabs/train_bert}{0.89}
\defR{tight-x0.75/timeslice/thr/train_bert}{55}
\defR{tight-x0.75/mps/attmin}{0.0}
\defR{tight-x0.75/mps/attmean}{0.0}
\defR{tight-x0.75/mps/thr}{80}
\defR{tight-x0.75/mps/n}{3}
\defR{tight-x0.75/mps/att/resnet50}{0.0}
\defR{tight-x0.75/mps/att/retinanet}{0.0}
\defR{tight-x0.75/mps/att/bertlarge}{0.0}
\defR{tight-x0.75/mps/p99/resnet50}{31307}
\defR{tight-x0.75/mps/p99/retinanet}{42486}
\defR{tight-x0.75/mps/p99/bertlarge}{6753}
\defR{tight-x0.75/mps/p99x/resnet50}{2336.8}
\defR{tight-x0.75/mps/p99x/retinanet}{1221.7}
\defR{tight-x0.75/mps/p99x/bertlarge}{93.3}
\defR{tight-x0.75/mps/thrabs/train_bert}{1.30}
\defR{tight-x0.75/mps/thr/train_bert}{80}
\defR{tight-x0.75/mpsprio/attmin}{0.0}
\defR{tight-x0.75/mpsprio/attmean}{31.4}
\defR{tight-x0.75/mpsprio/thr}{70}
\defR{tight-x0.75/mpsprio/n}{3}
\defR{tight-x0.75/mpsprio/att/resnet50}{0.0}
\defR{tight-x0.75/mpsprio/att/retinanet}{33.7}
\defR{tight-x0.75/mpsprio/att/bertlarge}{60.4}
\defR{tight-x0.75/mpsprio/p99/resnet50}{136}
\defR{tight-x0.75/mpsprio/p99/retinanet}{417}
\defR{tight-x0.75/mpsprio/p99/bertlarge}{261}
\defR{tight-x0.75/mpsprio/p99x/resnet50}{10.2}
\defR{tight-x0.75/mpsprio/p99x/retinanet}{12.0}
\defR{tight-x0.75/mpsprio/p99x/bertlarge}{3.6}
\defR{tight-x0.75/mpsprio/thrabs/train_bert}{1.13}
\defR{tight-x0.75/mpsprio/thr/train_bert}{70}
\defR{tight-x0.75/mpslimits/attmin}{83.8}
\defR{tight-x0.75/mpslimits/attmean}{90.4}
\defR{tight-x0.75/mpslimits/thr}{12}
\defR{tight-x0.75/mpslimits/n}{3}
\defR{tight-x0.75/mpslimits/att/resnet50}{93.0}
\defR{tight-x0.75/mpslimits/att/retinanet}{94.4}
\defR{tight-x0.75/mpslimits/att/bertlarge}{83.8}
\defR{tight-x0.75/mpslimits/p99/resnet50}{15.1}
\defR{tight-x0.75/mpslimits/p99/retinanet}{111}
\defR{tight-x0.75/mpslimits/p99/bertlarge}{171}
\defR{tight-x0.75/mpslimits/p99x/resnet50}{1.1}
\defR{tight-x0.75/mpslimits/p99x/retinanet}{3.2}
\defR{tight-x0.75/mpslimits/p99x/bertlarge}{2.4}
\defR{tight-x0.75/mpslimits/thrabs/train_bert}{0.19}
\defR{tight-x0.75/mpslimits/thr/train_bert}{12}
\defR{tight-x0.75/mpsparts/attmin}{83.3}
\defR{tight-x0.75/mpsparts/attmean}{91.3}
\defR{tight-x0.75/mpsparts/thr}{10}
\defR{tight-x0.75/mpsparts/n}{2}
\defR{tight-x0.75/mpsparts/att/resnet50}{93.1}
\defR{tight-x0.75/mpsparts/att/retinanet}{97.4}
\defR{tight-x0.75/mpsparts/att/bertlarge}{83.3}
\defR{tight-x0.75/mpsparts/p99/resnet50}{15.3}
\defR{tight-x0.75/mpsparts/p99/retinanet}{90.0}
\defR{tight-x0.75/mpsparts/p99/bertlarge}{172}
\defR{tight-x0.75/mpsparts/p99x/resnet50}{1.1}
\defR{tight-x0.75/mpsparts/p99x/retinanet}{2.6}
\defR{tight-x0.75/mpsparts/p99x/bertlarge}{2.4}
\defR{tight-x0.75/mpsparts/thrabs/train_bert}{0.16}
\defR{tight-x0.75/mpsparts/thr/train_bert}{10}
\defR{tight-x0.75/lithos/attmin}{83.8}
\defR{tight-x0.75/lithos/attmean}{90.4}
\defR{tight-x0.75/lithos/thr}{12}
\defR{tight-x0.75/lithos/n}{3}
\defR{tight-x0.75/lithos/att/resnet50}{91.9}
\defR{tight-x0.75/lithos/att/retinanet}{95.4}
\defR{tight-x0.75/lithos/att/bertlarge}{83.8}
\defR{tight-x0.75/lithos/p99/resnet50}{16.3}
\defR{tight-x0.75/lithos/p99/retinanet}{105}
\defR{tight-x0.75/lithos/p99/bertlarge}{170}
\defR{tight-x0.75/lithos/p99x/resnet50}{1.2}
\defR{tight-x0.75/lithos/p99x/retinanet}{3.0}
\defR{tight-x0.75/lithos/p99x/bertlarge}{2.3}
\defR{tight-x0.75/lithos/thrabs/train_bert}{0.19}
\defR{tight-x0.75/lithos/thr/train_bert}{12}
\defR{tight-x0.75/dysm/attmin}{0.0}
\defR{tight-x0.75/dysm/attmean}{26.7}
\defR{tight-x0.75/dysm/thr}{83}
\defR{tight-x0.75/dysm/n}{3}
\defR{tight-x0.75/dysm/att/resnet50}{0.0}
\defR{tight-x0.75/dysm/att/retinanet}{80.2}
\defR{tight-x0.75/dysm/att/bertlarge}{0.0}
\defR{tight-x0.75/dysm/p99/retinanet}{170}
\defR{tight-x0.75/dysm/p99x/retinanet}{4.9}
\defR{tight-x0.75/dysm/thrabs/train_bert}{1.35}
\defR{tight-x0.75/dysm/thr/train_bert}{83}
\defR{tight-x0.75/xsched/attmin}{74.5}
\defR{tight-x0.75/xsched/attmean}{90.9}
\defR{tight-x0.75/xsched/thr}{57}
\defR{tight-x0.75/xsched/n}{3}
\defR{tight-x0.75/xsched/att/resnet50}{74.5}
\defR{tight-x0.75/xsched/att/retinanet}{99.5}
\defR{tight-x0.75/xsched/att/bertlarge}{98.7}
\defR{tight-x0.75/xsched/p99/resnet50}{29.4}
\defR{tight-x0.75/xsched/p99/retinanet}{64.7}
\defR{tight-x0.75/xsched/p99/bertlarge}{101}
\defR{tight-x0.75/xsched/p99x/resnet50}{2.2}
\defR{tight-x0.75/xsched/p99x/retinanet}{1.9}
\defR{tight-x0.75/xsched/p99x/bertlarge}{1.4}
\defR{tight-x0.75/xsched/thrabs/train_bert}{0.92}
\defR{tight-x0.75/xsched/thr/train_bert}{57}
\defR{tight-x0.75/multistream/attmin}{0.0}
\defR{tight-x0.75/multistream/attmean}{32.6}
\defR{tight-x0.75/multistream/thr}{71}
\defR{tight-x0.75/multistream/n}{3}
\defR{tight-x0.75/multistream/att/resnet50}{0.0}
\defR{tight-x0.75/multistream/att/retinanet}{36.1}
\defR{tight-x0.75/multistream/att/bertlarge}{61.6}
\defR{tight-x0.75/multistream/p99/resnet50}{154}
\defR{tight-x0.75/multistream/p99/retinanet}{405}
\defR{tight-x0.75/multistream/p99/bertlarge}{254}
\defR{tight-x0.75/multistream/p99x/resnet50}{11.5}
\defR{tight-x0.75/multistream/p99x/retinanet}{11.6}
\defR{tight-x0.75/multistream/p99x/bertlarge}{3.5}
\defR{tight-x0.75/multistream/thrabs/train_bert}{1.15}
\defR{tight-x0.75/multistream/thr/train_bert}{71}
\defR{tight-x0.75/tgs/attmin}{91.7}
\defR{tight-x0.75/tgs/attmean}{97.2}
\defR{tight-x0.75/tgs/thr}{1}
\defR{tight-x0.75/tgs/n}{3}
\defR{tight-x0.75/tgs/att/resnet50}{91.7}
\defR{tight-x0.75/tgs/att/retinanet}{100.0}
\defR{tight-x0.75/tgs/att/bertlarge}{99.8}
\defR{tight-x0.75/tgs/p99/resnet50}{19.9}
\defR{tight-x0.75/tgs/p99/retinanet}{40.9}
\defR{tight-x0.75/tgs/p99/bertlarge}{78.3}
\defR{tight-x0.75/tgs/p99x/resnet50}{1.5}
\defR{tight-x0.75/tgs/p99x/retinanet}{1.2}
\defR{tight-x0.75/tgs/p99x/bertlarge}{1.1}
\defR{tight-x0.75/tgs/thrabs/train_bert}{0.02}
\defR{tight-x0.75/tgs/thr/train_bert}{1}
\defR{tight-x0.75/mig/attmin}{75.3}
\defR{tight-x0.75/mig/attmean}{86.9}
\defR{tight-x0.75/mig/thr}{0}
\defR{tight-x0.75/mig/n}{3}
\defR{tight-x0.75/mig/att/resnet50}{90.7}
\defR{tight-x0.75/mig/att/retinanet}{94.7}
\defR{tight-x0.75/mig/att/bertlarge}{75.3}
\defR{tight-x0.75/mig/p99/resnet50}{16.6}
\defR{tight-x0.75/mig/p99/retinanet}{109}
\defR{tight-x0.75/mig/p99/bertlarge}{207}
\defR{tight-x0.75/mig/p99x/resnet50}{1.2}
\defR{tight-x0.75/mig/p99x/retinanet}{3.1}
\defR{tight-x0.75/mig/p99x/bertlarge}{2.9}
\defR{tight-x0.75/mig/thrabs/train_bert}{0.00}
\defR{tight-x0.75/mig/thr/train_bert}{0}
\defR{tight-x0.75/tally/attmin}{0.0}
\defR{tight-x0.75/tally/attmean}{30.7}
\defR{tight-x0.75/tally/thr}{26}
\defR{tight-x0.75/tally/n}{3}
\defR{tight-x0.75/tally/att/resnet50}{0.0}
\defR{tight-x0.75/tally/att/retinanet}{2.7}
\defR{tight-x0.75/tally/att/bertlarge}{89.4}
\defR{tight-x0.75/tally/p99/resnet50}{94201}
\defR{tight-x0.75/tally/p99/retinanet}{3239}
\defR{tight-x0.75/tally/p99/bertlarge}{151}
\defR{tight-x0.75/tally/p99x/resnet50}{7031.3}
\defR{tight-x0.75/tally/p99x/retinanet}{93.1}
\defR{tight-x0.75/tally/p99x/bertlarge}{2.1}
\defR{tight-x0.75/tally/thrabs/train_bert}{0.43}
\defR{tight-x0.75/tally/thr/train_bert}{26}
\defR{tight-x0.75/orion/attmin}{0.0}
\defR{tight-x0.75/orion/attmean}{29.2}
\defR{tight-x0.75/orion/thr}{73}
\defR{tight-x0.75/orion/n}{3}
\defR{tight-x0.75/orion/att/resnet50}{85.8}
\defR{tight-x0.75/orion/att/retinanet}{0.0}
\defR{tight-x0.75/orion/att/bertlarge}{1.9}
\defR{tight-x0.75/orion/p99/resnet50}{27.0}
\defR{tight-x0.75/orion/p99/retinanet}{70455}
\defR{tight-x0.75/orion/p99/bertlarge}{1520}
\defR{tight-x0.75/orion/p99x/resnet50}{2.0}
\defR{tight-x0.75/orion/p99x/retinanet}{2025.9}
\defR{tight-x0.75/orion/p99x/bertlarge}{21.0}
\defR{tight-x0.75/orion/thrabs/train_bert}{1.18}
\defR{tight-x0.75/orion/thr/train_bert}{73}
\defR{tight-x1.50/berth-full/attmin}{100.0}
\defR{tight-x1.50/berth-full/attmean}{100.0}
\defR{tight-x1.50/berth-full/thr}{57}
\defR{tight-x1.50/berth-full/n}{3}
\defR{tight-x1.50/berth-full/att/resnet50}{100.0}
\defR{tight-x1.50/berth-full/att/retinanet}{100.0}
\defR{tight-x1.50/berth-full/att/bertlarge}{100.0}
\defR{tight-x1.50/berth-full/p99/resnet50}{14.3}
\defR{tight-x1.50/berth-full/p99/retinanet}{38.9}
\defR{tight-x1.50/berth-full/p99/bertlarge}{94.2}
\defR{tight-x1.50/berth-full/p99x/resnet50}{1.1}
\defR{tight-x1.50/berth-full/p99x/retinanet}{1.1}
\defR{tight-x1.50/berth-full/p99x/bertlarge}{1.3}
\defR{tight-x1.50/berth-full/thrabs/train_bert}{0.92}
\defR{tight-x1.50/berth-full/thr/train_bert}{57}
\defR{tight-x1.50/berth-nsr/attmin}{99.9}
\defR{tight-x1.50/berth-nsr/attmean}{100.0}
\defR{tight-x1.50/berth-nsr/thr}{56}
\defR{tight-x1.50/berth-nsr/n}{3}
\defR{tight-x1.50/berth-nsr/att/resnet50}{99.9}
\defR{tight-x1.50/berth-nsr/att/retinanet}{100.0}
\defR{tight-x1.50/berth-nsr/att/bertlarge}{100.0}
\defR{tight-x1.50/berth-nsr/p99/resnet50}{14.4}
\defR{tight-x1.50/berth-nsr/p99/retinanet}{39.5}
\defR{tight-x1.50/berth-nsr/p99/bertlarge}{90.0}
\defR{tight-x1.50/berth-nsr/p99x/resnet50}{1.1}
\defR{tight-x1.50/berth-nsr/p99x/retinanet}{1.1}
\defR{tight-x1.50/berth-nsr/p99x/bertlarge}{1.2}
\defR{tight-x1.50/berth-nsr/thrabs/train_bert}{0.91}
\defR{tight-x1.50/berth-nsr/thr/train_bert}{56}
\defR{tight-x1.50/berth-nba/attmin}{100.0}
\defR{tight-x1.50/berth-nba/attmean}{100.0}
\defR{tight-x1.50/berth-nba/thr}{57}
\defR{tight-x1.50/berth-nba/n}{3}
\defR{tight-x1.50/berth-nba/att/resnet50}{100.0}
\defR{tight-x1.50/berth-nba/att/retinanet}{100.0}
\defR{tight-x1.50/berth-nba/att/bertlarge}{100.0}
\defR{tight-x1.50/berth-nba/p99/resnet50}{13.9}
\defR{tight-x1.50/berth-nba/p99/retinanet}{39.0}
\defR{tight-x1.50/berth-nba/p99/bertlarge}{90.7}
\defR{tight-x1.50/berth-nba/p99x/resnet50}{1.0}
\defR{tight-x1.50/berth-nba/p99x/retinanet}{1.1}
\defR{tight-x1.50/berth-nba/p99x/bertlarge}{1.3}
\defR{tight-x1.50/berth-nba/thrabs/train_bert}{0.92}
\defR{tight-x1.50/berth-nba/thr/train_bert}{57}
\defR{tight-x1.50/berth-nsb/attmin}{100.0}
\defR{tight-x1.50/berth-nsb/attmean}{100.0}
\defR{tight-x1.50/berth-nsb/thr}{57}
\defR{tight-x1.50/berth-nsb/n}{3}
\defR{tight-x1.50/berth-nsb/att/resnet50}{100.0}
\defR{tight-x1.50/berth-nsb/att/retinanet}{100.0}
\defR{tight-x1.50/berth-nsb/att/bertlarge}{100.0}
\defR{tight-x1.50/berth-nsb/p99/resnet50}{13.8}
\defR{tight-x1.50/berth-nsb/p99/retinanet}{38.5}
\defR{tight-x1.50/berth-nsb/p99/bertlarge}{90.5}
\defR{tight-x1.50/berth-nsb/p99x/resnet50}{1.0}
\defR{tight-x1.50/berth-nsb/p99x/retinanet}{1.1}
\defR{tight-x1.50/berth-nsb/p99x/bertlarge}{1.3}
\defR{tight-x1.50/berth-nsb/thrabs/train_bert}{0.92}
\defR{tight-x1.50/berth-nsb/thr/train_bert}{57}
\defR{tight-x1.50/berth-ng/attmin}{99.8}
\defR{tight-x1.50/berth-ng/attmean}{99.9}
\defR{tight-x1.50/berth-ng/thr}{57}
\defR{tight-x1.50/berth-ng/n}{3}
\defR{tight-x1.50/berth-ng/att/resnet50}{99.8}
\defR{tight-x1.50/berth-ng/att/retinanet}{100.0}
\defR{tight-x1.50/berth-ng/att/bertlarge}{100.0}
\defR{tight-x1.50/berth-ng/p99/resnet50}{14.8}
\defR{tight-x1.50/berth-ng/p99/retinanet}{42.0}
\defR{tight-x1.50/berth-ng/p99/bertlarge}{97.2}
\defR{tight-x1.50/berth-ng/p99x/resnet50}{1.1}
\defR{tight-x1.50/berth-ng/p99x/retinanet}{1.2}
\defR{tight-x1.50/berth-ng/p99x/bertlarge}{1.3}
\defR{tight-x1.50/berth-ng/thrabs/train_bert}{0.92}
\defR{tight-x1.50/berth-ng/thr/train_bert}{57}
\defR{tight-x1.50/berth-nsng/attmin}{99.8}
\defR{tight-x1.50/berth-nsng/attmean}{99.9}
\defR{tight-x1.50/berth-nsng/thr}{57}
\defR{tight-x1.50/berth-nsng/n}{3}
\defR{tight-x1.50/berth-nsng/att/resnet50}{99.9}
\defR{tight-x1.50/berth-nsng/att/retinanet}{100.0}
\defR{tight-x1.50/berth-nsng/att/bertlarge}{99.8}
\defR{tight-x1.50/berth-nsng/p99/resnet50}{14.6}
\defR{tight-x1.50/berth-nsng/p99/retinanet}{40.5}
\defR{tight-x1.50/berth-nsng/p99/bertlarge}{103}
\defR{tight-x1.50/berth-nsng/p99x/resnet50}{1.1}
\defR{tight-x1.50/berth-nsng/p99x/retinanet}{1.2}
\defR{tight-x1.50/berth-nsng/p99x/bertlarge}{1.4}
\defR{tight-x1.50/berth-nsng/thrabs/train_bert}{0.93}
\defR{tight-x1.50/berth-nsng/thr/train_bert}{57}
\defR{tight-x1.50/berth-grouped/attmin}{99.2}
\defR{tight-x1.50/berth-grouped/attmean}{99.7}
\defR{tight-x1.50/berth-grouped/thr}{61}
\defR{tight-x1.50/berth-grouped/n}{3}
\defR{tight-x1.50/berth-grouped/att/resnet50}{99.2}
\defR{tight-x1.50/berth-grouped/att/retinanet}{100.0}
\defR{tight-x1.50/berth-grouped/att/bertlarge}{100.0}
\defR{tight-x1.50/berth-grouped/p99/resnet50}{21.7}
\defR{tight-x1.50/berth-grouped/p99/retinanet}{44.2}
\defR{tight-x1.50/berth-grouped/p99/bertlarge}{68.8}
\defR{tight-x1.50/berth-grouped/p99x/resnet50}{1.6}
\defR{tight-x1.50/berth-grouped/p99x/retinanet}{1.3}
\defR{tight-x1.50/berth-grouped/p99x/bertlarge}{1.0}
\defR{tight-x1.50/berth-grouped/thrabs/train_bert}{0.99}
\defR{tight-x1.50/berth-grouped/thr/train_bert}{61}
\defR{tight-x1.50/berth/attmin}{98.9}
\defR{tight-x1.50/berth/attmean}{99.6}
\defR{tight-x1.50/berth/thr}{58}
\defR{tight-x1.50/berth/n}{3}
\defR{tight-x1.50/berth/att/resnet50}{98.9}
\defR{tight-x1.50/berth/att/retinanet}{100.0}
\defR{tight-x1.50/berth/att/bertlarge}{100.0}
\defR{tight-x1.50/berth/p99/resnet50}{22.9}
\defR{tight-x1.50/berth/p99/retinanet}{51.8}
\defR{tight-x1.50/berth/p99/bertlarge}{103}
\defR{tight-x1.50/berth/p99x/resnet50}{1.7}
\defR{tight-x1.50/berth/p99x/retinanet}{1.5}
\defR{tight-x1.50/berth/p99x/bertlarge}{1.4}
\defR{tight-x1.50/berth/thrabs/train_bert}{0.94}
\defR{tight-x1.50/berth/thr/train_bert}{58}
\defR{tight-x1.50/timeslice/attmin}{86.7}
\defR{tight-x1.50/timeslice/attmean}{92.6}
\defR{tight-x1.50/timeslice/thr}{55}
\defR{tight-x1.50/timeslice/n}{3}
\defR{tight-x1.50/timeslice/att/resnet50}{86.7}
\defR{tight-x1.50/timeslice/att/retinanet}{99.4}
\defR{tight-x1.50/timeslice/att/bertlarge}{91.7}
\defR{tight-x1.50/timeslice/p99/resnet50}{41.3}
\defR{tight-x1.50/timeslice/p99/retinanet}{132}
\defR{tight-x1.50/timeslice/p99/bertlarge}{291}
\defR{tight-x1.50/timeslice/p99x/resnet50}{3.1}
\defR{tight-x1.50/timeslice/p99x/retinanet}{3.8}
\defR{tight-x1.50/timeslice/p99x/bertlarge}{4.0}
\defR{tight-x1.50/timeslice/thrabs/train_bert}{0.89}
\defR{tight-x1.50/timeslice/thr/train_bert}{55}
\defR{tight-x1.50/mps/attmin}{0.0}
\defR{tight-x1.50/mps/attmean}{0.2}
\defR{tight-x1.50/mps/thr}{80}
\defR{tight-x1.50/mps/n}{3}
\defR{tight-x1.50/mps/att/resnet50}{0.0}
\defR{tight-x1.50/mps/att/retinanet}{0.0}
\defR{tight-x1.50/mps/att/bertlarge}{0.6}
\defR{tight-x1.50/mps/p99/resnet50}{31770}
\defR{tight-x1.50/mps/p99/retinanet}{42506}
\defR{tight-x1.50/mps/p99/bertlarge}{6733}
\defR{tight-x1.50/mps/p99x/resnet50}{2371.3}
\defR{tight-x1.50/mps/p99x/retinanet}{1222.2}
\defR{tight-x1.50/mps/p99x/bertlarge}{93.1}
\defR{tight-x1.50/mps/thrabs/train_bert}{1.30}
\defR{tight-x1.50/mps/thr/train_bert}{80}
\defR{tight-x1.50/mpsprio/attmin}{25.1}
\defR{tight-x1.50/mpsprio/attmean}{62.2}
\defR{tight-x1.50/mpsprio/thr}{70}
\defR{tight-x1.50/mpsprio/n}{3}
\defR{tight-x1.50/mpsprio/att/resnet50}{25.1}
\defR{tight-x1.50/mpsprio/att/retinanet}{67.2}
\defR{tight-x1.50/mpsprio/att/bertlarge}{94.4}
\defR{tight-x1.50/mpsprio/p99/resnet50}{139}
\defR{tight-x1.50/mpsprio/p99/retinanet}{439}
\defR{tight-x1.50/mpsprio/p99/bertlarge}{265}
\defR{tight-x1.50/mpsprio/p99x/resnet50}{10.4}
\defR{tight-x1.50/mpsprio/p99x/retinanet}{12.6}
\defR{tight-x1.50/mpsprio/p99x/bertlarge}{3.7}
\defR{tight-x1.50/mpsprio/thrabs/train_bert}{1.13}
\defR{tight-x1.50/mpsprio/thr/train_bert}{70}
\defR{tight-x1.50/mpslimits/attmin}{98.1}
\defR{tight-x1.50/mpslimits/attmean}{99.2}
\defR{tight-x1.50/mpslimits/thr}{11}
\defR{tight-x1.50/mpslimits/n}{3}
\defR{tight-x1.50/mpslimits/att/resnet50}{100.0}
\defR{tight-x1.50/mpslimits/att/retinanet}{99.6}
\defR{tight-x1.50/mpslimits/att/bertlarge}{98.1}
\defR{tight-x1.50/mpslimits/p99/resnet50}{13.9}
\defR{tight-x1.50/mpslimits/p99/retinanet}{134}
\defR{tight-x1.50/mpslimits/p99/bertlarge}{214}
\defR{tight-x1.50/mpslimits/p99x/resnet50}{1.0}
\defR{tight-x1.50/mpslimits/p99x/retinanet}{3.9}
\defR{tight-x1.50/mpslimits/p99x/bertlarge}{3.0}
\defR{tight-x1.50/mpslimits/thrabs/train_bert}{0.19}
\defR{tight-x1.50/mpslimits/thr/train_bert}{11}
\defR{tight-x1.50/mpsparts/attmin}{98.0}
\defR{tight-x1.50/mpsparts/attmean}{98.9}
\defR{tight-x1.50/mpsparts/thr}{14}
\defR{tight-x1.50/mpsparts/n}{2}
\defR{tight-x1.50/mpsparts/att/resnet50}{100.0}
\defR{tight-x1.50/mpsparts/att/retinanet}{98.8}
\defR{tight-x1.50/mpsparts/att/bertlarge}{98.0}
\defR{tight-x1.50/mpsparts/p99/resnet50}{13.9}
\defR{tight-x1.50/mpsparts/p99/retinanet}{156}
\defR{tight-x1.50/mpsparts/p99/bertlarge}{211}
\defR{tight-x1.50/mpsparts/p99x/resnet50}{1.0}
\defR{tight-x1.50/mpsparts/p99x/retinanet}{4.5}
\defR{tight-x1.50/mpsparts/p99x/bertlarge}{2.9}
\defR{tight-x1.50/mpsparts/thrabs/train_bert}{0.22}
\defR{tight-x1.50/mpsparts/thr/train_bert}{14}
\defR{tight-x1.50/lithos/attmin}{97.7}
\defR{tight-x1.50/lithos/attmean}{99.0}
\defR{tight-x1.50/lithos/thr}{12}
\defR{tight-x1.50/lithos/n}{3}
\defR{tight-x1.50/lithos/att/resnet50}{99.9}
\defR{tight-x1.50/lithos/att/retinanet}{99.5}
\defR{tight-x1.50/lithos/att/bertlarge}{97.7}
\defR{tight-x1.50/lithos/p99/resnet50}{14.7}
\defR{tight-x1.50/lithos/p99/retinanet}{137}
\defR{tight-x1.50/lithos/p99/bertlarge}{218}
\defR{tight-x1.50/lithos/p99x/resnet50}{1.1}
\defR{tight-x1.50/lithos/p99x/retinanet}{3.9}
\defR{tight-x1.50/lithos/p99x/bertlarge}{3.0}
\defR{tight-x1.50/lithos/thrabs/train_bert}{0.19}
\defR{tight-x1.50/lithos/thr/train_bert}{12}
\defR{tight-x1.50/dysm/attmin}{0.0}
\defR{tight-x1.50/dysm/attmean}{44.4}
\defR{tight-x1.50/dysm/thr}{68}
\defR{tight-x1.50/dysm/n}{3}
\defR{tight-x1.50/dysm/att/resnet50}{0.0}
\defR{tight-x1.50/dysm/att/retinanet}{56.0}
\defR{tight-x1.50/dysm/att/bertlarge}{77.1}
\defR{tight-x1.50/dysm/p99/retinanet}{1091}
\defR{tight-x1.50/dysm/p99/bertlarge}{525}
\defR{tight-x1.50/dysm/p99x/retinanet}{31.4}
\defR{tight-x1.50/dysm/p99x/bertlarge}{7.3}
\defR{tight-x1.50/dysm/thrabs/train_bert}{1.10}
\defR{tight-x1.50/dysm/thr/train_bert}{68}
\defR{tight-x1.50/xsched/attmin}{96.9}
\defR{tight-x1.50/xsched/attmean}{99.0}
\defR{tight-x1.50/xsched/thr}{56}
\defR{tight-x1.50/xsched/n}{3}
\defR{tight-x1.50/xsched/att/resnet50}{96.9}
\defR{tight-x1.50/xsched/att/retinanet}{100.0}
\defR{tight-x1.50/xsched/att/bertlarge}{100.0}
\defR{tight-x1.50/xsched/p99/resnet50}{28.7}
\defR{tight-x1.50/xsched/p99/retinanet}{55.4}
\defR{tight-x1.50/xsched/p99/bertlarge}{107}
\defR{tight-x1.50/xsched/p99x/resnet50}{2.1}
\defR{tight-x1.50/xsched/p99x/retinanet}{1.6}
\defR{tight-x1.50/xsched/p99x/bertlarge}{1.5}
\defR{tight-x1.50/xsched/thrabs/train_bert}{0.91}
\defR{tight-x1.50/xsched/thr/train_bert}{56}
\defR{tight-x1.50/multistream/attmin}{21.5}
\defR{tight-x1.50/multistream/attmean}{63.2}
\defR{tight-x1.50/multistream/thr}{71}
\defR{tight-x1.50/multistream/n}{3}
\defR{tight-x1.50/multistream/att/resnet50}{21.5}
\defR{tight-x1.50/multistream/att/retinanet}{73.1}
\defR{tight-x1.50/multistream/att/bertlarge}{95.0}
\defR{tight-x1.50/multistream/p99/resnet50}{151}
\defR{tight-x1.50/multistream/p99/retinanet}{388}
\defR{tight-x1.50/multistream/p99/bertlarge}{258}
\defR{tight-x1.50/multistream/p99x/resnet50}{11.3}
\defR{tight-x1.50/multistream/p99x/retinanet}{11.2}
\defR{tight-x1.50/multistream/p99x/bertlarge}{3.6}
\defR{tight-x1.50/multistream/thrabs/train_bert}{1.14}
\defR{tight-x1.50/multistream/thr/train_bert}{71}
\defR{tight-x1.50/tgs/attmin}{99.5}
\defR{tight-x1.50/tgs/attmean}{99.8}
\defR{tight-x1.50/tgs/thr}{2}
\defR{tight-x1.50/tgs/n}{3}
\defR{tight-x1.50/tgs/att/resnet50}{99.5}
\defR{tight-x1.50/tgs/att/retinanet}{100.0}
\defR{tight-x1.50/tgs/att/bertlarge}{100.0}
\defR{tight-x1.50/tgs/p99/resnet50}{20.1}
\defR{tight-x1.50/tgs/p99/retinanet}{40.4}
\defR{tight-x1.50/tgs/p99/bertlarge}{81.7}
\defR{tight-x1.50/tgs/p99x/resnet50}{1.5}
\defR{tight-x1.50/tgs/p99x/retinanet}{1.2}
\defR{tight-x1.50/tgs/p99x/bertlarge}{1.1}
\defR{tight-x1.50/tgs/thrabs/train_bert}{0.03}
\defR{tight-x1.50/tgs/thr/train_bert}{2}
\defR{tight-x1.50/mig/attmin}{98.3}
\defR{tight-x1.50/mig/attmean}{99.4}
\defR{tight-x1.50/mig/thr}{0}
\defR{tight-x1.50/mig/n}{3}
\defR{tight-x1.50/mig/att/resnet50}{99.9}
\defR{tight-x1.50/mig/att/retinanet}{100.0}
\defR{tight-x1.50/mig/att/bertlarge}{98.3}
\defR{tight-x1.50/mig/p99/resnet50}{16.5}
\defR{tight-x1.50/mig/p99/retinanet}{109}
\defR{tight-x1.50/mig/p99/bertlarge}{207}
\defR{tight-x1.50/mig/p99x/resnet50}{1.2}
\defR{tight-x1.50/mig/p99x/retinanet}{3.1}
\defR{tight-x1.50/mig/p99x/bertlarge}{2.9}
\defR{tight-x1.50/mig/thrabs/train_bert}{0.00}
\defR{tight-x1.50/mig/thr/train_bert}{0}
\defR{tight-x1.50/tally/attmin}{0.0}
\defR{tight-x1.50/tally/attmean}{37.1}
\defR{tight-x1.50/tally/thr}{27}
\defR{tight-x1.50/tally/n}{3}
\defR{tight-x1.50/tally/att/resnet50}{0.0}
\defR{tight-x1.50/tally/att/retinanet}{11.3}
\defR{tight-x1.50/tally/att/bertlarge}{100.0}
\defR{tight-x1.50/tally/p99/resnet50}{94201}
\defR{tight-x1.50/tally/p99/retinanet}{3119}
\defR{tight-x1.50/tally/p99/bertlarge}{126}
\defR{tight-x1.50/tally/p99x/resnet50}{7031.3}
\defR{tight-x1.50/tally/p99x/retinanet}{89.7}
\defR{tight-x1.50/tally/p99x/bertlarge}{1.7}
\defR{tight-x1.50/tally/thrabs/train_bert}{0.44}
\defR{tight-x1.50/tally/thr/train_bert}{27}
\defR{tight-x1.50/orion/attmin}{0.0}
\defR{tight-x1.50/orion/attmean}{40.7}
\defR{tight-x1.50/orion/thr}{73}
\defR{tight-x1.50/orion/n}{3}
\defR{tight-x1.50/orion/att/resnet50}{98.4}
\defR{tight-x1.50/orion/att/retinanet}{0.0}
\defR{tight-x1.50/orion/att/bertlarge}{23.7}
\defR{tight-x1.50/orion/p99/resnet50}{25.6}
\defR{tight-x1.50/orion/p99/retinanet}{70293}
\defR{tight-x1.50/orion/p99/bertlarge}{1608}
\defR{tight-x1.50/orion/p99x/resnet50}{1.9}
\defR{tight-x1.50/orion/p99x/retinanet}{2021.2}
\defR{tight-x1.50/orion/p99x/bertlarge}{22.2}
\defR{tight-x1.50/orion/thrabs/train_bert}{1.18}
\defR{tight-x1.50/orion/thr/train_bert}{73}
\defR{tight-x2.00/berth-full/attmin}{100.0}
\defR{tight-x2.00/berth-full/attmean}{100.0}
\defR{tight-x2.00/berth-full/thr}{57}
\defR{tight-x2.00/berth-full/n}{3}
\defR{tight-x2.00/berth-full/att/resnet50}{100.0}
\defR{tight-x2.00/berth-full/att/retinanet}{100.0}
\defR{tight-x2.00/berth-full/att/bertlarge}{100.0}
\defR{tight-x2.00/berth-full/p99/resnet50}{13.7}
\defR{tight-x2.00/berth-full/p99/retinanet}{39.2}
\defR{tight-x2.00/berth-full/p99/bertlarge}{87.8}
\defR{tight-x2.00/berth-full/p99x/resnet50}{1.0}
\defR{tight-x2.00/berth-full/p99x/retinanet}{1.1}
\defR{tight-x2.00/berth-full/p99x/bertlarge}{1.2}
\defR{tight-x2.00/berth-full/thrabs/train_bert}{0.92}
\defR{tight-x2.00/berth-full/thr/train_bert}{57}
\defR{tight-x2.00/berth-nsr/attmin}{100.0}
\defR{tight-x2.00/berth-nsr/attmean}{100.0}
\defR{tight-x2.00/berth-nsr/thr}{57}
\defR{tight-x2.00/berth-nsr/n}{3}
\defR{tight-x2.00/berth-nsr/att/resnet50}{100.0}
\defR{tight-x2.00/berth-nsr/att/retinanet}{100.0}
\defR{tight-x2.00/berth-nsr/att/bertlarge}{100.0}
\defR{tight-x2.00/berth-nsr/p99/resnet50}{14.1}
\defR{tight-x2.00/berth-nsr/p99/retinanet}{40.4}
\defR{tight-x2.00/berth-nsr/p99/bertlarge}{92.7}
\defR{tight-x2.00/berth-nsr/p99x/resnet50}{1.1}
\defR{tight-x2.00/berth-nsr/p99x/retinanet}{1.2}
\defR{tight-x2.00/berth-nsr/p99x/bertlarge}{1.3}
\defR{tight-x2.00/berth-nsr/thrabs/train_bert}{0.92}
\defR{tight-x2.00/berth-nsr/thr/train_bert}{57}
\defR{tight-x2.00/berth-nba/attmin}{100.0}
\defR{tight-x2.00/berth-nba/attmean}{100.0}
\defR{tight-x2.00/berth-nba/thr}{57}
\defR{tight-x2.00/berth-nba/n}{3}
\defR{tight-x2.00/berth-nba/att/resnet50}{100.0}
\defR{tight-x2.00/berth-nba/att/retinanet}{100.0}
\defR{tight-x2.00/berth-nba/att/bertlarge}{100.0}
\defR{tight-x2.00/berth-nba/p99/resnet50}{14.2}
\defR{tight-x2.00/berth-nba/p99/retinanet}{39.3}
\defR{tight-x2.00/berth-nba/p99/bertlarge}{89.8}
\defR{tight-x2.00/berth-nba/p99x/resnet50}{1.1}
\defR{tight-x2.00/berth-nba/p99x/retinanet}{1.1}
\defR{tight-x2.00/berth-nba/p99x/bertlarge}{1.2}
\defR{tight-x2.00/berth-nba/thrabs/train_bert}{0.92}
\defR{tight-x2.00/berth-nba/thr/train_bert}{57}
\defR{tight-x2.00/berth-nsb/attmin}{100.0}
\defR{tight-x2.00/berth-nsb/attmean}{100.0}
\defR{tight-x2.00/berth-nsb/thr}{57}
\defR{tight-x2.00/berth-nsb/n}{3}
\defR{tight-x2.00/berth-nsb/att/resnet50}{100.0}
\defR{tight-x2.00/berth-nsb/att/retinanet}{100.0}
\defR{tight-x2.00/berth-nsb/att/bertlarge}{100.0}
\defR{tight-x2.00/berth-nsb/p99/resnet50}{13.9}
\defR{tight-x2.00/berth-nsb/p99/retinanet}{38.2}
\defR{tight-x2.00/berth-nsb/p99/bertlarge}{86.8}
\defR{tight-x2.00/berth-nsb/p99x/resnet50}{1.0}
\defR{tight-x2.00/berth-nsb/p99x/retinanet}{1.1}
\defR{tight-x2.00/berth-nsb/p99x/bertlarge}{1.2}
\defR{tight-x2.00/berth-nsb/thrabs/train_bert}{0.92}
\defR{tight-x2.00/berth-nsb/thr/train_bert}{57}
\defR{tight-x2.00/berth-ng/attmin}{100.0}
\defR{tight-x2.00/berth-ng/attmean}{100.0}
\defR{tight-x2.00/berth-ng/thr}{57}
\defR{tight-x2.00/berth-ng/n}{3}
\defR{tight-x2.00/berth-ng/att/resnet50}{100.0}
\defR{tight-x2.00/berth-ng/att/retinanet}{100.0}
\defR{tight-x2.00/berth-ng/att/bertlarge}{100.0}
\defR{tight-x2.00/berth-ng/p99/resnet50}{15.0}
\defR{tight-x2.00/berth-ng/p99/retinanet}{40.4}
\defR{tight-x2.00/berth-ng/p99/bertlarge}{100}
\defR{tight-x2.00/berth-ng/p99x/resnet50}{1.1}
\defR{tight-x2.00/berth-ng/p99x/retinanet}{1.2}
\defR{tight-x2.00/berth-ng/p99x/bertlarge}{1.4}
\defR{tight-x2.00/berth-ng/thrabs/train_bert}{0.93}
\defR{tight-x2.00/berth-ng/thr/train_bert}{57}
\defR{tight-x2.00/berth-nsng/attmin}{100.0}
\defR{tight-x2.00/berth-nsng/attmean}{100.0}
\defR{tight-x2.00/berth-nsng/thr}{57}
\defR{tight-x2.00/berth-nsng/n}{3}
\defR{tight-x2.00/berth-nsng/att/resnet50}{100.0}
\defR{tight-x2.00/berth-nsng/att/retinanet}{100.0}
\defR{tight-x2.00/berth-nsng/att/bertlarge}{100.0}
\defR{tight-x2.00/berth-nsng/p99/resnet50}{14.6}
\defR{tight-x2.00/berth-nsng/p99/retinanet}{41.6}
\defR{tight-x2.00/berth-nsng/p99/bertlarge}{103}
\defR{tight-x2.00/berth-nsng/p99x/resnet50}{1.1}
\defR{tight-x2.00/berth-nsng/p99x/retinanet}{1.2}
\defR{tight-x2.00/berth-nsng/p99x/bertlarge}{1.4}
\defR{tight-x2.00/berth-nsng/thrabs/train_bert}{0.92}
\defR{tight-x2.00/berth-nsng/thr/train_bert}{57}
\defR{tight-x2.00/berth-grouped/attmin}{99.9}
\defR{tight-x2.00/berth-grouped/attmean}{100.0}
\defR{tight-x2.00/berth-grouped/thr}{60}
\defR{tight-x2.00/berth-grouped/n}{3}
\defR{tight-x2.00/berth-grouped/att/resnet50}{99.9}
\defR{tight-x2.00/berth-grouped/att/retinanet}{100.0}
\defR{tight-x2.00/berth-grouped/att/bertlarge}{100.0}
\defR{tight-x2.00/berth-grouped/p99/resnet50}{21.7}
\defR{tight-x2.00/berth-grouped/p99/retinanet}{44.7}
\defR{tight-x2.00/berth-grouped/p99/bertlarge}{68.8}
\defR{tight-x2.00/berth-grouped/p99x/resnet50}{1.6}
\defR{tight-x2.00/berth-grouped/p99x/retinanet}{1.3}
\defR{tight-x2.00/berth-grouped/p99x/bertlarge}{1.0}
\defR{tight-x2.00/berth-grouped/thrabs/train_bert}{0.98}
\defR{tight-x2.00/berth-grouped/thr/train_bert}{60}
\defR{tight-x2.00/berth/attmin}{99.7}
\defR{tight-x2.00/berth/attmean}{99.9}
\defR{tight-x2.00/berth/thr}{58}
\defR{tight-x2.00/berth/n}{3}
\defR{tight-x2.00/berth/att/resnet50}{99.7}
\defR{tight-x2.00/berth/att/retinanet}{100.0}
\defR{tight-x2.00/berth/att/bertlarge}{100.0}
\defR{tight-x2.00/berth/p99/resnet50}{23.8}
\defR{tight-x2.00/berth/p99/retinanet}{53.5}
\defR{tight-x2.00/berth/p99/bertlarge}{99.2}
\defR{tight-x2.00/berth/p99x/resnet50}{1.8}
\defR{tight-x2.00/berth/p99x/retinanet}{1.5}
\defR{tight-x2.00/berth/p99x/bertlarge}{1.4}
\defR{tight-x2.00/berth/thrabs/train_bert}{0.94}
\defR{tight-x2.00/berth/thr/train_bert}{58}
\defR{tight-x2.00/timeslice/attmin}{93.9}
\defR{tight-x2.00/timeslice/attmean}{96.9}
\defR{tight-x2.00/timeslice/thr}{54}
\defR{tight-x2.00/timeslice/n}{3}
\defR{tight-x2.00/timeslice/att/resnet50}{93.9}
\defR{tight-x2.00/timeslice/att/retinanet}{99.9}
\defR{tight-x2.00/timeslice/att/bertlarge}{97.1}
\defR{tight-x2.00/timeslice/p99/resnet50}{43.3}
\defR{tight-x2.00/timeslice/p99/retinanet}{133}
\defR{tight-x2.00/timeslice/p99/bertlarge}{291}
\defR{tight-x2.00/timeslice/p99x/resnet50}{3.2}
\defR{tight-x2.00/timeslice/p99x/retinanet}{3.8}
\defR{tight-x2.00/timeslice/p99x/bertlarge}{4.0}
\defR{tight-x2.00/timeslice/thrabs/train_bert}{0.88}
\defR{tight-x2.00/timeslice/thr/train_bert}{54}
\defR{tight-x2.00/mps/attmin}{0.0}
\defR{tight-x2.00/mps/attmean}{0.5}
\defR{tight-x2.00/mps/thr}{80}
\defR{tight-x2.00/mps/n}{3}
\defR{tight-x2.00/mps/att/resnet50}{0.0}
\defR{tight-x2.00/mps/att/retinanet}{0.0}
\defR{tight-x2.00/mps/att/bertlarge}{1.4}
\defR{tight-x2.00/mps/p99/resnet50}{31465}
\defR{tight-x2.00/mps/p99/retinanet}{42458}
\defR{tight-x2.00/mps/p99/bertlarge}{6787}
\defR{tight-x2.00/mps/p99x/resnet50}{2348.6}
\defR{tight-x2.00/mps/p99x/retinanet}{1220.8}
\defR{tight-x2.00/mps/p99x/bertlarge}{93.8}
\defR{tight-x2.00/mps/thrabs/train_bert}{1.30}
\defR{tight-x2.00/mps/thr/train_bert}{80}
\defR{tight-x2.00/mpsprio/attmin}{48.4}
\defR{tight-x2.00/mpsprio/attmean}{76.9}
\defR{tight-x2.00/mpsprio/thr}{70}
\defR{tight-x2.00/mpsprio/n}{3}
\defR{tight-x2.00/mpsprio/att/resnet50}{48.4}
\defR{tight-x2.00/mpsprio/att/retinanet}{83.4}
\defR{tight-x2.00/mpsprio/att/bertlarge}{98.9}
\defR{tight-x2.00/mpsprio/p99/resnet50}{143}
\defR{tight-x2.00/mpsprio/p99/retinanet}{422}
\defR{tight-x2.00/mpsprio/p99/bertlarge}{263}
\defR{tight-x2.00/mpsprio/p99x/resnet50}{10.7}
\defR{tight-x2.00/mpsprio/p99x/retinanet}{12.1}
\defR{tight-x2.00/mpsprio/p99x/bertlarge}{3.6}
\defR{tight-x2.00/mpsprio/thrabs/train_bert}{1.13}
\defR{tight-x2.00/mpsprio/thr/train_bert}{70}
\defR{tight-x2.00/mpslimits/attmin}{99.9}
\defR{tight-x2.00/mpslimits/attmean}{100.0}
\defR{tight-x2.00/mpslimits/thr}{12}
\defR{tight-x2.00/mpslimits/n}{3}
\defR{tight-x2.00/mpslimits/att/resnet50}{100.0}
\defR{tight-x2.00/mpslimits/att/retinanet}{100.0}
\defR{tight-x2.00/mpslimits/att/bertlarge}{99.9}
\defR{tight-x2.00/mpslimits/p99/resnet50}{14.7}
\defR{tight-x2.00/mpslimits/p99/retinanet}{111}
\defR{tight-x2.00/mpslimits/p99/bertlarge}{219}
\defR{tight-x2.00/mpslimits/p99x/resnet50}{1.1}
\defR{tight-x2.00/mpslimits/p99x/retinanet}{3.2}
\defR{tight-x2.00/mpslimits/p99x/bertlarge}{3.0}
\defR{tight-x2.00/mpslimits/thrabs/train_bert}{0.19}
\defR{tight-x2.00/mpslimits/thr/train_bert}{12}
\defR{tight-x2.00/mpsparts/attmin}{99.9}
\defR{tight-x2.00/mpsparts/attmean}{100.0}
\defR{tight-x2.00/mpsparts/thr}{10}
\defR{tight-x2.00/mpsparts/n}{2}
\defR{tight-x2.00/mpsparts/att/resnet50}{100.0}
\defR{tight-x2.00/mpsparts/att/retinanet}{100.0}
\defR{tight-x2.00/mpsparts/att/bertlarge}{99.9}
\defR{tight-x2.00/mpsparts/p99/resnet50}{15.2}
\defR{tight-x2.00/mpsparts/p99/retinanet}{109}
\defR{tight-x2.00/mpsparts/p99/bertlarge}{172}
\defR{tight-x2.00/mpsparts/p99x/resnet50}{1.1}
\defR{tight-x2.00/mpsparts/p99x/retinanet}{3.1}
\defR{tight-x2.00/mpsparts/p99x/bertlarge}{2.4}
\defR{tight-x2.00/mpsparts/thrabs/train_bert}{0.16}
\defR{tight-x2.00/mpsparts/thr/train_bert}{10}
\defR{tight-x2.00/lithos/attmin}{99.9}
\defR{tight-x2.00/lithos/attmean}{100.0}
\defR{tight-x2.00/lithos/thr}{9}
\defR{tight-x2.00/lithos/n}{3}
\defR{tight-x2.00/lithos/att/resnet50}{100.0}
\defR{tight-x2.00/lithos/att/retinanet}{100.0}
\defR{tight-x2.00/lithos/att/bertlarge}{99.9}
\defR{tight-x2.00/lithos/p99/resnet50}{14.7}
\defR{tight-x2.00/lithos/p99/retinanet}{116}
\defR{tight-x2.00/lithos/p99/bertlarge}{209}
\defR{tight-x2.00/lithos/p99x/resnet50}{1.1}
\defR{tight-x2.00/lithos/p99x/retinanet}{3.3}
\defR{tight-x2.00/lithos/p99x/bertlarge}{2.9}
\defR{tight-x2.00/lithos/thrabs/train_bert}{0.15}
\defR{tight-x2.00/lithos/thr/train_bert}{9}
\defR{tight-x2.00/dysm/attmin}{0.0}
\defR{tight-x2.00/dysm/attmean}{46.3}
\defR{tight-x2.00/dysm/thr}{74}
\defR{tight-x2.00/dysm/n}{3}
\defR{tight-x2.00/dysm/att/resnet50}{81.0}
\defR{tight-x2.00/dysm/att/retinanet}{57.9}
\defR{tight-x2.00/dysm/att/bertlarge}{0.0}
\defR{tight-x2.00/dysm/p99/resnet50}{69.1}
\defR{tight-x2.00/dysm/p99/retinanet}{1884}
\defR{tight-x2.00/dysm/p99x/resnet50}{5.2}
\defR{tight-x2.00/dysm/p99x/retinanet}{54.2}
\defR{tight-x2.00/dysm/thrabs/train_bert}{1.19}
\defR{tight-x2.00/dysm/thr/train_bert}{74}
\defR{tight-x2.00/xsched/attmin}{99.1}
\defR{tight-x2.00/xsched/attmean}{99.7}
\defR{tight-x2.00/xsched/thr}{57}
\defR{tight-x2.00/xsched/n}{3}
\defR{tight-x2.00/xsched/att/resnet50}{99.1}
\defR{tight-x2.00/xsched/att/retinanet}{100.0}
\defR{tight-x2.00/xsched/att/bertlarge}{100.0}
\defR{tight-x2.00/xsched/p99/resnet50}{28.7}
\defR{tight-x2.00/xsched/p99/retinanet}{57.4}
\defR{tight-x2.00/xsched/p99/bertlarge}{102}
\defR{tight-x2.00/xsched/p99x/resnet50}{2.1}
\defR{tight-x2.00/xsched/p99x/retinanet}{1.6}
\defR{tight-x2.00/xsched/p99x/bertlarge}{1.4}
\defR{tight-x2.00/xsched/thrabs/train_bert}{0.92}
\defR{tight-x2.00/xsched/thr/train_bert}{57}
\defR{tight-x2.00/multistream/attmin}{45.5}
\defR{tight-x2.00/multistream/attmean}{76.6}
\defR{tight-x2.00/multistream/thr}{71}
\defR{tight-x2.00/multistream/n}{3}
\defR{tight-x2.00/multistream/att/resnet50}{45.5}
\defR{tight-x2.00/multistream/att/retinanet}{85.3}
\defR{tight-x2.00/multistream/att/bertlarge}{98.9}
\defR{tight-x2.00/multistream/p99/resnet50}{159}
\defR{tight-x2.00/multistream/p99/retinanet}{393}
\defR{tight-x2.00/multistream/p99/bertlarge}{260}
\defR{tight-x2.00/multistream/p99x/resnet50}{11.9}
\defR{tight-x2.00/multistream/p99x/retinanet}{11.3}
\defR{tight-x2.00/multistream/p99x/bertlarge}{3.6}
\defR{tight-x2.00/multistream/thrabs/train_bert}{1.14}
\defR{tight-x2.00/multistream/thr/train_bert}{71}
\defR{tight-x2.00/tgs/attmin}{99.9}
\defR{tight-x2.00/tgs/attmean}{100.0}
\defR{tight-x2.00/tgs/thr}{2}
\defR{tight-x2.00/tgs/n}{3}
\defR{tight-x2.00/tgs/att/resnet50}{99.9}
\defR{tight-x2.00/tgs/att/retinanet}{100.0}
\defR{tight-x2.00/tgs/att/bertlarge}{100.0}
\defR{tight-x2.00/tgs/p99/resnet50}{19.9}
\defR{tight-x2.00/tgs/p99/retinanet}{40.1}
\defR{tight-x2.00/tgs/p99/bertlarge}{82.2}
\defR{tight-x2.00/tgs/p99x/resnet50}{1.5}
\defR{tight-x2.00/tgs/p99x/retinanet}{1.2}
\defR{tight-x2.00/tgs/p99x/bertlarge}{1.1}
\defR{tight-x2.00/tgs/thrabs/train_bert}{0.03}
\defR{tight-x2.00/tgs/thr/train_bert}{2}
\defR{tight-x2.00/mig/attmin}{99.9}
\defR{tight-x2.00/mig/attmean}{100.0}
\defR{tight-x2.00/mig/thr}{0}
\defR{tight-x2.00/mig/n}{3}
\defR{tight-x2.00/mig/att/resnet50}{100.0}
\defR{tight-x2.00/mig/att/retinanet}{100.0}
\defR{tight-x2.00/mig/att/bertlarge}{99.9}
\defR{tight-x2.00/mig/p99/resnet50}{16.5}
\defR{tight-x2.00/mig/p99/retinanet}{109}
\defR{tight-x2.00/mig/p99/bertlarge}{207}
\defR{tight-x2.00/mig/p99x/resnet50}{1.2}
\defR{tight-x2.00/mig/p99x/retinanet}{3.1}
\defR{tight-x2.00/mig/p99x/bertlarge}{2.9}
\defR{tight-x2.00/mig/thrabs/train_bert}{0.00}
\defR{tight-x2.00/mig/thr/train_bert}{0}
\defR{tight-x2.00/tally/attmin}{0.0}
\defR{tight-x2.00/tally/attmean}{34.2}
\defR{tight-x2.00/tally/thr}{27}
\defR{tight-x2.00/tally/n}{3}
\defR{tight-x2.00/tally/att/resnet50}{0.0}
\defR{tight-x2.00/tally/att/retinanet}{2.6}
\defR{tight-x2.00/tally/att/bertlarge}{100.0}
\defR{tight-x2.00/tally/p99/resnet50}{94201}
\defR{tight-x2.00/tally/p99/retinanet}{4674}
\defR{tight-x2.00/tally/p99/bertlarge}{152}
\defR{tight-x2.00/tally/p99x/resnet50}{7031.3}
\defR{tight-x2.00/tally/p99x/retinanet}{134.4}
\defR{tight-x2.00/tally/p99x/bertlarge}{2.1}
\defR{tight-x2.00/tally/thrabs/train_bert}{0.43}
\defR{tight-x2.00/tally/thr/train_bert}{27}
\defR{tight-x2.00/orion/attmin}{0.0}
\defR{tight-x2.00/orion/attmean}{45.4}
\defR{tight-x2.00/orion/thr}{73}
\defR{tight-x2.00/orion/n}{3}
\defR{tight-x2.00/orion/att/resnet50}{99.5}
\defR{tight-x2.00/orion/att/retinanet}{0.0}
\defR{tight-x2.00/orion/att/bertlarge}{36.8}
\defR{tight-x2.00/orion/p99/resnet50}{27.1}
\defR{tight-x2.00/orion/p99/retinanet}{70359}
\defR{tight-x2.00/orion/p99/bertlarge}{1613}
\defR{tight-x2.00/orion/p99x/resnet50}{2.0}
\defR{tight-x2.00/orion/p99x/retinanet}{2023.1}
\defR{tight-x2.00/orion/p99x/bertlarge}{22.3}
\defR{tight-x2.00/orion/thrabs/train_bert}{1.18}
\defR{tight-x2.00/orion/thr/train_bert}{73}
\defR{tight-x3.00/berth-full/attmin}{100.0}
\defR{tight-x3.00/berth-full/attmean}{100.0}
\defR{tight-x3.00/berth-full/thr}{57}
\defR{tight-x3.00/berth-full/n}{3}
\defR{tight-x3.00/berth-full/att/resnet50}{100.0}
\defR{tight-x3.00/berth-full/att/retinanet}{100.0}
\defR{tight-x3.00/berth-full/att/bertlarge}{100.0}
\defR{tight-x3.00/berth-full/p99/resnet50}{14.4}
\defR{tight-x3.00/berth-full/p99/retinanet}{41.6}
\defR{tight-x3.00/berth-full/p99/bertlarge}{95.1}
\defR{tight-x3.00/berth-full/p99x/resnet50}{1.1}
\defR{tight-x3.00/berth-full/p99x/retinanet}{1.2}
\defR{tight-x3.00/berth-full/p99x/bertlarge}{1.3}
\defR{tight-x3.00/berth-full/thrabs/train_bert}{0.92}
\defR{tight-x3.00/berth-full/thr/train_bert}{57}
\defR{tight-x3.00/berth-nsr/attmin}{100.0}
\defR{tight-x3.00/berth-nsr/attmean}{100.0}
\defR{tight-x3.00/berth-nsr/thr}{56}
\defR{tight-x3.00/berth-nsr/n}{3}
\defR{tight-x3.00/berth-nsr/att/resnet50}{100.0}
\defR{tight-x3.00/berth-nsr/att/retinanet}{100.0}
\defR{tight-x3.00/berth-nsr/att/bertlarge}{100.0}
\defR{tight-x3.00/berth-nsr/p99/resnet50}{14.2}
\defR{tight-x3.00/berth-nsr/p99/retinanet}{38.1}
\defR{tight-x3.00/berth-nsr/p99/bertlarge}{93.4}
\defR{tight-x3.00/berth-nsr/p99x/resnet50}{1.1}
\defR{tight-x3.00/berth-nsr/p99x/retinanet}{1.1}
\defR{tight-x3.00/berth-nsr/p99x/bertlarge}{1.3}
\defR{tight-x3.00/berth-nsr/thrabs/train_bert}{0.91}
\defR{tight-x3.00/berth-nsr/thr/train_bert}{56}
\defR{tight-x3.00/berth-nba/attmin}{100.0}
\defR{tight-x3.00/berth-nba/attmean}{100.0}
\defR{tight-x3.00/berth-nba/thr}{57}
\defR{tight-x3.00/berth-nba/n}{3}
\defR{tight-x3.00/berth-nba/att/resnet50}{100.0}
\defR{tight-x3.00/berth-nba/att/retinanet}{100.0}
\defR{tight-x3.00/berth-nba/att/bertlarge}{100.0}
\defR{tight-x3.00/berth-nba/p99/resnet50}{14.0}
\defR{tight-x3.00/berth-nba/p99/retinanet}{38.6}
\defR{tight-x3.00/berth-nba/p99/bertlarge}{89.6}
\defR{tight-x3.00/berth-nba/p99x/resnet50}{1.0}
\defR{tight-x3.00/berth-nba/p99x/retinanet}{1.1}
\defR{tight-x3.00/berth-nba/p99x/bertlarge}{1.2}
\defR{tight-x3.00/berth-nba/thrabs/train_bert}{0.92}
\defR{tight-x3.00/berth-nba/thr/train_bert}{57}
\defR{tight-x3.00/berth-nsb/attmin}{100.0}
\defR{tight-x3.00/berth-nsb/attmean}{100.0}
\defR{tight-x3.00/berth-nsb/thr}{57}
\defR{tight-x3.00/berth-nsb/n}{3}
\defR{tight-x3.00/berth-nsb/att/resnet50}{100.0}
\defR{tight-x3.00/berth-nsb/att/retinanet}{100.0}
\defR{tight-x3.00/berth-nsb/att/bertlarge}{100.0}
\defR{tight-x3.00/berth-nsb/p99/resnet50}{14.0}
\defR{tight-x3.00/berth-nsb/p99/retinanet}{37.8}
\defR{tight-x3.00/berth-nsb/p99/bertlarge}{90.3}
\defR{tight-x3.00/berth-nsb/p99x/resnet50}{1.0}
\defR{tight-x3.00/berth-nsb/p99x/retinanet}{1.1}
\defR{tight-x3.00/berth-nsb/p99x/bertlarge}{1.2}
\defR{tight-x3.00/berth-nsb/thrabs/train_bert}{0.92}
\defR{tight-x3.00/berth-nsb/thr/train_bert}{57}
\defR{tight-x3.00/berth-ng/attmin}{100.0}
\defR{tight-x3.00/berth-ng/attmean}{100.0}
\defR{tight-x3.00/berth-ng/thr}{57}
\defR{tight-x3.00/berth-ng/n}{3}
\defR{tight-x3.00/berth-ng/att/resnet50}{100.0}
\defR{tight-x3.00/berth-ng/att/retinanet}{100.0}
\defR{tight-x3.00/berth-ng/att/bertlarge}{100.0}
\defR{tight-x3.00/berth-ng/p99/resnet50}{14.5}
\defR{tight-x3.00/berth-ng/p99/retinanet}{42.2}
\defR{tight-x3.00/berth-ng/p99/bertlarge}{111}
\defR{tight-x3.00/berth-ng/p99x/resnet50}{1.1}
\defR{tight-x3.00/berth-ng/p99x/retinanet}{1.2}
\defR{tight-x3.00/berth-ng/p99x/bertlarge}{1.5}
\defR{tight-x3.00/berth-ng/thrabs/train_bert}{0.92}
\defR{tight-x3.00/berth-ng/thr/train_bert}{57}
\defR{tight-x3.00/berth-nsng/attmin}{100.0}
\defR{tight-x3.00/berth-nsng/attmean}{100.0}
\defR{tight-x3.00/berth-nsng/thr}{57}
\defR{tight-x3.00/berth-nsng/n}{3}
\defR{tight-x3.00/berth-nsng/att/resnet50}{100.0}
\defR{tight-x3.00/berth-nsng/att/retinanet}{100.0}
\defR{tight-x3.00/berth-nsng/att/bertlarge}{100.0}
\defR{tight-x3.00/berth-nsng/p99/resnet50}{14.6}
\defR{tight-x3.00/berth-nsng/p99/retinanet}{41.4}
\defR{tight-x3.00/berth-nsng/p99/bertlarge}{108}
\defR{tight-x3.00/berth-nsng/p99x/resnet50}{1.1}
\defR{tight-x3.00/berth-nsng/p99x/retinanet}{1.2}
\defR{tight-x3.00/berth-nsng/p99x/bertlarge}{1.5}
\defR{tight-x3.00/berth-nsng/thrabs/train_bert}{0.92}
\defR{tight-x3.00/berth-nsng/thr/train_bert}{57}
\defR{tight-x3.00/berth-grouped/attmin}{100.0}
\defR{tight-x3.00/berth-grouped/attmean}{100.0}
\defR{tight-x3.00/berth-grouped/thr}{61}
\defR{tight-x3.00/berth-grouped/n}{3}
\defR{tight-x3.00/berth-grouped/att/resnet50}{100.0}
\defR{tight-x3.00/berth-grouped/att/retinanet}{100.0}
\defR{tight-x3.00/berth-grouped/att/bertlarge}{100.0}
\defR{tight-x3.00/berth-grouped/p99/resnet50}{22.1}
\defR{tight-x3.00/berth-grouped/p99/retinanet}{43.5}
\defR{tight-x3.00/berth-grouped/p99/bertlarge}{67.5}
\defR{tight-x3.00/berth-grouped/p99x/resnet50}{1.6}
\defR{tight-x3.00/berth-grouped/p99x/retinanet}{1.3}
\defR{tight-x3.00/berth-grouped/p99x/bertlarge}{0.9}
\defR{tight-x3.00/berth-grouped/thrabs/train_bert}{0.99}
\defR{tight-x3.00/berth-grouped/thr/train_bert}{61}
\defR{tight-x3.00/berth/attmin}{100.0}
\defR{tight-x3.00/berth/attmean}{100.0}
\defR{tight-x3.00/berth/thr}{58}
\defR{tight-x3.00/berth/n}{3}
\defR{tight-x3.00/berth/att/resnet50}{100.0}
\defR{tight-x3.00/berth/att/retinanet}{100.0}
\defR{tight-x3.00/berth/att/bertlarge}{100.0}
\defR{tight-x3.00/berth/p99/resnet50}{23.1}
\defR{tight-x3.00/berth/p99/retinanet}{51.4}
\defR{tight-x3.00/berth/p99/bertlarge}{101}
\defR{tight-x3.00/berth/p99x/resnet50}{1.7}
\defR{tight-x3.00/berth/p99x/retinanet}{1.5}
\defR{tight-x3.00/berth/p99x/bertlarge}{1.4}
\defR{tight-x3.00/berth/thrabs/train_bert}{0.94}
\defR{tight-x3.00/berth/thr/train_bert}{58}
\defR{tight-x3.00/timeslice/attmin}{99.5}
\defR{tight-x3.00/timeslice/attmean}{99.8}
\defR{tight-x3.00/timeslice/thr}{55}
\defR{tight-x3.00/timeslice/n}{3}
\defR{tight-x3.00/timeslice/att/resnet50}{99.5}
\defR{tight-x3.00/timeslice/att/retinanet}{100.0}
\defR{tight-x3.00/timeslice/att/bertlarge}{99.9}
\defR{tight-x3.00/timeslice/p99/resnet50}{40.7}
\defR{tight-x3.00/timeslice/p99/retinanet}{130}
\defR{tight-x3.00/timeslice/p99/bertlarge}{291}
\defR{tight-x3.00/timeslice/p99x/resnet50}{3.0}
\defR{tight-x3.00/timeslice/p99x/retinanet}{3.7}
\defR{tight-x3.00/timeslice/p99x/bertlarge}{4.0}
\defR{tight-x3.00/timeslice/thrabs/train_bert}{0.89}
\defR{tight-x3.00/timeslice/thr/train_bert}{55}
\defR{tight-x3.00/mps/attmin}{0.0}
\defR{tight-x3.00/mps/attmean}{0.7}
\defR{tight-x3.00/mps/thr}{80}
\defR{tight-x3.00/mps/n}{3}
\defR{tight-x3.00/mps/att/resnet50}{0.0}
\defR{tight-x3.00/mps/att/retinanet}{0.0}
\defR{tight-x3.00/mps/att/bertlarge}{2.0}
\defR{tight-x3.00/mps/p99/resnet50}{31456}
\defR{tight-x3.00/mps/p99/retinanet}{42381}
\defR{tight-x3.00/mps/p99/bertlarge}{6772}
\defR{tight-x3.00/mps/p99x/resnet50}{2347.9}
\defR{tight-x3.00/mps/p99x/retinanet}{1218.7}
\defR{tight-x3.00/mps/p99x/bertlarge}{93.6}
\defR{tight-x3.00/mps/thrabs/train_bert}{1.30}
\defR{tight-x3.00/mps/thr/train_bert}{80}
\defR{tight-x3.00/mpsprio/attmin}{69.8}
\defR{tight-x3.00/mpsprio/attmean}{88.5}
\defR{tight-x3.00/mpsprio/thr}{70}
\defR{tight-x3.00/mpsprio/n}{3}
\defR{tight-x3.00/mpsprio/att/resnet50}{69.8}
\defR{tight-x3.00/mpsprio/att/retinanet}{95.9}
\defR{tight-x3.00/mpsprio/att/bertlarge}{99.9}
\defR{tight-x3.00/mpsprio/p99/resnet50}{137}
\defR{tight-x3.00/mpsprio/p99/retinanet}{414}
\defR{tight-x3.00/mpsprio/p99/bertlarge}{263}
\defR{tight-x3.00/mpsprio/p99x/resnet50}{10.3}
\defR{tight-x3.00/mpsprio/p99x/retinanet}{11.9}
\defR{tight-x3.00/mpsprio/p99x/bertlarge}{3.6}
\defR{tight-x3.00/mpsprio/thrabs/train_bert}{1.14}
\defR{tight-x3.00/mpsprio/thr/train_bert}{70}
\defR{tight-x3.00/mpslimits/attmin}{99.9}
\defR{tight-x3.00/mpslimits/attmean}{100.0}
\defR{tight-x3.00/mpslimits/thr}{25}
\defR{tight-x3.00/mpslimits/n}{3}
\defR{tight-x3.00/mpslimits/att/resnet50}{100.0}
\defR{tight-x3.00/mpslimits/att/retinanet}{100.0}
\defR{tight-x3.00/mpslimits/att/bertlarge}{99.9}
\defR{tight-x3.00/mpslimits/p99/resnet50}{19.1}
\defR{tight-x3.00/mpslimits/p99/retinanet}{115}
\defR{tight-x3.00/mpslimits/p99/bertlarge}{274}
\defR{tight-x3.00/mpslimits/p99x/resnet50}{1.4}
\defR{tight-x3.00/mpslimits/p99x/retinanet}{3.3}
\defR{tight-x3.00/mpslimits/p99x/bertlarge}{3.8}
\defR{tight-x3.00/mpslimits/thrabs/train_bert}{0.40}
\defR{tight-x3.00/mpslimits/thr/train_bert}{25}
\defR{tight-x3.00/mpsparts/attmin}{99.9}
\defR{tight-x3.00/mpsparts/attmean}{100.0}
\defR{tight-x3.00/mpsparts/thr}{27}
\defR{tight-x3.00/mpsparts/n}{2}
\defR{tight-x3.00/mpsparts/att/resnet50}{100.0}
\defR{tight-x3.00/mpsparts/att/retinanet}{100.0}
\defR{tight-x3.00/mpsparts/att/bertlarge}{99.9}
\defR{tight-x3.00/mpsparts/p99/resnet50}{17.7}
\defR{tight-x3.00/mpsparts/p99/retinanet}{119}
\defR{tight-x3.00/mpsparts/p99/bertlarge}{272}
\defR{tight-x3.00/mpsparts/p99x/resnet50}{1.3}
\defR{tight-x3.00/mpsparts/p99x/retinanet}{3.4}
\defR{tight-x3.00/mpsparts/p99x/bertlarge}{3.8}
\defR{tight-x3.00/mpsparts/thrabs/train_bert}{0.43}
\defR{tight-x3.00/mpsparts/thr/train_bert}{27}
\defR{tight-x3.00/lithos/attmin}{99.9}
\defR{tight-x3.00/lithos/attmean}{100.0}
\defR{tight-x3.00/lithos/thr}{26}
\defR{tight-x3.00/lithos/n}{3}
\defR{tight-x3.00/lithos/att/resnet50}{100.0}
\defR{tight-x3.00/lithos/att/retinanet}{100.0}
\defR{tight-x3.00/lithos/att/bertlarge}{99.9}
\defR{tight-x3.00/lithos/p99/resnet50}{20.9}
\defR{tight-x3.00/lithos/p99/retinanet}{119}
\defR{tight-x3.00/lithos/p99/bertlarge}{267}
\defR{tight-x3.00/lithos/p99x/resnet50}{1.6}
\defR{tight-x3.00/lithos/p99x/retinanet}{3.4}
\defR{tight-x3.00/lithos/p99x/bertlarge}{3.7}
\defR{tight-x3.00/lithos/thrabs/train_bert}{0.43}
\defR{tight-x3.00/lithos/thr/train_bert}{26}
\defR{tight-x3.00/dysm/attmin}{53.7}
\defR{tight-x3.00/dysm/attmean}{63.6}
\defR{tight-x3.00/dysm/thr}{61}
\defR{tight-x3.00/dysm/n}{3}
\defR{tight-x3.00/dysm/att/resnet50}{73.6}
\defR{tight-x3.00/dysm/att/retinanet}{53.7}
\defR{tight-x3.00/dysm/att/bertlarge}{63.4}
\defR{tight-x3.00/dysm/p99/resnet50}{167}
\defR{tight-x3.00/dysm/p99/retinanet}{1917}
\defR{tight-x3.00/dysm/p99/bertlarge}{1520}
\defR{tight-x3.00/dysm/p99x/resnet50}{12.4}
\defR{tight-x3.00/dysm/p99x/retinanet}{55.1}
\defR{tight-x3.00/dysm/p99x/bertlarge}{21.0}
\defR{tight-x3.00/dysm/thrabs/train_bert}{0.99}
\defR{tight-x3.00/dysm/thr/train_bert}{61}
\defR{tight-x3.00/xsched/attmin}{99.9}
\defR{tight-x3.00/xsched/attmean}{100.0}
\defR{tight-x3.00/xsched/thr}{57}
\defR{tight-x3.00/xsched/n}{3}
\defR{tight-x3.00/xsched/att/resnet50}{99.9}
\defR{tight-x3.00/xsched/att/retinanet}{100.0}
\defR{tight-x3.00/xsched/att/bertlarge}{100.0}
\defR{tight-x3.00/xsched/p99/resnet50}{28.5}
\defR{tight-x3.00/xsched/p99/retinanet}{60.1}
\defR{tight-x3.00/xsched/p99/bertlarge}{98.1}
\defR{tight-x3.00/xsched/p99x/resnet50}{2.1}
\defR{tight-x3.00/xsched/p99x/retinanet}{1.7}
\defR{tight-x3.00/xsched/p99x/bertlarge}{1.4}
\defR{tight-x3.00/xsched/thrabs/train_bert}{0.92}
\defR{tight-x3.00/xsched/thr/train_bert}{57}
\defR{tight-x3.00/multistream/attmin}{64.5}
\defR{tight-x3.00/multistream/attmean}{87.1}
\defR{tight-x3.00/multistream/thr}{71}
\defR{tight-x3.00/multistream/n}{3}
\defR{tight-x3.00/multistream/att/resnet50}{64.5}
\defR{tight-x3.00/multistream/att/retinanet}{96.8}
\defR{tight-x3.00/multistream/att/bertlarge}{99.9}
\defR{tight-x3.00/multistream/p99/resnet50}{156}
\defR{tight-x3.00/multistream/p99/retinanet}{382}
\defR{tight-x3.00/multistream/p99/bertlarge}{256}
\defR{tight-x3.00/multistream/p99x/resnet50}{11.6}
\defR{tight-x3.00/multistream/p99x/retinanet}{11.0}
\defR{tight-x3.00/multistream/p99x/bertlarge}{3.5}
\defR{tight-x3.00/multistream/thrabs/train_bert}{1.15}
\defR{tight-x3.00/multistream/thr/train_bert}{71}
\defR{tight-x3.00/tgs/attmin}{100.0}
\defR{tight-x3.00/tgs/attmean}{100.0}
\defR{tight-x3.00/tgs/thr}{2}
\defR{tight-x3.00/tgs/n}{3}
\defR{tight-x3.00/tgs/att/resnet50}{100.0}
\defR{tight-x3.00/tgs/att/retinanet}{100.0}
\defR{tight-x3.00/tgs/att/bertlarge}{100.0}
\defR{tight-x3.00/tgs/p99/resnet50}{20.0}
\defR{tight-x3.00/tgs/p99/retinanet}{40.1}
\defR{tight-x3.00/tgs/p99/bertlarge}{81.6}
\defR{tight-x3.00/tgs/p99x/resnet50}{1.5}
\defR{tight-x3.00/tgs/p99x/retinanet}{1.2}
\defR{tight-x3.00/tgs/p99x/bertlarge}{1.1}
\defR{tight-x3.00/tgs/thrabs/train_bert}{0.03}
\defR{tight-x3.00/tgs/thr/train_bert}{2}
\defR{tight-x3.00/mig/attmin}{99.9}
\defR{tight-x3.00/mig/attmean}{100.0}
\defR{tight-x3.00/mig/thr}{0}
\defR{tight-x3.00/mig/n}{2}
\defR{tight-x3.00/mig/att/resnet50}{99.9}
\defR{tight-x3.00/mig/att/retinanet}{100.0}
\defR{tight-x3.00/mig/att/bertlarge}{100.0}
\defR{tight-x3.00/mig/p99/resnet50}{34.5}
\defR{tight-x3.00/mig/p99/retinanet}{109}
\defR{tight-x3.00/mig/p99/bertlarge}{207}
\defR{tight-x3.00/mig/p99x/resnet50}{2.6}
\defR{tight-x3.00/mig/p99x/retinanet}{3.1}
\defR{tight-x3.00/mig/p99x/bertlarge}{2.9}
\defR{tight-x3.00/mig/thrabs/train_bert}{0.00}
\defR{tight-x3.00/mig/thr/train_bert}{0}
\defR{tight-x3.00/tally/attmin}{0.0}
\defR{tight-x3.00/tally/attmean}{40.2}
\defR{tight-x3.00/tally/thr}{26}
\defR{tight-x3.00/tally/n}{3}
\defR{tight-x3.00/tally/att/resnet50}{0.0}
\defR{tight-x3.00/tally/att/retinanet}{20.7}
\defR{tight-x3.00/tally/att/bertlarge}{100.0}
\defR{tight-x3.00/tally/p99/resnet50}{94201}
\defR{tight-x3.00/tally/p99/retinanet}{2744}
\defR{tight-x3.00/tally/p99/bertlarge}{158}
\defR{tight-x3.00/tally/p99x/resnet50}{7031.3}
\defR{tight-x3.00/tally/p99x/retinanet}{78.9}
\defR{tight-x3.00/tally/p99x/bertlarge}{2.2}
\defR{tight-x3.00/tally/thrabs/train_bert}{0.41}
\defR{tight-x3.00/tally/thr/train_bert}{26}
\defR{tight-x3.00/orion/attmin}{0.0}
\defR{tight-x3.00/orion/attmean}{51.0}
\defR{tight-x3.00/orion/thr}{73}
\defR{tight-x3.00/orion/n}{3}
\defR{tight-x3.00/orion/att/resnet50}{100.0}
\defR{tight-x3.00/orion/att/retinanet}{0.0}
\defR{tight-x3.00/orion/att/bertlarge}{52.9}
\defR{tight-x3.00/orion/p99/resnet50}{27.7}
\defR{tight-x3.00/orion/p99/retinanet}{70517}
\defR{tight-x3.00/orion/p99/bertlarge}{1648}
\defR{tight-x3.00/orion/p99x/resnet50}{2.1}
\defR{tight-x3.00/orion/p99x/retinanet}{2027.7}
\defR{tight-x3.00/orion/p99x/bertlarge}{22.8}
\defR{tight-x3.00/orion/thrabs/train_bert}{1.18}
\defR{tight-x3.00/orion/thr/train_bert}{73}
\defR{pair-resnet50-load30-trainresnet/berth-full/attmin}{97.4}
\defR{pair-resnet50-load30-trainresnet/berth-full/attmean}{97.4}
\defR{pair-resnet50-load30-trainresnet/berth-full/thr}{64}
\defR{pair-resnet50-load30-trainresnet/berth-full/n}{3}
\defR{pair-resnet50-load30-trainresnet/berth-full/att/resnet50}{97.4}
\defR{pair-resnet50-load30-trainresnet/berth-full/p99/resnet50}{17.5}
\defR{pair-resnet50-load30-trainresnet/berth-full/p99x/resnet50}{1.0}
\defR{pair-resnet50-load30-trainresnet/berth-full/thrabs/train_resnet}{8.50}
\defR{pair-resnet50-load30-trainresnet/berth-full/thr/train_resnet}{64}
\defR{pair-resnet50-load30-trainresnet/berth-nsr/attmin}{97.9}
\defR{pair-resnet50-load30-trainresnet/berth-nsr/attmean}{97.9}
\defR{pair-resnet50-load30-trainresnet/berth-nsr/thr}{64}
\defR{pair-resnet50-load30-trainresnet/berth-nsr/n}{3}
\defR{pair-resnet50-load30-trainresnet/berth-nsr/att/resnet50}{97.9}
\defR{pair-resnet50-load30-trainresnet/berth-nsr/p99/resnet50}{16.9}
\defR{pair-resnet50-load30-trainresnet/berth-nsr/p99x/resnet50}{0.9}
\defR{pair-resnet50-load30-trainresnet/berth-nsr/thrabs/train_resnet}{8.56}
\defR{pair-resnet50-load30-trainresnet/berth-nsr/thr/train_resnet}{64}
\defR{pair-resnet50-load30-trainresnet/berth-nba/attmin}{97.8}
\defR{pair-resnet50-load30-trainresnet/berth-nba/attmean}{97.8}
\defR{pair-resnet50-load30-trainresnet/berth-nba/thr}{64}
\defR{pair-resnet50-load30-trainresnet/berth-nba/n}{3}
\defR{pair-resnet50-load30-trainresnet/berth-nba/att/resnet50}{97.8}
\defR{pair-resnet50-load30-trainresnet/berth-nba/p99/resnet50}{16.9}
\defR{pair-resnet50-load30-trainresnet/berth-nba/p99x/resnet50}{0.9}
\defR{pair-resnet50-load30-trainresnet/berth-nba/thrabs/train_resnet}{8.56}
\defR{pair-resnet50-load30-trainresnet/berth-nba/thr/train_resnet}{64}
\defR{pair-resnet50-load30-trainresnet/berth-nsb/attmin}{98.0}
\defR{pair-resnet50-load30-trainresnet/berth-nsb/attmean}{98.0}
\defR{pair-resnet50-load30-trainresnet/berth-nsb/thr}{65}
\defR{pair-resnet50-load30-trainresnet/berth-nsb/n}{3}
\defR{pair-resnet50-load30-trainresnet/berth-nsb/att/resnet50}{98.0}
\defR{pair-resnet50-load30-trainresnet/berth-nsb/p99/resnet50}{16.8}
\defR{pair-resnet50-load30-trainresnet/berth-nsb/p99x/resnet50}{0.9}
\defR{pair-resnet50-load30-trainresnet/berth-nsb/thrabs/train_resnet}{8.57}
\defR{pair-resnet50-load30-trainresnet/berth-nsb/thr/train_resnet}{65}
\defR{pair-resnet50-load30-trainresnet/berth-ng/attmin}{97.9}
\defR{pair-resnet50-load30-trainresnet/berth-ng/attmean}{97.9}
\defR{pair-resnet50-load30-trainresnet/berth-ng/thr}{66}
\defR{pair-resnet50-load30-trainresnet/berth-ng/n}{3}
\defR{pair-resnet50-load30-trainresnet/berth-ng/att/resnet50}{97.9}
\defR{pair-resnet50-load30-trainresnet/berth-ng/p99/resnet50}{16.8}
\defR{pair-resnet50-load30-trainresnet/berth-ng/p99x/resnet50}{0.9}
\defR{pair-resnet50-load30-trainresnet/berth-ng/thrabs/train_resnet}{8.83}
\defR{pair-resnet50-load30-trainresnet/berth-ng/thr/train_resnet}{66}
\defR{pair-resnet50-load30-trainresnet/berth-nsng/attmin}{98.0}
\defR{pair-resnet50-load30-trainresnet/berth-nsng/attmean}{98.0}
\defR{pair-resnet50-load30-trainresnet/berth-nsng/thr}{66}
\defR{pair-resnet50-load30-trainresnet/berth-nsng/n}{3}
\defR{pair-resnet50-load30-trainresnet/berth-nsng/att/resnet50}{98.0}
\defR{pair-resnet50-load30-trainresnet/berth-nsng/p99/resnet50}{16.6}
\defR{pair-resnet50-load30-trainresnet/berth-nsng/p99x/resnet50}{0.9}
\defR{pair-resnet50-load30-trainresnet/berth-nsng/thrabs/train_resnet}{8.81}
\defR{pair-resnet50-load30-trainresnet/berth-nsng/thr/train_resnet}{66}
\defR{pair-resnet50-load30-trainresnet/berth-grouped/attmin}{97.6}
\defR{pair-resnet50-load30-trainresnet/berth-grouped/attmean}{97.6}
\defR{pair-resnet50-load30-trainresnet/berth-grouped/thr}{67}
\defR{pair-resnet50-load30-trainresnet/berth-grouped/n}{3}
\defR{pair-resnet50-load30-trainresnet/berth-grouped/att/resnet50}{97.6}
\defR{pair-resnet50-load30-trainresnet/berth-grouped/p99/resnet50}{17.2}
\defR{pair-resnet50-load30-trainresnet/berth-grouped/p99x/resnet50}{0.9}
\defR{pair-resnet50-load30-trainresnet/berth-grouped/thrabs/train_resnet}{8.85}
\defR{pair-resnet50-load30-trainresnet/berth-grouped/thr/train_resnet}{67}
\defR{pair-resnet50-load30-trainresnet/berth/attmin}{97.5}
\defR{pair-resnet50-load30-trainresnet/berth/attmean}{97.5}
\defR{pair-resnet50-load30-trainresnet/berth/thr}{67}
\defR{pair-resnet50-load30-trainresnet/berth/n}{3}
\defR{pair-resnet50-load30-trainresnet/berth/att/resnet50}{97.5}
\defR{pair-resnet50-load30-trainresnet/berth/p99/resnet50}{17.3}
\defR{pair-resnet50-load30-trainresnet/berth/p99x/resnet50}{1.0}
\defR{pair-resnet50-load30-trainresnet/berth/thrabs/train_resnet}{8.85}
\defR{pair-resnet50-load30-trainresnet/berth/thr/train_resnet}{67}
\defR{pair-resnet50-load30-trainresnet/timeslice/attmin}{67.4}
\defR{pair-resnet50-load30-trainresnet/timeslice/attmean}{67.4}
\defR{pair-resnet50-load30-trainresnet/timeslice/thr}{70}
\defR{pair-resnet50-load30-trainresnet/timeslice/n}{3}
\defR{pair-resnet50-load30-trainresnet/timeslice/att/resnet50}{67.4}
\defR{pair-resnet50-load30-trainresnet/timeslice/p99/resnet50}{42.4}
\defR{pair-resnet50-load30-trainresnet/timeslice/p99x/resnet50}{2.3}
\defR{pair-resnet50-load30-trainresnet/timeslice/thrabs/train_resnet}{9.35}
\defR{pair-resnet50-load30-trainresnet/timeslice/thr/train_resnet}{70}
\defR{pair-resnet50-load30-trainresnet/mps/attmin}{6.5}
\defR{pair-resnet50-load30-trainresnet/mps/attmean}{6.5}
\defR{pair-resnet50-load30-trainresnet/mps/thr}{85}
\defR{pair-resnet50-load30-trainresnet/mps/n}{3}
\defR{pair-resnet50-load30-trainresnet/mps/att/resnet50}{6.5}
\defR{pair-resnet50-load30-trainresnet/mps/p99/resnet50}{284}
\defR{pair-resnet50-load30-trainresnet/mps/p99x/resnet50}{15.6}
\defR{pair-resnet50-load30-trainresnet/mps/thrabs/train_resnet}{11.31}
\defR{pair-resnet50-load30-trainresnet/mps/thr/train_resnet}{85}
\defR{pair-resnet50-load30-trainresnet/mpsprio/attmin}{60.7}
\defR{pair-resnet50-load30-trainresnet/mpsprio/attmean}{60.7}
\defR{pair-resnet50-load30-trainresnet/mpsprio/thr}{84}
\defR{pair-resnet50-load30-trainresnet/mpsprio/n}{3}
\defR{pair-resnet50-load30-trainresnet/mpsprio/att/resnet50}{60.7}
\defR{pair-resnet50-load30-trainresnet/mpsprio/p99/resnet50}{49.5}
\defR{pair-resnet50-load30-trainresnet/mpsprio/p99x/resnet50}{2.7}
\defR{pair-resnet50-load30-trainresnet/mpsprio/thrabs/train_resnet}{11.10}
\defR{pair-resnet50-load30-trainresnet/mpsprio/thr/train_resnet}{84}
\defR{pair-resnet50-load30-trainresnet/mpslimits/attmin}{98.5}
\defR{pair-resnet50-load30-trainresnet/mpslimits/attmean}{98.5}
\defR{pair-resnet50-load30-trainresnet/mpslimits/thr}{20}
\defR{pair-resnet50-load30-trainresnet/mpslimits/n}{3}
\defR{pair-resnet50-load30-trainresnet/mpslimits/att/resnet50}{98.5}
\defR{pair-resnet50-load30-trainresnet/mpslimits/p99/resnet50}{15.9}
\defR{pair-resnet50-load30-trainresnet/mpslimits/p99x/resnet50}{0.9}
\defR{pair-resnet50-load30-trainresnet/mpslimits/thrabs/train_resnet}{2.71}
\defR{pair-resnet50-load30-trainresnet/mpslimits/thr/train_resnet}{20}
\defR{pair-resnet50-load30-trainresnet/mpsparts/attmin}{98.2}
\defR{pair-resnet50-load30-trainresnet/mpsparts/attmean}{98.2}
\defR{pair-resnet50-load30-trainresnet/mpsparts/thr}{17}
\defR{pair-resnet50-load30-trainresnet/mpsparts/n}{2}
\defR{pair-resnet50-load30-trainresnet/mpsparts/att/resnet50}{98.2}
\defR{pair-resnet50-load30-trainresnet/mpsparts/p99/resnet50}{16.4}
\defR{pair-resnet50-load30-trainresnet/mpsparts/p99x/resnet50}{0.9}
\defR{pair-resnet50-load30-trainresnet/mpsparts/thrabs/train_resnet}{2.21}
\defR{pair-resnet50-load30-trainresnet/mpsparts/thr/train_resnet}{17}
\defR{pair-resnet50-load30-trainresnet/lithos/attmin}{87.7}
\defR{pair-resnet50-load30-trainresnet/lithos/attmean}{87.7}
\defR{pair-resnet50-load30-trainresnet/lithos/thr}{31}
\defR{pair-resnet50-load30-trainresnet/lithos/n}{3}
\defR{pair-resnet50-load30-trainresnet/lithos/att/resnet50}{87.7}
\defR{pair-resnet50-load30-trainresnet/lithos/p99/resnet50}{33.4}
\defR{pair-resnet50-load30-trainresnet/lithos/p99x/resnet50}{1.8}
\defR{pair-resnet50-load30-trainresnet/lithos/thrabs/train_resnet}{4.15}
\defR{pair-resnet50-load30-trainresnet/lithos/thr/train_resnet}{31}
\defR{pair-resnet50-load30-trainresnet/dysm/attmin}{0.0}
\defR{pair-resnet50-load30-trainresnet/dysm/attmean}{0.0}
\defR{pair-resnet50-load30-trainresnet/dysm/thr}{100}
\defR{pair-resnet50-load30-trainresnet/dysm/n}{3}
\defR{pair-resnet50-load30-trainresnet/dysm/att/resnet50}{0.0}
\defR{pair-resnet50-load30-trainresnet/dysm/thrabs/train_resnet}{13.25}
\defR{pair-resnet50-load30-trainresnet/dysm/thr/train_resnet}{100}
\defR{pair-resnet50-load30-trainresnet/xsched/attmin}{0.0}
\defR{pair-resnet50-load30-trainresnet/xsched/attmean}{0.0}
\defR{pair-resnet50-load30-trainresnet/xsched/thr}{0}
\defR{pair-resnet50-load30-trainresnet/xsched/n}{3}
\defR{pair-resnet50-load30-trainresnet/xsched/att/resnet50}{0.0}
\defR{pair-resnet50-load30-trainresnet/xsched/thrabs/train_resnet}{0.00}
\defR{pair-resnet50-load30-trainresnet/xsched/thr/train_resnet}{0}
\defR{pair-resnet50-load30-trainresnet/multistream/attmin}{62.7}
\defR{pair-resnet50-load30-trainresnet/multistream/attmean}{62.7}
\defR{pair-resnet50-load30-trainresnet/multistream/thr}{84}
\defR{pair-resnet50-load30-trainresnet/multistream/n}{3}
\defR{pair-resnet50-load30-trainresnet/multistream/att/resnet50}{62.7}
\defR{pair-resnet50-load30-trainresnet/multistream/p99/resnet50}{48.4}
\defR{pair-resnet50-load30-trainresnet/multistream/p99x/resnet50}{2.7}
\defR{pair-resnet50-load30-trainresnet/multistream/thrabs/train_resnet}{11.18}
\defR{pair-resnet50-load30-trainresnet/multistream/thr/train_resnet}{84}
\defR{pair-resnet50-load30-trainresnet/tgs/attmin}{97.7}
\defR{pair-resnet50-load30-trainresnet/tgs/attmean}{97.7}
\defR{pair-resnet50-load30-trainresnet/tgs/thr}{1}
\defR{pair-resnet50-load30-trainresnet/tgs/n}{3}
\defR{pair-resnet50-load30-trainresnet/tgs/att/resnet50}{97.7}
\defR{pair-resnet50-load30-trainresnet/tgs/p99/resnet50}{17.1}
\defR{pair-resnet50-load30-trainresnet/tgs/p99x/resnet50}{0.9}
\defR{pair-resnet50-load30-trainresnet/tgs/thrabs/train_resnet}{0.13}
\defR{pair-resnet50-load30-trainresnet/tgs/thr/train_resnet}{1}
\defR{pair-resnet50-load30-trainresnet/mig/attmin}{95.3}
\defR{pair-resnet50-load30-trainresnet/mig/attmean}{95.3}
\defR{pair-resnet50-load30-trainresnet/mig/thr}{51}
\defR{pair-resnet50-load30-trainresnet/mig/n}{3}
\defR{pair-resnet50-load30-trainresnet/mig/att/resnet50}{95.3}
\defR{pair-resnet50-load30-trainresnet/mig/p99/resnet50}{20.1}
\defR{pair-resnet50-load30-trainresnet/mig/p99x/resnet50}{1.1}
\defR{pair-resnet50-load30-trainresnet/mig/thrabs/train_resnet}{6.83}
\defR{pair-resnet50-load30-trainresnet/mig/thr/train_resnet}{51}
\defR{pair-resnet50-load30-trainresnet/tally/attmin}{0.0}
\defR{pair-resnet50-load30-trainresnet/tally/attmean}{0.0}
\defR{pair-resnet50-load30-trainresnet/tally/thr}{0}
\defR{pair-resnet50-load30-trainresnet/tally/n}{3}
\defR{pair-resnet50-load30-trainresnet/tally/att/resnet50}{0.0}
\defR{pair-resnet50-load30-trainresnet/tally/p99/resnet50}{63190}
\defR{pair-resnet50-load30-trainresnet/tally/p99x/resnet50}{3469.6}
\defR{pair-resnet50-load30-trainresnet/tally/thrabs/train_resnet}{0.00}
\defR{pair-resnet50-load30-trainresnet/tally/thr/train_resnet}{0}
\defR{pair-resnet50-load30-trainresnet/orion/attmin}{99.5}
\defR{pair-resnet50-load30-trainresnet/orion/attmean}{99.5}
\defR{pair-resnet50-load30-trainresnet/orion/thr}{71}
\defR{pair-resnet50-load30-trainresnet/orion/n}{3}
\defR{pair-resnet50-load30-trainresnet/orion/att/resnet50}{99.5}
\defR{pair-resnet50-load30-trainresnet/orion/p99/resnet50}{13.5}
\defR{pair-resnet50-load30-trainresnet/orion/p99x/resnet50}{0.7}
\defR{pair-resnet50-load30-trainresnet/orion/thrabs/train_resnet}{9.47}
\defR{pair-resnet50-load30-trainresnet/orion/thr/train_resnet}{71}
\defR{pair-resnet50-load30-gemm/berth-full/attmin}{95.1}
\defR{pair-resnet50-load30-gemm/berth-full/attmean}{95.1}
\defR{pair-resnet50-load30-gemm/berth-full/thr}{71}
\defR{pair-resnet50-load30-gemm/berth-full/n}{3}
\defR{pair-resnet50-load30-gemm/berth-full/att/resnet50}{95.1}
\defR{pair-resnet50-load30-gemm/berth-full/p99/resnet50}{20.4}
\defR{pair-resnet50-load30-gemm/berth-full/p99x/resnet50}{1.1}
\defR{pair-resnet50-load30-gemm/berth-full/thrabs/gemm}{11.30}
\defR{pair-resnet50-load30-gemm/berth-full/thr/gemm}{71}
\defR{pair-resnet50-load30-gemm/berth-nsr/attmin}{96.0}
\defR{pair-resnet50-load30-gemm/berth-nsr/attmean}{96.0}
\defR{pair-resnet50-load30-gemm/berth-nsr/thr}{72}
\defR{pair-resnet50-load30-gemm/berth-nsr/n}{3}
\defR{pair-resnet50-load30-gemm/berth-nsr/att/resnet50}{96.0}
\defR{pair-resnet50-load30-gemm/berth-nsr/p99/resnet50}{19.1}
\defR{pair-resnet50-load30-gemm/berth-nsr/p99x/resnet50}{1.0}
\defR{pair-resnet50-load30-gemm/berth-nsr/thrabs/gemm}{11.33}
\defR{pair-resnet50-load30-gemm/berth-nsr/thr/gemm}{72}
\defR{pair-resnet50-load30-gemm/berth-nba/attmin}{95.4}
\defR{pair-resnet50-load30-gemm/berth-nba/attmean}{95.4}
\defR{pair-resnet50-load30-gemm/berth-nba/thr}{72}
\defR{pair-resnet50-load30-gemm/berth-nba/n}{3}
\defR{pair-resnet50-load30-gemm/berth-nba/att/resnet50}{95.4}
\defR{pair-resnet50-load30-gemm/berth-nba/p99/resnet50}{19.7}
\defR{pair-resnet50-load30-gemm/berth-nba/p99x/resnet50}{1.1}
\defR{pair-resnet50-load30-gemm/berth-nba/thrabs/gemm}{11.33}
\defR{pair-resnet50-load30-gemm/berth-nba/thr/gemm}{72}
\defR{pair-resnet50-load30-gemm/berth-nsb/attmin}{95.8}
\defR{pair-resnet50-load30-gemm/berth-nsb/attmean}{95.8}
\defR{pair-resnet50-load30-gemm/berth-nsb/thr}{72}
\defR{pair-resnet50-load30-gemm/berth-nsb/n}{3}
\defR{pair-resnet50-load30-gemm/berth-nsb/att/resnet50}{95.8}
\defR{pair-resnet50-load30-gemm/berth-nsb/p99/resnet50}{19.4}
\defR{pair-resnet50-load30-gemm/berth-nsb/p99x/resnet50}{1.1}
\defR{pair-resnet50-load30-gemm/berth-nsb/thrabs/gemm}{11.36}
\defR{pair-resnet50-load30-gemm/berth-nsb/thr/gemm}{72}
\defR{pair-resnet50-load30-gemm/berth-ng/attmin}{95.6}
\defR{pair-resnet50-load30-gemm/berth-ng/attmean}{95.6}
\defR{pair-resnet50-load30-gemm/berth-ng/thr}{72}
\defR{pair-resnet50-load30-gemm/berth-ng/n}{3}
\defR{pair-resnet50-load30-gemm/berth-ng/att/resnet50}{95.6}
\defR{pair-resnet50-load30-gemm/berth-ng/p99/resnet50}{19.7}
\defR{pair-resnet50-load30-gemm/berth-ng/p99x/resnet50}{1.1}
\defR{pair-resnet50-load30-gemm/berth-ng/thrabs/gemm}{11.36}
\defR{pair-resnet50-load30-gemm/berth-ng/thr/gemm}{72}
\defR{pair-resnet50-load30-gemm/berth-nsng/attmin}{96.2}
\defR{pair-resnet50-load30-gemm/berth-nsng/attmean}{96.2}
\defR{pair-resnet50-load30-gemm/berth-nsng/thr}{72}
\defR{pair-resnet50-load30-gemm/berth-nsng/n}{3}
\defR{pair-resnet50-load30-gemm/berth-nsng/att/resnet50}{96.2}
\defR{pair-resnet50-load30-gemm/berth-nsng/p99/resnet50}{19.0}
\defR{pair-resnet50-load30-gemm/berth-nsng/p99x/resnet50}{1.0}
\defR{pair-resnet50-load30-gemm/berth-nsng/thrabs/gemm}{11.41}
\defR{pair-resnet50-load30-gemm/berth-nsng/thr/gemm}{72}
\defR{pair-resnet50-load30-gemm/berth-grouped/attmin}{66.4}
\defR{pair-resnet50-load30-gemm/berth-grouped/attmean}{66.4}
\defR{pair-resnet50-load30-gemm/berth-grouped/thr}{72}
\defR{pair-resnet50-load30-gemm/berth-grouped/n}{3}
\defR{pair-resnet50-load30-gemm/berth-grouped/att/resnet50}{66.4}
\defR{pair-resnet50-load30-gemm/berth-grouped/p99/resnet50}{42.7}
\defR{pair-resnet50-load30-gemm/berth-grouped/p99x/resnet50}{2.3}
\defR{pair-resnet50-load30-gemm/berth-grouped/thrabs/gemm}{11.35}
\defR{pair-resnet50-load30-gemm/berth-grouped/thr/gemm}{72}
\defR{pair-resnet50-load30-gemm/berth/attmin}{68.6}
\defR{pair-resnet50-load30-gemm/berth/attmean}{68.6}
\defR{pair-resnet50-load30-gemm/berth/thr}{72}
\defR{pair-resnet50-load30-gemm/berth/n}{3}
\defR{pair-resnet50-load30-gemm/berth/att/resnet50}{68.6}
\defR{pair-resnet50-load30-gemm/berth/p99/resnet50}{40.2}
\defR{pair-resnet50-load30-gemm/berth/p99x/resnet50}{2.2}
\defR{pair-resnet50-load30-gemm/berth/thrabs/gemm}{11.41}
\defR{pair-resnet50-load30-gemm/berth/thr/gemm}{72}
\defR{pair-resnet50-load30-gemm/timeslice/attmin}{67.0}
\defR{pair-resnet50-load30-gemm/timeslice/attmean}{67.0}
\defR{pair-resnet50-load30-gemm/timeslice/thr}{72}
\defR{pair-resnet50-load30-gemm/timeslice/n}{3}
\defR{pair-resnet50-load30-gemm/timeslice/att/resnet50}{67.0}
\defR{pair-resnet50-load30-gemm/timeslice/p99/resnet50}{41.9}
\defR{pair-resnet50-load30-gemm/timeslice/p99x/resnet50}{2.3}
\defR{pair-resnet50-load30-gemm/timeslice/thrabs/gemm}{11.41}
\defR{pair-resnet50-load30-gemm/timeslice/thr/gemm}{72}
\defR{pair-resnet50-load30-gemm/mps/attmin}{0.0}
\defR{pair-resnet50-load30-gemm/mps/attmean}{0.0}
\defR{pair-resnet50-load30-gemm/mps/thr}{100}
\defR{pair-resnet50-load30-gemm/mps/n}{3}
\defR{pair-resnet50-load30-gemm/mps/att/resnet50}{0.0}
\defR{pair-resnet50-load30-gemm/mps/p99/resnet50}{90624}
\defR{pair-resnet50-load30-gemm/mps/p99x/resnet50}{4976.0}
\defR{pair-resnet50-load30-gemm/mps/thrabs/gemm}{15.88}
\defR{pair-resnet50-load30-gemm/mps/thr/gemm}{100}
\defR{pair-resnet50-load30-gemm/mpsprio/attmin}{0.0}
\defR{pair-resnet50-load30-gemm/mpsprio/attmean}{0.0}
\defR{pair-resnet50-load30-gemm/mpsprio/thr}{100}
\defR{pair-resnet50-load30-gemm/mpsprio/n}{3}
\defR{pair-resnet50-load30-gemm/mpsprio/att/resnet50}{0.0}
\defR{pair-resnet50-load30-gemm/mpsprio/p99/resnet50}{89104}
\defR{pair-resnet50-load30-gemm/mpsprio/p99x/resnet50}{4892.5}
\defR{pair-resnet50-load30-gemm/mpsprio/thrabs/gemm}{15.73}
\defR{pair-resnet50-load30-gemm/mpsprio/thr/gemm}{100}
\defR{pair-resnet50-load30-gemm/mpslimits/attmin}{98.6}
\defR{pair-resnet50-load30-gemm/mpslimits/attmean}{98.6}
\defR{pair-resnet50-load30-gemm/mpslimits/thr}{10}
\defR{pair-resnet50-load30-gemm/mpslimits/n}{3}
\defR{pair-resnet50-load30-gemm/mpslimits/att/resnet50}{98.6}
\defR{pair-resnet50-load30-gemm/mpslimits/p99/resnet50}{15.7}
\defR{pair-resnet50-load30-gemm/mpslimits/p99x/resnet50}{0.9}
\defR{pair-resnet50-load30-gemm/mpslimits/thrabs/gemm}{1.54}
\defR{pair-resnet50-load30-gemm/mpslimits/thr/gemm}{10}
\defR{pair-resnet50-load30-gemm/mpsparts/attmin}{98.1}
\defR{pair-resnet50-load30-gemm/mpsparts/attmean}{98.1}
\defR{pair-resnet50-load30-gemm/mpsparts/thr}{8}
\defR{pair-resnet50-load30-gemm/mpsparts/n}{2}
\defR{pair-resnet50-load30-gemm/mpsparts/att/resnet50}{98.1}
\defR{pair-resnet50-load30-gemm/mpsparts/p99/resnet50}{16.5}
\defR{pair-resnet50-load30-gemm/mpsparts/p99x/resnet50}{0.9}
\defR{pair-resnet50-load30-gemm/mpsparts/thrabs/gemm}{1.27}
\defR{pair-resnet50-load30-gemm/mpsparts/thr/gemm}{8}
\defR{pair-resnet50-load30-gemm/lithos/attmin}{97.9}
\defR{pair-resnet50-load30-gemm/lithos/attmean}{97.9}
\defR{pair-resnet50-load30-gemm/lithos/thr}{8}
\defR{pair-resnet50-load30-gemm/lithos/n}{3}
\defR{pair-resnet50-load30-gemm/lithos/att/resnet50}{97.9}
\defR{pair-resnet50-load30-gemm/lithos/p99/resnet50}{17.1}
\defR{pair-resnet50-load30-gemm/lithos/p99x/resnet50}{0.9}
\defR{pair-resnet50-load30-gemm/lithos/thrabs/gemm}{1.27}
\defR{pair-resnet50-load30-gemm/lithos/thr/gemm}{8}
\defR{pair-resnet50-load30-gemm/dysm/attmin}{0.0}
\defR{pair-resnet50-load30-gemm/dysm/attmean}{0.0}
\defR{pair-resnet50-load30-gemm/dysm/thr}{101}
\defR{pair-resnet50-load30-gemm/dysm/n}{3}
\defR{pair-resnet50-load30-gemm/dysm/att/resnet50}{0.0}
\defR{pair-resnet50-load30-gemm/dysm/thrabs/gemm}{15.89}
\defR{pair-resnet50-load30-gemm/dysm/thr/gemm}{101}
\defR{pair-resnet50-load30-gemm/xsched/attmin}{43.6}
\defR{pair-resnet50-load30-gemm/xsched/attmean}{43.6}
\defR{pair-resnet50-load30-gemm/xsched/thr}{71}
\defR{pair-resnet50-load30-gemm/xsched/n}{3}
\defR{pair-resnet50-load30-gemm/xsched/att/resnet50}{43.6}
\defR{pair-resnet50-load30-gemm/xsched/p99/resnet50}{90.4}
\defR{pair-resnet50-load30-gemm/xsched/p99x/resnet50}{5.0}
\defR{pair-resnet50-load30-gemm/xsched/thrabs/gemm}{11.20}
\defR{pair-resnet50-load30-gemm/xsched/thr/gemm}{71}
\defR{pair-resnet50-load30-gemm/multistream/attmin}{0.0}
\defR{pair-resnet50-load30-gemm/multistream/attmean}{0.0}
\defR{pair-resnet50-load30-gemm/multistream/thr}{100}
\defR{pair-resnet50-load30-gemm/multistream/n}{3}
\defR{pair-resnet50-load30-gemm/multistream/att/resnet50}{0.0}
\defR{pair-resnet50-load30-gemm/multistream/p99/resnet50}{89091}
\defR{pair-resnet50-load30-gemm/multistream/p99x/resnet50}{4891.8}
\defR{pair-resnet50-load30-gemm/multistream/thrabs/gemm}{15.75}
\defR{pair-resnet50-load30-gemm/multistream/thr/gemm}{100}
\defR{pair-resnet50-load30-gemm/tgs/attmin}{66.9}
\defR{pair-resnet50-load30-gemm/tgs/attmean}{66.9}
\defR{pair-resnet50-load30-gemm/tgs/thr}{72}
\defR{pair-resnet50-load30-gemm/tgs/n}{3}
\defR{pair-resnet50-load30-gemm/tgs/att/resnet50}{66.9}
\defR{pair-resnet50-load30-gemm/tgs/p99/resnet50}{38.6}
\defR{pair-resnet50-load30-gemm/tgs/p99x/resnet50}{2.1}
\defR{pair-resnet50-load30-gemm/tgs/thrabs/gemm}{11.44}
\defR{pair-resnet50-load30-gemm/tgs/thr/gemm}{72}
\defR{pair-resnet50-load30-gemm/mig/attmin}{95.2}
\defR{pair-resnet50-load30-gemm/mig/attmean}{95.2}
\defR{pair-resnet50-load30-gemm/mig/thr}{42}
\defR{pair-resnet50-load30-gemm/mig/n}{3}
\defR{pair-resnet50-load30-gemm/mig/att/resnet50}{95.2}
\defR{pair-resnet50-load30-gemm/mig/p99/resnet50}{20.1}
\defR{pair-resnet50-load30-gemm/mig/p99x/resnet50}{1.1}
\defR{pair-resnet50-load30-gemm/mig/thrabs/gemm}{6.64}
\defR{pair-resnet50-load30-gemm/mig/thr/gemm}{42}
\defR{pair-resnet50-load30-gemm/tally/attmin}{0.0}
\defR{pair-resnet50-load30-gemm/tally/attmean}{0.0}
\defR{pair-resnet50-load30-gemm/tally/thr}{57}
\defR{pair-resnet50-load30-gemm/tally/n}{3}
\defR{pair-resnet50-load30-gemm/tally/att/resnet50}{0.0}
\defR{pair-resnet50-load30-gemm/tally/p99/resnet50}{79041}
\defR{pair-resnet50-load30-gemm/tally/p99x/resnet50}{4340.0}
\defR{pair-resnet50-load30-gemm/tally/thrabs/gemm}{8.98}
\defR{pair-resnet50-load30-gemm/tally/thr/gemm}{57}
\defR{pair-resnet50-load30-gemm/orion/attmin}{0.0}
\defR{pair-resnet50-load30-gemm/orion/attmean}{0.0}
\defR{pair-resnet50-load30-gemm/orion/thr}{95}
\defR{pair-resnet50-load30-gemm/orion/n}{3}
\defR{pair-resnet50-load30-gemm/orion/att/resnet50}{0.0}
\defR{pair-resnet50-load30-gemm/orion/p99/resnet50}{133368}
\defR{pair-resnet50-load30-gemm/orion/p99x/resnet50}{7323.0}
\defR{pair-resnet50-load30-gemm/orion/thrabs/gemm}{15.01}
\defR{pair-resnet50-load30-gemm/orion/thr/gemm}{95}
\defR{pair-bertlarge-load30-trainresnet/berth-full/attmin}{100.0}
\defR{pair-bertlarge-load30-trainresnet/berth-full/attmean}{100.0}
\defR{pair-bertlarge-load30-trainresnet/berth-full/thr}{63}
\defR{pair-bertlarge-load30-trainresnet/berth-full/n}{3}
\defR{pair-bertlarge-load30-trainresnet/berth-full/att/bertlarge}{100.0}
\defR{pair-bertlarge-load30-trainresnet/berth-full/p99/bertlarge}{69.0}
\defR{pair-bertlarge-load30-trainresnet/berth-full/p99x/bertlarge}{1.2}
\defR{pair-bertlarge-load30-trainresnet/berth-full/thrabs/train_resnet}{8.37}
\defR{pair-bertlarge-load30-trainresnet/berth-full/thr/train_resnet}{63}
\defR{pair-bertlarge-load30-trainresnet/berth-nsr/attmin}{100.0}
\defR{pair-bertlarge-load30-trainresnet/berth-nsr/attmean}{100.0}
\defR{pair-bertlarge-load30-trainresnet/berth-nsr/thr}{63}
\defR{pair-bertlarge-load30-trainresnet/berth-nsr/n}{3}
\defR{pair-bertlarge-load30-trainresnet/berth-nsr/att/bertlarge}{100.0}
\defR{pair-bertlarge-load30-trainresnet/berth-nsr/p99/bertlarge}{67.6}
\defR{pair-bertlarge-load30-trainresnet/berth-nsr/p99x/bertlarge}{1.2}
\defR{pair-bertlarge-load30-trainresnet/berth-nsr/thrabs/train_resnet}{8.34}
\defR{pair-bertlarge-load30-trainresnet/berth-nsr/thr/train_resnet}{63}
\defR{pair-bertlarge-load30-trainresnet/berth-nba/attmin}{100.0}
\defR{pair-bertlarge-load30-trainresnet/berth-nba/attmean}{100.0}
\defR{pair-bertlarge-load30-trainresnet/berth-nba/thr}{63}
\defR{pair-bertlarge-load30-trainresnet/berth-nba/n}{3}
\defR{pair-bertlarge-load30-trainresnet/berth-nba/att/bertlarge}{100.0}
\defR{pair-bertlarge-load30-trainresnet/berth-nba/p99/bertlarge}{67.5}
\defR{pair-bertlarge-load30-trainresnet/berth-nba/p99x/bertlarge}{1.2}
\defR{pair-bertlarge-load30-trainresnet/berth-nba/thrabs/train_resnet}{8.38}
\defR{pair-bertlarge-load30-trainresnet/berth-nba/thr/train_resnet}{63}
\defR{pair-bertlarge-load30-trainresnet/berth-nsb/attmin}{100.0}
\defR{pair-bertlarge-load30-trainresnet/berth-nsb/attmean}{100.0}
\defR{pair-bertlarge-load30-trainresnet/berth-nsb/thr}{63}
\defR{pair-bertlarge-load30-trainresnet/berth-nsb/n}{3}
\defR{pair-bertlarge-load30-trainresnet/berth-nsb/att/bertlarge}{100.0}
\defR{pair-bertlarge-load30-trainresnet/berth-nsb/p99/bertlarge}{67.9}
\defR{pair-bertlarge-load30-trainresnet/berth-nsb/p99x/bertlarge}{1.2}
\defR{pair-bertlarge-load30-trainresnet/berth-nsb/thrabs/train_resnet}{8.39}
\defR{pair-bertlarge-load30-trainresnet/berth-nsb/thr/train_resnet}{63}
\defR{pair-bertlarge-load30-trainresnet/berth-ng/attmin}{100.0}
\defR{pair-bertlarge-load30-trainresnet/berth-ng/attmean}{100.0}
\defR{pair-bertlarge-load30-trainresnet/berth-ng/thr}{63}
\defR{pair-bertlarge-load30-trainresnet/berth-ng/n}{3}
\defR{pair-bertlarge-load30-trainresnet/berth-ng/att/bertlarge}{100.0}
\defR{pair-bertlarge-load30-trainresnet/berth-ng/p99/bertlarge}{86.7}
\defR{pair-bertlarge-load30-trainresnet/berth-ng/p99x/bertlarge}{1.5}
\defR{pair-bertlarge-load30-trainresnet/berth-ng/thrabs/train_resnet}{8.32}
\defR{pair-bertlarge-load30-trainresnet/berth-ng/thr/train_resnet}{63}
\defR{pair-bertlarge-load30-trainresnet/berth-nsng/attmin}{100.0}
\defR{pair-bertlarge-load30-trainresnet/berth-nsng/attmean}{100.0}
\defR{pair-bertlarge-load30-trainresnet/berth-nsng/thr}{63}
\defR{pair-bertlarge-load30-trainresnet/berth-nsng/n}{3}
\defR{pair-bertlarge-load30-trainresnet/berth-nsng/att/bertlarge}{100.0}
\defR{pair-bertlarge-load30-trainresnet/berth-nsng/p99/bertlarge}{84.8}
\defR{pair-bertlarge-load30-trainresnet/berth-nsng/p99x/bertlarge}{1.5}
\defR{pair-bertlarge-load30-trainresnet/berth-nsng/thrabs/train_resnet}{8.32}
\defR{pair-bertlarge-load30-trainresnet/berth-nsng/thr/train_resnet}{63}
\defR{pair-bertlarge-load30-trainresnet/berth-grouped/attmin}{100.0}
\defR{pair-bertlarge-load30-trainresnet/berth-grouped/attmean}{100.0}
\defR{pair-bertlarge-load30-trainresnet/berth-grouped/thr}{64}
\defR{pair-bertlarge-load30-trainresnet/berth-grouped/n}{3}
\defR{pair-bertlarge-load30-trainresnet/berth-grouped/att/bertlarge}{100.0}
\defR{pair-bertlarge-load30-trainresnet/berth-grouped/p99/bertlarge}{89.2}
\defR{pair-bertlarge-load30-trainresnet/berth-grouped/p99x/bertlarge}{1.5}
\defR{pair-bertlarge-load30-trainresnet/berth-grouped/thrabs/train_resnet}{8.49}
\defR{pair-bertlarge-load30-trainresnet/berth-grouped/thr/train_resnet}{64}
\defR{pair-bertlarge-load30-trainresnet/berth/attmin}{100.0}
\defR{pair-bertlarge-load30-trainresnet/berth/attmean}{100.0}
\defR{pair-bertlarge-load30-trainresnet/berth/thr}{64}
\defR{pair-bertlarge-load30-trainresnet/berth/n}{3}
\defR{pair-bertlarge-load30-trainresnet/berth/att/bertlarge}{100.0}
\defR{pair-bertlarge-load30-trainresnet/berth/p99/bertlarge}{87.8}
\defR{pair-bertlarge-load30-trainresnet/berth/p99x/bertlarge}{1.5}
\defR{pair-bertlarge-load30-trainresnet/berth/thrabs/train_resnet}{8.51}
\defR{pair-bertlarge-load30-trainresnet/berth/thr/train_resnet}{64}
\defR{pair-bertlarge-load30-trainresnet/timeslice/attmin}{50.4}
\defR{pair-bertlarge-load30-trainresnet/timeslice/attmean}{50.4}
\defR{pair-bertlarge-load30-trainresnet/timeslice/thr}{59}
\defR{pair-bertlarge-load30-trainresnet/timeslice/n}{3}
\defR{pair-bertlarge-load30-trainresnet/timeslice/att/bertlarge}{50.4}
\defR{pair-bertlarge-load30-trainresnet/timeslice/p99/bertlarge}{861}
\defR{pair-bertlarge-load30-trainresnet/timeslice/p99x/bertlarge}{14.7}
\defR{pair-bertlarge-load30-trainresnet/timeslice/thrabs/train_resnet}{7.86}
\defR{pair-bertlarge-load30-trainresnet/timeslice/thr/train_resnet}{59}
\defR{pair-bertlarge-load30-trainresnet/mps/attmin}{54.5}
\defR{pair-bertlarge-load30-trainresnet/mps/attmean}{54.5}
\defR{pair-bertlarge-load30-trainresnet/mps/thr}{80}
\defR{pair-bertlarge-load30-trainresnet/mps/n}{3}
\defR{pair-bertlarge-load30-trainresnet/mps/att/bertlarge}{54.5}
\defR{pair-bertlarge-load30-trainresnet/mps/p99/bertlarge}{775}
\defR{pair-bertlarge-load30-trainresnet/mps/p99x/bertlarge}{13.3}
\defR{pair-bertlarge-load30-trainresnet/mps/thrabs/train_resnet}{10.59}
\defR{pair-bertlarge-load30-trainresnet/mps/thr/train_resnet}{80}
\defR{pair-bertlarge-load30-trainresnet/mpsprio/attmin}{85.1}
\defR{pair-bertlarge-load30-trainresnet/mpsprio/attmean}{85.1}
\defR{pair-bertlarge-load30-trainresnet/mpsprio/thr}{78}
\defR{pair-bertlarge-load30-trainresnet/mpsprio/n}{3}
\defR{pair-bertlarge-load30-trainresnet/mpsprio/att/bertlarge}{85.1}
\defR{pair-bertlarge-load30-trainresnet/mpsprio/p99/bertlarge}{350}
\defR{pair-bertlarge-load30-trainresnet/mpsprio/p99x/bertlarge}{6.0}
\defR{pair-bertlarge-load30-trainresnet/mpsprio/thrabs/train_resnet}{10.37}
\defR{pair-bertlarge-load30-trainresnet/mpsprio/thr/train_resnet}{78}
\defR{pair-bertlarge-load30-trainresnet/mpslimits/attmin}{99.5}
\defR{pair-bertlarge-load30-trainresnet/mpslimits/attmean}{99.5}
\defR{pair-bertlarge-load30-trainresnet/mpslimits/thr}{35}
\defR{pair-bertlarge-load30-trainresnet/mpslimits/n}{3}
\defR{pair-bertlarge-load30-trainresnet/mpslimits/att/bertlarge}{99.5}
\defR{pair-bertlarge-load30-trainresnet/mpslimits/p99/bertlarge}{115}
\defR{pair-bertlarge-load30-trainresnet/mpslimits/p99x/bertlarge}{2.0}
\defR{pair-bertlarge-load30-trainresnet/mpslimits/thrabs/train_resnet}{4.64}
\defR{pair-bertlarge-load30-trainresnet/mpslimits/thr/train_resnet}{35}
\defR{pair-bertlarge-load30-trainresnet/mpsparts/attmin}{99.7}
\defR{pair-bertlarge-load30-trainresnet/mpsparts/attmean}{99.7}
\defR{pair-bertlarge-load30-trainresnet/mpsparts/thr}{38}
\defR{pair-bertlarge-load30-trainresnet/mpsparts/n}{2}
\defR{pair-bertlarge-load30-trainresnet/mpsparts/att/bertlarge}{99.7}
\defR{pair-bertlarge-load30-trainresnet/mpsparts/p99/bertlarge}{113}
\defR{pair-bertlarge-load30-trainresnet/mpsparts/p99x/bertlarge}{1.9}
\defR{pair-bertlarge-load30-trainresnet/mpsparts/thrabs/train_resnet}{5.06}
\defR{pair-bertlarge-load30-trainresnet/mpsparts/thr/train_resnet}{38}
\defR{pair-bertlarge-load30-trainresnet/lithos/attmin}{98.3}
\defR{pair-bertlarge-load30-trainresnet/lithos/attmean}{98.3}
\defR{pair-bertlarge-load30-trainresnet/lithos/thr}{48}
\defR{pair-bertlarge-load30-trainresnet/lithos/n}{3}
\defR{pair-bertlarge-load30-trainresnet/lithos/att/bertlarge}{98.3}
\defR{pair-bertlarge-load30-trainresnet/lithos/p99/bertlarge}{148}
\defR{pair-bertlarge-load30-trainresnet/lithos/p99x/bertlarge}{2.5}
\defR{pair-bertlarge-load30-trainresnet/lithos/thrabs/train_resnet}{6.36}
\defR{pair-bertlarge-load30-trainresnet/lithos/thr/train_resnet}{48}
\defR{pair-bertlarge-load30-trainresnet/dysm/attmin}{69.8}
\defR{pair-bertlarge-load30-trainresnet/dysm/attmean}{69.8}
\defR{pair-bertlarge-load30-trainresnet/dysm/thr}{72}
\defR{pair-bertlarge-load30-trainresnet/dysm/n}{3}
\defR{pair-bertlarge-load30-trainresnet/dysm/att/bertlarge}{69.8}
\defR{pair-bertlarge-load30-trainresnet/dysm/p99/bertlarge}{498}
\defR{pair-bertlarge-load30-trainresnet/dysm/p99x/bertlarge}{8.5}
\defR{pair-bertlarge-load30-trainresnet/dysm/thrabs/train_resnet}{9.55}
\defR{pair-bertlarge-load30-trainresnet/dysm/thr/train_resnet}{72}
\defR{pair-bertlarge-load30-trainresnet/xsched/attmin}{0.0}
\defR{pair-bertlarge-load30-trainresnet/xsched/attmean}{0.0}
\defR{pair-bertlarge-load30-trainresnet/xsched/thr}{0}
\defR{pair-bertlarge-load30-trainresnet/xsched/n}{3}
\defR{pair-bertlarge-load30-trainresnet/xsched/att/bertlarge}{0.0}
\defR{pair-bertlarge-load30-trainresnet/xsched/thrabs/train_resnet}{0.00}
\defR{pair-bertlarge-load30-trainresnet/xsched/thr/train_resnet}{0}
\defR{pair-bertlarge-load30-trainresnet/multistream/attmin}{85.5}
\defR{pair-bertlarge-load30-trainresnet/multistream/attmean}{85.5}
\defR{pair-bertlarge-load30-trainresnet/multistream/thr}{78}
\defR{pair-bertlarge-load30-trainresnet/multistream/n}{3}
\defR{pair-bertlarge-load30-trainresnet/multistream/att/bertlarge}{85.5}
\defR{pair-bertlarge-load30-trainresnet/multistream/p99/bertlarge}{340}
\defR{pair-bertlarge-load30-trainresnet/multistream/p99x/bertlarge}{5.8}
\defR{pair-bertlarge-load30-trainresnet/multistream/thrabs/train_resnet}{10.36}
\defR{pair-bertlarge-load30-trainresnet/multistream/thr/train_resnet}{78}
\defR{pair-bertlarge-load30-trainresnet/tgs/attmin}{100.0}
\defR{pair-bertlarge-load30-trainresnet/tgs/attmean}{100.0}
\defR{pair-bertlarge-load30-trainresnet/tgs/thr}{0}
\defR{pair-bertlarge-load30-trainresnet/tgs/n}{3}
\defR{pair-bertlarge-load30-trainresnet/tgs/att/bertlarge}{100.0}
\defR{pair-bertlarge-load30-trainresnet/tgs/p99/bertlarge}{62.3}
\defR{pair-bertlarge-load30-trainresnet/tgs/p99x/bertlarge}{1.1}
\defR{pair-bertlarge-load30-trainresnet/tgs/thrabs/train_resnet}{0.04}
\defR{pair-bertlarge-load30-trainresnet/tgs/thr/train_resnet}{0}
\defR{pair-bertlarge-load30-trainresnet/mig/attmin}{97.6}
\defR{pair-bertlarge-load30-trainresnet/mig/attmean}{97.6}
\defR{pair-bertlarge-load30-trainresnet/mig/thr}{51}
\defR{pair-bertlarge-load30-trainresnet/mig/n}{3}
\defR{pair-bertlarge-load30-trainresnet/mig/att/bertlarge}{97.6}
\defR{pair-bertlarge-load30-trainresnet/mig/p99/bertlarge}{170}
\defR{pair-bertlarge-load30-trainresnet/mig/p99x/bertlarge}{2.9}
\defR{pair-bertlarge-load30-trainresnet/mig/thrabs/train_resnet}{6.77}
\defR{pair-bertlarge-load30-trainresnet/mig/thr/train_resnet}{51}
\defR{pair-bertlarge-load30-trainresnet/tally/attmin}{74.5}
\defR{pair-bertlarge-load30-trainresnet/tally/attmean}{74.5}
\defR{pair-bertlarge-load30-trainresnet/tally/thr}{4}
\defR{pair-bertlarge-load30-trainresnet/tally/n}{3}
\defR{pair-bertlarge-load30-trainresnet/tally/att/bertlarge}{74.5}
\defR{pair-bertlarge-load30-trainresnet/tally/p99/bertlarge}{547}
\defR{pair-bertlarge-load30-trainresnet/tally/p99x/bertlarge}{9.4}
\defR{pair-bertlarge-load30-trainresnet/tally/thrabs/train_resnet}{0.48}
\defR{pair-bertlarge-load30-trainresnet/tally/thr/train_resnet}{4}
\defR{pair-bertlarge-load30-trainresnet/orion/attmin}{100.0}
\defR{pair-bertlarge-load30-trainresnet/orion/attmean}{100.0}
\defR{pair-bertlarge-load30-trainresnet/orion/thr}{67}
\defR{pair-bertlarge-load30-trainresnet/orion/n}{3}
\defR{pair-bertlarge-load30-trainresnet/orion/att/bertlarge}{100.0}
\defR{pair-bertlarge-load30-trainresnet/orion/p99/bertlarge}{61.0}
\defR{pair-bertlarge-load30-trainresnet/orion/p99x/bertlarge}{1.0}
\defR{pair-bertlarge-load30-trainresnet/orion/thrabs/train_resnet}{8.90}
\defR{pair-bertlarge-load30-trainresnet/orion/thr/train_resnet}{67}
\defR{pair-bertlarge-load30-gemm/berth-full/attmin}{99.6}
\defR{pair-bertlarge-load30-gemm/berth-full/attmean}{99.6}
\defR{pair-bertlarge-load30-gemm/berth-full/thr}{63}
\defR{pair-bertlarge-load30-gemm/berth-full/n}{3}
\defR{pair-bertlarge-load30-gemm/berth-full/att/bertlarge}{99.6}
\defR{pair-bertlarge-load30-gemm/berth-full/p99/bertlarge}{112}
\defR{pair-bertlarge-load30-gemm/berth-full/p99x/bertlarge}{1.9}
\defR{pair-bertlarge-load30-gemm/berth-full/thrabs/gemm}{9.90}
\defR{pair-bertlarge-load30-gemm/berth-full/thr/gemm}{63}
\defR{pair-bertlarge-load30-gemm/berth-nsr/attmin}{99.1}
\defR{pair-bertlarge-load30-gemm/berth-nsr/attmean}{99.1}
\defR{pair-bertlarge-load30-gemm/berth-nsr/thr}{62}
\defR{pair-bertlarge-load30-gemm/berth-nsr/n}{3}
\defR{pair-bertlarge-load30-gemm/berth-nsr/att/bertlarge}{99.1}
\defR{pair-bertlarge-load30-gemm/berth-nsr/p99/bertlarge}{125}
\defR{pair-bertlarge-load30-gemm/berth-nsr/p99x/bertlarge}{2.1}
\defR{pair-bertlarge-load30-gemm/berth-nsr/thrabs/gemm}{9.84}
\defR{pair-bertlarge-load30-gemm/berth-nsr/thr/gemm}{62}
\defR{pair-bertlarge-load30-gemm/berth-nba/attmin}{99.3}
\defR{pair-bertlarge-load30-gemm/berth-nba/attmean}{99.3}
\defR{pair-bertlarge-load30-gemm/berth-nba/thr}{63}
\defR{pair-bertlarge-load30-gemm/berth-nba/n}{3}
\defR{pair-bertlarge-load30-gemm/berth-nba/att/bertlarge}{99.3}
\defR{pair-bertlarge-load30-gemm/berth-nba/p99/bertlarge}{118}
\defR{pair-bertlarge-load30-gemm/berth-nba/p99x/bertlarge}{2.0}
\defR{pair-bertlarge-load30-gemm/berth-nba/thrabs/gemm}{9.90}
\defR{pair-bertlarge-load30-gemm/berth-nba/thr/gemm}{63}
\defR{pair-bertlarge-load30-gemm/berth-nsb/attmin}{99.5}
\defR{pair-bertlarge-load30-gemm/berth-nsb/attmean}{99.5}
\defR{pair-bertlarge-load30-gemm/berth-nsb/thr}{63}
\defR{pair-bertlarge-load30-gemm/berth-nsb/n}{3}
\defR{pair-bertlarge-load30-gemm/berth-nsb/att/bertlarge}{99.5}
\defR{pair-bertlarge-load30-gemm/berth-nsb/p99/bertlarge}{111}
\defR{pair-bertlarge-load30-gemm/berth-nsb/p99x/bertlarge}{1.9}
\defR{pair-bertlarge-load30-gemm/berth-nsb/thrabs/gemm}{9.90}
\defR{pair-bertlarge-load30-gemm/berth-nsb/thr/gemm}{63}
\defR{pair-bertlarge-load30-gemm/berth-ng/attmin}{98.4}
\defR{pair-bertlarge-load30-gemm/berth-ng/attmean}{98.4}
\defR{pair-bertlarge-load30-gemm/berth-ng/thr}{62}
\defR{pair-bertlarge-load30-gemm/berth-ng/n}{3}
\defR{pair-bertlarge-load30-gemm/berth-ng/att/bertlarge}{98.4}
\defR{pair-bertlarge-load30-gemm/berth-ng/p99/bertlarge}{145}
\defR{pair-bertlarge-load30-gemm/berth-ng/p99x/bertlarge}{2.5}
\defR{pair-bertlarge-load30-gemm/berth-ng/thrabs/gemm}{9.81}
\defR{pair-bertlarge-load30-gemm/berth-ng/thr/gemm}{62}
\defR{pair-bertlarge-load30-gemm/berth-nsng/attmin}{98.5}
\defR{pair-bertlarge-load30-gemm/berth-nsng/attmean}{98.5}
\defR{pair-bertlarge-load30-gemm/berth-nsng/thr}{62}
\defR{pair-bertlarge-load30-gemm/berth-nsng/n}{3}
\defR{pair-bertlarge-load30-gemm/berth-nsng/att/bertlarge}{98.5}
\defR{pair-bertlarge-load30-gemm/berth-nsng/p99/bertlarge}{143}
\defR{pair-bertlarge-load30-gemm/berth-nsng/p99x/bertlarge}{2.4}
\defR{pair-bertlarge-load30-gemm/berth-nsng/thrabs/gemm}{9.81}
\defR{pair-bertlarge-load30-gemm/berth-nsng/thr/gemm}{62}
\defR{pair-bertlarge-load30-gemm/berth-grouped/attmin}{75.6}
\defR{pair-bertlarge-load30-gemm/berth-grouped/attmean}{75.6}
\defR{pair-bertlarge-load30-gemm/berth-grouped/thr}{64}
\defR{pair-bertlarge-load30-gemm/berth-grouped/n}{3}
\defR{pair-bertlarge-load30-gemm/berth-grouped/att/bertlarge}{75.6}
\defR{pair-bertlarge-load30-gemm/berth-grouped/p99/bertlarge}{417}
\defR{pair-bertlarge-load30-gemm/berth-grouped/p99x/bertlarge}{7.1}
\defR{pair-bertlarge-load30-gemm/berth-grouped/thrabs/gemm}{10.05}
\defR{pair-bertlarge-load30-gemm/berth-grouped/thr/gemm}{64}
\defR{pair-bertlarge-load30-gemm/berth/attmin}{75.5}
\defR{pair-bertlarge-load30-gemm/berth/attmean}{75.5}
\defR{pair-bertlarge-load30-gemm/berth/thr}{63}
\defR{pair-bertlarge-load30-gemm/berth/n}{3}
\defR{pair-bertlarge-load30-gemm/berth/att/bertlarge}{75.5}
\defR{pair-bertlarge-load30-gemm/berth/p99/bertlarge}{425}
\defR{pair-bertlarge-load30-gemm/berth/p99x/bertlarge}{7.3}
\defR{pair-bertlarge-load30-gemm/berth/thrabs/gemm}{10.03}
\defR{pair-bertlarge-load30-gemm/berth/thr/gemm}{63}
\defR{pair-bertlarge-load30-gemm/timeslice/attmin}{51.9}
\defR{pair-bertlarge-load30-gemm/timeslice/attmean}{51.9}
\defR{pair-bertlarge-load30-gemm/timeslice/thr}{62}
\defR{pair-bertlarge-load30-gemm/timeslice/n}{3}
\defR{pair-bertlarge-load30-gemm/timeslice/att/bertlarge}{51.9}
\defR{pair-bertlarge-load30-gemm/timeslice/p99/bertlarge}{844}
\defR{pair-bertlarge-load30-gemm/timeslice/p99x/bertlarge}{14.4}
\defR{pair-bertlarge-load30-gemm/timeslice/thrabs/gemm}{9.74}
\defR{pair-bertlarge-load30-gemm/timeslice/thr/gemm}{62}
\defR{pair-bertlarge-load30-gemm/mps/attmin}{0.0}
\defR{pair-bertlarge-load30-gemm/mps/attmean}{0.0}
\defR{pair-bertlarge-load30-gemm/mps/thr}{100}
\defR{pair-bertlarge-load30-gemm/mps/n}{3}
\defR{pair-bertlarge-load30-gemm/mps/att/bertlarge}{0.0}
\defR{pair-bertlarge-load30-gemm/mps/p99/bertlarge}{90612}
\defR{pair-bertlarge-load30-gemm/mps/p99x/bertlarge}{1549.9}
\defR{pair-bertlarge-load30-gemm/mps/thrabs/gemm}{15.88}
\defR{pair-bertlarge-load30-gemm/mps/thr/gemm}{100}
\defR{pair-bertlarge-load30-gemm/mpsprio/attmin}{0.0}
\defR{pair-bertlarge-load30-gemm/mpsprio/attmean}{0.0}
\defR{pair-bertlarge-load30-gemm/mpsprio/thr}{97}
\defR{pair-bertlarge-load30-gemm/mpsprio/n}{3}
\defR{pair-bertlarge-load30-gemm/mpsprio/att/bertlarge}{0.0}
\defR{pair-bertlarge-load30-gemm/mpsprio/p99/bertlarge}{84946}
\defR{pair-bertlarge-load30-gemm/mpsprio/p99x/bertlarge}{1453.0}
\defR{pair-bertlarge-load30-gemm/mpsprio/thrabs/gemm}{15.34}
\defR{pair-bertlarge-load30-gemm/mpsprio/thr/gemm}{97}
\defR{pair-bertlarge-load30-gemm/mpslimits/attmin}{99.9}
\defR{pair-bertlarge-load30-gemm/mpslimits/attmean}{99.9}
\defR{pair-bertlarge-load30-gemm/mpslimits/thr}{18}
\defR{pair-bertlarge-load30-gemm/mpslimits/n}{3}
\defR{pair-bertlarge-load30-gemm/mpslimits/att/bertlarge}{99.9}
\defR{pair-bertlarge-load30-gemm/mpslimits/p99/bertlarge}{97.7}
\defR{pair-bertlarge-load30-gemm/mpslimits/p99x/bertlarge}{1.7}
\defR{pair-bertlarge-load30-gemm/mpslimits/thrabs/gemm}{2.82}
\defR{pair-bertlarge-load30-gemm/mpslimits/thr/gemm}{18}
\defR{pair-bertlarge-load30-gemm/mpsparts/attmin}{100.0}
\defR{pair-bertlarge-load30-gemm/mpsparts/attmean}{100.0}
\defR{pair-bertlarge-load30-gemm/mpsparts/thr}{20}
\defR{pair-bertlarge-load30-gemm/mpsparts/n}{2}
\defR{pair-bertlarge-load30-gemm/mpsparts/att/bertlarge}{100.0}
\defR{pair-bertlarge-load30-gemm/mpsparts/p99/bertlarge}{96.8}
\defR{pair-bertlarge-load30-gemm/mpsparts/p99x/bertlarge}{1.7}
\defR{pair-bertlarge-load30-gemm/mpsparts/thrabs/gemm}{3.12}
\defR{pair-bertlarge-load30-gemm/mpsparts/thr/gemm}{20}
\defR{pair-bertlarge-load30-gemm/lithos/attmin}{99.4}
\defR{pair-bertlarge-load30-gemm/lithos/attmean}{99.4}
\defR{pair-bertlarge-load30-gemm/lithos/thr}{18}
\defR{pair-bertlarge-load30-gemm/lithos/n}{3}
\defR{pair-bertlarge-load30-gemm/lithos/att/bertlarge}{99.4}
\defR{pair-bertlarge-load30-gemm/lithos/p99/bertlarge}{118}
\defR{pair-bertlarge-load30-gemm/lithos/p99x/bertlarge}{2.0}
\defR{pair-bertlarge-load30-gemm/lithos/thrabs/gemm}{2.88}
\defR{pair-bertlarge-load30-gemm/lithos/thr/gemm}{18}
\defR{pair-bertlarge-load30-gemm/dysm/attmin}{68.7}
\defR{pair-bertlarge-load30-gemm/dysm/attmean}{68.7}
\defR{pair-bertlarge-load30-gemm/dysm/thr}{61}
\defR{pair-bertlarge-load30-gemm/dysm/n}{3}
\defR{pair-bertlarge-load30-gemm/dysm/att/bertlarge}{68.7}
\defR{pair-bertlarge-load30-gemm/dysm/p99/bertlarge}{479}
\defR{pair-bertlarge-load30-gemm/dysm/p99x/bertlarge}{8.2}
\defR{pair-bertlarge-load30-gemm/dysm/thrabs/gemm}{9.70}
\defR{pair-bertlarge-load30-gemm/dysm/thr/gemm}{61}
\defR{pair-bertlarge-load30-gemm/xsched/attmin}{43.2}
\defR{pair-bertlarge-load30-gemm/xsched/attmean}{43.2}
\defR{pair-bertlarge-load30-gemm/xsched/thr}{61}
\defR{pair-bertlarge-load30-gemm/xsched/n}{3}
\defR{pair-bertlarge-load30-gemm/xsched/att/bertlarge}{43.2}
\defR{pair-bertlarge-load30-gemm/xsched/p99/bertlarge}{1095}
\defR{pair-bertlarge-load30-gemm/xsched/p99x/bertlarge}{18.7}
\defR{pair-bertlarge-load30-gemm/xsched/thrabs/gemm}{9.71}
\defR{pair-bertlarge-load30-gemm/xsched/thr/gemm}{61}
\defR{pair-bertlarge-load30-gemm/multistream/attmin}{0.0}
\defR{pair-bertlarge-load30-gemm/multistream/attmean}{0.0}
\defR{pair-bertlarge-load30-gemm/multistream/thr}{97}
\defR{pair-bertlarge-load30-gemm/multistream/n}{3}
\defR{pair-bertlarge-load30-gemm/multistream/att/bertlarge}{0.0}
\defR{pair-bertlarge-load30-gemm/multistream/p99/bertlarge}{85006}
\defR{pair-bertlarge-load30-gemm/multistream/p99x/bertlarge}{1454.0}
\defR{pair-bertlarge-load30-gemm/multistream/thrabs/gemm}{15.36}
\defR{pair-bertlarge-load30-gemm/multistream/thr/gemm}{97}
\defR{pair-bertlarge-load30-gemm/tgs/attmin}{51.2}
\defR{pair-bertlarge-load30-gemm/tgs/attmean}{51.2}
\defR{pair-bertlarge-load30-gemm/tgs/thr}{61}
\defR{pair-bertlarge-load30-gemm/tgs/n}{3}
\defR{pair-bertlarge-load30-gemm/tgs/att/bertlarge}{51.2}
\defR{pair-bertlarge-load30-gemm/tgs/p99/bertlarge}{648}
\defR{pair-bertlarge-load30-gemm/tgs/p99x/bertlarge}{11.1}
\defR{pair-bertlarge-load30-gemm/tgs/thrabs/gemm}{9.61}
\defR{pair-bertlarge-load30-gemm/tgs/thr/gemm}{61}
\defR{pair-bertlarge-load30-gemm/mig/attmin}{97.7}
\defR{pair-bertlarge-load30-gemm/mig/attmean}{97.7}
\defR{pair-bertlarge-load30-gemm/mig/thr}{41}
\defR{pair-bertlarge-load30-gemm/mig/n}{3}
\defR{pair-bertlarge-load30-gemm/mig/att/bertlarge}{97.7}
\defR{pair-bertlarge-load30-gemm/mig/p99/bertlarge}{170}
\defR{pair-bertlarge-load30-gemm/mig/p99x/bertlarge}{2.9}
\defR{pair-bertlarge-load30-gemm/mig/thrabs/gemm}{6.49}
\defR{pair-bertlarge-load30-gemm/mig/thr/gemm}{41}
\defR{pair-bertlarge-load30-gemm/tally/attmin}{0.0}
\defR{pair-bertlarge-load30-gemm/tally/attmean}{0.0}
\defR{pair-bertlarge-load30-gemm/tally/thr}{75}
\defR{pair-bertlarge-load30-gemm/tally/n}{3}
\defR{pair-bertlarge-load30-gemm/tally/att/bertlarge}{0.0}
\defR{pair-bertlarge-load30-gemm/tally/p99/bertlarge}{19580}
\defR{pair-bertlarge-load30-gemm/tally/p99x/bertlarge}{334.9}
\defR{pair-bertlarge-load30-gemm/tally/thrabs/gemm}{11.91}
\defR{pair-bertlarge-load30-gemm/tally/thr/gemm}{75}
\defR{pair-bertlarge-load30-gemm/orion/attmin}{0.3}
\defR{pair-bertlarge-load30-gemm/orion/attmean}{0.3}
\defR{pair-bertlarge-load30-gemm/orion/thr}{75}
\defR{pair-bertlarge-load30-gemm/orion/n}{3}
\defR{pair-bertlarge-load30-gemm/orion/att/bertlarge}{0.3}
\defR{pair-bertlarge-load30-gemm/orion/p99/bertlarge}{14591}
\defR{pair-bertlarge-load30-gemm/orion/p99x/bertlarge}{249.6}
\defR{pair-bertlarge-load30-gemm/orion/thrabs/gemm}{11.90}
\defR{pair-bertlarge-load30-gemm/orion/thr/gemm}{75}
\defR{mix-load40-trainresnet/berth-full/attmin}{99.7}
\defR{mix-load40-trainresnet/berth-full/attmean}{99.9}
\defR{mix-load40-trainresnet/berth-full/thr}{53}
\defR{mix-load40-trainresnet/berth-full/n}{3}
\defR{mix-load40-trainresnet/berth-full/att/resnet50}{99.7}
\defR{mix-load40-trainresnet/berth-full/att/retinanet}{100.0}
\defR{mix-load40-trainresnet/berth-full/att/bertlarge}{100.0}
\defR{mix-load40-trainresnet/berth-full/p99/resnet50}{13.1}
\defR{mix-load40-trainresnet/berth-full/p99/retinanet}{36.9}
\defR{mix-load40-trainresnet/berth-full/p99/bertlarge}{81.9}
\defR{mix-load40-trainresnet/berth-full/p99x/resnet50}{1.0}
\defR{mix-load40-trainresnet/berth-full/p99x/retinanet}{1.1}
\defR{mix-load40-trainresnet/berth-full/p99x/bertlarge}{1.1}
\defR{mix-load40-trainresnet/berth-full/thrabs/train_resnet}{7.07}
\defR{mix-load40-trainresnet/berth-full/thr/train_resnet}{53}
\defR{mix-load40-trainresnet/berth-nsr/attmin}{99.7}
\defR{mix-load40-trainresnet/berth-nsr/attmean}{99.9}
\defR{mix-load40-trainresnet/berth-nsr/thr}{53}
\defR{mix-load40-trainresnet/berth-nsr/n}{3}
\defR{mix-load40-trainresnet/berth-nsr/att/resnet50}{99.7}
\defR{mix-load40-trainresnet/berth-nsr/att/retinanet}{100.0}
\defR{mix-load40-trainresnet/berth-nsr/att/bertlarge}{99.9}
\defR{mix-load40-trainresnet/berth-nsr/p99/resnet50}{13.3}
\defR{mix-load40-trainresnet/berth-nsr/p99/retinanet}{36.9}
\defR{mix-load40-trainresnet/berth-nsr/p99/bertlarge}{82.8}
\defR{mix-load40-trainresnet/berth-nsr/p99x/resnet50}{1.0}
\defR{mix-load40-trainresnet/berth-nsr/p99x/retinanet}{1.1}
\defR{mix-load40-trainresnet/berth-nsr/p99x/bertlarge}{1.1}
\defR{mix-load40-trainresnet/berth-nsr/thrabs/train_resnet}{7.06}
\defR{mix-load40-trainresnet/berth-nsr/thr/train_resnet}{53}
\defR{mix-load40-trainresnet/berth-nba/attmin}{99.7}
\defR{mix-load40-trainresnet/berth-nba/attmean}{99.9}
\defR{mix-load40-trainresnet/berth-nba/thr}{53}
\defR{mix-load40-trainresnet/berth-nba/n}{3}
\defR{mix-load40-trainresnet/berth-nba/att/resnet50}{99.7}
\defR{mix-load40-trainresnet/berth-nba/att/retinanet}{100.0}
\defR{mix-load40-trainresnet/berth-nba/att/bertlarge}{99.9}
\defR{mix-load40-trainresnet/berth-nba/p99/resnet50}{13.0}
\defR{mix-load40-trainresnet/berth-nba/p99/retinanet}{35.5}
\defR{mix-load40-trainresnet/berth-nba/p99/bertlarge}{83.5}
\defR{mix-load40-trainresnet/berth-nba/p99x/resnet50}{1.0}
\defR{mix-load40-trainresnet/berth-nba/p99x/retinanet}{1.0}
\defR{mix-load40-trainresnet/berth-nba/p99x/bertlarge}{1.2}
\defR{mix-load40-trainresnet/berth-nba/thrabs/train_resnet}{7.09}
\defR{mix-load40-trainresnet/berth-nba/thr/train_resnet}{53}
\defR{mix-load40-trainresnet/berth-nsb/attmin}{99.7}
\defR{mix-load40-trainresnet/berth-nsb/attmean}{99.9}
\defR{mix-load40-trainresnet/berth-nsb/thr}{53}
\defR{mix-load40-trainresnet/berth-nsb/n}{3}
\defR{mix-load40-trainresnet/berth-nsb/att/resnet50}{99.7}
\defR{mix-load40-trainresnet/berth-nsb/att/retinanet}{100.0}
\defR{mix-load40-trainresnet/berth-nsb/att/bertlarge}{100.0}
\defR{mix-load40-trainresnet/berth-nsb/p99/resnet50}{13.1}
\defR{mix-load40-trainresnet/berth-nsb/p99/retinanet}{37.2}
\defR{mix-load40-trainresnet/berth-nsb/p99/bertlarge}{82.1}
\defR{mix-load40-trainresnet/berth-nsb/p99x/resnet50}{1.0}
\defR{mix-load40-trainresnet/berth-nsb/p99x/retinanet}{1.1}
\defR{mix-load40-trainresnet/berth-nsb/p99x/bertlarge}{1.1}
\defR{mix-load40-trainresnet/berth-nsb/thrabs/train_resnet}{7.10}
\defR{mix-load40-trainresnet/berth-nsb/thr/train_resnet}{53}
\defR{mix-load40-trainresnet/berth-ng/attmin}{99.7}
\defR{mix-load40-trainresnet/berth-ng/attmean}{99.9}
\defR{mix-load40-trainresnet/berth-ng/thr}{54}
\defR{mix-load40-trainresnet/berth-ng/n}{3}
\defR{mix-load40-trainresnet/berth-ng/att/resnet50}{99.7}
\defR{mix-load40-trainresnet/berth-ng/att/retinanet}{100.0}
\defR{mix-load40-trainresnet/berth-ng/att/bertlarge}{99.9}
\defR{mix-load40-trainresnet/berth-ng/p99/resnet50}{13.0}
\defR{mix-load40-trainresnet/berth-ng/p99/retinanet}{35.8}
\defR{mix-load40-trainresnet/berth-ng/p99/bertlarge}{86.9}
\defR{mix-load40-trainresnet/berth-ng/p99x/resnet50}{1.0}
\defR{mix-load40-trainresnet/berth-ng/p99x/retinanet}{1.0}
\defR{mix-load40-trainresnet/berth-ng/p99x/bertlarge}{1.2}
\defR{mix-load40-trainresnet/berth-ng/thrabs/train_resnet}{7.22}
\defR{mix-load40-trainresnet/berth-ng/thr/train_resnet}{54}
\defR{mix-load40-trainresnet/berth-nsng/attmin}{99.6}
\defR{mix-load40-trainresnet/berth-nsng/attmean}{99.8}
\defR{mix-load40-trainresnet/berth-nsng/thr}{54}
\defR{mix-load40-trainresnet/berth-nsng/n}{3}
\defR{mix-load40-trainresnet/berth-nsng/att/resnet50}{99.6}
\defR{mix-load40-trainresnet/berth-nsng/att/retinanet}{100.0}
\defR{mix-load40-trainresnet/berth-nsng/att/bertlarge}{99.9}
\defR{mix-load40-trainresnet/berth-nsng/p99/resnet50}{13.0}
\defR{mix-load40-trainresnet/berth-nsng/p99/retinanet}{36.1}
\defR{mix-load40-trainresnet/berth-nsng/p99/bertlarge}{87.8}
\defR{mix-load40-trainresnet/berth-nsng/p99x/resnet50}{1.0}
\defR{mix-load40-trainresnet/berth-nsng/p99x/retinanet}{1.0}
\defR{mix-load40-trainresnet/berth-nsng/p99x/bertlarge}{1.2}
\defR{mix-load40-trainresnet/berth-nsng/thrabs/train_resnet}{7.22}
\defR{mix-load40-trainresnet/berth-nsng/thr/train_resnet}{54}
\defR{mix-load40-trainresnet/berth-grouped/attmin}{97.1}
\defR{mix-load40-trainresnet/berth-grouped/attmean}{99.0}
\defR{mix-load40-trainresnet/berth-grouped/thr}{57}
\defR{mix-load40-trainresnet/berth-grouped/n}{3}
\defR{mix-load40-trainresnet/berth-grouped/att/resnet50}{97.1}
\defR{mix-load40-trainresnet/berth-grouped/att/retinanet}{100.0}
\defR{mix-load40-trainresnet/berth-grouped/att/bertlarge}{100.0}
\defR{mix-load40-trainresnet/berth-grouped/p99/resnet50}{18.8}
\defR{mix-load40-trainresnet/berth-grouped/p99/retinanet}{38.6}
\defR{mix-load40-trainresnet/berth-grouped/p99/bertlarge}{63.5}
\defR{mix-load40-trainresnet/berth-grouped/p99x/resnet50}{1.4}
\defR{mix-load40-trainresnet/berth-grouped/p99x/retinanet}{1.1}
\defR{mix-load40-trainresnet/berth-grouped/p99x/bertlarge}{0.9}
\defR{mix-load40-trainresnet/berth-grouped/thrabs/train_resnet}{7.58}
\defR{mix-load40-trainresnet/berth-grouped/thr/train_resnet}{57}
\defR{mix-load40-trainresnet/berth/attmin}{95.5}
\defR{mix-load40-trainresnet/berth/attmean}{98.5}
\defR{mix-load40-trainresnet/berth/thr}{54}
\defR{mix-load40-trainresnet/berth/n}{3}
\defR{mix-load40-trainresnet/berth/att/resnet50}{95.5}
\defR{mix-load40-trainresnet/berth/att/retinanet}{100.0}
\defR{mix-load40-trainresnet/berth/att/bertlarge}{99.9}
\defR{mix-load40-trainresnet/berth/p99/resnet50}{20.8}
\defR{mix-load40-trainresnet/berth/p99/retinanet}{50.8}
\defR{mix-load40-trainresnet/berth/p99/bertlarge}{90.0}
\defR{mix-load40-trainresnet/berth/p99x/resnet50}{1.5}
\defR{mix-load40-trainresnet/berth/p99x/retinanet}{1.5}
\defR{mix-load40-trainresnet/berth/p99x/bertlarge}{1.2}
\defR{mix-load40-trainresnet/berth/thrabs/train_resnet}{7.22}
\defR{mix-load40-trainresnet/berth/thr/train_resnet}{54}
\defR{mix-load40-trainresnet/timeslice/attmin}{68.5}
\defR{mix-load40-trainresnet/timeslice/attmean}{80.7}
\defR{mix-load40-trainresnet/timeslice/thr}{54}
\defR{mix-load40-trainresnet/timeslice/n}{3}
\defR{mix-load40-trainresnet/timeslice/att/resnet50}{68.5}
\defR{mix-load40-trainresnet/timeslice/att/retinanet}{97.3}
\defR{mix-load40-trainresnet/timeslice/att/bertlarge}{76.3}
\defR{mix-load40-trainresnet/timeslice/p99/resnet50}{41.0}
\defR{mix-load40-trainresnet/timeslice/p99/retinanet}{125}
\defR{mix-load40-trainresnet/timeslice/p99/bertlarge}{284}
\defR{mix-load40-trainresnet/timeslice/p99x/resnet50}{3.1}
\defR{mix-load40-trainresnet/timeslice/p99x/retinanet}{3.6}
\defR{mix-load40-trainresnet/timeslice/p99x/bertlarge}{3.9}
\defR{mix-load40-trainresnet/timeslice/thrabs/train_resnet}{7.13}
\defR{mix-load40-trainresnet/timeslice/thr/train_resnet}{54}
\defR{mix-load40-trainresnet/mps/attmin}{9.4}
\defR{mix-load40-trainresnet/mps/attmean}{55.4}
\defR{mix-load40-trainresnet/mps/thr}{74}
\defR{mix-load40-trainresnet/mps/n}{3}
\defR{mix-load40-trainresnet/mps/att/resnet50}{9.4}
\defR{mix-load40-trainresnet/mps/att/retinanet}{68.9}
\defR{mix-load40-trainresnet/mps/att/bertlarge}{88.0}
\defR{mix-load40-trainresnet/mps/p99/resnet50}{99.9}
\defR{mix-load40-trainresnet/mps/p99/retinanet}{273}
\defR{mix-load40-trainresnet/mps/p99/bertlarge}{216}
\defR{mix-load40-trainresnet/mps/p99x/resnet50}{7.5}
\defR{mix-load40-trainresnet/mps/p99x/retinanet}{7.9}
\defR{mix-load40-trainresnet/mps/p99x/bertlarge}{3.0}
\defR{mix-load40-trainresnet/mps/thrabs/train_resnet}{9.90}
\defR{mix-load40-trainresnet/mps/thr/train_resnet}{74}
\defR{mix-load40-trainresnet/mpsprio/attmin}{72.0}
\defR{mix-load40-trainresnet/mpsprio/attmean}{89.9}
\defR{mix-load40-trainresnet/mpsprio/thr}{72}
\defR{mix-load40-trainresnet/mpsprio/n}{3}
\defR{mix-load40-trainresnet/mpsprio/att/resnet50}{72.0}
\defR{mix-load40-trainresnet/mpsprio/att/retinanet}{99.1}
\defR{mix-load40-trainresnet/mpsprio/att/bertlarge}{98.8}
\defR{mix-load40-trainresnet/mpsprio/p99/resnet50}{35.9}
\defR{mix-load40-trainresnet/mpsprio/p99/retinanet}{97.9}
\defR{mix-load40-trainresnet/mpsprio/p99/bertlarge}{132}
\defR{mix-load40-trainresnet/mpsprio/p99x/resnet50}{2.7}
\defR{mix-load40-trainresnet/mpsprio/p99x/retinanet}{2.8}
\defR{mix-load40-trainresnet/mpsprio/p99x/bertlarge}{1.8}
\defR{mix-load40-trainresnet/mpsprio/thrabs/train_resnet}{9.61}
\defR{mix-load40-trainresnet/mpsprio/thr/train_resnet}{72}
\defR{mix-load40-trainresnet/mpslimits/attmin}{87.0}
\defR{mix-load40-trainresnet/mpslimits/attmean}{93.7}
\defR{mix-load40-trainresnet/mpslimits/thr}{20}
\defR{mix-load40-trainresnet/mpslimits/n}{3}
\defR{mix-load40-trainresnet/mpslimits/att/resnet50}{98.7}
\defR{mix-load40-trainresnet/mpslimits/att/retinanet}{95.5}
\defR{mix-load40-trainresnet/mpslimits/att/bertlarge}{87.0}
\defR{mix-load40-trainresnet/mpslimits/p99/resnet50}{15.6}
\defR{mix-load40-trainresnet/mpslimits/p99/retinanet}{142}
\defR{mix-load40-trainresnet/mpslimits/p99/bertlarge}{223}
\defR{mix-load40-trainresnet/mpslimits/p99x/resnet50}{1.2}
\defR{mix-load40-trainresnet/mpslimits/p99x/retinanet}{4.1}
\defR{mix-load40-trainresnet/mpslimits/p99x/bertlarge}{3.1}
\defR{mix-load40-trainresnet/mpslimits/thrabs/train_resnet}{2.70}
\defR{mix-load40-trainresnet/mpslimits/thr/train_resnet}{20}
\defR{mix-load40-trainresnet/mpsparts/attmin}{84.0}
\defR{mix-load40-trainresnet/mpsparts/attmean}{90.6}
\defR{mix-load40-trainresnet/mpsparts/thr}{31}
\defR{mix-load40-trainresnet/mpsparts/n}{2}
\defR{mix-load40-trainresnet/mpsparts/att/resnet50}{97.9}
\defR{mix-load40-trainresnet/mpsparts/att/retinanet}{89.8}
\defR{mix-load40-trainresnet/mpsparts/att/bertlarge}{84.0}
\defR{mix-load40-trainresnet/mpsparts/p99/resnet50}{16.9}
\defR{mix-load40-trainresnet/mpsparts/p99/retinanet}{179}
\defR{mix-load40-trainresnet/mpsparts/p99/bertlarge}{227}
\defR{mix-load40-trainresnet/mpsparts/p99x/resnet50}{1.3}
\defR{mix-load40-trainresnet/mpsparts/p99x/retinanet}{5.2}
\defR{mix-load40-trainresnet/mpsparts/p99x/bertlarge}{3.1}
\defR{mix-load40-trainresnet/mpsparts/thrabs/train_resnet}{4.12}
\defR{mix-load40-trainresnet/mpsparts/thr/train_resnet}{31}
\defR{mix-load40-trainresnet/lithos/attmin}{67.8}
\defR{mix-load40-trainresnet/lithos/attmean}{71.7}
\defR{mix-load40-trainresnet/lithos/thr}{47}
\defR{mix-load40-trainresnet/lithos/n}{3}
\defR{mix-load40-trainresnet/lithos/att/resnet50}{68.7}
\defR{mix-load40-trainresnet/lithos/att/retinanet}{78.4}
\defR{mix-load40-trainresnet/lithos/att/bertlarge}{67.8}
\defR{mix-load40-trainresnet/lithos/p99/resnet50}{40.5}
\defR{mix-load40-trainresnet/lithos/p99/retinanet}{219}
\defR{mix-load40-trainresnet/lithos/p99/bertlarge}{283}
\defR{mix-load40-trainresnet/lithos/p99x/resnet50}{3.0}
\defR{mix-load40-trainresnet/lithos/p99x/retinanet}{6.3}
\defR{mix-load40-trainresnet/lithos/p99x/bertlarge}{3.9}
\defR{mix-load40-trainresnet/lithos/thrabs/train_resnet}{6.27}
\defR{mix-load40-trainresnet/lithos/thr/train_resnet}{47}
\defR{mix-load40-trainresnet/dysm/attmin}{0.0}
\defR{mix-load40-trainresnet/dysm/attmean}{24.8}
\defR{mix-load40-trainresnet/dysm/thr}{86}
\defR{mix-load40-trainresnet/dysm/n}{3}
\defR{mix-load40-trainresnet/dysm/att/resnet50}{0.0}
\defR{mix-load40-trainresnet/dysm/att/retinanet}{74.4}
\defR{mix-load40-trainresnet/dysm/att/bertlarge}{0.0}
\defR{mix-load40-trainresnet/dysm/p99/retinanet}{264}
\defR{mix-load40-trainresnet/dysm/p99x/retinanet}{7.6}
\defR{mix-load40-trainresnet/dysm/thrabs/train_resnet}{11.36}
\defR{mix-load40-trainresnet/dysm/thr/train_resnet}{86}
\defR{mix-load40-trainresnet/xsched/attmin}{0.0}
\defR{mix-load40-trainresnet/xsched/attmean}{0.0}
\defR{mix-load40-trainresnet/xsched/thr}{0}
\defR{mix-load40-trainresnet/xsched/n}{3}
\defR{mix-load40-trainresnet/xsched/att/resnet50}{0.0}
\defR{mix-load40-trainresnet/xsched/att/retinanet}{0.0}
\defR{mix-load40-trainresnet/xsched/att/bertlarge}{0.0}
\defR{mix-load40-trainresnet/xsched/thrabs/train_resnet}{0.00}
\defR{mix-load40-trainresnet/xsched/thr/train_resnet}{0}
\defR{mix-load40-trainresnet/multistream/attmin}{71.1}
\defR{mix-load40-trainresnet/multistream/attmean}{89.9}
\defR{mix-load40-trainresnet/multistream/thr}{73}
\defR{mix-load40-trainresnet/multistream/n}{3}
\defR{mix-load40-trainresnet/multistream/att/resnet50}{71.1}
\defR{mix-load40-trainresnet/multistream/att/retinanet}{99.5}
\defR{mix-load40-trainresnet/multistream/att/bertlarge}{99.0}
\defR{mix-load40-trainresnet/multistream/p99/resnet50}{36.5}
\defR{mix-load40-trainresnet/multistream/p99/retinanet}{86.2}
\defR{mix-load40-trainresnet/multistream/p99/bertlarge}{130}
\defR{mix-load40-trainresnet/multistream/p99x/resnet50}{2.7}
\defR{mix-load40-trainresnet/multistream/p99x/retinanet}{2.5}
\defR{mix-load40-trainresnet/multistream/p99x/bertlarge}{1.8}
\defR{mix-load40-trainresnet/multistream/thrabs/train_resnet}{9.73}
\defR{mix-load40-trainresnet/multistream/thr/train_resnet}{73}
\defR{mix-load40-trainresnet/tgs/attmin}{96.5}
\defR{mix-load40-trainresnet/tgs/attmean}{98.8}
\defR{mix-load40-trainresnet/tgs/thr}{0}
\defR{mix-load40-trainresnet/tgs/n}{3}
\defR{mix-load40-trainresnet/tgs/att/resnet50}{96.5}
\defR{mix-load40-trainresnet/tgs/att/retinanet}{100.0}
\defR{mix-load40-trainresnet/tgs/att/bertlarge}{99.9}
\defR{mix-load40-trainresnet/tgs/p99/resnet50}{19.8}
\defR{mix-load40-trainresnet/tgs/p99/retinanet}{39.8}
\defR{mix-load40-trainresnet/tgs/p99/bertlarge}{77.5}
\defR{mix-load40-trainresnet/tgs/p99x/resnet50}{1.5}
\defR{mix-load40-trainresnet/tgs/p99x/retinanet}{1.1}
\defR{mix-load40-trainresnet/tgs/p99x/bertlarge}{1.1}
\defR{mix-load40-trainresnet/tgs/thrabs/train_resnet}{0.04}
\defR{mix-load40-trainresnet/tgs/thr/train_resnet}{0}
\defR{mix-load40-trainresnet/mig/attmin}{90.6}
\defR{mix-load40-trainresnet/mig/attmean}{95.8}
\defR{mix-load40-trainresnet/mig/thr}{0}
\defR{mix-load40-trainresnet/mig/n}{3}
\defR{mix-load40-trainresnet/mig/att/resnet50}{98.5}
\defR{mix-load40-trainresnet/mig/att/retinanet}{98.4}
\defR{mix-load40-trainresnet/mig/att/bertlarge}{90.6}
\defR{mix-load40-trainresnet/mig/p99/resnet50}{16.5}
\defR{mix-load40-trainresnet/mig/p99/retinanet}{109}
\defR{mix-load40-trainresnet/mig/p99/bertlarge}{207}
\defR{mix-load40-trainresnet/mig/p99x/resnet50}{1.2}
\defR{mix-load40-trainresnet/mig/p99x/retinanet}{3.1}
\defR{mix-load40-trainresnet/mig/p99x/bertlarge}{2.9}
\defR{mix-load40-trainresnet/mig/thrabs/train_resnet}{0.00}
\defR{mix-load40-trainresnet/mig/thr/train_resnet}{0}
\defR{mix-load40-trainresnet/tally/attmin}{0.0}
\defR{mix-load40-trainresnet/tally/attmean}{43.7}
\defR{mix-load40-trainresnet/tally/thr}{2}
\defR{mix-load40-trainresnet/tally/n}{3}
\defR{mix-load40-trainresnet/tally/att/resnet50}{0.0}
\defR{mix-load40-trainresnet/tally/att/retinanet}{35.1}
\defR{mix-load40-trainresnet/tally/att/bertlarge}{96.1}
\defR{mix-load40-trainresnet/tally/p99/resnet50}{89719}
\defR{mix-load40-trainresnet/tally/p99/retinanet}{549}
\defR{mix-load40-trainresnet/tally/p99/bertlarge}{162}
\defR{mix-load40-trainresnet/tally/p99x/resnet50}{6696.7}
\defR{mix-load40-trainresnet/tally/p99x/retinanet}{15.8}
\defR{mix-load40-trainresnet/tally/p99x/bertlarge}{2.2}
\defR{mix-load40-trainresnet/tally/thrabs/train_resnet}{0.20}
\defR{mix-load40-trainresnet/tally/thr/train_resnet}{2}
\defR{mix-load40-trainresnet/orion/attmin}{0.3}
\defR{mix-load40-trainresnet/orion/attmean}{57.3}
\defR{mix-load40-trainresnet/orion/thr}{62}
\defR{mix-load40-trainresnet/orion/n}{3}
\defR{mix-load40-trainresnet/orion/att/resnet50}{99.4}
\defR{mix-load40-trainresnet/orion/att/retinanet}{0.3}
\defR{mix-load40-trainresnet/orion/att/bertlarge}{72.2}
\defR{mix-load40-trainresnet/orion/p99/resnet50}{13.6}
\defR{mix-load40-trainresnet/orion/p99/retinanet}{9814}
\defR{mix-load40-trainresnet/orion/p99/bertlarge}{293}
\defR{mix-load40-trainresnet/orion/p99x/resnet50}{1.0}
\defR{mix-load40-trainresnet/orion/p99x/retinanet}{282.2}
\defR{mix-load40-trainresnet/orion/p99x/bertlarge}{4.0}
\defR{mix-load40-trainresnet/orion/thrabs/train_resnet}{8.20}
\defR{mix-load40-trainresnet/orion/thr/train_resnet}{62}
\defR{mix-load40-gemm/berth-full/attmin}{96.4}
\defR{mix-load40-gemm/berth-full/attmean}{98.0}
\defR{mix-load40-gemm/berth-full/thr}{55}
\defR{mix-load40-gemm/berth-full/n}{3}
\defR{mix-load40-gemm/berth-full/att/resnet50}{96.4}
\defR{mix-load40-gemm/berth-full/att/retinanet}{100.0}
\defR{mix-load40-gemm/berth-full/att/bertlarge}{97.5}
\defR{mix-load40-gemm/berth-full/p99/resnet50}{19.6}
\defR{mix-load40-gemm/berth-full/p99/retinanet}{53.5}
\defR{mix-load40-gemm/berth-full/p99/bertlarge}{149}
\defR{mix-load40-gemm/berth-full/p99x/resnet50}{1.5}
\defR{mix-load40-gemm/berth-full/p99x/retinanet}{1.5}
\defR{mix-load40-gemm/berth-full/p99x/bertlarge}{2.1}
\defR{mix-load40-gemm/berth-full/thrabs/gemm}{8.75}
\defR{mix-load40-gemm/berth-full/thr/gemm}{55}
\defR{mix-load40-gemm/berth-nsr/attmin}{96.4}
\defR{mix-load40-gemm/berth-nsr/attmean}{97.9}
\defR{mix-load40-gemm/berth-nsr/thr}{55}
\defR{mix-load40-gemm/berth-nsr/n}{3}
\defR{mix-load40-gemm/berth-nsr/att/resnet50}{96.4}
\defR{mix-load40-gemm/berth-nsr/att/retinanet}{100.0}
\defR{mix-load40-gemm/berth-nsr/att/bertlarge}{97.4}
\defR{mix-load40-gemm/berth-nsr/p99/resnet50}{19.4}
\defR{mix-load40-gemm/berth-nsr/p99/retinanet}{54.5}
\defR{mix-load40-gemm/berth-nsr/p99/bertlarge}{149}
\defR{mix-load40-gemm/berth-nsr/p99x/resnet50}{1.4}
\defR{mix-load40-gemm/berth-nsr/p99x/retinanet}{1.6}
\defR{mix-load40-gemm/berth-nsr/p99x/bertlarge}{2.1}
\defR{mix-load40-gemm/berth-nsr/thrabs/gemm}{8.70}
\defR{mix-load40-gemm/berth-nsr/thr/gemm}{55}
\defR{mix-load40-gemm/berth-nba/attmin}{96.8}
\defR{mix-load40-gemm/berth-nba/attmean}{98.2}
\defR{mix-load40-gemm/berth-nba/thr}{55}
\defR{mix-load40-gemm/berth-nba/n}{3}
\defR{mix-load40-gemm/berth-nba/att/resnet50}{96.8}
\defR{mix-load40-gemm/berth-nba/att/retinanet}{100.0}
\defR{mix-load40-gemm/berth-nba/att/bertlarge}{97.7}
\defR{mix-load40-gemm/berth-nba/p99/resnet50}{18.8}
\defR{mix-load40-gemm/berth-nba/p99/retinanet}{54.2}
\defR{mix-load40-gemm/berth-nba/p99/bertlarge}{148}
\defR{mix-load40-gemm/berth-nba/p99x/resnet50}{1.4}
\defR{mix-load40-gemm/berth-nba/p99x/retinanet}{1.6}
\defR{mix-load40-gemm/berth-nba/p99x/bertlarge}{2.0}
\defR{mix-load40-gemm/berth-nba/thrabs/gemm}{8.76}
\defR{mix-load40-gemm/berth-nba/thr/gemm}{55}
\defR{mix-load40-gemm/berth-nsb/attmin}{96.5}
\defR{mix-load40-gemm/berth-nsb/attmean}{98.1}
\defR{mix-load40-gemm/berth-nsb/thr}{56}
\defR{mix-load40-gemm/berth-nsb/n}{3}
\defR{mix-load40-gemm/berth-nsb/att/resnet50}{96.5}
\defR{mix-load40-gemm/berth-nsb/att/retinanet}{100.0}
\defR{mix-load40-gemm/berth-nsb/att/bertlarge}{97.9}
\defR{mix-load40-gemm/berth-nsb/p99/resnet50}{18.9}
\defR{mix-load40-gemm/berth-nsb/p99/retinanet}{54.6}
\defR{mix-load40-gemm/berth-nsb/p99/bertlarge}{149}
\defR{mix-load40-gemm/berth-nsb/p99x/resnet50}{1.4}
\defR{mix-load40-gemm/berth-nsb/p99x/retinanet}{1.6}
\defR{mix-load40-gemm/berth-nsb/p99x/bertlarge}{2.1}
\defR{mix-load40-gemm/berth-nsb/thrabs/gemm}{8.77}
\defR{mix-load40-gemm/berth-nsb/thr/gemm}{56}
\defR{mix-load40-gemm/berth-ng/attmin}{95.8}
\defR{mix-load40-gemm/berth-ng/attmean}{97.6}
\defR{mix-load40-gemm/berth-ng/thr}{55}
\defR{mix-load40-gemm/berth-ng/n}{3}
\defR{mix-load40-gemm/berth-ng/att/resnet50}{95.8}
\defR{mix-load40-gemm/berth-ng/att/retinanet}{100.0}
\defR{mix-load40-gemm/berth-ng/att/bertlarge}{96.9}
\defR{mix-load40-gemm/berth-ng/p99/resnet50}{20.0}
\defR{mix-load40-gemm/berth-ng/p99/retinanet}{56.3}
\defR{mix-load40-gemm/berth-ng/p99/bertlarge}{157}
\defR{mix-load40-gemm/berth-ng/p99x/resnet50}{1.5}
\defR{mix-load40-gemm/berth-ng/p99x/retinanet}{1.6}
\defR{mix-load40-gemm/berth-ng/p99x/bertlarge}{2.2}
\defR{mix-load40-gemm/berth-ng/thrabs/gemm}{8.72}
\defR{mix-load40-gemm/berth-ng/thr/gemm}{55}
\defR{mix-load40-gemm/berth-nsng/attmin}{96.2}
\defR{mix-load40-gemm/berth-nsng/attmean}{97.7}
\defR{mix-load40-gemm/berth-nsng/thr}{55}
\defR{mix-load40-gemm/berth-nsng/n}{3}
\defR{mix-load40-gemm/berth-nsng/att/resnet50}{96.2}
\defR{mix-load40-gemm/berth-nsng/att/retinanet}{100.0}
\defR{mix-load40-gemm/berth-nsng/att/bertlarge}{96.9}
\defR{mix-load40-gemm/berth-nsng/p99/resnet50}{19.6}
\defR{mix-load40-gemm/berth-nsng/p99/retinanet}{54.1}
\defR{mix-load40-gemm/berth-nsng/p99/bertlarge}{162}
\defR{mix-load40-gemm/berth-nsng/p99x/resnet50}{1.5}
\defR{mix-load40-gemm/berth-nsng/p99x/retinanet}{1.6}
\defR{mix-load40-gemm/berth-nsng/p99x/bertlarge}{2.2}
\defR{mix-load40-gemm/berth-nsng/thrabs/gemm}{8.73}
\defR{mix-load40-gemm/berth-nsng/thr/gemm}{55}
\defR{mix-load40-gemm/berth-grouped/attmin}{48.5}
\defR{mix-load40-gemm/berth-grouped/attmean}{80.2}
\defR{mix-load40-gemm/berth-grouped/thr}{59}
\defR{mix-load40-gemm/berth-grouped/n}{3}
\defR{mix-load40-gemm/berth-grouped/att/resnet50}{48.5}
\defR{mix-load40-gemm/berth-grouped/att/retinanet}{97.3}
\defR{mix-load40-gemm/berth-grouped/att/bertlarge}{94.8}
\defR{mix-load40-gemm/berth-grouped/p99/resnet50}{65.9}
\defR{mix-load40-gemm/berth-grouped/p99/retinanet}{114}
\defR{mix-load40-gemm/berth-grouped/p99/bertlarge}{181}
\defR{mix-load40-gemm/berth-grouped/p99x/resnet50}{4.9}
\defR{mix-load40-gemm/berth-grouped/p99x/retinanet}{3.3}
\defR{mix-load40-gemm/berth-grouped/p99x/bertlarge}{2.5}
\defR{mix-load40-gemm/berth-grouped/thrabs/gemm}{9.39}
\defR{mix-load40-gemm/berth-grouped/thr/gemm}{59}
\defR{mix-load40-gemm/berth/attmin}{70.3}
\defR{mix-load40-gemm/berth/attmean}{86.4}
\defR{mix-load40-gemm/berth/thr}{57}
\defR{mix-load40-gemm/berth/n}{3}
\defR{mix-load40-gemm/berth/att/resnet50}{70.3}
\defR{mix-load40-gemm/berth/att/retinanet}{98.8}
\defR{mix-load40-gemm/berth/att/bertlarge}{89.9}
\defR{mix-load40-gemm/berth/p99/resnet50}{36.7}
\defR{mix-load40-gemm/berth/p99/retinanet}{102}
\defR{mix-load40-gemm/berth/p99/bertlarge}{210}
\defR{mix-load40-gemm/berth/p99x/resnet50}{2.7}
\defR{mix-load40-gemm/berth/p99x/retinanet}{2.9}
\defR{mix-load40-gemm/berth/p99x/bertlarge}{2.9}
\defR{mix-load40-gemm/berth/thrabs/gemm}{8.99}
\defR{mix-load40-gemm/berth/thr/gemm}{57}
\defR{mix-load40-gemm/timeslice/attmin}{62.7}
\defR{mix-load40-gemm/timeslice/attmean}{79.6}
\defR{mix-load40-gemm/timeslice/thr}{56}
\defR{mix-load40-gemm/timeslice/n}{3}
\defR{mix-load40-gemm/timeslice/att/resnet50}{62.7}
\defR{mix-load40-gemm/timeslice/att/retinanet}{97.7}
\defR{mix-load40-gemm/timeslice/att/bertlarge}{78.3}
\defR{mix-load40-gemm/timeslice/p99/resnet50}{40.6}
\defR{mix-load40-gemm/timeslice/p99/retinanet}{122}
\defR{mix-load40-gemm/timeslice/p99/bertlarge}{277}
\defR{mix-load40-gemm/timeslice/p99x/resnet50}{3.0}
\defR{mix-load40-gemm/timeslice/p99x/retinanet}{3.5}
\defR{mix-load40-gemm/timeslice/p99x/bertlarge}{3.8}
\defR{mix-load40-gemm/timeslice/thrabs/gemm}{8.88}
\defR{mix-load40-gemm/timeslice/thr/gemm}{56}
\defR{mix-load40-gemm/mps/attmin}{0.0}
\defR{mix-load40-gemm/mps/attmean}{0.0}
\defR{mix-load40-gemm/mps/thr}{100}
\defR{mix-load40-gemm/mps/n}{3}
\defR{mix-load40-gemm/mps/att/resnet50}{0.0}
\defR{mix-load40-gemm/mps/att/retinanet}{0.0}
\defR{mix-load40-gemm/mps/att/bertlarge}{0.0}
\defR{mix-load40-gemm/mps/p99/resnet50}{90452}
\defR{mix-load40-gemm/mps/p99/retinanet}{90981}
\defR{mix-load40-gemm/mps/p99/bertlarge}{90484}
\defR{mix-load40-gemm/mps/p99x/resnet50}{6751.4}
\defR{mix-load40-gemm/mps/p99x/retinanet}{2616.1}
\defR{mix-load40-gemm/mps/p99x/bertlarge}{1250.7}
\defR{mix-load40-gemm/mps/thrabs/gemm}{15.86}
\defR{mix-load40-gemm/mps/thr/gemm}{100}
\defR{mix-load40-gemm/mpsprio/attmin}{0.0}
\defR{mix-load40-gemm/mpsprio/attmean}{0.0}
\defR{mix-load40-gemm/mpsprio/thr}{92}
\defR{mix-load40-gemm/mpsprio/n}{3}
\defR{mix-load40-gemm/mpsprio/att/resnet50}{0.0}
\defR{mix-load40-gemm/mpsprio/att/retinanet}{0.0}
\defR{mix-load40-gemm/mpsprio/att/bertlarge}{0.0}
\defR{mix-load40-gemm/mpsprio/p99/resnet50}{73321}
\defR{mix-load40-gemm/mpsprio/p99/retinanet}{77266}
\defR{mix-load40-gemm/mpsprio/p99/bertlarge}{68005}
\defR{mix-load40-gemm/mpsprio/p99x/resnet50}{5472.8}
\defR{mix-load40-gemm/mpsprio/p99x/retinanet}{2221.7}
\defR{mix-load40-gemm/mpsprio/p99x/bertlarge}{940.0}
\defR{mix-load40-gemm/mpsprio/thrabs/gemm}{14.58}
\defR{mix-load40-gemm/mpsprio/thr/gemm}{92}
\defR{mix-load40-gemm/mpslimits/attmin}{88.9}
\defR{mix-load40-gemm/mpslimits/attmean}{95.0}
\defR{mix-load40-gemm/mpslimits/thr}{10}
\defR{mix-load40-gemm/mpslimits/n}{3}
\defR{mix-load40-gemm/mpslimits/att/resnet50}{98.9}
\defR{mix-load40-gemm/mpslimits/att/retinanet}{97.2}
\defR{mix-load40-gemm/mpslimits/att/bertlarge}{88.9}
\defR{mix-load40-gemm/mpslimits/p99/resnet50}{14.9}
\defR{mix-load40-gemm/mpslimits/p99/retinanet}{127}
\defR{mix-load40-gemm/mpslimits/p99/bertlarge}{215}
\defR{mix-load40-gemm/mpslimits/p99x/resnet50}{1.1}
\defR{mix-load40-gemm/mpslimits/p99x/retinanet}{3.6}
\defR{mix-load40-gemm/mpslimits/p99x/bertlarge}{3.0}
\defR{mix-load40-gemm/mpslimits/thrabs/gemm}{1.56}
\defR{mix-load40-gemm/mpslimits/thr/gemm}{10}
\defR{mix-load40-gemm/mpsparts/attmin}{88.5}
\defR{mix-load40-gemm/mpsparts/attmean}{94.2}
\defR{mix-load40-gemm/mpsparts/thr}{16}
\defR{mix-load40-gemm/mpsparts/n}{2}
\defR{mix-load40-gemm/mpsparts/att/resnet50}{99.1}
\defR{mix-load40-gemm/mpsparts/att/retinanet}{94.8}
\defR{mix-load40-gemm/mpsparts/att/bertlarge}{88.5}
\defR{mix-load40-gemm/mpsparts/p99/resnet50}{15.1}
\defR{mix-load40-gemm/mpsparts/p99/retinanet}{147}
\defR{mix-load40-gemm/mpsparts/p99/bertlarge}{212}
\defR{mix-load40-gemm/mpsparts/p99x/resnet50}{1.1}
\defR{mix-load40-gemm/mpsparts/p99x/retinanet}{4.2}
\defR{mix-load40-gemm/mpsparts/p99x/bertlarge}{2.9}
\defR{mix-load40-gemm/mpsparts/thrabs/gemm}{2.50}
\defR{mix-load40-gemm/mpsparts/thr/gemm}{16}
\defR{mix-load40-gemm/lithos/attmin}{88.2}
\defR{mix-load40-gemm/lithos/attmean}{94.8}
\defR{mix-load40-gemm/lithos/thr}{12}
\defR{mix-load40-gemm/lithos/n}{3}
\defR{mix-load40-gemm/lithos/att/resnet50}{99.3}
\defR{mix-load40-gemm/lithos/att/retinanet}{96.9}
\defR{mix-load40-gemm/lithos/att/bertlarge}{88.2}
\defR{mix-load40-gemm/lithos/p99/resnet50}{14.1}
\defR{mix-load40-gemm/lithos/p99/retinanet}{131}
\defR{mix-load40-gemm/lithos/p99/bertlarge}{216}
\defR{mix-load40-gemm/lithos/p99x/resnet50}{1.0}
\defR{mix-load40-gemm/lithos/p99x/retinanet}{3.8}
\defR{mix-load40-gemm/lithos/p99x/bertlarge}{3.0}
\defR{mix-load40-gemm/lithos/thrabs/gemm}{1.91}
\defR{mix-load40-gemm/lithos/thr/gemm}{12}
\defR{mix-load40-gemm/dysm/attmin}{0.0}
\defR{mix-load40-gemm/dysm/attmean}{24.2}
\defR{mix-load40-gemm/dysm/thr}{87}
\defR{mix-load40-gemm/dysm/n}{3}
\defR{mix-load40-gemm/dysm/att/resnet50}{0.0}
\defR{mix-load40-gemm/dysm/att/retinanet}{72.6}
\defR{mix-load40-gemm/dysm/att/bertlarge}{0.0}
\defR{mix-load40-gemm/dysm/p99/retinanet}{313}
\defR{mix-load40-gemm/dysm/p99x/retinanet}{9.0}
\defR{mix-load40-gemm/dysm/thrabs/gemm}{13.72}
\defR{mix-load40-gemm/dysm/thr/gemm}{87}
\defR{mix-load40-gemm/xsched/attmin}{46.8}
\defR{mix-load40-gemm/xsched/attmean}{74.1}
\defR{mix-load40-gemm/xsched/thr}{55}
\defR{mix-load40-gemm/xsched/n}{3}
\defR{mix-load40-gemm/xsched/att/resnet50}{46.8}
\defR{mix-load40-gemm/xsched/att/retinanet}{94.7}
\defR{mix-load40-gemm/xsched/att/bertlarge}{80.9}
\defR{mix-load40-gemm/xsched/p99/resnet50}{54.5}
\defR{mix-load40-gemm/xsched/p99/retinanet}{145}
\defR{mix-load40-gemm/xsched/p99/bertlarge}{258}
\defR{mix-load40-gemm/xsched/p99x/resnet50}{4.1}
\defR{mix-load40-gemm/xsched/p99x/retinanet}{4.2}
\defR{mix-load40-gemm/xsched/p99x/bertlarge}{3.6}
\defR{mix-load40-gemm/xsched/thrabs/gemm}{8.68}
\defR{mix-load40-gemm/xsched/thr/gemm}{55}
\defR{mix-load40-gemm/multistream/attmin}{0.0}
\defR{mix-load40-gemm/multistream/attmean}{0.0}
\defR{mix-load40-gemm/multistream/thr}{92}
\defR{mix-load40-gemm/multistream/n}{3}
\defR{mix-load40-gemm/multistream/att/resnet50}{0.0}
\defR{mix-load40-gemm/multistream/att/retinanet}{0.0}
\defR{mix-load40-gemm/multistream/att/bertlarge}{0.0}
\defR{mix-load40-gemm/multistream/p99/resnet50}{73241}
\defR{mix-load40-gemm/multistream/p99/retinanet}{77447}
\defR{mix-load40-gemm/multistream/p99/bertlarge}{67857}
\defR{mix-load40-gemm/multistream/p99x/resnet50}{5466.8}
\defR{mix-load40-gemm/multistream/p99x/retinanet}{2227.0}
\defR{mix-load40-gemm/multistream/p99x/bertlarge}{937.9}
\defR{mix-load40-gemm/multistream/thrabs/gemm}{14.58}
\defR{mix-load40-gemm/multistream/thr/gemm}{92}
\defR{mix-load40-gemm/tgs/attmin}{62.7}
\defR{mix-load40-gemm/tgs/attmean}{78.7}
\defR{mix-load40-gemm/tgs/thr}{56}
\defR{mix-load40-gemm/tgs/n}{3}
\defR{mix-load40-gemm/tgs/att/resnet50}{62.7}
\defR{mix-load40-gemm/tgs/att/retinanet}{97.6}
\defR{mix-load40-gemm/tgs/att/bertlarge}{75.7}
\defR{mix-load40-gemm/tgs/p99/resnet50}{43.0}
\defR{mix-load40-gemm/tgs/p99/retinanet}{127}
\defR{mix-load40-gemm/tgs/p99/bertlarge}{289}
\defR{mix-load40-gemm/tgs/p99x/resnet50}{3.2}
\defR{mix-load40-gemm/tgs/p99x/retinanet}{3.6}
\defR{mix-load40-gemm/tgs/p99x/bertlarge}{4.0}
\defR{mix-load40-gemm/tgs/thrabs/gemm}{8.93}
\defR{mix-load40-gemm/tgs/thr/gemm}{56}
\defR{mix-load40-gemm/mig/attmin}{90.6}
\defR{mix-load40-gemm/mig/attmean}{95.8}
\defR{mix-load40-gemm/mig/thr}{0}
\defR{mix-load40-gemm/mig/n}{3}
\defR{mix-load40-gemm/mig/att/resnet50}{98.5}
\defR{mix-load40-gemm/mig/att/retinanet}{98.4}
\defR{mix-load40-gemm/mig/att/bertlarge}{90.6}
\defR{mix-load40-gemm/mig/p99/resnet50}{16.5}
\defR{mix-load40-gemm/mig/p99/retinanet}{109}
\defR{mix-load40-gemm/mig/p99/bertlarge}{207}
\defR{mix-load40-gemm/mig/p99x/resnet50}{1.2}
\defR{mix-load40-gemm/mig/p99x/retinanet}{3.1}
\defR{mix-load40-gemm/mig/p99x/bertlarge}{2.9}
\defR{mix-load40-gemm/mig/thrabs/gemm}{0.00}
\defR{mix-load40-gemm/mig/thr/gemm}{0}
\defR{mix-load40-gemm/tally/attmin}{0.0}
\defR{mix-load40-gemm/tally/attmean}{20.4}
\defR{mix-load40-gemm/tally/thr}{83}
\defR{mix-load40-gemm/tally/n}{3}
\defR{mix-load40-gemm/tally/att/resnet50}{0.0}
\defR{mix-load40-gemm/tally/att/retinanet}{0.0}
\defR{mix-load40-gemm/tally/att/bertlarge}{61.1}
\defR{mix-load40-gemm/tally/p99/resnet50}{94201}
\defR{mix-load40-gemm/tally/p99/retinanet}{94181}
\defR{mix-load40-gemm/tally/p99/bertlarge}{397}
\defR{mix-load40-gemm/tally/p99x/resnet50}{7031.3}
\defR{mix-load40-gemm/tally/p99x/retinanet}{2708.1}
\defR{mix-load40-gemm/tally/p99x/bertlarge}{5.5}
\defR{mix-load40-gemm/tally/thrabs/gemm}{13.14}
\defR{mix-load40-gemm/tally/thr/gemm}{83}
\defR{mix-load40-gemm/orion/attmin}{0.0}
\defR{mix-load40-gemm/orion/attmean}{0.0}
\defR{mix-load40-gemm/orion/thr}{89}
\defR{mix-load40-gemm/orion/n}{3}
\defR{mix-load40-gemm/orion/att/resnet50}{0.0}
\defR{mix-load40-gemm/orion/att/retinanet}{0.0}
\defR{mix-load40-gemm/orion/att/bertlarge}{0.0}
\defR{mix-load40-gemm/orion/p99/resnet50}{4836}
\defR{mix-load40-gemm/orion/p99/bertlarge}{82071}
\defR{mix-load40-gemm/orion/p99x/resnet50}{360.9}
\defR{mix-load40-gemm/orion/p99x/bertlarge}{1134.4}
\defR{mix-load40-gemm/orion/thrabs/gemm}{14.02}
\defR{mix-load40-gemm/orion/thr/gemm}{89}
\defR{bursty-resnet50-load30/berth-full/attmin}{86.0}
\defR{bursty-resnet50-load30/berth-full/attmean}{86.0}
\defR{bursty-resnet50-load30/berth-full/thr}{66}
\defR{bursty-resnet50-load30/berth-full/n}{3}
\defR{bursty-resnet50-load30/berth-full/att/resnet50}{86.0}
\defR{bursty-resnet50-load30/berth-full/p99/resnet50}{74.6}
\defR{bursty-resnet50-load30/berth-full/p99x/resnet50}{1.1}
\defR{bursty-resnet50-load30/berth-full/thrabs/train_bert}{1.07}
\defR{bursty-resnet50-load30/berth-full/thr/train_bert}{66}
\defR{bursty-resnet50-load30/berth-nsr/attmin}{86.3}
\defR{bursty-resnet50-load30/berth-nsr/attmean}{86.3}
\defR{bursty-resnet50-load30/berth-nsr/thr}{66}
\defR{bursty-resnet50-load30/berth-nsr/n}{3}
\defR{bursty-resnet50-load30/berth-nsr/att/resnet50}{86.3}
\defR{bursty-resnet50-load30/berth-nsr/p99/resnet50}{68.9}
\defR{bursty-resnet50-load30/berth-nsr/p99x/resnet50}{1.0}
\defR{bursty-resnet50-load30/berth-nsr/thrabs/train_bert}{1.08}
\defR{bursty-resnet50-load30/berth-nsr/thr/train_bert}{66}
\defR{bursty-resnet50-load30/berth-nba/attmin}{86.9}
\defR{bursty-resnet50-load30/berth-nba/attmean}{86.9}
\defR{bursty-resnet50-load30/berth-nba/thr}{67}
\defR{bursty-resnet50-load30/berth-nba/n}{3}
\defR{bursty-resnet50-load30/berth-nba/att/resnet50}{86.9}
\defR{bursty-resnet50-load30/berth-nba/p99/resnet50}{64.4}
\defR{bursty-resnet50-load30/berth-nba/p99x/resnet50}{1.0}
\defR{bursty-resnet50-load30/berth-nba/thrabs/train_bert}{1.09}
\defR{bursty-resnet50-load30/berth-nba/thr/train_bert}{67}
\defR{bursty-resnet50-load30/berth-nsb/attmin}{86.3}
\defR{bursty-resnet50-load30/berth-nsb/attmean}{86.3}
\defR{bursty-resnet50-load30/berth-nsb/thr}{66}
\defR{bursty-resnet50-load30/berth-nsb/n}{3}
\defR{bursty-resnet50-load30/berth-nsb/att/resnet50}{86.3}
\defR{bursty-resnet50-load30/berth-nsb/p99/resnet50}{69.6}
\defR{bursty-resnet50-load30/berth-nsb/p99x/resnet50}{1.0}
\defR{bursty-resnet50-load30/berth-nsb/thrabs/train_bert}{1.08}
\defR{bursty-resnet50-load30/berth-nsb/thr/train_bert}{66}
\defR{bursty-resnet50-load30/berth-ng/attmin}{84.6}
\defR{bursty-resnet50-load30/berth-ng/attmean}{84.6}
\defR{bursty-resnet50-load30/berth-ng/thr}{68}
\defR{bursty-resnet50-load30/berth-ng/n}{3}
\defR{bursty-resnet50-load30/berth-ng/att/resnet50}{84.6}
\defR{bursty-resnet50-load30/berth-ng/p99/resnet50}{77.9}
\defR{bursty-resnet50-load30/berth-ng/p99x/resnet50}{1.2}
\defR{bursty-resnet50-load30/berth-ng/thrabs/train_bert}{1.11}
\defR{bursty-resnet50-load30/berth-ng/thr/train_bert}{68}
\defR{bursty-resnet50-load30/berth-nsng/attmin}{83.6}
\defR{bursty-resnet50-load30/berth-nsng/attmean}{83.6}
\defR{bursty-resnet50-load30/berth-nsng/thr}{68}
\defR{bursty-resnet50-load30/berth-nsng/n}{3}
\defR{bursty-resnet50-load30/berth-nsng/att/resnet50}{83.6}
\defR{bursty-resnet50-load30/berth-nsng/p99/resnet50}{80.0}
\defR{bursty-resnet50-load30/berth-nsng/p99x/resnet50}{1.2}
\defR{bursty-resnet50-load30/berth-nsng/thrabs/train_bert}{1.10}
\defR{bursty-resnet50-load30/berth-nsng/thr/train_bert}{68}
\defR{bursty-resnet50-load30/berth-grouped/attmin}{83.9}
\defR{bursty-resnet50-load30/berth-grouped/attmean}{83.9}
\defR{bursty-resnet50-load30/berth-grouped/thr}{68}
\defR{bursty-resnet50-load30/berth-grouped/n}{3}
\defR{bursty-resnet50-load30/berth-grouped/att/resnet50}{83.9}
\defR{bursty-resnet50-load30/berth-grouped/p99/resnet50}{72.0}
\defR{bursty-resnet50-load30/berth-grouped/p99x/resnet50}{1.1}
\defR{bursty-resnet50-load30/berth-grouped/thrabs/train_bert}{1.11}
\defR{bursty-resnet50-load30/berth-grouped/thr/train_bert}{68}
\defR{bursty-resnet50-load30/berth/attmin}{83.7}
\defR{bursty-resnet50-load30/berth/attmean}{83.7}
\defR{bursty-resnet50-load30/berth/thr}{68}
\defR{bursty-resnet50-load30/berth/n}{3}
\defR{bursty-resnet50-load30/berth/att/resnet50}{83.7}
\defR{bursty-resnet50-load30/berth/p99/resnet50}{89.0}
\defR{bursty-resnet50-load30/berth/p99x/resnet50}{1.3}
\defR{bursty-resnet50-load30/berth/thrabs/train_bert}{1.11}
\defR{bursty-resnet50-load30/berth/thr/train_bert}{68}
\defR{bursty-resnet50-load30/timeslice/attmin}{33.7}
\defR{bursty-resnet50-load30/timeslice/attmean}{33.7}
\defR{bursty-resnet50-load30/timeslice/thr}{68}
\defR{bursty-resnet50-load30/timeslice/n}{3}
\defR{bursty-resnet50-load30/timeslice/att/resnet50}{33.7}
\defR{bursty-resnet50-load30/timeslice/p99/resnet50}{935}
\defR{bursty-resnet50-load30/timeslice/p99x/resnet50}{14.0}
\defR{bursty-resnet50-load30/timeslice/thrabs/train_bert}{1.11}
\defR{bursty-resnet50-load30/timeslice/thr/train_bert}{68}
\defR{bursty-resnet50-load30/mps/attmin}{0.0}
\defR{bursty-resnet50-load30/mps/attmean}{0.0}
\defR{bursty-resnet50-load30/mps/thr}{93}
\defR{bursty-resnet50-load30/mps/n}{3}
\defR{bursty-resnet50-load30/mps/att/resnet50}{0.0}
\defR{bursty-resnet50-load30/mps/p99/resnet50}{68595}
\defR{bursty-resnet50-load30/mps/p99x/resnet50}{1028.5}
\defR{bursty-resnet50-load30/mps/thrabs/train_bert}{1.51}
\defR{bursty-resnet50-load30/mps/thr/train_bert}{93}
\defR{bursty-resnet50-load30/mpsprio/attmin}{0.0}
\defR{bursty-resnet50-load30/mpsprio/attmean}{0.0}
\defR{bursty-resnet50-load30/mpsprio/thr}{83}
\defR{bursty-resnet50-load30/mpsprio/n}{3}
\defR{bursty-resnet50-load30/mpsprio/att/resnet50}{0.0}
\defR{bursty-resnet50-load30/mpsprio/p99/resnet50}{28265}
\defR{bursty-resnet50-load30/mpsprio/p99x/resnet50}{423.8}
\defR{bursty-resnet50-load30/mpsprio/thrabs/train_bert}{1.35}
\defR{bursty-resnet50-load30/mpsprio/thr/train_bert}{83}
\defR{bursty-resnet50-load30/mpslimits/attmin}{89.1}
\defR{bursty-resnet50-load30/mpslimits/attmean}{89.1}
\defR{bursty-resnet50-load30/mpslimits/thr}{12}
\defR{bursty-resnet50-load30/mpslimits/n}{3}
\defR{bursty-resnet50-load30/mpslimits/att/resnet50}{89.1}
\defR{bursty-resnet50-load30/mpslimits/p99/resnet50}{62.3}
\defR{bursty-resnet50-load30/mpslimits/p99x/resnet50}{0.9}
\defR{bursty-resnet50-load30/mpslimits/thrabs/train_bert}{0.19}
\defR{bursty-resnet50-load30/mpslimits/thr/train_bert}{12}
\defR{bursty-resnet50-load30/mpsparts/attmin}{89.0}
\defR{bursty-resnet50-load30/mpsparts/attmean}{89.0}
\defR{bursty-resnet50-load30/mpsparts/thr}{10}
\defR{bursty-resnet50-load30/mpsparts/n}{2}
\defR{bursty-resnet50-load30/mpsparts/att/resnet50}{89.0}
\defR{bursty-resnet50-load30/mpsparts/p99/resnet50}{66.5}
\defR{bursty-resnet50-load30/mpsparts/p99x/resnet50}{1.0}
\defR{bursty-resnet50-load30/mpsparts/thrabs/train_bert}{0.16}
\defR{bursty-resnet50-load30/mpsparts/thr/train_bert}{10}
\defR{bursty-resnet50-load30/lithos/attmin}{87.2}
\defR{bursty-resnet50-load30/lithos/attmean}{87.2}
\defR{bursty-resnet50-load30/lithos/thr}{9}
\defR{bursty-resnet50-load30/lithos/n}{3}
\defR{bursty-resnet50-load30/lithos/att/resnet50}{87.2}
\defR{bursty-resnet50-load30/lithos/p99/resnet50}{74.0}
\defR{bursty-resnet50-load30/lithos/p99x/resnet50}{1.1}
\defR{bursty-resnet50-load30/lithos/thrabs/train_bert}{0.14}
\defR{bursty-resnet50-load30/lithos/thr/train_bert}{9}
\defR{bursty-resnet50-load30/dysm/attmin}{0.0}
\defR{bursty-resnet50-load30/dysm/attmean}{0.0}
\defR{bursty-resnet50-load30/dysm/thr}{100}
\defR{bursty-resnet50-load30/dysm/n}{3}
\defR{bursty-resnet50-load30/dysm/att/resnet50}{0.0}
\defR{bursty-resnet50-load30/dysm/thrabs/train_bert}{1.62}
\defR{bursty-resnet50-load30/dysm/thr/train_bert}{100}
\defR{bursty-resnet50-load30/xsched/attmin}{71.6}
\defR{bursty-resnet50-load30/xsched/attmean}{71.6}
\defR{bursty-resnet50-load30/xsched/thr}{67}
\defR{bursty-resnet50-load30/xsched/n}{3}
\defR{bursty-resnet50-load30/xsched/att/resnet50}{71.6}
\defR{bursty-resnet50-load30/xsched/p99/resnet50}{254}
\defR{bursty-resnet50-load30/xsched/p99x/resnet50}{3.8}
\defR{bursty-resnet50-load30/xsched/thrabs/train_bert}{1.09}
\defR{bursty-resnet50-load30/xsched/thr/train_bert}{67}
\defR{bursty-resnet50-load30/multistream/attmin}{0.0}
\defR{bursty-resnet50-load30/multistream/attmean}{0.0}
\defR{bursty-resnet50-load30/multistream/thr}{84}
\defR{bursty-resnet50-load30/multistream/n}{3}
\defR{bursty-resnet50-load30/multistream/att/resnet50}{0.0}
\defR{bursty-resnet50-load30/multistream/p99/resnet50}{33262}
\defR{bursty-resnet50-load30/multistream/p99x/resnet50}{498.7}
\defR{bursty-resnet50-load30/multistream/thrabs/train_bert}{1.37}
\defR{bursty-resnet50-load30/multistream/thr/train_bert}{84}
\defR{bursty-resnet50-load30/tgs/attmin}{88.3}
\defR{bursty-resnet50-load30/tgs/attmean}{88.3}
\defR{bursty-resnet50-load30/tgs/thr}{2}
\defR{bursty-resnet50-load30/tgs/n}{3}
\defR{bursty-resnet50-load30/tgs/att/resnet50}{88.3}
\defR{bursty-resnet50-load30/tgs/p99/resnet50}{63.5}
\defR{bursty-resnet50-load30/tgs/p99x/resnet50}{1.0}
\defR{bursty-resnet50-load30/tgs/thrabs/train_bert}{0.03}
\defR{bursty-resnet50-load30/tgs/thr/train_bert}{2}
\defR{bursty-resnet50-load30/mig/attmin}{80.6}
\defR{bursty-resnet50-load30/mig/attmean}{80.6}
\defR{bursty-resnet50-load30/mig/thr}{45}
\defR{bursty-resnet50-load30/mig/n}{3}
\defR{bursty-resnet50-load30/mig/att/resnet50}{80.6}
\defR{bursty-resnet50-load30/mig/p99/resnet50}{173}
\defR{bursty-resnet50-load30/mig/p99x/resnet50}{2.6}
\defR{bursty-resnet50-load30/mig/thrabs/train_bert}{0.72}
\defR{bursty-resnet50-load30/mig/thr/train_bert}{45}
\defR{bursty-resnet50-load30/tally/attmin}{0.0}
\defR{bursty-resnet50-load30/tally/attmean}{0.0}
\defR{bursty-resnet50-load30/tally/thr}{1}
\defR{bursty-resnet50-load30/tally/n}{3}
\defR{bursty-resnet50-load30/tally/att/resnet50}{0.0}
\defR{bursty-resnet50-load30/tally/p99/resnet50}{52201}
\defR{bursty-resnet50-load30/tally/p99x/resnet50}{782.7}
\defR{bursty-resnet50-load30/tally/thrabs/train_bert}{0.01}
\defR{bursty-resnet50-load30/tally/thr/train_bert}{1}
\defR{bursty-resnet50-load30/orion/attmin}{84.1}
\defR{bursty-resnet50-load30/orion/attmean}{84.1}
\defR{bursty-resnet50-load30/orion/thr}{72}
\defR{bursty-resnet50-load30/orion/n}{3}
\defR{bursty-resnet50-load30/orion/att/resnet50}{84.1}
\defR{bursty-resnet50-load30/orion/p99/resnet50}{60.4}
\defR{bursty-resnet50-load30/orion/p99x/resnet50}{0.9}
\defR{bursty-resnet50-load30/orion/thrabs/train_bert}{1.17}
\defR{bursty-resnet50-load30/orion/thr/train_bert}{72}
\defR{bursty-mix-load40/berth-full/attmin}{84.8}
\defR{bursty-mix-load40/berth-full/attmean}{92.9}
\defR{bursty-mix-load40/berth-full/thr}{35}
\defR{bursty-mix-load40/berth-full/n}{3}
\defR{bursty-mix-load40/berth-full/att/resnet50}{95.0}
\defR{bursty-mix-load40/berth-full/att/retinanet}{98.8}
\defR{bursty-mix-load40/berth-full/att/bertlarge}{84.8}
\defR{bursty-mix-load40/berth-full/p99/resnet50}{22.0}
\defR{bursty-mix-load40/berth-full/p99/retinanet}{103}
\defR{bursty-mix-load40/berth-full/p99/bertlarge}{351}
\defR{bursty-mix-load40/berth-full/p99x/resnet50}{1.0}
\defR{bursty-mix-load40/berth-full/p99x/retinanet}{2.1}
\defR{bursty-mix-load40/berth-full/p99x/bertlarge}{3.5}
\defR{bursty-mix-load40/berth-full/thrabs/train_bert}{0.56}
\defR{bursty-mix-load40/berth-full/thr/train_bert}{35}
\defR{bursty-mix-load40/berth-nsr/attmin}{83.9}
\defR{bursty-mix-load40/berth-nsr/attmean}{92.6}
\defR{bursty-mix-load40/berth-nsr/thr}{34}
\defR{bursty-mix-load40/berth-nsr/n}{3}
\defR{bursty-mix-load40/berth-nsr/att/resnet50}{95.2}
\defR{bursty-mix-load40/berth-nsr/att/retinanet}{98.7}
\defR{bursty-mix-load40/berth-nsr/att/bertlarge}{83.9}
\defR{bursty-mix-load40/berth-nsr/p99/resnet50}{21.9}
\defR{bursty-mix-load40/berth-nsr/p99/retinanet}{105}
\defR{bursty-mix-load40/berth-nsr/p99/bertlarge}{367}
\defR{bursty-mix-load40/berth-nsr/p99x/resnet50}{1.0}
\defR{bursty-mix-load40/berth-nsr/p99x/retinanet}{2.2}
\defR{bursty-mix-load40/berth-nsr/p99x/bertlarge}{3.6}
\defR{bursty-mix-load40/berth-nsr/thrabs/train_bert}{0.55}
\defR{bursty-mix-load40/berth-nsr/thr/train_bert}{34}
\defR{bursty-mix-load40/berth-nba/attmin}{85.4}
\defR{bursty-mix-load40/berth-nba/attmean}{93.2}
\defR{bursty-mix-load40/berth-nba/thr}{35}
\defR{bursty-mix-load40/berth-nba/n}{3}
\defR{bursty-mix-load40/berth-nba/att/resnet50}{95.4}
\defR{bursty-mix-load40/berth-nba/att/retinanet}{99.0}
\defR{bursty-mix-load40/berth-nba/att/bertlarge}{85.4}
\defR{bursty-mix-load40/berth-nba/p99/resnet50}{21.7}
\defR{bursty-mix-load40/berth-nba/p99/retinanet}{98.8}
\defR{bursty-mix-load40/berth-nba/p99/bertlarge}{339}
\defR{bursty-mix-load40/berth-nba/p99x/resnet50}{1.0}
\defR{bursty-mix-load40/berth-nba/p99x/retinanet}{2.0}
\defR{bursty-mix-load40/berth-nba/p99x/bertlarge}{3.4}
\defR{bursty-mix-load40/berth-nba/thrabs/train_bert}{0.56}
\defR{bursty-mix-load40/berth-nba/thr/train_bert}{35}
\defR{bursty-mix-load40/berth-nsb/attmin}{85.6}
\defR{bursty-mix-load40/berth-nsb/attmean}{93.4}
\defR{bursty-mix-load40/berth-nsb/thr}{35}
\defR{bursty-mix-load40/berth-nsb/n}{3}
\defR{bursty-mix-load40/berth-nsb/att/resnet50}{95.4}
\defR{bursty-mix-load40/berth-nsb/att/retinanet}{99.1}
\defR{bursty-mix-load40/berth-nsb/att/bertlarge}{85.6}
\defR{bursty-mix-load40/berth-nsb/p99/resnet50}{21.8}
\defR{bursty-mix-load40/berth-nsb/p99/retinanet}{97.5}
\defR{bursty-mix-load40/berth-nsb/p99/bertlarge}{341}
\defR{bursty-mix-load40/berth-nsb/p99x/resnet50}{1.0}
\defR{bursty-mix-load40/berth-nsb/p99x/retinanet}{2.0}
\defR{bursty-mix-load40/berth-nsb/p99x/bertlarge}{3.4}
\defR{bursty-mix-load40/berth-nsb/thrabs/train_bert}{0.56}
\defR{bursty-mix-load40/berth-nsb/thr/train_bert}{35}
\defR{bursty-mix-load40/berth-ng/attmin}{76.7}
\defR{bursty-mix-load40/berth-ng/attmean}{90.1}
\defR{bursty-mix-load40/berth-ng/thr}{34}
\defR{bursty-mix-load40/berth-ng/n}{3}
\defR{bursty-mix-load40/berth-ng/att/resnet50}{94.7}
\defR{bursty-mix-load40/berth-ng/att/retinanet}{98.9}
\defR{bursty-mix-load40/berth-ng/att/bertlarge}{76.7}
\defR{bursty-mix-load40/berth-ng/p99/resnet50}{22.6}
\defR{bursty-mix-load40/berth-ng/p99/retinanet}{102}
\defR{bursty-mix-load40/berth-ng/p99/bertlarge}{403}
\defR{bursty-mix-load40/berth-ng/p99x/resnet50}{1.0}
\defR{bursty-mix-load40/berth-ng/p99x/retinanet}{2.1}
\defR{bursty-mix-load40/berth-ng/p99x/bertlarge}{4.0}
\defR{bursty-mix-load40/berth-ng/thrabs/train_bert}{0.56}
\defR{bursty-mix-load40/berth-ng/thr/train_bert}{34}
\defR{bursty-mix-load40/berth-nsng/attmin}{77.3}
\defR{bursty-mix-load40/berth-nsng/attmean}{90.2}
\defR{bursty-mix-load40/berth-nsng/thr}{34}
\defR{bursty-mix-load40/berth-nsng/n}{3}
\defR{bursty-mix-load40/berth-nsng/att/resnet50}{94.6}
\defR{bursty-mix-load40/berth-nsng/att/retinanet}{98.7}
\defR{bursty-mix-load40/berth-nsng/att/bertlarge}{77.3}
\defR{bursty-mix-load40/berth-nsng/p99/resnet50}{22.4}
\defR{bursty-mix-load40/berth-nsng/p99/retinanet}{105}
\defR{bursty-mix-load40/berth-nsng/p99/bertlarge}{407}
\defR{bursty-mix-load40/berth-nsng/p99x/resnet50}{1.0}
\defR{bursty-mix-load40/berth-nsng/p99x/retinanet}{2.2}
\defR{bursty-mix-load40/berth-nsng/p99x/bertlarge}{4.0}
\defR{bursty-mix-load40/berth-nsng/thrabs/train_bert}{0.56}
\defR{bursty-mix-load40/berth-nsng/thr/train_bert}{34}
\defR{bursty-mix-load40/berth-grouped/attmin}{82.8}
\defR{bursty-mix-load40/berth-grouped/attmean}{91.2}
\defR{bursty-mix-load40/berth-grouped/thr}{41}
\defR{bursty-mix-load40/berth-grouped/n}{3}
\defR{bursty-mix-load40/berth-grouped/att/resnet50}{82.8}
\defR{bursty-mix-load40/berth-grouped/att/retinanet}{97.7}
\defR{bursty-mix-load40/berth-grouped/att/bertlarge}{93.3}
\defR{bursty-mix-load40/berth-grouped/p99/resnet50}{38.5}
\defR{bursty-mix-load40/berth-grouped/p99/retinanet}{135}
\defR{bursty-mix-load40/berth-grouped/p99/bertlarge}{213}
\defR{bursty-mix-load40/berth-grouped/p99x/resnet50}{1.8}
\defR{bursty-mix-load40/berth-grouped/p99x/retinanet}{2.8}
\defR{bursty-mix-load40/berth-grouped/p99x/bertlarge}{2.1}
\defR{bursty-mix-load40/berth-grouped/thrabs/train_bert}{0.66}
\defR{bursty-mix-load40/berth-grouped/thr/train_bert}{41}
\defR{bursty-mix-load40/berth/attmin}{78.0}
\defR{bursty-mix-load40/berth/attmean}{83.3}
\defR{bursty-mix-load40/berth/thr}{34}
\defR{bursty-mix-load40/berth/n}{3}
\defR{bursty-mix-load40/berth/att/resnet50}{78.3}
\defR{bursty-mix-load40/berth/att/retinanet}{93.6}
\defR{bursty-mix-load40/berth/att/bertlarge}{78.0}
\defR{bursty-mix-load40/berth/p99/resnet50}{55.8}
\defR{bursty-mix-load40/berth/p99/retinanet}{252}
\defR{bursty-mix-load40/berth/p99/bertlarge}{407}
\defR{bursty-mix-load40/berth/p99x/resnet50}{2.6}
\defR{bursty-mix-load40/berth/p99x/retinanet}{5.2}
\defR{bursty-mix-load40/berth/p99x/bertlarge}{4.0}
\defR{bursty-mix-load40/berth/thrabs/train_bert}{0.55}
\defR{bursty-mix-load40/berth/thr/train_bert}{34}
\defR{bursty-mix-load40/timeslice/attmin}{6.2}
\defR{bursty-mix-load40/timeslice/attmean}{23.3}
\defR{bursty-mix-load40/timeslice/thr}{32}
\defR{bursty-mix-load40/timeslice/n}{3}
\defR{bursty-mix-load40/timeslice/att/resnet50}{28.3}
\defR{bursty-mix-load40/timeslice/att/retinanet}{35.3}
\defR{bursty-mix-load40/timeslice/att/bertlarge}{6.2}
\defR{bursty-mix-load40/timeslice/p99/resnet50}{711}
\defR{bursty-mix-load40/timeslice/p99/retinanet}{4367}
\defR{bursty-mix-load40/timeslice/p99/bertlarge}{12904}
\defR{bursty-mix-load40/timeslice/p99x/resnet50}{32.6}
\defR{bursty-mix-load40/timeslice/p99x/retinanet}{90.5}
\defR{bursty-mix-load40/timeslice/p99x/bertlarge}{127.4}
\defR{bursty-mix-load40/timeslice/thrabs/train_bert}{0.53}
\defR{bursty-mix-load40/timeslice/thr/train_bert}{32}
\defR{bursty-mix-load40/mps/attmin}{0.0}
\defR{bursty-mix-load40/mps/attmean}{0.0}
\defR{bursty-mix-load40/mps/thr}{80}
\defR{bursty-mix-load40/mps/n}{3}
\defR{bursty-mix-load40/mps/att/resnet50}{0.0}
\defR{bursty-mix-load40/mps/att/retinanet}{0.0}
\defR{bursty-mix-load40/mps/att/bertlarge}{0.0}
\defR{bursty-mix-load40/mps/p99/resnet50}{56280}
\defR{bursty-mix-load40/mps/p99/retinanet}{63214}
\defR{bursty-mix-load40/mps/p99/bertlarge}{34232}
\defR{bursty-mix-load40/mps/p99x/resnet50}{2578.7}
\defR{bursty-mix-load40/mps/p99x/retinanet}{1310.0}
\defR{bursty-mix-load40/mps/p99x/bertlarge}{337.9}
\defR{bursty-mix-load40/mps/thrabs/train_bert}{1.30}
\defR{bursty-mix-load40/mps/thr/train_bert}{80}
\defR{bursty-mix-load40/mpsprio/attmin}{0.0}
\defR{bursty-mix-load40/mpsprio/attmean}{7.3}
\defR{bursty-mix-load40/mpsprio/thr}{56}
\defR{bursty-mix-load40/mpsprio/n}{3}
\defR{bursty-mix-load40/mpsprio/att/resnet50}{0.0}
\defR{bursty-mix-load40/mpsprio/att/retinanet}{0.0}
\defR{bursty-mix-load40/mpsprio/att/bertlarge}{21.8}
\defR{bursty-mix-load40/mpsprio/p99/resnet50}{3676}
\defR{bursty-mix-load40/mpsprio/p99/retinanet}{18917}
\defR{bursty-mix-load40/mpsprio/p99/bertlarge}{6574}
\defR{bursty-mix-load40/mpsprio/p99x/resnet50}{168.4}
\defR{bursty-mix-load40/mpsprio/p99x/retinanet}{392.0}
\defR{bursty-mix-load40/mpsprio/p99x/bertlarge}{64.9}
\defR{bursty-mix-load40/mpsprio/thrabs/train_bert}{0.91}
\defR{bursty-mix-load40/mpsprio/thr/train_bert}{56}
\defR{bursty-mix-load40/mpslimits/attmin}{34.5}
\defR{bursty-mix-load40/mpslimits/attmean}{59.0}
\defR{bursty-mix-load40/mpslimits/thr}{12}
\defR{bursty-mix-load40/mpslimits/n}{3}
\defR{bursty-mix-load40/mpslimits/att/resnet50}{90.7}
\defR{bursty-mix-load40/mpslimits/att/retinanet}{51.9}
\defR{bursty-mix-load40/mpslimits/att/bertlarge}{34.5}
\defR{bursty-mix-load40/mpslimits/p99/resnet50}{30.2}
\defR{bursty-mix-load40/mpslimits/p99/retinanet}{2762}
\defR{bursty-mix-load40/mpslimits/p99/bertlarge}{4256}
\defR{bursty-mix-load40/mpslimits/p99x/resnet50}{1.4}
\defR{bursty-mix-load40/mpslimits/p99x/retinanet}{57.2}
\defR{bursty-mix-load40/mpslimits/p99x/bertlarge}{42.0}
\defR{bursty-mix-load40/mpslimits/thrabs/train_bert}{0.19}
\defR{bursty-mix-load40/mpslimits/thr/train_bert}{12}
\defR{bursty-mix-load40/mpsparts/attmin}{33.5}
\defR{bursty-mix-load40/mpsparts/attmean}{54.6}
\defR{bursty-mix-load40/mpsparts/thr}{18}
\defR{bursty-mix-load40/mpsparts/n}{2}
\defR{bursty-mix-load40/mpsparts/att/resnet50}{90.8}
\defR{bursty-mix-load40/mpsparts/att/retinanet}{39.5}
\defR{bursty-mix-load40/mpsparts/att/bertlarge}{33.5}
\defR{bursty-mix-load40/mpsparts/p99/resnet50}{30.8}
\defR{bursty-mix-load40/mpsparts/p99/retinanet}{3727}
\defR{bursty-mix-load40/mpsparts/p99/bertlarge}{4416}
\defR{bursty-mix-load40/mpsparts/p99x/resnet50}{1.4}
\defR{bursty-mix-load40/mpsparts/p99x/retinanet}{77.2}
\defR{bursty-mix-load40/mpsparts/p99x/bertlarge}{43.6}
\defR{bursty-mix-load40/mpsparts/thrabs/train_bert}{0.29}
\defR{bursty-mix-load40/mpsparts/thr/train_bert}{18}
\defR{bursty-mix-load40/lithos/attmin}{31.6}
\defR{bursty-mix-load40/lithos/attmean}{57.1}
\defR{bursty-mix-load40/lithos/thr}{13}
\defR{bursty-mix-load40/lithos/n}{3}
\defR{bursty-mix-load40/lithos/att/resnet50}{91.4}
\defR{bursty-mix-load40/lithos/att/retinanet}{48.2}
\defR{bursty-mix-load40/lithos/att/bertlarge}{31.6}
\defR{bursty-mix-load40/lithos/p99/resnet50}{29.5}
\defR{bursty-mix-load40/lithos/p99/retinanet}{3092}
\defR{bursty-mix-load40/lithos/p99/bertlarge}{4618}
\defR{bursty-mix-load40/lithos/p99x/resnet50}{1.3}
\defR{bursty-mix-load40/lithos/p99x/retinanet}{64.1}
\defR{bursty-mix-load40/lithos/p99x/bertlarge}{45.6}
\defR{bursty-mix-load40/lithos/thrabs/train_bert}{0.21}
\defR{bursty-mix-load40/lithos/thr/train_bert}{13}
\defR{bursty-mix-load40/dysm/attmin}{0.0}
\defR{bursty-mix-load40/dysm/attmean}{20.3}
\defR{bursty-mix-load40/dysm/thr}{78}
\defR{bursty-mix-load40/dysm/n}{3}
\defR{bursty-mix-load40/dysm/att/resnet50}{0.0}
\defR{bursty-mix-load40/dysm/att/retinanet}{60.9}
\defR{bursty-mix-load40/dysm/att/bertlarge}{0.0}
\defR{bursty-mix-load40/dysm/p99/retinanet}{899}
\defR{bursty-mix-load40/dysm/p99x/retinanet}{18.6}
\defR{bursty-mix-load40/dysm/thrabs/train_bert}{1.26}
\defR{bursty-mix-load40/dysm/thr/train_bert}{78}
\defR{bursty-mix-load40/xsched/attmin}{66.5}
\defR{bursty-mix-load40/xsched/attmean}{78.2}
\defR{bursty-mix-load40/xsched/thr}{32}
\defR{bursty-mix-load40/xsched/n}{3}
\defR{bursty-mix-load40/xsched/att/resnet50}{66.5}
\defR{bursty-mix-load40/xsched/att/retinanet}{91.4}
\defR{bursty-mix-load40/xsched/att/bertlarge}{76.6}
\defR{bursty-mix-load40/xsched/p99/resnet50}{81.8}
\defR{bursty-mix-load40/xsched/p99/retinanet}{321}
\defR{bursty-mix-load40/xsched/p99/bertlarge}{457}
\defR{bursty-mix-load40/xsched/p99x/resnet50}{3.7}
\defR{bursty-mix-load40/xsched/p99x/retinanet}{6.7}
\defR{bursty-mix-load40/xsched/p99x/bertlarge}{4.5}
\defR{bursty-mix-load40/xsched/thrabs/train_bert}{0.52}
\defR{bursty-mix-load40/xsched/thr/train_bert}{32}
\defR{bursty-mix-load40/multistream/attmin}{0.0}
\defR{bursty-mix-load40/multistream/attmean}{7.6}
\defR{bursty-mix-load40/multistream/thr}{57}
\defR{bursty-mix-load40/multistream/n}{3}
\defR{bursty-mix-load40/multistream/att/resnet50}{0.0}
\defR{bursty-mix-load40/multistream/att/retinanet}{0.6}
\defR{bursty-mix-load40/multistream/att/bertlarge}{22.3}
\defR{bursty-mix-load40/multistream/p99/resnet50}{4124}
\defR{bursty-mix-load40/multistream/p99/retinanet}{17133}
\defR{bursty-mix-load40/multistream/p99/bertlarge}{6414}
\defR{bursty-mix-load40/multistream/p99x/resnet50}{189.0}
\defR{bursty-mix-load40/multistream/p99x/retinanet}{355.1}
\defR{bursty-mix-load40/multistream/p99x/bertlarge}{63.3}
\defR{bursty-mix-load40/multistream/thrabs/train_bert}{0.92}
\defR{bursty-mix-load40/multistream/thr/train_bert}{57}
\defR{bursty-mix-load40/tgs/attmin}{85.3}
\defR{bursty-mix-load40/tgs/attmean}{94.8}
\defR{bursty-mix-load40/tgs/thr}{1}
\defR{bursty-mix-load40/tgs/n}{3}
\defR{bursty-mix-load40/tgs/att/resnet50}{85.3}
\defR{bursty-mix-load40/tgs/att/retinanet}{100.0}
\defR{bursty-mix-load40/tgs/att/bertlarge}{99.1}
\defR{bursty-mix-load40/tgs/p99/resnet50}{47.3}
\defR{bursty-mix-load40/tgs/p99/retinanet}{60.9}
\defR{bursty-mix-load40/tgs/p99/bertlarge}{125}
\defR{bursty-mix-load40/tgs/p99x/resnet50}{2.2}
\defR{bursty-mix-load40/tgs/p99x/retinanet}{1.3}
\defR{bursty-mix-load40/tgs/p99x/bertlarge}{1.2}
\defR{bursty-mix-load40/tgs/thrabs/train_bert}{0.01}
\defR{bursty-mix-load40/tgs/thr/train_bert}{1}
\defR{bursty-mix-load40/mig/attmin}{37.9}
\defR{bursty-mix-load40/mig/attmean}{65.2}
\defR{bursty-mix-load40/mig/thr}{0}
\defR{bursty-mix-load40/mig/n}{3}
\defR{bursty-mix-load40/mig/att/resnet50}{90.4}
\defR{bursty-mix-load40/mig/att/retinanet}{67.2}
\defR{bursty-mix-load40/mig/att/bertlarge}{37.9}
\defR{bursty-mix-load40/mig/p99/resnet50}{32.8}
\defR{bursty-mix-load40/mig/p99/retinanet}{1403}
\defR{bursty-mix-load40/mig/p99/bertlarge}{3456}
\defR{bursty-mix-load40/mig/p99x/resnet50}{1.5}
\defR{bursty-mix-load40/mig/p99x/retinanet}{29.1}
\defR{bursty-mix-load40/mig/p99x/bertlarge}{34.1}
\defR{bursty-mix-load40/mig/thrabs/train_bert}{0.00}
\defR{bursty-mix-load40/mig/thr/train_bert}{0}
\defR{bursty-mix-load40/tally/attmin}{0.0}
\defR{bursty-mix-load40/tally/attmean}{17.5}
\defR{bursty-mix-load40/tally/thr}{15}
\defR{bursty-mix-load40/tally/n}{3}
\defR{bursty-mix-load40/tally/att/resnet50}{0.0}
\defR{bursty-mix-load40/tally/att/retinanet}{0.0}
\defR{bursty-mix-load40/tally/att/bertlarge}{52.5}
\defR{bursty-mix-load40/tally/p99/resnet50}{94122}
\defR{bursty-mix-load40/tally/p99/retinanet}{56422}
\defR{bursty-mix-load40/tally/p99/bertlarge}{1374}
\defR{bursty-mix-load40/tally/p99x/resnet50}{4312.7}
\defR{bursty-mix-load40/tally/p99x/retinanet}{1169.3}
\defR{bursty-mix-load40/tally/p99x/bertlarge}{13.6}
\defR{bursty-mix-load40/tally/thrabs/train_bert}{0.24}
\defR{bursty-mix-load40/tally/thr/train_bert}{15}
\defR{bursty-mix-load40/orion/attmin}{0.0}
\defR{bursty-mix-load40/orion/attmean}{29.4}
\defR{bursty-mix-load40/orion/thr}{48}
\defR{bursty-mix-load40/orion/n}{3}
\defR{bursty-mix-load40/orion/att/resnet50}{86.4}
\defR{bursty-mix-load40/orion/att/retinanet}{0.0}
\defR{bursty-mix-load40/orion/att/bertlarge}{1.8}
\defR{bursty-mix-load40/orion/p99/resnet50}{34.9}
\defR{bursty-mix-load40/orion/p99/retinanet}{55566}
\defR{bursty-mix-load40/orion/p99/bertlarge}{10301}
\defR{bursty-mix-load40/orion/p99x/resnet50}{1.6}
\defR{bursty-mix-load40/orion/p99x/retinanet}{1151.5}
\defR{bursty-mix-load40/orion/p99x/bertlarge}{101.7}
\defR{bursty-mix-load40/orion/thrabs/train_bert}{0.79}
\defR{bursty-mix-load40/orion/thr/train_bert}{48}
\defR{churn-load40/berth-full/attmin}{96.6}
\defR{churn-load40/berth-full/attmean}{98.5}
\defR{churn-load40/berth-full/thr}{47}
\defR{churn-load40/berth-full/n}{3}
\defR{churn-load40/berth-full/att/resnet50}{96.6}
\defR{churn-load40/berth-full/att/bertlarge}{98.2}
\defR{churn-load40/berth-full/att/retinanet}{99.8}
\defR{churn-load40/berth-full/att/perception}{99.3}
\defR{churn-load40/berth-full/p99/resnet50}{18.2}
\defR{churn-load40/berth-full/p99/bertlarge}{148}
\defR{churn-load40/berth-full/p99/retinanet}{74.3}
\defR{churn-load40/berth-full/p99/perception}{30.7}
\defR{churn-load40/berth-full/p99x/resnet50}{1.4}
\defR{churn-load40/berth-full/p99x/bertlarge}{2.0}
\defR{churn-load40/berth-full/p99x/retinanet}{2.1}
\defR{churn-load40/berth-full/p99x/perception}{2.2}
\defR{churn-load40/berth-full/thrabs/train_bert}{0.76}
\defR{churn-load40/berth-full/thr/train_bert}{47}
\defR{churn-load40/berth-nsr/attmin}{96.8}
\defR{churn-load40/berth-nsr/attmean}{98.6}
\defR{churn-load40/berth-nsr/thr}{47}
\defR{churn-load40/berth-nsr/n}{3}
\defR{churn-load40/berth-nsr/att/resnet50}{96.8}
\defR{churn-load40/berth-nsr/att/bertlarge}{98.4}
\defR{churn-load40/berth-nsr/att/retinanet}{99.8}
\defR{churn-load40/berth-nsr/att/perception}{99.4}
\defR{churn-load40/berth-nsr/p99/resnet50}{18.2}
\defR{churn-load40/berth-nsr/p99/bertlarge}{148}
\defR{churn-load40/berth-nsr/p99/retinanet}{72.3}
\defR{churn-load40/berth-nsr/p99/perception}{30.1}
\defR{churn-load40/berth-nsr/p99x/resnet50}{1.4}
\defR{churn-load40/berth-nsr/p99x/bertlarge}{2.0}
\defR{churn-load40/berth-nsr/p99x/retinanet}{2.1}
\defR{churn-load40/berth-nsr/p99x/perception}{2.2}
\defR{churn-load40/berth-nsr/thrabs/train_bert}{0.76}
\defR{churn-load40/berth-nsr/thr/train_bert}{47}
\defR{churn-load40/berth-nba/attmin}{97.2}
\defR{churn-load40/berth-nba/attmean}{98.6}
\defR{churn-load40/berth-nba/thr}{47}
\defR{churn-load40/berth-nba/n}{3}
\defR{churn-load40/berth-nba/att/resnet50}{97.2}
\defR{churn-load40/berth-nba/att/bertlarge}{98.4}
\defR{churn-load40/berth-nba/att/retinanet}{99.3}
\defR{churn-load40/berth-nba/att/perception}{99.5}
\defR{churn-load40/berth-nba/p99/resnet50}{17.6}
\defR{churn-load40/berth-nba/p99/bertlarge}{140}
\defR{churn-load40/berth-nba/p99/retinanet}{89.7}
\defR{churn-load40/berth-nba/p99/perception}{29.1}
\defR{churn-load40/berth-nba/p99x/resnet50}{1.3}
\defR{churn-load40/berth-nba/p99x/bertlarge}{1.9}
\defR{churn-load40/berth-nba/p99x/retinanet}{2.6}
\defR{churn-load40/berth-nba/p99x/perception}{2.1}
\defR{churn-load40/berth-nba/thrabs/train_bert}{0.77}
\defR{churn-load40/berth-nba/thr/train_bert}{47}
\defR{churn-load40/berth-nsb/attmin}{96.9}
\defR{churn-load40/berth-nsb/attmean}{98.5}
\defR{churn-load40/berth-nsb/thr}{47}
\defR{churn-load40/berth-nsb/n}{3}
\defR{churn-load40/berth-nsb/att/resnet50}{96.9}
\defR{churn-load40/berth-nsb/att/bertlarge}{98.2}
\defR{churn-load40/berth-nsb/att/retinanet}{99.5}
\defR{churn-load40/berth-nsb/att/perception}{99.2}
\defR{churn-load40/berth-nsb/p99/resnet50}{19.1}
\defR{churn-load40/berth-nsb/p99/bertlarge}{154}
\defR{churn-load40/berth-nsb/p99/retinanet}{79.0}
\defR{churn-load40/berth-nsb/p99/perception}{30.6}
\defR{churn-load40/berth-nsb/p99x/resnet50}{1.4}
\defR{churn-load40/berth-nsb/p99x/bertlarge}{2.1}
\defR{churn-load40/berth-nsb/p99x/retinanet}{2.3}
\defR{churn-load40/berth-nsb/p99x/perception}{2.2}
\defR{churn-load40/berth-nsb/thrabs/train_bert}{0.77}
\defR{churn-load40/berth-nsb/thr/train_bert}{47}
\defR{churn-load40/berth-ng/attmin}{96.0}
\defR{churn-load40/berth-ng/attmean}{97.7}
\defR{churn-load40/berth-ng/thr}{48}
\defR{churn-load40/berth-ng/n}{3}
\defR{churn-load40/berth-ng/att/resnet50}{96.0}
\defR{churn-load40/berth-ng/att/bertlarge}{97.1}
\defR{churn-load40/berth-ng/att/retinanet}{98.6}
\defR{churn-load40/berth-ng/att/perception}{99.2}
\defR{churn-load40/berth-ng/p99/resnet50}{18.6}
\defR{churn-load40/berth-ng/p99/bertlarge}{172}
\defR{churn-load40/berth-ng/p99/retinanet}{105}
\defR{churn-load40/berth-ng/p99/perception}{31.0}
\defR{churn-load40/berth-ng/p99x/resnet50}{1.4}
\defR{churn-load40/berth-ng/p99x/bertlarge}{2.4}
\defR{churn-load40/berth-ng/p99x/retinanet}{3.0}
\defR{churn-load40/berth-ng/p99x/perception}{2.3}
\defR{churn-load40/berth-ng/thrabs/train_bert}{0.78}
\defR{churn-load40/berth-ng/thr/train_bert}{48}
\defR{churn-load40/berth-nsng/attmin}{96.1}
\defR{churn-load40/berth-nsng/attmean}{98.0}
\defR{churn-load40/berth-nsng/thr}{48}
\defR{churn-load40/berth-nsng/n}{3}
\defR{churn-load40/berth-nsng/att/resnet50}{96.1}
\defR{churn-load40/berth-nsng/att/bertlarge}{97.4}
\defR{churn-load40/berth-nsng/att/retinanet}{99.3}
\defR{churn-load40/berth-nsng/att/perception}{99.3}
\defR{churn-load40/berth-nsng/p99/resnet50}{18.8}
\defR{churn-load40/berth-nsng/p99/bertlarge}{161}
\defR{churn-load40/berth-nsng/p99/retinanet}{86.0}
\defR{churn-load40/berth-nsng/p99/perception}{31.3}
\defR{churn-load40/berth-nsng/p99x/resnet50}{1.4}
\defR{churn-load40/berth-nsng/p99x/bertlarge}{2.2}
\defR{churn-load40/berth-nsng/p99x/retinanet}{2.5}
\defR{churn-load40/berth-nsng/p99x/perception}{2.3}
\defR{churn-load40/berth-nsng/thrabs/train_bert}{0.78}
\defR{churn-load40/berth-nsng/thr/train_bert}{48}
\defR{churn-load40/berth-grouped/attmin}{83.9}
\defR{churn-load40/berth-grouped/attmean}{95.6}
\defR{churn-load40/berth-grouped/thr}{52}
\defR{churn-load40/berth-grouped/n}{3}
\defR{churn-load40/berth-grouped/att/resnet50}{83.9}
\defR{churn-load40/berth-grouped/att/bertlarge}{99.9}
\defR{churn-load40/berth-grouped/att/retinanet}{99.8}
\defR{churn-load40/berth-grouped/att/perception}{98.8}
\defR{churn-load40/berth-grouped/p99/resnet50}{32.7}
\defR{churn-load40/berth-grouped/p99/bertlarge}{89.1}
\defR{churn-load40/berth-grouped/p99/retinanet}{69.8}
\defR{churn-load40/berth-grouped/p99/perception}{34.5}
\defR{churn-load40/berth-grouped/p99x/resnet50}{2.4}
\defR{churn-load40/berth-grouped/p99x/bertlarge}{1.2}
\defR{churn-load40/berth-grouped/p99x/retinanet}{2.0}
\defR{churn-load40/berth-grouped/p99x/perception}{2.5}
\defR{churn-load40/berth-grouped/thrabs/train_bert}{0.84}
\defR{churn-load40/berth-grouped/thr/train_bert}{52}
\defR{churn-load40/berth/attmin}{85.8}
\defR{churn-load40/berth/attmean}{93.0}
\defR{churn-load40/berth/thr}{47}
\defR{churn-load40/berth/n}{3}
\defR{churn-load40/berth/att/resnet50}{90.2}
\defR{churn-load40/berth/att/bertlarge}{96.8}
\defR{churn-load40/berth/att/retinanet}{99.4}
\defR{churn-load40/berth/att/perception}{85.8}
\defR{churn-load40/berth/p99/resnet50}{29.1}
\defR{churn-load40/berth/p99/bertlarge}{167}
\defR{churn-load40/berth/p99/retinanet}{92.9}
\defR{churn-load40/berth/p99/perception}{69.6}
\defR{churn-load40/berth/p99x/resnet50}{2.2}
\defR{churn-load40/berth/p99x/bertlarge}{2.3}
\defR{churn-load40/berth/p99x/retinanet}{2.7}
\defR{churn-load40/berth/p99x/perception}{5.1}
\defR{churn-load40/berth/thrabs/train_bert}{0.77}
\defR{churn-load40/berth/thr/train_bert}{47}
\defR{churn-load40/timeslice/attmin}{0.0}
\defR{churn-load40/timeslice/attmean}{49.3}
\defR{churn-load40/timeslice/thr}{48}
\defR{churn-load40/timeslice/n}{3}
\defR{churn-load40/timeslice/att/resnet50}{56.9}
\defR{churn-load40/timeslice/att/bertlarge}{60.9}
\defR{churn-load40/timeslice/att/retinanet}{79.5}
\defR{churn-load40/timeslice/att/perception}{0.0}
\defR{churn-load40/timeslice/p99/resnet50}{52.8}
\defR{churn-load40/timeslice/p99/bertlarge}{686}
\defR{churn-load40/timeslice/p99/retinanet}{227}
\defR{churn-load40/timeslice/p99/perception}{8704}
\defR{churn-load40/timeslice/p99x/resnet50}{3.9}
\defR{churn-load40/timeslice/p99x/bertlarge}{9.5}
\defR{churn-load40/timeslice/p99x/retinanet}{6.5}
\defR{churn-load40/timeslice/p99x/perception}{635.2}
\defR{churn-load40/timeslice/thrabs/train_bert}{0.77}
\defR{churn-load40/timeslice/thr/train_bert}{48}
\defR{churn-load40/mps/attmin}{0.0}
\defR{churn-load40/mps/attmean}{0.1}
\defR{churn-load40/mps/thr}{79}
\defR{churn-load40/mps/n}{3}
\defR{churn-load40/mps/att/resnet50}{0.0}
\defR{churn-load40/mps/att/bertlarge}{0.5}
\defR{churn-load40/mps/att/retinanet}{0.0}
\defR{churn-load40/mps/att/perception}{0.0}
\defR{churn-load40/mps/p99/resnet50}{34700}
\defR{churn-load40/mps/p99/bertlarge}{8459}
\defR{churn-load40/mps/p99/retinanet}{18680}
\defR{churn-load40/mps/p99/perception}{27556}
\defR{churn-load40/mps/p99x/resnet50}{2590.0}
\defR{churn-load40/mps/p99x/bertlarge}{116.9}
\defR{churn-load40/mps/p99x/retinanet}{537.1}
\defR{churn-load40/mps/p99x/perception}{2011.1}
\defR{churn-load40/mps/thrabs/train_bert}{1.27}
\defR{churn-load40/mps/thr/train_bert}{79}
\defR{churn-load40/mpsprio/attmin}{0.0}
\defR{churn-load40/mpsprio/attmean}{25.7}
\defR{churn-load40/mpsprio/thr}{62}
\defR{churn-load40/mpsprio/n}{3}
\defR{churn-load40/mpsprio/att/resnet50}{0.0}
\defR{churn-load40/mpsprio/att/bertlarge}{73.9}
\defR{churn-load40/mpsprio/att/retinanet}{29.0}
\defR{churn-load40/mpsprio/att/perception}{0.0}
\defR{churn-load40/mpsprio/p99/resnet50}{210}
\defR{churn-load40/mpsprio/p99/bertlarge}{305}
\defR{churn-load40/mpsprio/p99/retinanet}{983}
\defR{churn-load40/mpsprio/p99/perception}{1624}
\defR{churn-load40/mpsprio/p99x/resnet50}{15.7}
\defR{churn-load40/mpsprio/p99x/bertlarge}{4.2}
\defR{churn-load40/mpsprio/p99x/retinanet}{28.3}
\defR{churn-load40/mpsprio/p99x/perception}{118.5}
\defR{churn-load40/mpsprio/thrabs/train_bert}{1.01}
\defR{churn-load40/mpsprio/thr/train_bert}{62}
\defR{churn-load40/mpslimits/attmin}{0.0}
\defR{churn-load40/mpslimits/attmean}{55.4}
\defR{churn-load40/mpslimits/thr}{12}
\defR{churn-load40/mpslimits/n}{3}
\defR{churn-load40/mpslimits/att/resnet50}{95.4}
\defR{churn-load40/mpslimits/att/bertlarge}{50.1}
\defR{churn-load40/mpslimits/att/retinanet}{76.2}
\defR{churn-load40/mpslimits/att/perception}{0.0}
\defR{churn-load40/mpslimits/p99/resnet50}{18.8}
\defR{churn-load40/mpslimits/p99/bertlarge}{457}
\defR{churn-load40/mpslimits/p99/retinanet}{211}
\defR{churn-load40/mpslimits/p99/perception}{1654}
\defR{churn-load40/mpslimits/p99x/resnet50}{1.4}
\defR{churn-load40/mpslimits/p99x/bertlarge}{6.3}
\defR{churn-load40/mpslimits/p99x/retinanet}{6.1}
\defR{churn-load40/mpslimits/p99x/perception}{120.7}
\defR{churn-load40/mpslimits/thrabs/train_bert}{0.19}
\defR{churn-load40/mpslimits/thr/train_bert}{12}
\defR{churn-load40/mpsparts/attmin}{0.0}
\defR{churn-load40/mpsparts/attmean}{65.1}
\defR{churn-load40/mpsparts/thr}{9}
\defR{churn-load40/mpsparts/n}{2}
\defR{churn-load40/mpsparts/att/resnet50}{96.9}
\defR{churn-load40/mpsparts/att/bertlarge}{75.1}
\defR{churn-load40/mpsparts/att/retinanet}{88.3}
\defR{churn-load40/mpsparts/att/perception}{0.0}
\defR{churn-load40/mpsparts/p99/resnet50}{18.0}
\defR{churn-load40/mpsparts/p99/bertlarge}{314}
\defR{churn-load40/mpsparts/p99/retinanet}{159}
\defR{churn-load40/mpsparts/p99/perception}{1654}
\defR{churn-load40/mpsparts/p99x/resnet50}{1.3}
\defR{churn-load40/mpsparts/p99x/bertlarge}{4.3}
\defR{churn-load40/mpsparts/p99x/retinanet}{4.6}
\defR{churn-load40/mpsparts/p99x/perception}{120.7}
\defR{churn-load40/mpsparts/thrabs/train_bert}{0.14}
\defR{churn-load40/mpsparts/thr/train_bert}{9}
\defR{churn-load40/lithos/attmin}{0.0}
\defR{churn-load40/lithos/attmean}{56.5}
\defR{churn-load40/lithos/thr}{17}
\defR{churn-load40/lithos/n}{3}
\defR{churn-load40/lithos/att/resnet50}{94.3}
\defR{churn-load40/lithos/att/bertlarge}{59.2}
\defR{churn-load40/lithos/att/retinanet}{72.6}
\defR{churn-load40/lithos/att/perception}{0.0}
\defR{churn-load40/lithos/p99/resnet50}{20.9}
\defR{churn-load40/lithos/p99/bertlarge}{384}
\defR{churn-load40/lithos/p99/retinanet}{225}
\defR{churn-load40/lithos/p99/perception}{3836}
\defR{churn-load40/lithos/p99x/resnet50}{1.6}
\defR{churn-load40/lithos/p99x/bertlarge}{5.3}
\defR{churn-load40/lithos/p99x/retinanet}{6.5}
\defR{churn-load40/lithos/p99x/perception}{280.0}
\defR{churn-load40/lithos/thrabs/train_bert}{0.28}
\defR{churn-load40/lithos/thr/train_bert}{17}
\defR{churn-load40/dysm/attmin}{0.0}
\defR{churn-load40/dysm/attmean}{43.4}
\defR{churn-load40/dysm/thr}{38}
\defR{churn-load40/dysm/n}{3}
\defR{churn-load40/dysm/att/resnet50}{0.0}
\defR{churn-load40/dysm/att/bertlarge}{0.0}
\defR{churn-load40/dysm/att/retinanet}{74.9}
\defR{churn-load40/dysm/att/perception}{98.6}
\defR{churn-load40/dysm/p99/retinanet}{331}
\defR{churn-load40/dysm/p99/perception}{34.9}
\defR{churn-load40/dysm/p99x/retinanet}{9.5}
\defR{churn-load40/dysm/p99x/perception}{2.5}
\defR{churn-load40/dysm/thrabs/train_bert}{0.61}
\defR{churn-load40/dysm/thr/train_bert}{38}
\defR{churn-load40/xsched/attmin}{80.2}
\defR{churn-load40/xsched/attmean}{89.6}
\defR{churn-load40/xsched/thr}{46}
\defR{churn-load40/xsched/n}{3}
\defR{churn-load40/xsched/att/resnet50}{80.2}
\defR{churn-load40/xsched/att/bertlarge}{96.8}
\defR{churn-load40/xsched/att/retinanet}{97.1}
\defR{churn-load40/xsched/att/perception}{84.4}
\defR{churn-load40/xsched/p99/resnet50}{38.4}
\defR{churn-load40/xsched/p99/bertlarge}{178}
\defR{churn-load40/xsched/p99/retinanet}{115}
\defR{churn-load40/xsched/p99/perception}{73.0}
\defR{churn-load40/xsched/p99x/resnet50}{2.9}
\defR{churn-load40/xsched/p99x/bertlarge}{2.5}
\defR{churn-load40/xsched/p99x/retinanet}{3.3}
\defR{churn-load40/xsched/p99x/perception}{5.3}
\defR{churn-load40/xsched/thrabs/train_bert}{0.75}
\defR{churn-load40/xsched/thr/train_bert}{46}
\defR{churn-load40/multistream/attmin}{0.0}
\defR{churn-load40/multistream/attmean}{25.3}
\defR{churn-load40/multistream/thr}{63}
\defR{churn-load40/multistream/n}{3}
\defR{churn-load40/multistream/att/resnet50}{0.1}
\defR{churn-load40/multistream/att/bertlarge}{75.6}
\defR{churn-load40/multistream/att/retinanet}{25.5}
\defR{churn-load40/multistream/att/perception}{0.0}
\defR{churn-load40/multistream/p99/resnet50}{194}
\defR{churn-load40/multistream/p99/bertlarge}{299}
\defR{churn-load40/multistream/p99/retinanet}{1100}
\defR{churn-load40/multistream/p99/perception}{1485}
\defR{churn-load40/multistream/p99x/resnet50}{14.5}
\defR{churn-load40/multistream/p99x/bertlarge}{4.1}
\defR{churn-load40/multistream/p99x/retinanet}{31.6}
\defR{churn-load40/multistream/p99x/perception}{108.4}
\defR{churn-load40/multistream/thrabs/train_bert}{1.02}
\defR{churn-load40/multistream/thr/train_bert}{63}
\defR{churn-load40/tgs/attmin}{90.3}
\defR{churn-load40/tgs/attmean}{95.3}
\defR{churn-load40/tgs/thr}{1}
\defR{churn-load40/tgs/n}{3}
\defR{churn-load40/tgs/att/resnet50}{93.5}
\defR{churn-load40/tgs/att/bertlarge}{98.0}
\defR{churn-load40/tgs/att/retinanet}{99.2}
\defR{churn-load40/tgs/att/perception}{90.3}
\defR{churn-load40/tgs/p99/resnet50}{26.9}
\defR{churn-load40/tgs/p99/bertlarge}{151}
\defR{churn-load40/tgs/p99/retinanet}{91.5}
\defR{churn-load40/tgs/p99/perception}{81.2}
\defR{churn-load40/tgs/p99x/resnet50}{2.0}
\defR{churn-load40/tgs/p99x/bertlarge}{2.1}
\defR{churn-load40/tgs/p99x/retinanet}{2.6}
\defR{churn-load40/tgs/p99x/perception}{5.9}
\defR{churn-load40/tgs/thrabs/train_bert}{0.02}
\defR{churn-load40/tgs/thr/train_bert}{1}
\defR{churn-load40/mig/attmin}{0.0}
\defR{churn-load40/mig/attmean}{55.3}
\defR{churn-load40/mig/thr}{0}
\defR{churn-load40/mig/n}{3}
\defR{churn-load40/mig/att/resnet50}{0.0}
\defR{churn-load40/mig/att/bertlarge}{23.4}
\defR{churn-load40/mig/att/retinanet}{98.5}
\defR{churn-load40/mig/att/perception}{99.3}
\defR{churn-load40/mig/p99/resnet50}{161}
\defR{churn-load40/mig/p99/bertlarge}{1007}
\defR{churn-load40/mig/p99/retinanet}{103}
\defR{churn-load40/mig/p99/perception}{24.1}
\defR{churn-load40/mig/p99x/resnet50}{12.1}
\defR{churn-load40/mig/p99x/bertlarge}{13.9}
\defR{churn-load40/mig/p99x/retinanet}{3.0}
\defR{churn-load40/mig/p99x/perception}{1.8}
\defR{churn-load40/mig/thrabs/train_bert}{0.00}
\defR{churn-load40/mig/thr/train_bert}{0}
\defR{churn-load40/tally/attmin}{0.0}
\defR{churn-load40/tally/attmean}{27.1}
\defR{churn-load40/tally/thr}{31}
\defR{churn-load40/tally/n}{3}
\defR{churn-load40/tally/att/resnet50}{0.0}
\defR{churn-load40/tally/att/bertlarge}{3.2}
\defR{churn-load40/tally/att/retinanet}{9.3}
\defR{churn-load40/tally/att/perception}{95.9}
\defR{churn-load40/tally/p99/resnet50}{94201}
\defR{churn-load40/tally/p99/bertlarge}{43493}
\defR{churn-load40/tally/p99/retinanet}{17520}
\defR{churn-load40/tally/p99/perception}{90.1}
\defR{churn-load40/tally/p99x/resnet50}{7031.3}
\defR{churn-load40/tally/p99x/bertlarge}{601.2}
\defR{churn-load40/tally/p99x/retinanet}{503.8}
\defR{churn-load40/tally/p99x/perception}{6.6}
\defR{churn-load40/tally/thrabs/train_bert}{0.50}
\defR{churn-load40/tally/thr/train_bert}{31}
\defR{churn-load40/orion/attmin}{0.0}
\defR{churn-load40/orion/attmean}{0.0}
\defR{churn-load40/orion/thr}{0}
\defR{churn-load40/orion/n}{3}
\defR{churn-load40/orion/att/resnet50}{0.0}
\defR{churn-load40/orion/att/bertlarge}{0.0}
\defR{churn-load40/orion/att/retinanet}{0.0}
\defR{churn-load40/orion/att/perception}{0.0}
\defR{churn-load40/orion/thrabs/train_bert}{0.00}
\defR{churn-load40/orion/thr/train_bert}{0}
\defR{scale-5/berth-full/attmin}{95.1}
\defR{scale-5/berth-full/attmean}{97.4}
\defR{scale-5/berth-full/thr}{29}
\defR{scale-5/berth-full/n}{3}
\defR{scale-5/berth-full/att/resnet50}{95.4}
\defR{scale-5/berth-full/att/retinanet}{99.7}
\defR{scale-5/berth-full/att/bertlarge}{97.2}
\defR{scale-5/berth-full/att/perception}{99.7}
\defR{scale-5/berth-full/att/resnet50_b}{95.1}
\defR{scale-5/berth-full/p99/resnet50}{19.0}
\defR{scale-5/berth-full/p99/retinanet}{85.7}
\defR{scale-5/berth-full/p99/bertlarge}{154}
\defR{scale-5/berth-full/p99/perception}{29.1}
\defR{scale-5/berth-full/p99/resnet50_b}{19.3}
\defR{scale-5/berth-full/p99x/resnet50}{1.5}
\defR{scale-5/berth-full/p99x/retinanet}{2.6}
\defR{scale-5/berth-full/p99x/bertlarge}{2.2}
\defR{scale-5/berth-full/p99x/perception}{2.1}
\defR{scale-5/berth-full/p99x/resnet50_b}{1.5}
\defR{scale-5/berth-full/thrabs/train_bert}{0.47}
\defR{scale-5/berth-full/thr/train_bert}{29}
\defR{scale-5/berth-nsr/attmin}{94.6}
\defR{scale-5/berth-nsr/attmean}{97.2}
\defR{scale-5/berth-nsr/thr}{29}
\defR{scale-5/berth-nsr/n}{3}
\defR{scale-5/berth-nsr/att/resnet50}{95.5}
\defR{scale-5/berth-nsr/att/retinanet}{99.4}
\defR{scale-5/berth-nsr/att/bertlarge}{97.1}
\defR{scale-5/berth-nsr/att/perception}{99.6}
\defR{scale-5/berth-nsr/att/resnet50_b}{94.6}
\defR{scale-5/berth-nsr/p99/resnet50}{19.6}
\defR{scale-5/berth-nsr/p99/retinanet}{89.1}
\defR{scale-5/berth-nsr/p99/bertlarge}{156}
\defR{scale-5/berth-nsr/p99/perception}{30.0}
\defR{scale-5/berth-nsr/p99/resnet50_b}{20.9}
\defR{scale-5/berth-nsr/p99x/resnet50}{1.6}
\defR{scale-5/berth-nsr/p99x/retinanet}{2.7}
\defR{scale-5/berth-nsr/p99x/bertlarge}{2.2}
\defR{scale-5/berth-nsr/p99x/perception}{2.2}
\defR{scale-5/berth-nsr/p99x/resnet50_b}{1.7}
\defR{scale-5/berth-nsr/thrabs/train_bert}{0.47}
\defR{scale-5/berth-nsr/thr/train_bert}{29}
\defR{scale-5/berth-nba/attmin}{94.4}
\defR{scale-5/berth-nba/attmean}{97.1}
\defR{scale-5/berth-nba/thr}{29}
\defR{scale-5/berth-nba/n}{3}
\defR{scale-5/berth-nba/att/resnet50}{94.9}
\defR{scale-5/berth-nba/att/retinanet}{99.6}
\defR{scale-5/berth-nba/att/bertlarge}{96.9}
\defR{scale-5/berth-nba/att/perception}{99.6}
\defR{scale-5/berth-nba/att/resnet50_b}{94.4}
\defR{scale-5/berth-nba/p99/resnet50}{20.0}
\defR{scale-5/berth-nba/p99/retinanet}{84.9}
\defR{scale-5/berth-nba/p99/bertlarge}{166}
\defR{scale-5/berth-nba/p99/perception}{30.2}
\defR{scale-5/berth-nba/p99/resnet50_b}{20.5}
\defR{scale-5/berth-nba/p99x/resnet50}{1.6}
\defR{scale-5/berth-nba/p99x/retinanet}{2.6}
\defR{scale-5/berth-nba/p99x/bertlarge}{2.4}
\defR{scale-5/berth-nba/p99x/perception}{2.2}
\defR{scale-5/berth-nba/p99x/resnet50_b}{1.6}
\defR{scale-5/berth-nba/thrabs/train_bert}{0.47}
\defR{scale-5/berth-nba/thr/train_bert}{29}
\defR{scale-5/berth-nsb/attmin}{95.2}
\defR{scale-5/berth-nsb/attmean}{97.2}
\defR{scale-5/berth-nsb/thr}{29}
\defR{scale-5/berth-nsb/n}{3}
\defR{scale-5/berth-nsb/att/resnet50}{95.2}
\defR{scale-5/berth-nsb/att/retinanet}{99.5}
\defR{scale-5/berth-nsb/att/bertlarge}{96.6}
\defR{scale-5/berth-nsb/att/perception}{99.7}
\defR{scale-5/berth-nsb/att/resnet50_b}{95.2}
\defR{scale-5/berth-nsb/p99/resnet50}{19.9}
\defR{scale-5/berth-nsb/p99/retinanet}{87.3}
\defR{scale-5/berth-nsb/p99/bertlarge}{163}
\defR{scale-5/berth-nsb/p99/perception}{29.5}
\defR{scale-5/berth-nsb/p99/resnet50_b}{20.3}
\defR{scale-5/berth-nsb/p99x/resnet50}{1.6}
\defR{scale-5/berth-nsb/p99x/retinanet}{2.7}
\defR{scale-5/berth-nsb/p99x/bertlarge}{2.3}
\defR{scale-5/berth-nsb/p99x/perception}{2.2}
\defR{scale-5/berth-nsb/p99x/resnet50_b}{1.6}
\defR{scale-5/berth-nsb/thrabs/train_bert}{0.47}
\defR{scale-5/berth-nsb/thr/train_bert}{29}
\defR{scale-5/berth-ng/attmin}{93.9}
\defR{scale-5/berth-ng/attmean}{96.2}
\defR{scale-5/berth-ng/thr}{29}
\defR{scale-5/berth-ng/n}{3}
\defR{scale-5/berth-ng/att/resnet50}{94.2}
\defR{scale-5/berth-ng/att/retinanet}{98.6}
\defR{scale-5/berth-ng/att/bertlarge}{94.6}
\defR{scale-5/berth-ng/att/perception}{99.5}
\defR{scale-5/berth-ng/att/resnet50_b}{93.9}
\defR{scale-5/berth-ng/p99/resnet50}{21.2}
\defR{scale-5/berth-ng/p99/retinanet}{103}
\defR{scale-5/berth-ng/p99/bertlarge}{185}
\defR{scale-5/berth-ng/p99/perception}{30.8}
\defR{scale-5/berth-ng/p99/resnet50_b}{21.3}
\defR{scale-5/berth-ng/p99x/resnet50}{1.7}
\defR{scale-5/berth-ng/p99x/retinanet}{3.2}
\defR{scale-5/berth-ng/p99x/bertlarge}{2.6}
\defR{scale-5/berth-ng/p99x/perception}{2.2}
\defR{scale-5/berth-ng/p99x/resnet50_b}{1.7}
\defR{scale-5/berth-ng/thrabs/train_bert}{0.48}
\defR{scale-5/berth-ng/thr/train_bert}{29}
\defR{scale-5/berth-nsng/attmin}{93.6}
\defR{scale-5/berth-nsng/attmean}{96.2}
\defR{scale-5/berth-nsng/thr}{29}
\defR{scale-5/berth-nsng/n}{3}
\defR{scale-5/berth-nsng/att/resnet50}{93.6}
\defR{scale-5/berth-nsng/att/retinanet}{98.8}
\defR{scale-5/berth-nsng/att/bertlarge}{95.1}
\defR{scale-5/berth-nsng/att/perception}{99.5}
\defR{scale-5/berth-nsng/att/resnet50_b}{94.1}
\defR{scale-5/berth-nsng/p99/resnet50}{21.6}
\defR{scale-5/berth-nsng/p99/retinanet}{102}
\defR{scale-5/berth-nsng/p99/bertlarge}{191}
\defR{scale-5/berth-nsng/p99/perception}{31.4}
\defR{scale-5/berth-nsng/p99/resnet50_b}{20.9}
\defR{scale-5/berth-nsng/p99x/resnet50}{1.7}
\defR{scale-5/berth-nsng/p99x/retinanet}{3.1}
\defR{scale-5/berth-nsng/p99x/bertlarge}{2.7}
\defR{scale-5/berth-nsng/p99x/perception}{2.3}
\defR{scale-5/berth-nsng/p99x/resnet50_b}{1.7}
\defR{scale-5/berth-nsng/thrabs/train_bert}{0.47}
\defR{scale-5/berth-nsng/thr/train_bert}{29}
\defR{scale-5/berth-grouped/attmin}{72.1}
\defR{scale-5/berth-grouped/attmean}{89.0}
\defR{scale-5/berth-grouped/thr}{36}
\defR{scale-5/berth-grouped/n}{3}
\defR{scale-5/berth-grouped/att/resnet50}{74.2}
\defR{scale-5/berth-grouped/att/retinanet}{99.8}
\defR{scale-5/berth-grouped/att/bertlarge}{100.0}
\defR{scale-5/berth-grouped/att/perception}{99.0}
\defR{scale-5/berth-grouped/att/resnet50_b}{72.1}
\defR{scale-5/berth-grouped/p99/resnet50}{34.9}
\defR{scale-5/berth-grouped/p99/retinanet}{66.7}
\defR{scale-5/berth-grouped/p99/bertlarge}{88.4}
\defR{scale-5/berth-grouped/p99/perception}{33.1}
\defR{scale-5/berth-grouped/p99/resnet50_b}{35.6}
\defR{scale-5/berth-grouped/p99x/resnet50}{2.8}
\defR{scale-5/berth-grouped/p99x/retinanet}{2.1}
\defR{scale-5/berth-grouped/p99x/bertlarge}{1.3}
\defR{scale-5/berth-grouped/p99x/perception}{2.4}
\defR{scale-5/berth-grouped/p99x/resnet50_b}{2.8}
\defR{scale-5/berth-grouped/thrabs/train_bert}{0.58}
\defR{scale-5/berth-grouped/thr/train_bert}{36}
\defR{scale-5/berth/attmin}{84.0}
\defR{scale-5/berth/attmean}{89.9}
\defR{scale-5/berth/thr}{28}
\defR{scale-5/berth/n}{3}
\defR{scale-5/berth/att/resnet50}{84.0}
\defR{scale-5/berth/att/retinanet}{99.6}
\defR{scale-5/berth/att/bertlarge}{95.7}
\defR{scale-5/berth/att/perception}{86.1}
\defR{scale-5/berth/att/resnet50_b}{84.1}
\defR{scale-5/berth/p99/resnet50}{30.5}
\defR{scale-5/berth/p99/retinanet}{86.0}
\defR{scale-5/berth/p99/bertlarge}{179}
\defR{scale-5/berth/p99/perception}{57.6}
\defR{scale-5/berth/p99/resnet50_b}{29.8}
\defR{scale-5/berth/p99x/resnet50}{2.4}
\defR{scale-5/berth/p99x/retinanet}{2.6}
\defR{scale-5/berth/p99x/bertlarge}{2.5}
\defR{scale-5/berth/p99x/perception}{4.2}
\defR{scale-5/berth/p99x/resnet50_b}{2.4}
\defR{scale-5/berth/thrabs/train_bert}{0.46}
\defR{scale-5/berth/thr/train_bert}{28}
\defR{scale-5/timeslice/attmin}{0.0}
\defR{scale-5/timeslice/attmean}{46.2}
\defR{scale-5/timeslice/thr}{31}
\defR{scale-5/timeslice/n}{3}
\defR{scale-5/timeslice/att/resnet50}{40.1}
\defR{scale-5/timeslice/att/retinanet}{89.9}
\defR{scale-5/timeslice/att/bertlarge}{59.0}
\defR{scale-5/timeslice/att/perception}{0.0}
\defR{scale-5/timeslice/att/resnet50_b}{41.9}
\defR{scale-5/timeslice/p99/resnet50}{49.6}
\defR{scale-5/timeslice/p99/retinanet}{190}
\defR{scale-5/timeslice/p99/bertlarge}{429}
\defR{scale-5/timeslice/p99/perception}{18368}
\defR{scale-5/timeslice/p99/resnet50_b}{51.1}
\defR{scale-5/timeslice/p99x/resnet50}{3.9}
\defR{scale-5/timeslice/p99x/retinanet}{5.8}
\defR{scale-5/timeslice/p99x/bertlarge}{6.1}
\defR{scale-5/timeslice/p99x/perception}{1340.6}
\defR{scale-5/timeslice/p99x/resnet50_b}{4.1}
\defR{scale-5/timeslice/thrabs/train_bert}{0.50}
\defR{scale-5/timeslice/thr/train_bert}{31}
\defR{scale-5/mps/attmin}{0.0}
\defR{scale-5/mps/attmean}{0.0}
\defR{scale-5/mps/thr}{71}
\defR{scale-5/mps/n}{3}
\defR{scale-5/mps/att/resnet50}{0.0}
\defR{scale-5/mps/att/retinanet}{0.0}
\defR{scale-5/mps/att/bertlarge}{0.2}
\defR{scale-5/mps/att/perception}{0.0}
\defR{scale-5/mps/att/resnet50_b}{0.0}
\defR{scale-5/mps/p99/resnet50}{18046}
\defR{scale-5/mps/p99/retinanet}{32236}
\defR{scale-5/mps/p99/bertlarge}{1094}
\defR{scale-5/mps/p99/perception}{59261}
\defR{scale-5/mps/p99/resnet50_b}{17527}
\defR{scale-5/mps/p99x/resnet50}{1433.4}
\defR{scale-5/mps/p99x/retinanet}{991.3}
\defR{scale-5/mps/p99x/bertlarge}{15.5}
\defR{scale-5/mps/p99x/perception}{4325.1}
\defR{scale-5/mps/p99x/resnet50_b}{1392.2}
\defR{scale-5/mps/thrabs/train_bert}{1.16}
\defR{scale-5/mps/thr/train_bert}{71}
\defR{scale-5/mpsprio/attmin}{0.0}
\defR{scale-5/mpsprio/attmean}{29.4}
\defR{scale-5/mpsprio/thr}{47}
\defR{scale-5/mpsprio/n}{3}
\defR{scale-5/mpsprio/att/resnet50}{0.0}
\defR{scale-5/mpsprio/att/retinanet}{60.3}
\defR{scale-5/mpsprio/att/bertlarge}{85.6}
\defR{scale-5/mpsprio/att/perception}{1.0}
\defR{scale-5/mpsprio/att/resnet50_b}{0.0}
\defR{scale-5/mpsprio/p99/resnet50}{118}
\defR{scale-5/mpsprio/p99/retinanet}{308}
\defR{scale-5/mpsprio/p99/bertlarge}{219}
\defR{scale-5/mpsprio/p99/perception}{946}
\defR{scale-5/mpsprio/p99/resnet50_b}{116}
\defR{scale-5/mpsprio/p99x/resnet50}{9.4}
\defR{scale-5/mpsprio/p99x/retinanet}{9.5}
\defR{scale-5/mpsprio/p99x/bertlarge}{3.1}
\defR{scale-5/mpsprio/p99x/perception}{69.0}
\defR{scale-5/mpsprio/p99x/resnet50_b}{9.2}
\defR{scale-5/mpsprio/thrabs/train_bert}{0.76}
\defR{scale-5/mpsprio/thr/train_bert}{47}
\defR{scale-5/mpslimits/attmin}{0.0}
\defR{scale-5/mpslimits/attmean}{50.6}
\defR{scale-5/mpslimits/thr}{12}
\defR{scale-5/mpslimits/n}{3}
\defR{scale-5/mpslimits/att/resnet50}{91.4}
\defR{scale-5/mpslimits/att/retinanet}{19.5}
\defR{scale-5/mpslimits/att/bertlarge}{51.3}
\defR{scale-5/mpslimits/att/perception}{0.0}
\defR{scale-5/mpslimits/att/resnet50_b}{90.7}
\defR{scale-5/mpslimits/p99/resnet50}{21.4}
\defR{scale-5/mpslimits/p99/retinanet}{766}
\defR{scale-5/mpslimits/p99/bertlarge}{464}
\defR{scale-5/mpslimits/p99/perception}{28903}
\defR{scale-5/mpslimits/p99/resnet50_b}{22.5}
\defR{scale-5/mpslimits/p99x/resnet50}{1.7}
\defR{scale-5/mpslimits/p99x/retinanet}{23.6}
\defR{scale-5/mpslimits/p99x/bertlarge}{6.6}
\defR{scale-5/mpslimits/p99x/perception}{2109.4}
\defR{scale-5/mpslimits/p99x/resnet50_b}{1.8}
\defR{scale-5/mpslimits/thrabs/train_bert}{0.19}
\defR{scale-5/mpslimits/thr/train_bert}{12}
\defR{scale-5/mpsparts/attmin}{0.0}
\defR{scale-5/mpsparts/attmean}{61.9}
\defR{scale-5/mpsparts/thr}{3}
\defR{scale-5/mpsparts/n}{2}
\defR{scale-5/mpsparts/att/resnet50}{92.9}
\defR{scale-5/mpsparts/att/retinanet}{53.8}
\defR{scale-5/mpsparts/att/bertlarge}{69.4}
\defR{scale-5/mpsparts/att/perception}{0.0}
\defR{scale-5/mpsparts/att/resnet50_b}{93.2}
\defR{scale-5/mpsparts/p99/resnet50}{20.2}
\defR{scale-5/mpsparts/p99/retinanet}{355}
\defR{scale-5/mpsparts/p99/bertlarge}{333}
\defR{scale-5/mpsparts/p99/perception}{18857}
\defR{scale-5/mpsparts/p99/resnet50_b}{20.9}
\defR{scale-5/mpsparts/p99x/resnet50}{1.6}
\defR{scale-5/mpsparts/p99x/retinanet}{10.9}
\defR{scale-5/mpsparts/p99x/bertlarge}{4.7}
\defR{scale-5/mpsparts/p99x/perception}{1376.2}
\defR{scale-5/mpsparts/p99x/resnet50_b}{1.7}
\defR{scale-5/mpsparts/thrabs/train_bert}{0.06}
\defR{scale-5/mpsparts/thr/train_bert}{3}
\defR{scale-5/lithos/attmin}{0.0}
\defR{scale-5/lithos/attmean}{48.9}
\defR{scale-5/lithos/thr}{13}
\defR{scale-5/lithos/n}{3}
\defR{scale-5/lithos/att/resnet50}{90.6}
\defR{scale-5/lithos/att/retinanet}{48.8}
\defR{scale-5/lithos/att/bertlarge}{15.7}
\defR{scale-5/lithos/att/perception}{0.0}
\defR{scale-5/lithos/att/resnet50_b}{89.7}
\defR{scale-5/lithos/p99/resnet50}{23.0}
\defR{scale-5/lithos/p99/retinanet}{390}
\defR{scale-5/lithos/p99/bertlarge}{501}
\defR{scale-5/lithos/p99/perception}{30579}
\defR{scale-5/lithos/p99/resnet50_b}{24.8}
\defR{scale-5/lithos/p99x/resnet50}{1.8}
\defR{scale-5/lithos/p99x/retinanet}{12.0}
\defR{scale-5/lithos/p99x/bertlarge}{7.1}
\defR{scale-5/lithos/p99x/perception}{2231.7}
\defR{scale-5/lithos/p99x/resnet50_b}{2.0}
\defR{scale-5/lithos/thrabs/train_bert}{0.22}
\defR{scale-5/lithos/thr/train_bert}{13}
\defR{scale-5/dysm/attmin}{0.0}
\defR{scale-5/dysm/attmean}{36.0}
\defR{scale-5/dysm/thr}{51}
\defR{scale-5/dysm/n}{3}
\defR{scale-5/dysm/att/resnet50}{0.0}
\defR{scale-5/dysm/att/retinanet}{81.4}
\defR{scale-5/dysm/att/bertlarge}{0.0}
\defR{scale-5/dysm/att/perception}{98.7}
\defR{scale-5/dysm/att/resnet50_b}{0.0}
\defR{scale-5/dysm/p99/retinanet}{214}
\defR{scale-5/dysm/p99/perception}{34.5}
\defR{scale-5/dysm/p99x/retinanet}{6.6}
\defR{scale-5/dysm/p99x/perception}{2.5}
\defR{scale-5/dysm/thrabs/train_bert}{0.83}
\defR{scale-5/dysm/thr/train_bert}{51}
\defR{scale-5/xsched/attmin}{69.3}
\defR{scale-5/xsched/attmean}{83.7}
\defR{scale-5/xsched/thr}{27}
\defR{scale-5/xsched/n}{3}
\defR{scale-5/xsched/att/resnet50}{70.1}
\defR{scale-5/xsched/att/retinanet}{99.1}
\defR{scale-5/xsched/att/bertlarge}{95.3}
\defR{scale-5/xsched/att/perception}{84.4}
\defR{scale-5/xsched/att/resnet50_b}{69.3}
\defR{scale-5/xsched/p99/resnet50}{39.9}
\defR{scale-5/xsched/p99/retinanet}{99.8}
\defR{scale-5/xsched/p99/bertlarge}{179}
\defR{scale-5/xsched/p99/perception}{57.9}
\defR{scale-5/xsched/p99/resnet50_b}{37.8}
\defR{scale-5/xsched/p99x/resnet50}{3.2}
\defR{scale-5/xsched/p99x/retinanet}{3.1}
\defR{scale-5/xsched/p99x/bertlarge}{2.5}
\defR{scale-5/xsched/p99x/perception}{4.2}
\defR{scale-5/xsched/p99x/resnet50_b}{3.0}
\defR{scale-5/xsched/thrabs/train_bert}{0.44}
\defR{scale-5/xsched/thr/train_bert}{27}
\defR{scale-5/multistream/attmin}{0.0}
\defR{scale-5/multistream/attmean}{31.0}
\defR{scale-5/multistream/thr}{47}
\defR{scale-5/multistream/n}{3}
\defR{scale-5/multistream/att/resnet50}{0.0}
\defR{scale-5/multistream/att/retinanet}{57.7}
\defR{scale-5/multistream/att/bertlarge}{85.4}
\defR{scale-5/multistream/att/perception}{11.7}
\defR{scale-5/multistream/att/resnet50_b}{0.0}
\defR{scale-5/multistream/p99/resnet50}{123}
\defR{scale-5/multistream/p99/retinanet}{322}
\defR{scale-5/multistream/p99/bertlarge}{214}
\defR{scale-5/multistream/p99/perception}{528}
\defR{scale-5/multistream/p99/resnet50_b}{144}
\defR{scale-5/multistream/p99x/resnet50}{9.8}
\defR{scale-5/multistream/p99x/retinanet}{9.9}
\defR{scale-5/multistream/p99x/bertlarge}{3.0}
\defR{scale-5/multistream/p99x/perception}{38.5}
\defR{scale-5/multistream/p99x/resnet50_b}{11.5}
\defR{scale-5/multistream/thrabs/train_bert}{0.77}
\defR{scale-5/multistream/thr/train_bert}{47}
\defR{scale-5/tgs/attmin}{88.2}
\defR{scale-5/tgs/attmean}{92.4}
\defR{scale-5/tgs/thr}{2}
\defR{scale-5/tgs/n}{3}
\defR{scale-5/tgs/att/resnet50}{88.2}
\defR{scale-5/tgs/att/retinanet}{99.7}
\defR{scale-5/tgs/att/bertlarge}{95.7}
\defR{scale-5/tgs/att/perception}{90.2}
\defR{scale-5/tgs/att/resnet50_b}{88.3}
\defR{scale-5/tgs/p99/resnet50}{28.6}
\defR{scale-5/tgs/p99/retinanet}{76.8}
\defR{scale-5/tgs/p99/bertlarge}{173}
\defR{scale-5/tgs/p99/perception}{64.7}
\defR{scale-5/tgs/p99/resnet50_b}{30.2}
\defR{scale-5/tgs/p99x/resnet50}{2.3}
\defR{scale-5/tgs/p99x/retinanet}{2.4}
\defR{scale-5/tgs/p99x/bertlarge}{2.5}
\defR{scale-5/tgs/p99x/perception}{4.7}
\defR{scale-5/tgs/p99x/resnet50_b}{2.4}
\defR{scale-5/tgs/thrabs/train_bert}{0.03}
\defR{scale-5/tgs/thr/train_bert}{2}
\defR{scale-5/mig/attmin}{0.0}
\defR{scale-5/mig/attmean}{45.8}
\defR{scale-5/mig/thr}{0}
\defR{scale-5/mig/n}{3}
\defR{scale-5/mig/att/resnet50}{0.0}
\defR{scale-5/mig/att/retinanet}{71.6}
\defR{scale-5/mig/att/bertlarge}{57.3}
\defR{scale-5/mig/att/perception}{100.0}
\defR{scale-5/mig/att/resnet50_b}{0.0}
\defR{scale-5/mig/p99/resnet50}{84.4}
\defR{scale-5/mig/p99/retinanet}{257}
\defR{scale-5/mig/p99/bertlarge}{426}
\defR{scale-5/mig/p99/perception}{23.9}
\defR{scale-5/mig/p99/resnet50_b}{84.5}
\defR{scale-5/mig/p99x/resnet50}{6.7}
\defR{scale-5/mig/p99x/retinanet}{7.9}
\defR{scale-5/mig/p99x/bertlarge}{6.0}
\defR{scale-5/mig/p99x/perception}{1.7}
\defR{scale-5/mig/p99x/resnet50_b}{6.7}
\defR{scale-5/mig/thrabs/train_bert}{0.00}
\defR{scale-5/mig/thr/train_bert}{0}
\defR{scale-5/tally/attmin}{0.0}
\defR{scale-5/tally/attmean}{0.0}
\defR{scale-5/tally/thr}{2}
\defR{scale-5/tally/n}{3}
\defR{scale-5/tally/att/resnet50}{0.0}
\defR{scale-5/tally/att/retinanet}{0.0}
\defR{scale-5/tally/att/bertlarge}{0.0}
\defR{scale-5/tally/att/perception}{0.0}
\defR{scale-5/tally/att/resnet50_b}{0.0}
\defR{scale-5/tally/p99/resnet50}{94058}
\defR{scale-5/tally/p99/retinanet}{90067}
\defR{scale-5/tally/p99/bertlarge}{22853}
\defR{scale-5/tally/p99/perception}{91417}
\defR{scale-5/tally/p99/resnet50_b}{201}
\defR{scale-5/tally/p99x/resnet50}{7471.1}
\defR{scale-5/tally/p99x/retinanet}{2769.7}
\defR{scale-5/tally/p99x/bertlarge}{324.1}
\defR{scale-5/tally/p99x/perception}{6672.0}
\defR{scale-5/tally/p99x/resnet50_b}{16.0}
\defR{scale-5/tally/thrabs/train_bert}{0.03}
\defR{scale-5/tally/thr/train_bert}{2}
\defR{scale-5/orion/attmin}{0.0}
\defR{scale-5/orion/attmean}{20.7}
\defR{scale-5/orion/thr}{68}
\defR{scale-5/orion/n}{3}
\defR{scale-5/orion/att/resnet50}{0.0}
\defR{scale-5/orion/att/retinanet}{0.0}
\defR{scale-5/orion/att/bertlarge}{6.2}
\defR{scale-5/orion/att/perception}{0.0}
\defR{scale-5/orion/att/resnet50_b}{97.3}
\defR{scale-5/orion/p99/resnet50}{53047}
\defR{scale-5/orion/p99/retinanet}{46297}
\defR{scale-5/orion/p99/bertlarge}{854}
\defR{scale-5/orion/p99/perception}{79487}
\defR{scale-5/orion/p99/resnet50_b}{21.4}
\defR{scale-5/orion/p99x/resnet50}{4213.5}
\defR{scale-5/orion/p99x/retinanet}{1423.7}
\defR{scale-5/orion/p99x/bertlarge}{12.1}
\defR{scale-5/orion/p99x/perception}{5801.3}
\defR{scale-5/orion/p99x/resnet50_b}{1.7}
\defR{scale-5/orion/thrabs/train_bert}{1.10}
\defR{scale-5/orion/thr/train_bert}{68}
\defR{scale-10/berth-full/attmin}{88.3}
\defR{scale-10/berth-full/attmean}{94.7}
\defR{scale-10/berth-full/thr}{18}
\defR{scale-10/berth-full/n}{3}
\defR{scale-10/berth-full/att/resnet50}{94.2}
\defR{scale-10/berth-full/att/retinanet}{99.6}
\defR{scale-10/berth-full/att/bertlarge}{90.4}
\defR{scale-10/berth-full/att/perception}{99.5}
\defR{scale-10/berth-full/att/resnet50_b}{93.9}
\defR{scale-10/berth-full/att/retinanet_b}{99.5}
\defR{scale-10/berth-full/att/bertlarge_b}{88.8}
\defR{scale-10/berth-full/att/resnet50_c}{93.8}
\defR{scale-10/berth-full/att/bertlarge_c}{88.3}
\defR{scale-10/berth-full/att/retinanet_c}{99.4}
\defR{scale-10/berth-full/p99/resnet50}{22.2}
\defR{scale-10/berth-full/p99/retinanet}{88.4}
\defR{scale-10/berth-full/p99/bertlarge}{215}
\defR{scale-10/berth-full/p99/perception}{31.1}
\defR{scale-10/berth-full/p99/resnet50_b}{20.2}
\defR{scale-10/berth-full/p99/retinanet_b}{87.9}
\defR{scale-10/berth-full/p99/bertlarge_b}{218}
\defR{scale-10/berth-full/p99/resnet50_c}{22.6}
\defR{scale-10/berth-full/p99/bertlarge_c}{245}
\defR{scale-10/berth-full/p99/retinanet_c}{83.6}
\defR{scale-10/berth-full/p99x/resnet50}{1.8}
\defR{scale-10/berth-full/p99x/retinanet}{2.8}
\defR{scale-10/berth-full/p99x/bertlarge}{3.0}
\defR{scale-10/berth-full/p99x/perception}{2.3}
\defR{scale-10/berth-full/p99x/resnet50_b}{1.7}
\defR{scale-10/berth-full/p99x/retinanet_b}{2.7}
\defR{scale-10/berth-full/p99x/bertlarge_b}{3.1}
\defR{scale-10/berth-full/p99x/resnet50_c}{1.8}
\defR{scale-10/berth-full/p99x/bertlarge_c}{3.4}
\defR{scale-10/berth-full/p99x/retinanet_c}{2.6}
\defR{scale-10/berth-full/thrabs/train_bert}{0.30}
\defR{scale-10/berth-full/thr/train_bert}{18}
\defR{scale-10/berth-nsr/attmin}{87.5}
\defR{scale-10/berth-nsr/attmean}{93.9}
\defR{scale-10/berth-nsr/thr}{18}
\defR{scale-10/berth-nsr/n}{3}
\defR{scale-10/berth-nsr/att/resnet50}{91.8}
\defR{scale-10/berth-nsr/att/retinanet}{98.9}
\defR{scale-10/berth-nsr/att/bertlarge}{90.0}
\defR{scale-10/berth-nsr/att/perception}{99.0}
\defR{scale-10/berth-nsr/att/resnet50_b}{92.7}
\defR{scale-10/berth-nsr/att/retinanet_b}{99.5}
\defR{scale-10/berth-nsr/att/bertlarge_b}{87.8}
\defR{scale-10/berth-nsr/att/resnet50_c}{93.3}
\defR{scale-10/berth-nsr/att/bertlarge_c}{87.5}
\defR{scale-10/berth-nsr/att/retinanet_c}{99.0}
\defR{scale-10/berth-nsr/p99/resnet50}{24.4}
\defR{scale-10/berth-nsr/p99/retinanet}{99.3}
\defR{scale-10/berth-nsr/p99/bertlarge}{232}
\defR{scale-10/berth-nsr/p99/perception}{32.8}
\defR{scale-10/berth-nsr/p99/resnet50_b}{22.5}
\defR{scale-10/berth-nsr/p99/retinanet_b}{89.3}
\defR{scale-10/berth-nsr/p99/bertlarge_b}{225}
\defR{scale-10/berth-nsr/p99/resnet50_c}{23.5}
\defR{scale-10/berth-nsr/p99/bertlarge_c}{248}
\defR{scale-10/berth-nsr/p99/retinanet_c}{92.2}
\defR{scale-10/berth-nsr/p99x/resnet50}{2.0}
\defR{scale-10/berth-nsr/p99x/retinanet}{3.1}
\defR{scale-10/berth-nsr/p99x/bertlarge}{3.3}
\defR{scale-10/berth-nsr/p99x/perception}{2.4}
\defR{scale-10/berth-nsr/p99x/resnet50_b}{1.8}
\defR{scale-10/berth-nsr/p99x/retinanet_b}{2.8}
\defR{scale-10/berth-nsr/p99x/bertlarge_b}{3.2}
\defR{scale-10/berth-nsr/p99x/resnet50_c}{1.9}
\defR{scale-10/berth-nsr/p99x/bertlarge_c}{3.5}
\defR{scale-10/berth-nsr/p99x/retinanet_c}{2.9}
\defR{scale-10/berth-nsr/thrabs/train_bert}{0.29}
\defR{scale-10/berth-nsr/thr/train_bert}{18}
\defR{scale-10/berth-nba/attmin}{87.8}
\defR{scale-10/berth-nba/attmean}{94.7}
\defR{scale-10/berth-nba/thr}{18}
\defR{scale-10/berth-nba/n}{3}
\defR{scale-10/berth-nba/att/resnet50}{93.6}
\defR{scale-10/berth-nba/att/retinanet}{99.5}
\defR{scale-10/berth-nba/att/bertlarge}{90.2}
\defR{scale-10/berth-nba/att/perception}{99.5}
\defR{scale-10/berth-nba/att/resnet50_b}{94.1}
\defR{scale-10/berth-nba/att/retinanet_b}{99.4}
\defR{scale-10/berth-nba/att/bertlarge_b}{89.9}
\defR{scale-10/berth-nba/att/resnet50_c}{93.9}
\defR{scale-10/berth-nba/att/bertlarge_c}{87.8}
\defR{scale-10/berth-nba/att/retinanet_c}{99.2}
\defR{scale-10/berth-nba/p99/resnet50}{22.3}
\defR{scale-10/berth-nba/p99/retinanet}{89.0}
\defR{scale-10/berth-nba/p99/bertlarge}{208}
\defR{scale-10/berth-nba/p99/perception}{30.7}
\defR{scale-10/berth-nba/p99/resnet50_b}{20.4}
\defR{scale-10/berth-nba/p99/retinanet_b}{92.7}
\defR{scale-10/berth-nba/p99/bertlarge_b}{233}
\defR{scale-10/berth-nba/p99/resnet50_c}{21.0}
\defR{scale-10/berth-nba/p99/bertlarge_c}{241}
\defR{scale-10/berth-nba/p99/retinanet_c}{88.5}
\defR{scale-10/berth-nba/p99x/resnet50}{1.8}
\defR{scale-10/berth-nba/p99x/retinanet}{2.8}
\defR{scale-10/berth-nba/p99x/bertlarge}{2.9}
\defR{scale-10/berth-nba/p99x/perception}{2.2}
\defR{scale-10/berth-nba/p99x/resnet50_b}{1.7}
\defR{scale-10/berth-nba/p99x/retinanet_b}{2.9}
\defR{scale-10/berth-nba/p99x/bertlarge_b}{3.3}
\defR{scale-10/berth-nba/p99x/resnet50_c}{1.7}
\defR{scale-10/berth-nba/p99x/bertlarge_c}{3.4}
\defR{scale-10/berth-nba/p99x/retinanet_c}{2.8}
\defR{scale-10/berth-nba/thrabs/train_bert}{0.30}
\defR{scale-10/berth-nba/thr/train_bert}{18}
\defR{scale-10/berth-nsb/attmin}{88.7}
\defR{scale-10/berth-nsb/attmean}{94.9}
\defR{scale-10/berth-nsb/thr}{18}
\defR{scale-10/berth-nsb/n}{3}
\defR{scale-10/berth-nsb/att/resnet50}{93.7}
\defR{scale-10/berth-nsb/att/retinanet}{99.3}
\defR{scale-10/berth-nsb/att/bertlarge}{92.1}
\defR{scale-10/berth-nsb/att/perception}{99.4}
\defR{scale-10/berth-nsb/att/resnet50_b}{93.8}
\defR{scale-10/berth-nsb/att/retinanet_b}{99.5}
\defR{scale-10/berth-nsb/att/bertlarge_b}{89.5}
\defR{scale-10/berth-nsb/att/resnet50_c}{93.3}
\defR{scale-10/berth-nsb/att/bertlarge_c}{88.7}
\defR{scale-10/berth-nsb/att/retinanet_c}{99.5}
\defR{scale-10/berth-nsb/p99/resnet50}{23.0}
\defR{scale-10/berth-nsb/p99/retinanet}{90.4}
\defR{scale-10/berth-nsb/p99/bertlarge}{218}
\defR{scale-10/berth-nsb/p99/perception}{31.1}
\defR{scale-10/berth-nsb/p99/resnet50_b}{20.7}
\defR{scale-10/berth-nsb/p99/retinanet_b}{92.8}
\defR{scale-10/berth-nsb/p99/bertlarge_b}{220}
\defR{scale-10/berth-nsb/p99/resnet50_c}{22.6}
\defR{scale-10/berth-nsb/p99/bertlarge_c}{245}
\defR{scale-10/berth-nsb/p99/retinanet_c}{88.8}
\defR{scale-10/berth-nsb/p99x/resnet50}{1.9}
\defR{scale-10/berth-nsb/p99x/retinanet}{2.8}
\defR{scale-10/berth-nsb/p99x/bertlarge}{3.1}
\defR{scale-10/berth-nsb/p99x/perception}{2.3}
\defR{scale-10/berth-nsb/p99x/resnet50_b}{1.7}
\defR{scale-10/berth-nsb/p99x/retinanet_b}{2.9}
\defR{scale-10/berth-nsb/p99x/bertlarge_b}{3.1}
\defR{scale-10/berth-nsb/p99x/resnet50_c}{1.9}
\defR{scale-10/berth-nsb/p99x/bertlarge_c}{3.5}
\defR{scale-10/berth-nsb/p99x/retinanet_c}{2.8}
\defR{scale-10/berth-nsb/thrabs/train_bert}{0.30}
\defR{scale-10/berth-nsb/thr/train_bert}{18}
\defR{scale-10/berth-ng/attmin}{86.5}
\defR{scale-10/berth-ng/attmean}{93.5}
\defR{scale-10/berth-ng/thr}{19}
\defR{scale-10/berth-ng/n}{3}
\defR{scale-10/berth-ng/att/resnet50}{92.6}
\defR{scale-10/berth-ng/att/retinanet}{99.0}
\defR{scale-10/berth-ng/att/bertlarge}{87.3}
\defR{scale-10/berth-ng/att/perception}{98.9}
\defR{scale-10/berth-ng/att/resnet50_b}{92.4}
\defR{scale-10/berth-ng/att/retinanet_b}{98.8}
\defR{scale-10/berth-ng/att/bertlarge_b}{87.6}
\defR{scale-10/berth-ng/att/resnet50_c}{93.2}
\defR{scale-10/berth-ng/att/bertlarge_c}{86.5}
\defR{scale-10/berth-ng/att/retinanet_c}{98.8}
\defR{scale-10/berth-ng/p99/resnet50}{24.1}
\defR{scale-10/berth-ng/p99/retinanet}{99.8}
\defR{scale-10/berth-ng/p99/bertlarge}{243}
\defR{scale-10/berth-ng/p99/perception}{34.5}
\defR{scale-10/berth-ng/p99/resnet50_b}{22.9}
\defR{scale-10/berth-ng/p99/retinanet_b}{108}
\defR{scale-10/berth-ng/p99/bertlarge_b}{214}
\defR{scale-10/berth-ng/p99/resnet50_c}{23.2}
\defR{scale-10/berth-ng/p99/bertlarge_c}{279}
\defR{scale-10/berth-ng/p99/retinanet_c}{96.2}
\defR{scale-10/berth-ng/p99x/resnet50}{2.0}
\defR{scale-10/berth-ng/p99x/retinanet}{3.1}
\defR{scale-10/berth-ng/p99x/bertlarge}{3.4}
\defR{scale-10/berth-ng/p99x/perception}{2.5}
\defR{scale-10/berth-ng/p99x/resnet50_b}{1.9}
\defR{scale-10/berth-ng/p99x/retinanet_b}{3.4}
\defR{scale-10/berth-ng/p99x/bertlarge_b}{3.0}
\defR{scale-10/berth-ng/p99x/resnet50_c}{1.9}
\defR{scale-10/berth-ng/p99x/bertlarge_c}{3.9}
\defR{scale-10/berth-ng/p99x/retinanet_c}{3.0}
\defR{scale-10/berth-ng/thrabs/train_bert}{0.31}
\defR{scale-10/berth-ng/thr/train_bert}{19}
\defR{scale-10/berth-nsng/attmin}{85.5}
\defR{scale-10/berth-nsng/attmean}{93.4}
\defR{scale-10/berth-nsng/thr}{19}
\defR{scale-10/berth-nsng/n}{3}
\defR{scale-10/berth-nsng/att/resnet50}{93.4}
\defR{scale-10/berth-nsng/att/retinanet}{98.4}
\defR{scale-10/berth-nsng/att/bertlarge}{87.4}
\defR{scale-10/berth-nsng/att/perception}{99.0}
\defR{scale-10/berth-nsng/att/resnet50_b}{93.6}
\defR{scale-10/berth-nsng/att/retinanet_b}{99.3}
\defR{scale-10/berth-nsng/att/bertlarge_b}{86.1}
\defR{scale-10/berth-nsng/att/resnet50_c}{93.1}
\defR{scale-10/berth-nsng/att/bertlarge_c}{85.5}
\defR{scale-10/berth-nsng/att/retinanet_c}{98.5}
\defR{scale-10/berth-nsng/p99/resnet50}{23.9}
\defR{scale-10/berth-nsng/p99/retinanet}{108}
\defR{scale-10/berth-nsng/p99/bertlarge}{230}
\defR{scale-10/berth-nsng/p99/perception}{33.9}
\defR{scale-10/berth-nsng/p99/resnet50_b}{21.4}
\defR{scale-10/berth-nsng/p99/retinanet_b}{93.1}
\defR{scale-10/berth-nsng/p99/bertlarge_b}{254}
\defR{scale-10/berth-nsng/p99/resnet50_c}{22.7}
\defR{scale-10/berth-nsng/p99/bertlarge_c}{300}
\defR{scale-10/berth-nsng/p99/retinanet_c}{106}
\defR{scale-10/berth-nsng/p99x/resnet50}{2.0}
\defR{scale-10/berth-nsng/p99x/retinanet}{3.4}
\defR{scale-10/berth-nsng/p99x/bertlarge}{3.2}
\defR{scale-10/berth-nsng/p99x/perception}{2.5}
\defR{scale-10/berth-nsng/p99x/resnet50_b}{1.8}
\defR{scale-10/berth-nsng/p99x/retinanet_b}{2.9}
\defR{scale-10/berth-nsng/p99x/bertlarge_b}{3.6}
\defR{scale-10/berth-nsng/p99x/resnet50_c}{1.9}
\defR{scale-10/berth-nsng/p99x/bertlarge_c}{4.2}
\defR{scale-10/berth-nsng/p99x/retinanet_c}{3.3}
\defR{scale-10/berth-nsng/thrabs/train_bert}{0.30}
\defR{scale-10/berth-nsng/thr/train_bert}{19}
\defR{scale-10/berth-grouped/attmin}{59.3}
\defR{scale-10/berth-grouped/attmean}{87.3}
\defR{scale-10/berth-grouped/thr}{25}
\defR{scale-10/berth-grouped/n}{3}
\defR{scale-10/berth-grouped/att/resnet50}{59.3}
\defR{scale-10/berth-grouped/att/retinanet}{98.8}
\defR{scale-10/berth-grouped/att/bertlarge}{99.6}
\defR{scale-10/berth-grouped/att/perception}{96.9}
\defR{scale-10/berth-grouped/att/resnet50_b}{61.0}
\defR{scale-10/berth-grouped/att/retinanet_b}{99.3}
\defR{scale-10/berth-grouped/att/bertlarge_b}{99.2}
\defR{scale-10/berth-grouped/att/resnet50_c}{60.9}
\defR{scale-10/berth-grouped/att/bertlarge_c}{98.9}
\defR{scale-10/berth-grouped/att/retinanet_c}{99.1}
\defR{scale-10/berth-grouped/p99/resnet50}{47.9}
\defR{scale-10/berth-grouped/p99/retinanet}{103}
\defR{scale-10/berth-grouped/p99/bertlarge}{113}
\defR{scale-10/berth-grouped/p99/perception}{38.4}
\defR{scale-10/berth-grouped/p99/resnet50_b}{44.6}
\defR{scale-10/berth-grouped/p99/retinanet_b}{89.6}
\defR{scale-10/berth-grouped/p99/bertlarge_b}{114}
\defR{scale-10/berth-grouped/p99/resnet50_c}{46.3}
\defR{scale-10/berth-grouped/p99/bertlarge_c}{120}
\defR{scale-10/berth-grouped/p99/retinanet_c}{95.9}
\defR{scale-10/berth-grouped/p99x/resnet50}{3.9}
\defR{scale-10/berth-grouped/p99x/retinanet}{3.2}
\defR{scale-10/berth-grouped/p99x/bertlarge}{1.6}
\defR{scale-10/berth-grouped/p99x/perception}{2.8}
\defR{scale-10/berth-grouped/p99x/resnet50_b}{3.7}
\defR{scale-10/berth-grouped/p99x/retinanet_b}{2.8}
\defR{scale-10/berth-grouped/p99x/bertlarge_b}{1.6}
\defR{scale-10/berth-grouped/p99x/resnet50_c}{3.8}
\defR{scale-10/berth-grouped/p99x/bertlarge_c}{1.7}
\defR{scale-10/berth-grouped/p99x/retinanet_c}{3.0}
\defR{scale-10/berth-grouped/thrabs/train_bert}{0.41}
\defR{scale-10/berth-grouped/thr/train_bert}{25}
\defR{scale-10/berth/attmin}{49.5}
\defR{scale-10/berth/attmean}{80.2}
\defR{scale-10/berth/thr}{12}
\defR{scale-10/berth/n}{3}
\defR{scale-10/berth/att/resnet50}{68.3}
\defR{scale-10/berth/att/retinanet}{97.6}
\defR{scale-10/berth/att/bertlarge}{85.5}
\defR{scale-10/berth/att/perception}{49.5}
\defR{scale-10/berth/att/resnet50_b}{67.3}
\defR{scale-10/berth/att/retinanet_b}{98.8}
\defR{scale-10/berth/att/bertlarge_b}{84.4}
\defR{scale-10/berth/att/resnet50_c}{67.4}
\defR{scale-10/berth/att/bertlarge_c}{84.7}
\defR{scale-10/berth/att/retinanet_c}{98.2}
\defR{scale-10/berth/p99/resnet50}{38.1}
\defR{scale-10/berth/p99/retinanet}{117}
\defR{scale-10/berth/p99/bertlarge}{243}
\defR{scale-10/berth/p99/perception}{273}
\defR{scale-10/berth/p99/resnet50_b}{40.5}
\defR{scale-10/berth/p99/retinanet_b}{101}
\defR{scale-10/berth/p99/bertlarge_b}{271}
\defR{scale-10/berth/p99/resnet50_c}{41.7}
\defR{scale-10/berth/p99/bertlarge_c}{255}
\defR{scale-10/berth/p99/retinanet_c}{109}
\defR{scale-10/berth/p99x/resnet50}{3.1}
\defR{scale-10/berth/p99x/retinanet}{3.6}
\defR{scale-10/berth/p99x/bertlarge}{3.4}
\defR{scale-10/berth/p99x/perception}{19.9}
\defR{scale-10/berth/p99x/resnet50_b}{3.3}
\defR{scale-10/berth/p99x/retinanet_b}{3.1}
\defR{scale-10/berth/p99x/bertlarge_b}{3.8}
\defR{scale-10/berth/p99x/resnet50_c}{3.4}
\defR{scale-10/berth/p99x/bertlarge_c}{3.6}
\defR{scale-10/berth/p99x/retinanet_c}{3.4}
\defR{scale-10/berth/thrabs/train_bert}{0.19}
\defR{scale-10/berth/thr/train_bert}{12}
\defR{scale-10/timeslice/attmin}{0.0}
\defR{scale-10/timeslice/attmean}{37.9}
\defR{scale-10/timeslice/thr}{24}
\defR{scale-10/timeslice/n}{3}
\defR{scale-10/timeslice/att/resnet50}{19.7}
\defR{scale-10/timeslice/att/retinanet}{84.0}
\defR{scale-10/timeslice/att/bertlarge}{22.9}
\defR{scale-10/timeslice/att/perception}{0.0}
\defR{scale-10/timeslice/att/resnet50_b}{19.5}
\defR{scale-10/timeslice/att/retinanet_b}{84.1}
\defR{scale-10/timeslice/att/bertlarge_b}{21.8}
\defR{scale-10/timeslice/att/resnet50_c}{19.6}
\defR{scale-10/timeslice/att/bertlarge_c}{22.3}
\defR{scale-10/timeslice/att/retinanet_c}{84.8}
\defR{scale-10/timeslice/p99/resnet50}{62.1}
\defR{scale-10/timeslice/p99/retinanet}{223}
\defR{scale-10/timeslice/p99/bertlarge}{485}
\defR{scale-10/timeslice/p99/perception}{35481}
\defR{scale-10/timeslice/p99/resnet50_b}{66.9}
\defR{scale-10/timeslice/p99/retinanet_b}{194}
\defR{scale-10/timeslice/p99/bertlarge_b}{521}
\defR{scale-10/timeslice/p99/resnet50_c}{66.9}
\defR{scale-10/timeslice/p99/bertlarge_c}{516}
\defR{scale-10/timeslice/p99/retinanet_c}{196}
\defR{scale-10/timeslice/p99x/resnet50}{5.1}
\defR{scale-10/timeslice/p99x/retinanet}{6.9}
\defR{scale-10/timeslice/p99x/bertlarge}{6.8}
\defR{scale-10/timeslice/p99x/perception}{2589.5}
\defR{scale-10/timeslice/p99x/resnet50_b}{5.5}
\defR{scale-10/timeslice/p99x/retinanet_b}{6.0}
\defR{scale-10/timeslice/p99x/bertlarge_b}{7.3}
\defR{scale-10/timeslice/p99x/resnet50_c}{5.5}
\defR{scale-10/timeslice/p99x/bertlarge_c}{7.3}
\defR{scale-10/timeslice/p99x/retinanet_c}{6.1}
\defR{scale-10/timeslice/thrabs/train_bert}{0.39}
\defR{scale-10/timeslice/thr/train_bert}{24}
\defR{scale-10/mps/attmin}{0.0}
\defR{scale-10/mps/attmean}{0.0}
\defR{scale-10/mps/thr}{62}
\defR{scale-10/mps/n}{3}
\defR{scale-10/mps/att/resnet50}{0.0}
\defR{scale-10/mps/att/retinanet}{0.0}
\defR{scale-10/mps/att/bertlarge}{0.0}
\defR{scale-10/mps/att/perception}{0.0}
\defR{scale-10/mps/att/resnet50_b}{0.0}
\defR{scale-10/mps/att/retinanet_b}{0.0}
\defR{scale-10/mps/att/bertlarge_b}{0.0}
\defR{scale-10/mps/att/resnet50_c}{0.0}
\defR{scale-10/mps/att/bertlarge_c}{0.0}
\defR{scale-10/mps/att/retinanet_c}{0.0}
\defR{scale-10/mps/p99/resnet50}{7324}
\defR{scale-10/mps/p99/retinanet}{23796}
\defR{scale-10/mps/p99/bertlarge}{966}
\defR{scale-10/mps/p99/perception}{68829}
\defR{scale-10/mps/p99/resnet50_b}{7485}
\defR{scale-10/mps/p99/retinanet_b}{24201}
\defR{scale-10/mps/p99/bertlarge_b}{908}
\defR{scale-10/mps/p99/resnet50_c}{9985}
\defR{scale-10/mps/p99/bertlarge_c}{846}
\defR{scale-10/mps/p99/retinanet_c}{22759}
\defR{scale-10/mps/p99x/resnet50}{600.4}
\defR{scale-10/mps/p99x/retinanet}{740.9}
\defR{scale-10/mps/p99x/bertlarge}{13.6}
\defR{scale-10/mps/p99x/perception}{5023.4}
\defR{scale-10/mps/p99x/resnet50_b}{613.6}
\defR{scale-10/mps/p99x/retinanet_b}{753.5}
\defR{scale-10/mps/p99x/bertlarge_b}{12.8}
\defR{scale-10/mps/p99x/resnet50_c}{818.6}
\defR{scale-10/mps/p99x/bertlarge_c}{11.9}
\defR{scale-10/mps/p99x/retinanet_c}{708.6}
\defR{scale-10/mps/thrabs/train_bert}{1.01}
\defR{scale-10/mps/thr/train_bert}{62}
\defR{scale-10/mpsprio/attmin}{0.0}
\defR{scale-10/mpsprio/attmean}{48.3}
\defR{scale-10/mpsprio/thr}{36}
\defR{scale-10/mpsprio/n}{3}
\defR{scale-10/mpsprio/att/resnet50}{0.0}
\defR{scale-10/mpsprio/att/retinanet}{72.5}
\defR{scale-10/mpsprio/att/bertlarge}{88.4}
\defR{scale-10/mpsprio/att/perception}{0.0}
\defR{scale-10/mpsprio/att/resnet50_b}{0.0}
\defR{scale-10/mpsprio/att/retinanet_b}{72.0}
\defR{scale-10/mpsprio/att/bertlarge_b}{88.6}
\defR{scale-10/mpsprio/att/resnet50_c}{0.0}
\defR{scale-10/mpsprio/att/bertlarge_c}{88.1}
\defR{scale-10/mpsprio/att/retinanet_c}{73.1}
\defR{scale-10/mpsprio/p99/resnet50}{96.5}
\defR{scale-10/mpsprio/p99/retinanet}{262}
\defR{scale-10/mpsprio/p99/bertlarge}{183}
\defR{scale-10/mpsprio/p99/perception}{3979}
\defR{scale-10/mpsprio/p99/resnet50_b}{91.8}
\defR{scale-10/mpsprio/p99/retinanet_b}{228}
\defR{scale-10/mpsprio/p99/bertlarge_b}{206}
\defR{scale-10/mpsprio/p99/resnet50_c}{89.8}
\defR{scale-10/mpsprio/p99/bertlarge_c}{189}
\defR{scale-10/mpsprio/p99/retinanet_c}{234}
\defR{scale-10/mpsprio/p99x/resnet50}{7.9}
\defR{scale-10/mpsprio/p99x/retinanet}{8.2}
\defR{scale-10/mpsprio/p99x/bertlarge}{2.6}
\defR{scale-10/mpsprio/p99x/perception}{290.4}
\defR{scale-10/mpsprio/p99x/resnet50_b}{7.5}
\defR{scale-10/mpsprio/p99x/retinanet_b}{7.1}
\defR{scale-10/mpsprio/p99x/bertlarge_b}{2.9}
\defR{scale-10/mpsprio/p99x/resnet50_c}{7.4}
\defR{scale-10/mpsprio/p99x/bertlarge_c}{2.7}
\defR{scale-10/mpsprio/p99x/retinanet_c}{7.3}
\defR{scale-10/mpsprio/thrabs/train_bert}{0.59}
\defR{scale-10/mpsprio/thr/train_bert}{36}
\defR{scale-10/mpslimits/attmin}{0.0}
\defR{scale-10/mpslimits/attmean}{0.5}
\defR{scale-10/mpslimits/thr}{20}
\defR{scale-10/mpslimits/n}{3}
\defR{scale-10/mpslimits/att/resnet50}{1.6}
\defR{scale-10/mpslimits/att/retinanet}{0.0}
\defR{scale-10/mpslimits/att/bertlarge}{0.0}
\defR{scale-10/mpslimits/att/perception}{0.0}
\defR{scale-10/mpslimits/att/resnet50_b}{2.0}
\defR{scale-10/mpslimits/att/retinanet_b}{0.0}
\defR{scale-10/mpslimits/att/bertlarge_b}{0.0}
\defR{scale-10/mpslimits/att/resnet50_c}{1.3}
\defR{scale-10/mpslimits/att/bertlarge_c}{0.0}
\defR{scale-10/mpslimits/att/retinanet_c}{0.0}
\defR{scale-10/mpslimits/p99/resnet50}{34.7}
\defR{scale-10/mpslimits/p99/retinanet}{1049}
\defR{scale-10/mpslimits/p99/bertlarge}{1629}
\defR{scale-10/mpslimits/p99/perception}{68324}
\defR{scale-10/mpslimits/p99/resnet50_b}{36.2}
\defR{scale-10/mpslimits/p99/retinanet_b}{1234}
\defR{scale-10/mpslimits/p99/bertlarge_b}{1496}
\defR{scale-10/mpslimits/p99/resnet50_c}{34.7}
\defR{scale-10/mpslimits/p99/bertlarge_c}{1340}
\defR{scale-10/mpslimits/p99/retinanet_c}{946}
\defR{scale-10/mpslimits/p99x/resnet50}{2.8}
\defR{scale-10/mpslimits/p99x/retinanet}{32.7}
\defR{scale-10/mpslimits/p99x/bertlarge}{22.9}
\defR{scale-10/mpslimits/p99x/perception}{4986.6}
\defR{scale-10/mpslimits/p99x/resnet50_b}{3.0}
\defR{scale-10/mpslimits/p99x/retinanet_b}{38.4}
\defR{scale-10/mpslimits/p99x/bertlarge_b}{21.1}
\defR{scale-10/mpslimits/p99x/resnet50_c}{2.8}
\defR{scale-10/mpslimits/p99x/bertlarge_c}{18.9}
\defR{scale-10/mpslimits/p99x/retinanet_c}{29.4}
\defR{scale-10/mpslimits/thrabs/train_bert}{0.32}
\defR{scale-10/mpslimits/thr/train_bert}{20}
\defR{scale-10/mpsparts/attmin}{0.0}
\defR{scale-10/mpsparts/attmean}{0.8}
\defR{scale-10/mpsparts/thr}{18}
\defR{scale-10/mpsparts/n}{2}
\defR{scale-10/mpsparts/att/resnet50}{2.2}
\defR{scale-10/mpsparts/att/retinanet}{0.0}
\defR{scale-10/mpsparts/att/bertlarge}{0.0}
\defR{scale-10/mpsparts/att/perception}{0.0}
\defR{scale-10/mpsparts/att/resnet50_b}{2.8}
\defR{scale-10/mpsparts/att/retinanet_b}{0.0}
\defR{scale-10/mpsparts/att/bertlarge_b}{0.0}
\defR{scale-10/mpsparts/att/resnet50_c}{2.9}
\defR{scale-10/mpsparts/att/bertlarge_c}{0.0}
\defR{scale-10/mpsparts/att/retinanet_c}{0.0}
\defR{scale-10/mpsparts/p99/resnet50}{34.5}
\defR{scale-10/mpsparts/p99/retinanet}{1114}
\defR{scale-10/mpsparts/p99/bertlarge}{870}
\defR{scale-10/mpsparts/p99/perception}{58640}
\defR{scale-10/mpsparts/p99/resnet50_b}{37.9}
\defR{scale-10/mpsparts/p99/retinanet_b}{1132}
\defR{scale-10/mpsparts/p99/bertlarge_b}{734}
\defR{scale-10/mpsparts/p99/resnet50_c}{33.4}
\defR{scale-10/mpsparts/p99/bertlarge_c}{782}
\defR{scale-10/mpsparts/p99/retinanet_c}{994}
\defR{scale-10/mpsparts/p99x/resnet50}{2.8}
\defR{scale-10/mpsparts/p99x/retinanet}{34.7}
\defR{scale-10/mpsparts/p99x/bertlarge}{12.2}
\defR{scale-10/mpsparts/p99x/perception}{4279.7}
\defR{scale-10/mpsparts/p99x/resnet50_b}{3.1}
\defR{scale-10/mpsparts/p99x/retinanet_b}{35.2}
\defR{scale-10/mpsparts/p99x/bertlarge_b}{10.3}
\defR{scale-10/mpsparts/p99x/resnet50_c}{2.7}
\defR{scale-10/mpsparts/p99x/bertlarge_c}{11.0}
\defR{scale-10/mpsparts/p99x/retinanet_c}{31.0}
\defR{scale-10/mpsparts/thrabs/train_bert}{0.29}
\defR{scale-10/mpsparts/thr/train_bert}{18}
\defR{scale-10/lithos/attmin}{0.0}
\defR{scale-10/lithos/attmean}{0.0}
\defR{scale-10/lithos/thr}{23}
\defR{scale-10/lithos/n}{3}
\defR{scale-10/lithos/att/resnet50}{0.0}
\defR{scale-10/lithos/att/retinanet}{0.0}
\defR{scale-10/lithos/att/bertlarge}{0.0}
\defR{scale-10/lithos/att/perception}{0.0}
\defR{scale-10/lithos/att/resnet50_b}{0.0}
\defR{scale-10/lithos/att/retinanet_b}{0.0}
\defR{scale-10/lithos/att/bertlarge_b}{0.0}
\defR{scale-10/lithos/att/resnet50_c}{0.0}
\defR{scale-10/lithos/att/bertlarge_c}{0.0}
\defR{scale-10/lithos/att/retinanet_c}{0.0}
\defR{scale-10/lithos/p99/resnet50}{39.5}
\defR{scale-10/lithos/p99/retinanet}{1265}
\defR{scale-10/lithos/p99/bertlarge}{2152}
\defR{scale-10/lithos/p99/perception}{60720}
\defR{scale-10/lithos/p99/resnet50_b}{40.6}
\defR{scale-10/lithos/p99/retinanet_b}{1582}
\defR{scale-10/lithos/p99/bertlarge_b}{2285}
\defR{scale-10/lithos/p99/resnet50_c}{37.6}
\defR{scale-10/lithos/p99/bertlarge_c}{2146}
\defR{scale-10/lithos/p99/retinanet_c}{1264}
\defR{scale-10/lithos/p99x/resnet50}{3.2}
\defR{scale-10/lithos/p99x/retinanet}{39.4}
\defR{scale-10/lithos/p99x/bertlarge}{30.3}
\defR{scale-10/lithos/p99x/perception}{4431.6}
\defR{scale-10/lithos/p99x/resnet50_b}{3.3}
\defR{scale-10/lithos/p99x/retinanet_b}{49.3}
\defR{scale-10/lithos/p99x/bertlarge_b}{32.2}
\defR{scale-10/lithos/p99x/resnet50_c}{3.1}
\defR{scale-10/lithos/p99x/bertlarge_c}{30.2}
\defR{scale-10/lithos/p99x/retinanet_c}{39.3}
\defR{scale-10/lithos/thrabs/train_bert}{0.38}
\defR{scale-10/lithos/thr/train_bert}{23}
\defR{scale-10/dysm/attmin}{0.0}
\defR{scale-10/dysm/attmean}{18.5}
\defR{scale-10/dysm/thr}{53}
\defR{scale-10/dysm/n}{3}
\defR{scale-10/dysm/att/resnet50}{0.0}
\defR{scale-10/dysm/att/retinanet}{56.2}
\defR{scale-10/dysm/att/bertlarge}{0.0}
\defR{scale-10/dysm/att/perception}{99.0}
\defR{scale-10/dysm/att/resnet50_b}{0.0}
\defR{scale-10/dysm/att/retinanet_b}{0.0}
\defR{scale-10/dysm/att/bertlarge_b}{0.0}
\defR{scale-10/dysm/att/resnet50_c}{0.0}
\defR{scale-10/dysm/att/bertlarge_c}{0.0}
\defR{scale-10/dysm/att/retinanet_c}{29.9}
\defR{scale-10/dysm/p99/retinanet}{188}
\defR{scale-10/dysm/p99/perception}{33.3}
\defR{scale-10/dysm/p99/retinanet_c}{140}
\defR{scale-10/dysm/p99x/retinanet}{5.8}
\defR{scale-10/dysm/p99x/perception}{2.4}
\defR{scale-10/dysm/p99x/retinanet_c}{4.4}
\defR{scale-10/dysm/thrabs/train_bert}{0.86}
\defR{scale-10/dysm/thr/train_bert}{53}
\defR{scale-10/xsched/attmin}{43.5}
\defR{scale-10/xsched/attmean}{71.7}
\defR{scale-10/xsched/thr}{11}
\defR{scale-10/xsched/n}{3}
\defR{scale-10/xsched/att/resnet50}{46.5}
\defR{scale-10/xsched/att/retinanet}{95.7}
\defR{scale-10/xsched/att/bertlarge}{83.9}
\defR{scale-10/xsched/att/perception}{43.5}
\defR{scale-10/xsched/att/resnet50_b}{45.9}
\defR{scale-10/xsched/att/retinanet_b}{96.6}
\defR{scale-10/xsched/att/bertlarge_b}{80.6}
\defR{scale-10/xsched/att/resnet50_c}{45.9}
\defR{scale-10/xsched/att/bertlarge_c}{82.2}
\defR{scale-10/xsched/att/retinanet_c}{96.1}
\defR{scale-10/xsched/p99/resnet50}{53.1}
\defR{scale-10/xsched/p99/retinanet}{147}
\defR{scale-10/xsched/p99/bertlarge}{265}
\defR{scale-10/xsched/p99/perception}{326}
\defR{scale-10/xsched/p99/resnet50_b}{55.2}
\defR{scale-10/xsched/p99/retinanet_b}{126}
\defR{scale-10/xsched/p99/bertlarge_b}{285}
\defR{scale-10/xsched/p99/resnet50_c}{54.8}
\defR{scale-10/xsched/p99/bertlarge_c}{272}
\defR{scale-10/xsched/p99/retinanet_c}{137}
\defR{scale-10/xsched/p99x/resnet50}{4.4}
\defR{scale-10/xsched/p99x/retinanet}{4.6}
\defR{scale-10/xsched/p99x/bertlarge}{3.7}
\defR{scale-10/xsched/p99x/perception}{23.8}
\defR{scale-10/xsched/p99x/resnet50_b}{4.5}
\defR{scale-10/xsched/p99x/retinanet_b}{3.9}
\defR{scale-10/xsched/p99x/bertlarge_b}{4.0}
\defR{scale-10/xsched/p99x/resnet50_c}{4.5}
\defR{scale-10/xsched/p99x/bertlarge_c}{3.8}
\defR{scale-10/xsched/p99x/retinanet_c}{4.3}
\defR{scale-10/xsched/thrabs/train_bert}{0.18}
\defR{scale-10/xsched/thr/train_bert}{11}
\defR{scale-10/multistream/attmin}{0.0}
\defR{scale-10/multistream/attmean}{43.8}
\defR{scale-10/multistream/thr}{38}
\defR{scale-10/multistream/n}{3}
\defR{scale-10/multistream/att/resnet50}{0.0}
\defR{scale-10/multistream/att/retinanet}{59.0}
\defR{scale-10/multistream/att/bertlarge}{87.0}
\defR{scale-10/multistream/att/perception}{0.0}
\defR{scale-10/multistream/att/resnet50_b}{0.0}
\defR{scale-10/multistream/att/retinanet_b}{59.8}
\defR{scale-10/multistream/att/bertlarge_b}{86.3}
\defR{scale-10/multistream/att/resnet50_c}{0.0}
\defR{scale-10/multistream/att/bertlarge_c}{86.5}
\defR{scale-10/multistream/att/retinanet_c}{59.8}
\defR{scale-10/multistream/p99/resnet50}{531}
\defR{scale-10/multistream/p99/retinanet}{655}
\defR{scale-10/multistream/p99/bertlarge}{380}
\defR{scale-10/multistream/p99/perception}{6106}
\defR{scale-10/multistream/p99/resnet50_b}{480}
\defR{scale-10/multistream/p99/retinanet_b}{434}
\defR{scale-10/multistream/p99/bertlarge_b}{532}
\defR{scale-10/multistream/p99/resnet50_c}{377}
\defR{scale-10/multistream/p99/bertlarge_c}{528}
\defR{scale-10/multistream/p99/retinanet_c}{496}
\defR{scale-10/multistream/p99x/resnet50}{43.6}
\defR{scale-10/multistream/p99x/retinanet}{20.4}
\defR{scale-10/multistream/p99x/bertlarge}{5.4}
\defR{scale-10/multistream/p99x/perception}{445.6}
\defR{scale-10/multistream/p99x/resnet50_b}{39.3}
\defR{scale-10/multistream/p99x/retinanet_b}{13.5}
\defR{scale-10/multistream/p99x/bertlarge_b}{7.5}
\defR{scale-10/multistream/p99x/resnet50_c}{30.9}
\defR{scale-10/multistream/p99x/bertlarge_c}{7.4}
\defR{scale-10/multistream/p99x/retinanet_c}{15.4}
\defR{scale-10/multistream/thrabs/train_bert}{0.61}
\defR{scale-10/multistream/thr/train_bert}{38}
\defR{scale-10/tgs/attmin}{59.4}
\defR{scale-10/tgs/attmean}{83.8}
\defR{scale-10/tgs/thr}{1}
\defR{scale-10/tgs/n}{3}
\defR{scale-10/tgs/att/resnet50}{73.4}
\defR{scale-10/tgs/att/retinanet}{99.4}
\defR{scale-10/tgs/att/bertlarge}{87.5}
\defR{scale-10/tgs/att/perception}{59.4}
\defR{scale-10/tgs/att/resnet50_b}{71.9}
\defR{scale-10/tgs/att/retinanet_b}{98.0}
\defR{scale-10/tgs/att/bertlarge_b}{85.5}
\defR{scale-10/tgs/att/resnet50_c}{74.5}
\defR{scale-10/tgs/att/bertlarge_c}{89.2}
\defR{scale-10/tgs/att/retinanet_c}{99.4}
\defR{scale-10/tgs/p99/resnet50}{38.6}
\defR{scale-10/tgs/p99/retinanet}{86.8}
\defR{scale-10/tgs/p99/bertlarge}{228}
\defR{scale-10/tgs/p99/perception}{284}
\defR{scale-10/tgs/p99/resnet50_b}{39.3}
\defR{scale-10/tgs/p99/retinanet_b}{120}
\defR{scale-10/tgs/p99/bertlarge_b}{275}
\defR{scale-10/tgs/p99/resnet50_c}{33.6}
\defR{scale-10/tgs/p99/bertlarge_c}{218}
\defR{scale-10/tgs/p99/retinanet_c}{90.9}
\defR{scale-10/tgs/p99x/resnet50}{3.2}
\defR{scale-10/tgs/p99x/retinanet}{2.7}
\defR{scale-10/tgs/p99x/bertlarge}{3.2}
\defR{scale-10/tgs/p99x/perception}{20.7}
\defR{scale-10/tgs/p99x/resnet50_b}{3.2}
\defR{scale-10/tgs/p99x/retinanet_b}{3.7}
\defR{scale-10/tgs/p99x/bertlarge_b}{3.9}
\defR{scale-10/tgs/p99x/resnet50_c}{2.8}
\defR{scale-10/tgs/p99x/bertlarge_c}{3.1}
\defR{scale-10/tgs/p99x/retinanet_c}{2.8}
\defR{scale-10/tgs/thrabs/train_bert}{0.01}
\defR{scale-10/tgs/thr/train_bert}{1}
\defR{scale-10/mig/attmin}{0.0}
\defR{scale-10/mig/attmean}{46.9}
\defR{scale-10/mig/thr}{0}
\defR{scale-10/mig/n}{3}
\defR{scale-10/mig/att/resnet50}{0.0}
\defR{scale-10/mig/att/retinanet}{83.7}
\defR{scale-10/mig/att/bertlarge}{74.9}
\defR{scale-10/mig/att/perception}{0.0}
\defR{scale-10/mig/att/resnet50_b}{0.0}
\defR{scale-10/mig/att/retinanet_b}{84.0}
\defR{scale-10/mig/att/bertlarge_b}{70.1}
\defR{scale-10/mig/att/resnet50_c}{0.0}
\defR{scale-10/mig/att/bertlarge_c}{72.2}
\defR{scale-10/mig/att/retinanet_c}{84.2}
\defR{scale-10/mig/p99/retinanet}{176}
\defR{scale-10/mig/p99/bertlarge}{314}
\defR{scale-10/mig/p99/retinanet_b}{167}
\defR{scale-10/mig/p99/bertlarge_b}{329}
\defR{scale-10/mig/p99/resnet50_c}{59.0}
\defR{scale-10/mig/p99/bertlarge_c}{323}
\defR{scale-10/mig/p99/retinanet_c}{162}
\defR{scale-10/mig/p99x/retinanet}{5.5}
\defR{scale-10/mig/p99x/bertlarge}{4.4}
\defR{scale-10/mig/p99x/retinanet_b}{5.2}
\defR{scale-10/mig/p99x/bertlarge_b}{4.6}
\defR{scale-10/mig/p99x/resnet50_c}{4.8}
\defR{scale-10/mig/p99x/bertlarge_c}{4.6}
\defR{scale-10/mig/p99x/retinanet_c}{5.0}
\defR{scale-10/mig/thrabs/train_bert}{0.00}
\defR{scale-10/mig/thr/train_bert}{0}
\defR{scale-10/tally/attmin}{0.0}
\defR{scale-10/tally/attmean}{18.6}
\defR{scale-10/tally/thr}{3}
\defR{scale-10/tally/n}{3}
\defR{scale-10/tally/att/resnet50}{0.0}
\defR{scale-10/tally/att/retinanet}{0.0}
\defR{scale-10/tally/att/bertlarge}{0.0}
\defR{scale-10/tally/att/perception}{0.0}
\defR{scale-10/tally/att/resnet50_b}{0.0}
\defR{scale-10/tally/att/retinanet_b}{42.3}
\defR{scale-10/tally/att/bertlarge_b}{16.2}
\defR{scale-10/tally/att/resnet50_c}{0.0}
\defR{scale-10/tally/att/bertlarge_c}{27.1}
\defR{scale-10/tally/att/retinanet_c}{100.0}
\defR{scale-10/tally/p99/resnet50}{94109}
\defR{scale-10/tally/p99/retinanet}{85208}
\defR{scale-10/tally/p99/bertlarge}{24897}
\defR{scale-10/tally/p99/perception}{94067}
\defR{scale-10/tally/p99/resnet50_b}{64648}
\defR{scale-10/tally/p99/retinanet_b}{410}
\defR{scale-10/tally/p99/bertlarge_b}{490}
\defR{scale-10/tally/p99/resnet50_c}{19060}
\defR{scale-10/tally/p99/bertlarge_c}{456}
\defR{scale-10/tally/p99/retinanet_c}{55.3}
\defR{scale-10/tally/p99x/resnet50}{7715.1}
\defR{scale-10/tally/p99x/retinanet}{2652.9}
\defR{scale-10/tally/p99x/bertlarge}{350.5}
\defR{scale-10/tally/p99x/perception}{6865.4}
\defR{scale-10/tally/p99x/resnet50_b}{5299.8}
\defR{scale-10/tally/p99x/retinanet_b}{12.8}
\defR{scale-10/tally/p99x/bertlarge_b}{6.9}
\defR{scale-10/tally/p99x/resnet50_c}{1562.5}
\defR{scale-10/tally/p99x/bertlarge_c}{6.4}
\defR{scale-10/tally/p99x/retinanet_c}{1.7}
\defR{scale-10/tally/thrabs/train_bert}{0.05}
\defR{scale-10/tally/thr/train_bert}{3}
\defR{scale-10/orion/attmin}{0.0}
\defR{scale-10/orion/attmean}{10.3}
\defR{scale-10/orion/thr}{53}
\defR{scale-10/orion/n}{3}
\defR{scale-10/orion/att/resnet50}{0.0}
\defR{scale-10/orion/att/retinanet}{0.0}
\defR{scale-10/orion/att/bertlarge}{1.3}
\defR{scale-10/orion/att/perception}{0.0}
\defR{scale-10/orion/att/resnet50_b}{0.0}
\defR{scale-10/orion/att/retinanet_b}{0.0}
\defR{scale-10/orion/att/bertlarge_b}{2.4}
\defR{scale-10/orion/att/resnet50_c}{96.6}
\defR{scale-10/orion/att/bertlarge_c}{2.3}
\defR{scale-10/orion/att/retinanet_c}{0.0}
\defR{scale-10/orion/p99/resnet50}{35447}
\defR{scale-10/orion/p99/retinanet}{27333}
\defR{scale-10/orion/p99/bertlarge}{736}
\defR{scale-10/orion/p99/perception}{84668}
\defR{scale-10/orion/p99/resnet50_b}{31671}
\defR{scale-10/orion/p99/retinanet_b}{24303}
\defR{scale-10/orion/p99/bertlarge_b}{772}
\defR{scale-10/orion/p99/resnet50_c}{18.5}
\defR{scale-10/orion/p99/bertlarge_c}{809}
\defR{scale-10/orion/p99/retinanet_c}{27851}
\defR{scale-10/orion/p99x/resnet50}{2905.9}
\defR{scale-10/orion/p99x/retinanet}{851.0}
\defR{scale-10/orion/p99x/bertlarge}{10.4}
\defR{scale-10/orion/p99x/perception}{6179.4}
\defR{scale-10/orion/p99x/resnet50_b}{2596.4}
\defR{scale-10/orion/p99x/retinanet_b}{756.7}
\defR{scale-10/orion/p99x/bertlarge_b}{10.9}
\defR{scale-10/orion/p99x/resnet50_c}{1.5}
\defR{scale-10/orion/p99x/bertlarge_c}{11.4}
\defR{scale-10/orion/p99x/retinanet_c}{867.1}
\defR{scale-10/orion/thrabs/train_bert}{0.86}
\defR{scale-10/orion/thr/train_bert}{53}
\defR{twobatch-mix-load40/berth-full/attmin}{89.3}
\defR{twobatch-mix-load40/berth-full/attmean}{93.2}
\defR{twobatch-mix-load40/berth-full/thr}{19}
\defR{twobatch-mix-load40/berth-full/n}{3}
\defR{twobatch-mix-load40/berth-full/att/resnet50}{89.3}
\defR{twobatch-mix-load40/berth-full/att/retinanet}{99.9}
\defR{twobatch-mix-load40/berth-full/att/bertlarge}{90.4}
\defR{twobatch-mix-load40/berth-full/p99/resnet50}{23.5}
\defR{twobatch-mix-load40/berth-full/p99/retinanet}{70.9}
\defR{twobatch-mix-load40/berth-full/p99/bertlarge}{220}
\defR{twobatch-mix-load40/berth-full/p99x/resnet50}{1.8}
\defR{twobatch-mix-load40/berth-full/p99x/retinanet}{2.0}
\defR{twobatch-mix-load40/berth-full/p99x/bertlarge}{3.0}
\defR{twobatch-mix-load40/berth-full/thrabs/train_bert}{0.40}
\defR{twobatch-mix-load40/berth-full/thrabs/train_resnet}{1.68}
\defR{twobatch-mix-load40/berth-full/thr/train_bert}{25}
\defR{twobatch-mix-load40/berth-full/thr/train_resnet}{13}
\defR{twobatch-mix-load40/berth-nsr/attmin}{89.9}
\defR{twobatch-mix-load40/berth-nsr/attmean}{93.5}
\defR{twobatch-mix-load40/berth-nsr/thr}{19}
\defR{twobatch-mix-load40/berth-nsr/n}{3}
\defR{twobatch-mix-load40/berth-nsr/att/resnet50}{89.9}
\defR{twobatch-mix-load40/berth-nsr/att/retinanet}{99.8}
\defR{twobatch-mix-load40/berth-nsr/att/bertlarge}{90.8}
\defR{twobatch-mix-load40/berth-nsr/p99/resnet50}{23.9}
\defR{twobatch-mix-load40/berth-nsr/p99/retinanet}{73.5}
\defR{twobatch-mix-load40/berth-nsr/p99/bertlarge}{223}
\defR{twobatch-mix-load40/berth-nsr/p99x/resnet50}{1.8}
\defR{twobatch-mix-load40/berth-nsr/p99x/retinanet}{2.1}
\defR{twobatch-mix-load40/berth-nsr/p99x/bertlarge}{3.1}
\defR{twobatch-mix-load40/berth-nsr/thrabs/train_bert}{0.41}
\defR{twobatch-mix-load40/berth-nsr/thrabs/train_resnet}{1.67}
\defR{twobatch-mix-load40/berth-nsr/thr/train_bert}{25}
\defR{twobatch-mix-load40/berth-nsr/thr/train_resnet}{13}
\defR{twobatch-mix-load40/berth-nba/attmin}{89.8}
\defR{twobatch-mix-load40/berth-nba/attmean}{93.4}
\defR{twobatch-mix-load40/berth-nba/thr}{19}
\defR{twobatch-mix-load40/berth-nba/n}{3}
\defR{twobatch-mix-load40/berth-nba/att/resnet50}{89.8}
\defR{twobatch-mix-load40/berth-nba/att/retinanet}{99.9}
\defR{twobatch-mix-load40/berth-nba/att/bertlarge}{90.5}
\defR{twobatch-mix-load40/berth-nba/p99/resnet50}{24.4}
\defR{twobatch-mix-load40/berth-nba/p99/retinanet}{71.5}
\defR{twobatch-mix-load40/berth-nba/p99/bertlarge}{211}
\defR{twobatch-mix-load40/berth-nba/p99x/resnet50}{1.8}
\defR{twobatch-mix-load40/berth-nba/p99x/retinanet}{2.1}
\defR{twobatch-mix-load40/berth-nba/p99x/bertlarge}{2.9}
\defR{twobatch-mix-load40/berth-nba/thrabs/train_bert}{0.40}
\defR{twobatch-mix-load40/berth-nba/thrabs/train_resnet}{1.69}
\defR{twobatch-mix-load40/berth-nba/thr/train_bert}{25}
\defR{twobatch-mix-load40/berth-nba/thr/train_resnet}{13}
\defR{twobatch-mix-load40/berth-nsb/attmin}{83.1}
\defR{twobatch-mix-load40/berth-nsb/attmean}{89.5}
\defR{twobatch-mix-load40/berth-nsb/thr}{18}
\defR{twobatch-mix-load40/berth-nsb/n}{3}
\defR{twobatch-mix-load40/berth-nsb/att/resnet50}{83.1}
\defR{twobatch-mix-load40/berth-nsb/att/retinanet}{99.8}
\defR{twobatch-mix-load40/berth-nsb/att/bertlarge}{85.6}
\defR{twobatch-mix-load40/berth-nsb/p99/resnet50}{26.5}
\defR{twobatch-mix-load40/berth-nsb/p99/retinanet}{80.7}
\defR{twobatch-mix-load40/berth-nsb/p99/bertlarge}{255}
\defR{twobatch-mix-load40/berth-nsb/p99x/resnet50}{2.0}
\defR{twobatch-mix-load40/berth-nsb/p99x/retinanet}{2.3}
\defR{twobatch-mix-load40/berth-nsb/p99x/bertlarge}{3.5}
\defR{twobatch-mix-load40/berth-nsb/thrabs/train_bert}{0.32}
\defR{twobatch-mix-load40/berth-nsb/thrabs/train_resnet}{2.16}
\defR{twobatch-mix-load40/berth-nsb/thr/train_bert}{20}
\defR{twobatch-mix-load40/berth-nsb/thr/train_resnet}{16}
\defR{twobatch-mix-load40/berth-ng/attmin}{82.2}
\defR{twobatch-mix-load40/berth-ng/attmean}{88.4}
\defR{twobatch-mix-load40/berth-ng/thr}{18}
\defR{twobatch-mix-load40/berth-ng/n}{3}
\defR{twobatch-mix-load40/berth-ng/att/resnet50}{82.2}
\defR{twobatch-mix-load40/berth-ng/att/retinanet}{99.6}
\defR{twobatch-mix-load40/berth-ng/att/bertlarge}{83.5}
\defR{twobatch-mix-load40/berth-ng/p99/resnet50}{27.5}
\defR{twobatch-mix-load40/berth-ng/p99/retinanet}{82.3}
\defR{twobatch-mix-load40/berth-ng/p99/bertlarge}{248}
\defR{twobatch-mix-load40/berth-ng/p99x/resnet50}{2.1}
\defR{twobatch-mix-load40/berth-ng/p99x/retinanet}{2.4}
\defR{twobatch-mix-load40/berth-ng/p99x/bertlarge}{3.4}
\defR{twobatch-mix-load40/berth-ng/thrabs/train_bert}{0.32}
\defR{twobatch-mix-load40/berth-ng/thrabs/train_resnet}{2.15}
\defR{twobatch-mix-load40/berth-ng/thr/train_bert}{20}
\defR{twobatch-mix-load40/berth-ng/thr/train_resnet}{16}
\defR{twobatch-mix-load40/berth-nsng/attmin}{81.9}
\defR{twobatch-mix-load40/berth-nsng/attmean}{88.8}
\defR{twobatch-mix-load40/berth-nsng/thr}{18}
\defR{twobatch-mix-load40/berth-nsng/n}{3}
\defR{twobatch-mix-load40/berth-nsng/att/resnet50}{81.9}
\defR{twobatch-mix-load40/berth-nsng/att/retinanet}{99.7}
\defR{twobatch-mix-load40/berth-nsng/att/bertlarge}{84.7}
\defR{twobatch-mix-load40/berth-nsng/p99/resnet50}{27.6}
\defR{twobatch-mix-load40/berth-nsng/p99/retinanet}{83.7}
\defR{twobatch-mix-load40/berth-nsng/p99/bertlarge}{255}
\defR{twobatch-mix-load40/berth-nsng/p99x/resnet50}{2.1}
\defR{twobatch-mix-load40/berth-nsng/p99x/retinanet}{2.4}
\defR{twobatch-mix-load40/berth-nsng/p99x/bertlarge}{3.5}
\defR{twobatch-mix-load40/berth-nsng/thrabs/train_bert}{0.31}
\defR{twobatch-mix-load40/berth-nsng/thrabs/train_resnet}{2.15}
\defR{twobatch-mix-load40/berth-nsng/thr/train_bert}{19}
\defR{twobatch-mix-load40/berth-nsng/thr/train_resnet}{16}
\defR{twobatch-mix-load40/berth-grouped/attmin}{46.0}
\defR{twobatch-mix-load40/berth-grouped/attmean}{79.9}
\defR{twobatch-mix-load40/berth-grouped/thr}{28}
\defR{twobatch-mix-load40/berth-grouped/n}{3}
\defR{twobatch-mix-load40/berth-grouped/att/resnet50}{46.0}
\defR{twobatch-mix-load40/berth-grouped/att/retinanet}{97.1}
\defR{twobatch-mix-load40/berth-grouped/att/bertlarge}{96.7}
\defR{twobatch-mix-load40/berth-grouped/p99/resnet50}{73.3}
\defR{twobatch-mix-load40/berth-grouped/p99/retinanet}{122}
\defR{twobatch-mix-load40/berth-grouped/p99/bertlarge}{155}
\defR{twobatch-mix-load40/berth-grouped/p99x/resnet50}{5.5}
\defR{twobatch-mix-load40/berth-grouped/p99x/retinanet}{3.5}
\defR{twobatch-mix-load40/berth-grouped/p99x/bertlarge}{2.1}
\defR{twobatch-mix-load40/berth-grouped/thrabs/train_bert}{0.51}
\defR{twobatch-mix-load40/berth-grouped/thrabs/train_resnet}{3.24}
\defR{twobatch-mix-load40/berth-grouped/thr/train_bert}{31}
\defR{twobatch-mix-load40/berth-grouped/thr/train_resnet}{24}
\defR{twobatch-mix-load40/berth/attmin}{59.8}
\defR{twobatch-mix-load40/berth/attmean}{83.1}
\defR{twobatch-mix-load40/berth/thr}{26}
\defR{twobatch-mix-load40/berth/n}{3}
\defR{twobatch-mix-load40/berth/att/resnet50}{59.8}
\defR{twobatch-mix-load40/berth/att/retinanet}{97.8}
\defR{twobatch-mix-load40/berth/att/bertlarge}{91.8}
\defR{twobatch-mix-load40/berth/p99/resnet50}{48.3}
\defR{twobatch-mix-load40/berth/p99/retinanet}{112}
\defR{twobatch-mix-load40/berth/p99/bertlarge}{191}
\defR{twobatch-mix-load40/berth/p99x/resnet50}{3.6}
\defR{twobatch-mix-load40/berth/p99x/retinanet}{3.2}
\defR{twobatch-mix-load40/berth/p99x/bertlarge}{2.6}
\defR{twobatch-mix-load40/berth/thrabs/train_bert}{0.48}
\defR{twobatch-mix-load40/berth/thrabs/train_resnet}{3.04}
\defR{twobatch-mix-load40/berth/thr/train_bert}{29}
\defR{twobatch-mix-load40/berth/thr/train_resnet}{23}
\defR{twobatch-mix-load40/timeslice/attmin}{22.3}
\defR{twobatch-mix-load40/timeslice/attmean}{36.7}
\defR{twobatch-mix-load40/timeslice/thr}{26}
\defR{twobatch-mix-load40/timeslice/n}{3}
\defR{twobatch-mix-load40/timeslice/att/resnet50}{22.3}
\defR{twobatch-mix-load40/timeslice/att/retinanet}{61.0}
\defR{twobatch-mix-load40/timeslice/att/bertlarge}{27.0}
\defR{twobatch-mix-load40/timeslice/p99/resnet50}{84.0}
\defR{twobatch-mix-load40/timeslice/p99/retinanet}{349}
\defR{twobatch-mix-load40/timeslice/p99/bertlarge}{751}
\defR{twobatch-mix-load40/timeslice/p99x/resnet50}{6.3}
\defR{twobatch-mix-load40/timeslice/p99x/retinanet}{10.0}
\defR{twobatch-mix-load40/timeslice/p99x/bertlarge}{10.4}
\defR{twobatch-mix-load40/timeslice/thrabs/train_bert}{0.43}
\defR{twobatch-mix-load40/timeslice/thrabs/train_resnet}{3.37}
\defR{twobatch-mix-load40/timeslice/thr/train_bert}{26}
\defR{twobatch-mix-load40/timeslice/thr/train_resnet}{25}
\defR{twobatch-mix-load40/mps/attmin}{0.0}
\defR{twobatch-mix-load40/mps/attmean}{0.0}
\defR{twobatch-mix-load40/mps/thr}{44}
\defR{twobatch-mix-load40/mps/n}{3}
\defR{twobatch-mix-load40/mps/att/resnet50}{0.0}
\defR{twobatch-mix-load40/mps/att/retinanet}{0.0}
\defR{twobatch-mix-load40/mps/att/bertlarge}{0.0}
\defR{twobatch-mix-load40/mps/p99/resnet50}{52451}
\defR{twobatch-mix-load40/mps/p99/retinanet}{59028}
\defR{twobatch-mix-load40/mps/p99/bertlarge}{13482}
\defR{twobatch-mix-load40/mps/p99x/resnet50}{3915.0}
\defR{twobatch-mix-load40/mps/p99x/retinanet}{1697.3}
\defR{twobatch-mix-load40/mps/p99x/bertlarge}{186.4}
\defR{twobatch-mix-load40/mps/thrabs/train_bert}{1.10}
\defR{twobatch-mix-load40/mps/thrabs/train_resnet}{2.68}
\defR{twobatch-mix-load40/mps/thr/train_bert}{68}
\defR{twobatch-mix-load40/mps/thr/train_resnet}{20}
\defR{twobatch-mix-load40/mpsprio/attmin}{0.7}
\defR{twobatch-mix-load40/mpsprio/attmean}{45.5}
\defR{twobatch-mix-load40/mpsprio/thr}{37}
\defR{twobatch-mix-load40/mpsprio/n}{3}
\defR{twobatch-mix-load40/mpsprio/att/resnet50}{0.7}
\defR{twobatch-mix-load40/mpsprio/att/retinanet}{57.0}
\defR{twobatch-mix-load40/mpsprio/att/bertlarge}{78.7}
\defR{twobatch-mix-load40/mpsprio/p99/resnet50}{118}
\defR{twobatch-mix-load40/mpsprio/p99/retinanet}{356}
\defR{twobatch-mix-load40/mpsprio/p99/bertlarge}{317}
\defR{twobatch-mix-load40/mpsprio/p99x/resnet50}{8.8}
\defR{twobatch-mix-load40/mpsprio/p99x/retinanet}{10.2}
\defR{twobatch-mix-load40/mpsprio/p99x/bertlarge}{4.4}
\defR{twobatch-mix-load40/mpsprio/thrabs/train_bert}{0.91}
\defR{twobatch-mix-load40/mpsprio/thrabs/train_resnet}{2.48}
\defR{twobatch-mix-load40/mpsprio/thr/train_bert}{56}
\defR{twobatch-mix-load40/mpsprio/thr/train_resnet}{19}
\defR{twobatch-mix-load40/mpslimits/attmin}{84.4}
\defR{twobatch-mix-load40/mpslimits/attmean}{91.4}
\defR{twobatch-mix-load40/mpslimits/thr}{16}
\defR{twobatch-mix-load40/mpslimits/n}{3}
\defR{twobatch-mix-load40/mpslimits/att/resnet50}{97.0}
\defR{twobatch-mix-load40/mpslimits/att/retinanet}{92.7}
\defR{twobatch-mix-load40/mpslimits/att/bertlarge}{84.4}
\defR{twobatch-mix-load40/mpslimits/p99/resnet50}{18.4}
\defR{twobatch-mix-load40/mpslimits/p99/retinanet}{151}
\defR{twobatch-mix-load40/mpslimits/p99/bertlarge}{272}
\defR{twobatch-mix-load40/mpslimits/p99x/resnet50}{1.4}
\defR{twobatch-mix-load40/mpslimits/p99x/retinanet}{4.3}
\defR{twobatch-mix-load40/mpslimits/p99x/bertlarge}{3.8}
\defR{twobatch-mix-load40/mpslimits/thrabs/train_bert}{0.19}
\defR{twobatch-mix-load40/mpslimits/thrabs/train_resnet}{2.64}
\defR{twobatch-mix-load40/mpslimits/thr/train_bert}{12}
\defR{twobatch-mix-load40/mpslimits/thr/train_resnet}{20}
\defR{twobatch-mix-load40/mpsparts/attmin}{86.4}
\defR{twobatch-mix-load40/mpsparts/attmean}{91.3}
\defR{twobatch-mix-load40/mpsparts/thr}{13}
\defR{twobatch-mix-load40/mpsparts/n}{2}
\defR{twobatch-mix-load40/mpsparts/att/resnet50}{98.4}
\defR{twobatch-mix-load40/mpsparts/att/retinanet}{89.1}
\defR{twobatch-mix-load40/mpsparts/att/bertlarge}{86.4}
\defR{twobatch-mix-load40/mpsparts/p99/resnet50}{16.4}
\defR{twobatch-mix-load40/mpsparts/p99/retinanet}{171}
\defR{twobatch-mix-load40/mpsparts/p99/bertlarge}{268}
\defR{twobatch-mix-load40/mpsparts/p99x/resnet50}{1.2}
\defR{twobatch-mix-load40/mpsparts/p99x/retinanet}{4.9}
\defR{twobatch-mix-load40/mpsparts/p99x/bertlarge}{3.7}
\defR{twobatch-mix-load40/mpsparts/thrabs/train_bert}{0.14}
\defR{twobatch-mix-load40/mpsparts/thrabs/train_resnet}{2.18}
\defR{twobatch-mix-load40/mpsparts/thr/train_bert}{9}
\defR{twobatch-mix-load40/mpsparts/thr/train_resnet}{16}
\defR{twobatch-mix-load40/lithos/attmin}{83.3}
\defR{twobatch-mix-load40/lithos/attmean}{89.0}
\defR{twobatch-mix-load40/lithos/thr}{11}
\defR{twobatch-mix-load40/lithos/n}{3}
\defR{twobatch-mix-load40/lithos/att/resnet50}{93.4}
\defR{twobatch-mix-load40/lithos/att/retinanet}{90.3}
\defR{twobatch-mix-load40/lithos/att/bertlarge}{83.3}
\defR{twobatch-mix-load40/lithos/p99/resnet50}{27.5}
\defR{twobatch-mix-load40/lithos/p99/retinanet}{167}
\defR{twobatch-mix-load40/lithos/p99/bertlarge}{283}
\defR{twobatch-mix-load40/lithos/p99x/resnet50}{2.1}
\defR{twobatch-mix-load40/lithos/p99x/retinanet}{4.8}
\defR{twobatch-mix-load40/lithos/p99x/bertlarge}{3.9}
\defR{twobatch-mix-load40/lithos/thrabs/train_bert}{0.19}
\defR{twobatch-mix-load40/lithos/thrabs/train_resnet}{1.34}
\defR{twobatch-mix-load40/lithos/thr/train_bert}{12}
\defR{twobatch-mix-load40/lithos/thr/train_resnet}{10}
\defR{twobatch-mix-load40/dysm/attmin}{0.0}
\defR{twobatch-mix-load40/dysm/attmean}{24.0}
\defR{twobatch-mix-load40/dysm/thr}{45}
\defR{twobatch-mix-load40/dysm/n}{3}
\defR{twobatch-mix-load40/dysm/att/resnet50}{0.0}
\defR{twobatch-mix-load40/dysm/att/retinanet}{71.9}
\defR{twobatch-mix-load40/dysm/att/bertlarge}{0.0}
\defR{twobatch-mix-load40/dysm/p99/retinanet}{444}
\defR{twobatch-mix-load40/dysm/p99x/retinanet}{12.8}
\defR{twobatch-mix-load40/dysm/thrabs/train_bert}{1.08}
\defR{twobatch-mix-load40/dysm/thrabs/train_resnet}{3.07}
\defR{twobatch-mix-load40/dysm/thr/train_bert}{66}
\defR{twobatch-mix-load40/dysm/thr/train_resnet}{23}
\defR{twobatch-mix-load40/xsched/attmin}{0.0}
\defR{twobatch-mix-load40/xsched/attmean}{0.0}
\defR{twobatch-mix-load40/xsched/thr}{0}
\defR{twobatch-mix-load40/xsched/n}{2}
\defR{twobatch-mix-load40/xsched/att/resnet50}{0.0}
\defR{twobatch-mix-load40/xsched/att/retinanet}{0.0}
\defR{twobatch-mix-load40/xsched/att/bertlarge}{0.0}
\defR{twobatch-mix-load40/xsched/thrabs/train_bert}{0.00}
\defR{twobatch-mix-load40/xsched/thrabs/train_resnet}{0.00}
\defR{twobatch-mix-load40/xsched/thr/train_bert}{0}
\defR{twobatch-mix-load40/xsched/thr/train_resnet}{0}
\defR{twobatch-mix-load40/multistream/attmin}{0.7}
\defR{twobatch-mix-load40/multistream/attmean}{45.6}
\defR{twobatch-mix-load40/multistream/thr}{37}
\defR{twobatch-mix-load40/multistream/n}{3}
\defR{twobatch-mix-load40/multistream/att/resnet50}{0.7}
\defR{twobatch-mix-load40/multistream/att/retinanet}{55.8}
\defR{twobatch-mix-load40/multistream/att/bertlarge}{80.1}
\defR{twobatch-mix-load40/multistream/p99/resnet50}{133}
\defR{twobatch-mix-load40/multistream/p99/retinanet}{369}
\defR{twobatch-mix-load40/multistream/p99/bertlarge}{306}
\defR{twobatch-mix-load40/multistream/p99x/resnet50}{9.9}
\defR{twobatch-mix-load40/multistream/p99x/retinanet}{10.6}
\defR{twobatch-mix-load40/multistream/p99x/bertlarge}{4.2}
\defR{twobatch-mix-load40/multistream/thrabs/train_bert}{0.93}
\defR{twobatch-mix-load40/multistream/thrabs/train_resnet}{2.31}
\defR{twobatch-mix-load40/multistream/thr/train_bert}{58}
\defR{twobatch-mix-load40/multistream/thr/train_resnet}{17}
\defR{twobatch-mix-load40/tgs/attmin}{0.0}
\defR{twobatch-mix-load40/tgs/attmean}{0.0}
\defR{twobatch-mix-load40/tgs/thr}{0}
\defR{twobatch-mix-load40/tgs/n}{3}
\defR{twobatch-mix-load40/tgs/att/resnet50}{0.0}
\defR{twobatch-mix-load40/tgs/att/retinanet}{0.0}
\defR{twobatch-mix-load40/tgs/att/bertlarge}{0.0}
\defR{twobatch-mix-load40/tgs/thrabs/train_bert}{0.00}
\defR{twobatch-mix-load40/tgs/thrabs/train_resnet}{0.00}
\defR{twobatch-mix-load40/tgs/thr/train_bert}{0}
\defR{twobatch-mix-load40/tgs/thr/train_resnet}{0}
\defR{twobatch-mix-load40/mig/attmin}{90.7}
\defR{twobatch-mix-load40/mig/attmean}{96.0}
\defR{twobatch-mix-load40/mig/thr}{0}
\defR{twobatch-mix-load40/mig/n}{3}
\defR{twobatch-mix-load40/mig/att/resnet50}{98.6}
\defR{twobatch-mix-load40/mig/att/retinanet}{98.8}
\defR{twobatch-mix-load40/mig/att/bertlarge}{90.7}
\defR{twobatch-mix-load40/mig/p99/resnet50}{16.0}
\defR{twobatch-mix-load40/mig/p99/retinanet}{103}
\defR{twobatch-mix-load40/mig/p99/bertlarge}{228}
\defR{twobatch-mix-load40/mig/p99x/resnet50}{1.2}
\defR{twobatch-mix-load40/mig/p99x/retinanet}{3.0}
\defR{twobatch-mix-load40/mig/p99x/bertlarge}{3.1}
\defR{twobatch-mix-load40/mig/thrabs/train_bert}{0.00}
\defR{twobatch-mix-load40/mig/thrabs/train_resnet}{0.00}
\defR{twobatch-mix-load40/mig/thr/train_bert}{0}
\defR{twobatch-mix-load40/mig/thr/train_resnet}{0}
\defR{twobatch-mix-load40/tally/attmin}{0.0}
\defR{twobatch-mix-load40/tally/attmean}{36.2}
\defR{twobatch-mix-load40/tally/thr}{9}
\defR{twobatch-mix-load40/tally/n}{3}
\defR{twobatch-mix-load40/tally/att/resnet50}{0.0}
\defR{twobatch-mix-load40/tally/att/retinanet}{14.8}
\defR{twobatch-mix-load40/tally/att/bertlarge}{93.6}
\defR{twobatch-mix-load40/tally/p99/resnet50}{93933}
\defR{twobatch-mix-load40/tally/p99/retinanet}{922}
\defR{twobatch-mix-load40/tally/p99/bertlarge}{193}
\defR{twobatch-mix-load40/tally/p99x/resnet50}{7011.2}
\defR{twobatch-mix-load40/tally/p99x/retinanet}{26.5}
\defR{twobatch-mix-load40/tally/p99x/bertlarge}{2.7}
\defR{twobatch-mix-load40/tally/thrabs/train_bert}{0.30}
\defR{twobatch-mix-load40/tally/thrabs/train_resnet}{0.09}
\defR{twobatch-mix-load40/tally/thr/train_bert}{18}
\defR{twobatch-mix-load40/tally/thr/train_resnet}{1}
\defR{twobatch-mix-load40/orion/attmin}{0.0}
\defR{twobatch-mix-load40/orion/attmean}{0.0}
\defR{twobatch-mix-load40/orion/thr}{0}
\defR{twobatch-mix-load40/orion/n}{2}
\defR{twobatch-mix-load40/orion/att/resnet50}{0.0}
\defR{twobatch-mix-load40/orion/att/retinanet}{0.0}
\defR{twobatch-mix-load40/orion/att/bertlarge}{0.0}
\defR{twobatch-mix-load40/orion/thrabs/train_bert}{0.00}
\defR{twobatch-mix-load40/orion/thrabs/train_resnet}{0.00}
\defR{twobatch-mix-load40/orion/thr/train_bert}{0}
\defR{twobatch-mix-load40/orion/thr/train_resnet}{0}
\defR{twojobs-mix-load40-gemm/berth-full/attmin}{81.1}
\defR{twojobs-mix-load40-gemm/berth-full/attmean}{88.0}
\defR{twojobs-mix-load40-gemm/berth-full/thr}{19}
\defR{twojobs-mix-load40-gemm/berth-full/n}{3}
\defR{twojobs-mix-load40-gemm/berth-full/att/resnet50}{81.1}
\defR{twojobs-mix-load40-gemm/berth-full/att/retinanet}{99.6}
\defR{twojobs-mix-load40-gemm/berth-full/att/bertlarge}{83.2}
\defR{twojobs-mix-load40-gemm/berth-full/p99/resnet50}{27.6}
\defR{twojobs-mix-load40-gemm/berth-full/p99/retinanet}{81.1}
\defR{twojobs-mix-load40-gemm/berth-full/p99/bertlarge}{295}
\defR{twojobs-mix-load40-gemm/berth-full/p99x/resnet50}{2.1}
\defR{twojobs-mix-load40-gemm/berth-full/p99x/retinanet}{2.3}
\defR{twojobs-mix-load40-gemm/berth-full/p99x/bertlarge}{4.1}
\defR{twojobs-mix-load40-gemm/berth-full/thrabs/train_bert}{0.31}
\defR{twojobs-mix-load40-gemm/berth-full/thrabs/gemm}{3.12}
\defR{twojobs-mix-load40-gemm/berth-full/thr/train_bert}{19}
\defR{twojobs-mix-load40-gemm/berth-full/thr/gemm}{20}
\defR{twojobs-mix-load40-gemm/berth-nsr/attmin}{81.6}
\defR{twojobs-mix-load40-gemm/berth-nsr/attmean}{88.4}
\defR{twojobs-mix-load40-gemm/berth-nsr/thr}{19}
\defR{twojobs-mix-load40-gemm/berth-nsr/n}{3}
\defR{twojobs-mix-load40-gemm/berth-nsr/att/resnet50}{81.6}
\defR{twojobs-mix-load40-gemm/berth-nsr/att/retinanet}{99.6}
\defR{twojobs-mix-load40-gemm/berth-nsr/att/bertlarge}{83.9}
\defR{twojobs-mix-load40-gemm/berth-nsr/p99/resnet50}{27.5}
\defR{twojobs-mix-load40-gemm/berth-nsr/p99/retinanet}{86.2}
\defR{twojobs-mix-load40-gemm/berth-nsr/p99/bertlarge}{275}
\defR{twojobs-mix-load40-gemm/berth-nsr/p99x/resnet50}{2.1}
\defR{twojobs-mix-load40-gemm/berth-nsr/p99x/retinanet}{2.5}
\defR{twojobs-mix-load40-gemm/berth-nsr/p99x/bertlarge}{3.8}
\defR{twojobs-mix-load40-gemm/berth-nsr/thrabs/train_bert}{0.31}
\defR{twojobs-mix-load40-gemm/berth-nsr/thrabs/gemm}{3.08}
\defR{twojobs-mix-load40-gemm/berth-nsr/thr/train_bert}{19}
\defR{twojobs-mix-load40-gemm/berth-nsr/thr/gemm}{19}
\defR{twojobs-mix-load40-gemm/berth-nba/attmin}{82.1}
\defR{twojobs-mix-load40-gemm/berth-nba/attmean}{88.1}
\defR{twojobs-mix-load40-gemm/berth-nba/thr}{19}
\defR{twojobs-mix-load40-gemm/berth-nba/n}{3}
\defR{twojobs-mix-load40-gemm/berth-nba/att/resnet50}{82.1}
\defR{twojobs-mix-load40-gemm/berth-nba/att/retinanet}{99.6}
\defR{twojobs-mix-load40-gemm/berth-nba/att/bertlarge}{82.6}
\defR{twojobs-mix-load40-gemm/berth-nba/p99/resnet50}{27.6}
\defR{twojobs-mix-load40-gemm/berth-nba/p99/retinanet}{86.1}
\defR{twojobs-mix-load40-gemm/berth-nba/p99/bertlarge}{280}
\defR{twojobs-mix-load40-gemm/berth-nba/p99x/resnet50}{2.1}
\defR{twojobs-mix-load40-gemm/berth-nba/p99x/retinanet}{2.5}
\defR{twojobs-mix-load40-gemm/berth-nba/p99x/bertlarge}{3.9}
\defR{twojobs-mix-load40-gemm/berth-nba/thrabs/train_bert}{0.31}
\defR{twojobs-mix-load40-gemm/berth-nba/thrabs/gemm}{3.06}
\defR{twojobs-mix-load40-gemm/berth-nba/thr/train_bert}{19}
\defR{twojobs-mix-load40-gemm/berth-nba/thr/gemm}{19}
\defR{twojobs-mix-load40-gemm/berth-nsb/attmin}{80.9}
\defR{twojobs-mix-load40-gemm/berth-nsb/attmean}{87.3}
\defR{twojobs-mix-load40-gemm/berth-nsb/thr}{19}
\defR{twojobs-mix-load40-gemm/berth-nsb/n}{3}
\defR{twojobs-mix-load40-gemm/berth-nsb/att/resnet50}{80.9}
\defR{twojobs-mix-load40-gemm/berth-nsb/att/retinanet}{99.6}
\defR{twojobs-mix-load40-gemm/berth-nsb/att/bertlarge}{81.3}
\defR{twojobs-mix-load40-gemm/berth-nsb/p99/resnet50}{27.8}
\defR{twojobs-mix-load40-gemm/berth-nsb/p99/retinanet}{88.6}
\defR{twojobs-mix-load40-gemm/berth-nsb/p99/bertlarge}{304}
\defR{twojobs-mix-load40-gemm/berth-nsb/p99x/resnet50}{2.1}
\defR{twojobs-mix-load40-gemm/berth-nsb/p99x/retinanet}{2.5}
\defR{twojobs-mix-load40-gemm/berth-nsb/p99x/bertlarge}{4.2}
\defR{twojobs-mix-load40-gemm/berth-nsb/thrabs/train_bert}{0.31}
\defR{twojobs-mix-load40-gemm/berth-nsb/thrabs/gemm}{2.97}
\defR{twojobs-mix-load40-gemm/berth-nsb/thr/train_bert}{19}
\defR{twojobs-mix-load40-gemm/berth-nsb/thr/gemm}{19}
\defR{twojobs-mix-load40-gemm/berth-ng/attmin}{80.3}
\defR{twojobs-mix-load40-gemm/berth-ng/attmean}{87.2}
\defR{twojobs-mix-load40-gemm/berth-ng/thr}{19}
\defR{twojobs-mix-load40-gemm/berth-ng/n}{3}
\defR{twojobs-mix-load40-gemm/berth-ng/att/resnet50}{80.3}
\defR{twojobs-mix-load40-gemm/berth-ng/att/retinanet}{99.6}
\defR{twojobs-mix-load40-gemm/berth-ng/att/bertlarge}{81.7}
\defR{twojobs-mix-load40-gemm/berth-ng/p99/resnet50}{28.0}
\defR{twojobs-mix-load40-gemm/berth-ng/p99/retinanet}{83.1}
\defR{twojobs-mix-load40-gemm/berth-ng/p99/bertlarge}{314}
\defR{twojobs-mix-load40-gemm/berth-ng/p99x/resnet50}{2.1}
\defR{twojobs-mix-load40-gemm/berth-ng/p99x/retinanet}{2.4}
\defR{twojobs-mix-load40-gemm/berth-ng/p99x/bertlarge}{4.3}
\defR{twojobs-mix-load40-gemm/berth-ng/thrabs/train_bert}{0.31}
\defR{twojobs-mix-load40-gemm/berth-ng/thrabs/gemm}{2.99}
\defR{twojobs-mix-load40-gemm/berth-ng/thr/train_bert}{19}
\defR{twojobs-mix-load40-gemm/berth-ng/thr/gemm}{19}
\defR{twojobs-mix-load40-gemm/berth-nsng/attmin}{79.4}
\defR{twojobs-mix-load40-gemm/berth-nsng/attmean}{86.8}
\defR{twojobs-mix-load40-gemm/berth-nsng/thr}{19}
\defR{twojobs-mix-load40-gemm/berth-nsng/n}{3}
\defR{twojobs-mix-load40-gemm/berth-nsng/att/resnet50}{79.4}
\defR{twojobs-mix-load40-gemm/berth-nsng/att/retinanet}{99.4}
\defR{twojobs-mix-load40-gemm/berth-nsng/att/bertlarge}{81.7}
\defR{twojobs-mix-load40-gemm/berth-nsng/p99/resnet50}{28.6}
\defR{twojobs-mix-load40-gemm/berth-nsng/p99/retinanet}{85.5}
\defR{twojobs-mix-load40-gemm/berth-nsng/p99/bertlarge}{304}
\defR{twojobs-mix-load40-gemm/berth-nsng/p99x/resnet50}{2.1}
\defR{twojobs-mix-load40-gemm/berth-nsng/p99x/retinanet}{2.5}
\defR{twojobs-mix-load40-gemm/berth-nsng/p99x/bertlarge}{4.2}
\defR{twojobs-mix-load40-gemm/berth-nsng/thrabs/train_bert}{0.31}
\defR{twojobs-mix-load40-gemm/berth-nsng/thrabs/gemm}{3.00}
\defR{twojobs-mix-load40-gemm/berth-nsng/thr/train_bert}{19}
\defR{twojobs-mix-load40-gemm/berth-nsng/thr/gemm}{19}
\defR{twojobs-mix-load40-gemm/berth-grouped/attmin}{28.1}
\defR{twojobs-mix-load40-gemm/berth-grouped/attmean}{67.5}
\defR{twojobs-mix-load40-gemm/berth-grouped/thr}{28}
\defR{twojobs-mix-load40-gemm/berth-grouped/n}{3}
\defR{twojobs-mix-load40-gemm/berth-grouped/att/resnet50}{28.1}
\defR{twojobs-mix-load40-gemm/berth-grouped/att/retinanet}{88.3}
\defR{twojobs-mix-load40-gemm/berth-grouped/att/bertlarge}{86.1}
\defR{twojobs-mix-load40-gemm/berth-grouped/p99/resnet50}{86.4}
\defR{twojobs-mix-load40-gemm/berth-grouped/p99/retinanet}{166}
\defR{twojobs-mix-load40-gemm/berth-grouped/p99/bertlarge}{214}
\defR{twojobs-mix-load40-gemm/berth-grouped/p99x/resnet50}{6.5}
\defR{twojobs-mix-load40-gemm/berth-grouped/p99x/retinanet}{4.8}
\defR{twojobs-mix-load40-gemm/berth-grouped/p99x/bertlarge}{3.0}
\defR{twojobs-mix-load40-gemm/berth-grouped/thrabs/train_bert}{0.37}
\defR{twojobs-mix-load40-gemm/berth-grouped/thrabs/gemm}{5.24}
\defR{twojobs-mix-load40-gemm/berth-grouped/thr/train_bert}{23}
\defR{twojobs-mix-load40-gemm/berth-grouped/thr/gemm}{33}
\defR{twojobs-mix-load40-gemm/berth/attmin}{51.8}
\defR{twojobs-mix-load40-gemm/berth/attmean}{73.7}
\defR{twojobs-mix-load40-gemm/berth/thr}{27}
\defR{twojobs-mix-load40-gemm/berth/n}{3}
\defR{twojobs-mix-load40-gemm/berth/att/resnet50}{51.8}
\defR{twojobs-mix-load40-gemm/berth/att/retinanet}{93.7}
\defR{twojobs-mix-load40-gemm/berth/att/bertlarge}{75.6}
\defR{twojobs-mix-load40-gemm/berth/p99/resnet50}{54.0}
\defR{twojobs-mix-load40-gemm/berth/p99/retinanet}{148}
\defR{twojobs-mix-load40-gemm/berth/p99/bertlarge}{302}
\defR{twojobs-mix-load40-gemm/berth/p99x/resnet50}{4.0}
\defR{twojobs-mix-load40-gemm/berth/p99x/retinanet}{4.3}
\defR{twojobs-mix-load40-gemm/berth/p99x/bertlarge}{4.2}
\defR{twojobs-mix-load40-gemm/berth/thrabs/train_bert}{0.36}
\defR{twojobs-mix-load40-gemm/berth/thrabs/gemm}{4.84}
\defR{twojobs-mix-load40-gemm/berth/thr/train_bert}{22}
\defR{twojobs-mix-load40-gemm/berth/thr/gemm}{31}
\defR{twojobs-mix-load40-gemm/timeslice/attmin}{19.1}
\defR{twojobs-mix-load40-gemm/timeslice/attmean}{35.4}
\defR{twojobs-mix-load40-gemm/timeslice/thr}{26}
\defR{twojobs-mix-load40-gemm/timeslice/n}{3}
\defR{twojobs-mix-load40-gemm/timeslice/att/resnet50}{19.1}
\defR{twojobs-mix-load40-gemm/timeslice/att/retinanet}{59.8}
\defR{twojobs-mix-load40-gemm/timeslice/att/bertlarge}{27.3}
\defR{twojobs-mix-load40-gemm/timeslice/p99/resnet50}{87.0}
\defR{twojobs-mix-load40-gemm/timeslice/p99/retinanet}{344}
\defR{twojobs-mix-load40-gemm/timeslice/p99/bertlarge}{755}
\defR{twojobs-mix-load40-gemm/timeslice/p99x/resnet50}{6.5}
\defR{twojobs-mix-load40-gemm/timeslice/p99x/retinanet}{9.9}
\defR{twojobs-mix-load40-gemm/timeslice/p99x/bertlarge}{10.4}
\defR{twojobs-mix-load40-gemm/timeslice/thrabs/train_bert}{0.43}
\defR{twojobs-mix-load40-gemm/timeslice/thrabs/gemm}{4.11}
\defR{twojobs-mix-load40-gemm/timeslice/thr/train_bert}{26}
\defR{twojobs-mix-load40-gemm/timeslice/thr/gemm}{26}
\defR{twojobs-mix-load40-gemm/mps/attmin}{0.0}
\defR{twojobs-mix-load40-gemm/mps/attmean}{0.0}
\defR{twojobs-mix-load40-gemm/mps/thr}{--}
\defR{twojobs-mix-load40-gemm/mps/n}{3}
\defR{twojobs-mix-load40-gemm/mps/att/resnet50}{0.0}
\defR{twojobs-mix-load40-gemm/mps/att/retinanet}{0.0}
\defR{twojobs-mix-load40-gemm/mps/att/bertlarge}{0.0}
\defR{twojobs-mix-load40-gemm/mps/p99/resnet50}{91611}
\defR{twojobs-mix-load40-gemm/mps/p99/retinanet}{92238}
\defR{twojobs-mix-load40-gemm/mps/p99x/resnet50}{6837.9}
\defR{twojobs-mix-load40-gemm/mps/p99x/retinanet}{2652.2}
\defR{twojobs-mix-load40-gemm/mps/thrabs/train_bert}{0.00}
\defR{twojobs-mix-load40-gemm/mps/thrabs/gemm}{15.69}
\defR{twojobs-mix-load40-gemm/mps/thr/train_bert}{--}
\defR{twojobs-mix-load40-gemm/mps/thr/gemm}{--}
\defR{twojobs-mix-load40-gemm/mpsprio/attmin}{0.0}
\defR{twojobs-mix-load40-gemm/mpsprio/attmean}{0.0}
\defR{twojobs-mix-load40-gemm/mpsprio/thr}{45}
\defR{twojobs-mix-load40-gemm/mpsprio/n}{3}
\defR{twojobs-mix-load40-gemm/mpsprio/att/resnet50}{0.0}
\defR{twojobs-mix-load40-gemm/mpsprio/att/retinanet}{0.0}
\defR{twojobs-mix-load40-gemm/mpsprio/att/bertlarge}{0.0}
\defR{twojobs-mix-load40-gemm/mpsprio/p99/resnet50}{73730}
\defR{twojobs-mix-load40-gemm/mpsprio/p99/retinanet}{78042}
\defR{twojobs-mix-load40-gemm/mpsprio/p99/bertlarge}{64711}
\defR{twojobs-mix-load40-gemm/mpsprio/p99x/resnet50}{5503.3}
\defR{twojobs-mix-load40-gemm/mpsprio/p99x/retinanet}{2244.0}
\defR{twojobs-mix-load40-gemm/mpsprio/p99x/bertlarge}{894.4}
\defR{twojobs-mix-load40-gemm/mpsprio/thrabs/train_bert}{0.00}
\defR{twojobs-mix-load40-gemm/mpsprio/thrabs/gemm}{14.28}
\defR{twojobs-mix-load40-gemm/mpsprio/thr/train_bert}{0}
\defR{twojobs-mix-load40-gemm/mpsprio/thr/gemm}{90}
\defR{twojobs-mix-load40-gemm/mpslimits/attmin}{87.9}
\defR{twojobs-mix-load40-gemm/mpslimits/attmean}{93.9}
\defR{twojobs-mix-load40-gemm/mpslimits/thr}{11}
\defR{twojobs-mix-load40-gemm/mpslimits/n}{3}
\defR{twojobs-mix-load40-gemm/mpslimits/att/resnet50}{98.6}
\defR{twojobs-mix-load40-gemm/mpslimits/att/retinanet}{95.1}
\defR{twojobs-mix-load40-gemm/mpslimits/att/bertlarge}{87.9}
\defR{twojobs-mix-load40-gemm/mpslimits/p99/resnet50}{15.5}
\defR{twojobs-mix-load40-gemm/mpslimits/p99/retinanet}{139}
\defR{twojobs-mix-load40-gemm/mpslimits/p99/bertlarge}{249}
\defR{twojobs-mix-load40-gemm/mpslimits/p99x/resnet50}{1.2}
\defR{twojobs-mix-load40-gemm/mpslimits/p99x/retinanet}{4.0}
\defR{twojobs-mix-load40-gemm/mpslimits/p99x/bertlarge}{3.4}
\defR{twojobs-mix-load40-gemm/mpslimits/thrabs/train_bert}{0.19}
\defR{twojobs-mix-load40-gemm/mpslimits/thrabs/gemm}{1.54}
\defR{twojobs-mix-load40-gemm/mpslimits/thr/train_bert}{12}
\defR{twojobs-mix-load40-gemm/mpslimits/thr/gemm}{10}
\defR{twojobs-mix-load40-gemm/mpsparts/attmin}{0.0}
\defR{twojobs-mix-load40-gemm/mpsparts/attmean}{0.0}
\defR{twojobs-mix-load40-gemm/mpsparts/thr}{0}
\defR{twojobs-mix-load40-gemm/mpsparts/n}{0}
\defR{twojobs-mix-load40-gemm/mpsparts/att/resnet50}{0.0}
\defR{twojobs-mix-load40-gemm/mpsparts/att/retinanet}{0.0}
\defR{twojobs-mix-load40-gemm/mpsparts/att/bertlarge}{0.0}
\defR{twojobs-mix-load40-gemm/mpsparts/thrabs/train_bert}{0.00}
\defR{twojobs-mix-load40-gemm/mpsparts/thrabs/gemm}{0.00}
\defR{twojobs-mix-load40-gemm/mpsparts/thr/train_bert}{0}
\defR{twojobs-mix-load40-gemm/mpsparts/thr/gemm}{0}
\defR{twojobs-mix-load40-gemm/lithos/attmin}{89.0}
\defR{twojobs-mix-load40-gemm/lithos/attmean}{94.6}
\defR{twojobs-mix-load40-gemm/lithos/thr}{6}
\defR{twojobs-mix-load40-gemm/lithos/n}{3}
\defR{twojobs-mix-load40-gemm/lithos/att/resnet50}{99.4}
\defR{twojobs-mix-load40-gemm/lithos/att/retinanet}{95.4}
\defR{twojobs-mix-load40-gemm/lithos/att/bertlarge}{89.0}
\defR{twojobs-mix-load40-gemm/lithos/p99/resnet50}{13.7}
\defR{twojobs-mix-load40-gemm/lithos/p99/retinanet}{136}
\defR{twojobs-mix-load40-gemm/lithos/p99/bertlarge}{242}
\defR{twojobs-mix-load40-gemm/lithos/p99x/resnet50}{1.0}
\defR{twojobs-mix-load40-gemm/lithos/p99x/retinanet}{3.9}
\defR{twojobs-mix-load40-gemm/lithos/p99x/bertlarge}{3.3}
\defR{twojobs-mix-load40-gemm/lithos/thrabs/train_bert}{0.00}
\defR{twojobs-mix-load40-gemm/lithos/thrabs/gemm}{1.89}
\defR{twojobs-mix-load40-gemm/lithos/thr/train_bert}{0}
\defR{twojobs-mix-load40-gemm/lithos/thr/gemm}{12}
\defR{twojobs-mix-load40-gemm/dysm/attmin}{0.0}
\defR{twojobs-mix-load40-gemm/dysm/attmean}{24.2}
\defR{twojobs-mix-load40-gemm/dysm/thr}{42}
\defR{twojobs-mix-load40-gemm/dysm/n}{3}
\defR{twojobs-mix-load40-gemm/dysm/att/resnet50}{0.0}
\defR{twojobs-mix-load40-gemm/dysm/att/retinanet}{72.6}
\defR{twojobs-mix-load40-gemm/dysm/att/bertlarge}{0.0}
\defR{twojobs-mix-load40-gemm/dysm/p99/retinanet}{422}
\defR{twojobs-mix-load40-gemm/dysm/p99x/retinanet}{12.1}
\defR{twojobs-mix-load40-gemm/dysm/thrabs/train_bert}{0.00}
\defR{twojobs-mix-load40-gemm/dysm/thrabs/gemm}{13.36}
\defR{twojobs-mix-load40-gemm/dysm/thr/train_bert}{0}
\defR{twojobs-mix-load40-gemm/dysm/thr/gemm}{85}
\defR{twojobs-mix-load40-gemm/xsched/attmin}{44.8}
\defR{twojobs-mix-load40-gemm/xsched/attmean}{72.9}
\defR{twojobs-mix-load40-gemm/xsched/thr}{26}
\defR{twojobs-mix-load40-gemm/xsched/n}{3}
\defR{twojobs-mix-load40-gemm/xsched/att/resnet50}{44.8}
\defR{twojobs-mix-load40-gemm/xsched/att/retinanet}{94.9}
\defR{twojobs-mix-load40-gemm/xsched/att/bertlarge}{79.1}
\defR{twojobs-mix-load40-gemm/xsched/p99/resnet50}{57.5}
\defR{twojobs-mix-load40-gemm/xsched/p99/retinanet}{143}
\defR{twojobs-mix-load40-gemm/xsched/p99/bertlarge}{300}
\defR{twojobs-mix-load40-gemm/xsched/p99x/resnet50}{4.3}
\defR{twojobs-mix-load40-gemm/xsched/p99x/retinanet}{4.1}
\defR{twojobs-mix-load40-gemm/xsched/p99x/bertlarge}{4.1}
\defR{twojobs-mix-load40-gemm/xsched/thrabs/train_bert}{0.12}
\defR{twojobs-mix-load40-gemm/xsched/thrabs/gemm}{6.96}
\defR{twojobs-mix-load40-gemm/xsched/thr/train_bert}{7}
\defR{twojobs-mix-load40-gemm/xsched/thr/gemm}{44}
\defR{twojobs-mix-load40-gemm/multistream/attmin}{0.0}
\defR{twojobs-mix-load40-gemm/multistream/attmean}{0.0}
\defR{twojobs-mix-load40-gemm/multistream/thr}{46}
\defR{twojobs-mix-load40-gemm/multistream/n}{3}
\defR{twojobs-mix-load40-gemm/multistream/att/resnet50}{0.0}
\defR{twojobs-mix-load40-gemm/multistream/att/retinanet}{0.0}
\defR{twojobs-mix-load40-gemm/multistream/att/bertlarge}{0.0}
\defR{twojobs-mix-load40-gemm/multistream/p99/resnet50}{78694}
\defR{twojobs-mix-load40-gemm/multistream/p99/retinanet}{83162}
\defR{twojobs-mix-load40-gemm/multistream/p99/bertlarge}{66134}
\defR{twojobs-mix-load40-gemm/multistream/p99x/resnet50}{5873.8}
\defR{twojobs-mix-load40-gemm/multistream/p99x/retinanet}{2391.3}
\defR{twojobs-mix-load40-gemm/multistream/p99x/bertlarge}{914.1}
\defR{twojobs-mix-load40-gemm/multistream/thrabs/train_bert}{0.00}
\defR{twojobs-mix-load40-gemm/multistream/thrabs/gemm}{14.46}
\defR{twojobs-mix-load40-gemm/multistream/thr/train_bert}{0}
\defR{twojobs-mix-load40-gemm/multistream/thr/gemm}{91}
\defR{twojobs-mix-load40-gemm/tgs/attmin}{0.0}
\defR{twojobs-mix-load40-gemm/tgs/attmean}{0.0}
\defR{twojobs-mix-load40-gemm/tgs/thr}{0}
\defR{twojobs-mix-load40-gemm/tgs/n}{3}
\defR{twojobs-mix-load40-gemm/tgs/att/resnet50}{0.0}
\defR{twojobs-mix-load40-gemm/tgs/att/retinanet}{0.0}
\defR{twojobs-mix-load40-gemm/tgs/att/bertlarge}{0.0}
\defR{twojobs-mix-load40-gemm/tgs/thrabs/train_bert}{0.00}
\defR{twojobs-mix-load40-gemm/tgs/thrabs/gemm}{0.00}
\defR{twojobs-mix-load40-gemm/tgs/thr/train_bert}{0}
\defR{twojobs-mix-load40-gemm/tgs/thr/gemm}{0}
\defR{twojobs-mix-load40-gemm/mig/attmin}{90.7}
\defR{twojobs-mix-load40-gemm/mig/attmean}{96.0}
\defR{twojobs-mix-load40-gemm/mig/thr}{0}
\defR{twojobs-mix-load40-gemm/mig/n}{3}
\defR{twojobs-mix-load40-gemm/mig/att/resnet50}{98.6}
\defR{twojobs-mix-load40-gemm/mig/att/retinanet}{98.8}
\defR{twojobs-mix-load40-gemm/mig/att/bertlarge}{90.7}
\defR{twojobs-mix-load40-gemm/mig/p99/resnet50}{15.9}
\defR{twojobs-mix-load40-gemm/mig/p99/retinanet}{103}
\defR{twojobs-mix-load40-gemm/mig/p99/bertlarge}{228}
\defR{twojobs-mix-load40-gemm/mig/p99x/resnet50}{1.2}
\defR{twojobs-mix-load40-gemm/mig/p99x/retinanet}{3.0}
\defR{twojobs-mix-load40-gemm/mig/p99x/bertlarge}{3.1}
\defR{twojobs-mix-load40-gemm/mig/thrabs/train_bert}{0.00}
\defR{twojobs-mix-load40-gemm/mig/thrabs/gemm}{0.00}
\defR{twojobs-mix-load40-gemm/mig/thr/train_bert}{0}
\defR{twojobs-mix-load40-gemm/mig/thr/gemm}{0}
\defR{twojobs-mix-load40-gemm/tally/attmin}{0.0}
\defR{twojobs-mix-load40-gemm/tally/attmean}{19.2}
\defR{twojobs-mix-load40-gemm/tally/thr}{42}
\defR{twojobs-mix-load40-gemm/tally/n}{3}
\defR{twojobs-mix-load40-gemm/tally/att/resnet50}{0.0}
\defR{twojobs-mix-load40-gemm/tally/att/retinanet}{0.0}
\defR{twojobs-mix-load40-gemm/tally/att/bertlarge}{57.6}
\defR{twojobs-mix-load40-gemm/tally/p99/resnet50}{93933}
\defR{twojobs-mix-load40-gemm/tally/p99/retinanet}{94229}
\defR{twojobs-mix-load40-gemm/tally/p99/bertlarge}{434}
\defR{twojobs-mix-load40-gemm/tally/p99x/resnet50}{7011.2}
\defR{twojobs-mix-load40-gemm/tally/p99x/retinanet}{2709.5}
\defR{twojobs-mix-load40-gemm/tally/p99x/bertlarge}{6.0}
\defR{twojobs-mix-load40-gemm/tally/thrabs/train_bert}{0.00}
\defR{twojobs-mix-load40-gemm/tally/thrabs/gemm}{13.20}
\defR{twojobs-mix-load40-gemm/tally/thr/train_bert}{0}
\defR{twojobs-mix-load40-gemm/tally/thr/gemm}{84}
\defR{twojobs-mix-load40-gemm/orion/attmin}{0.0}
\defR{twojobs-mix-load40-gemm/orion/attmean}{0.0}
\defR{twojobs-mix-load40-gemm/orion/thr}{0}
\defR{twojobs-mix-load40-gemm/orion/n}{3}
\defR{twojobs-mix-load40-gemm/orion/att/resnet50}{0.0}
\defR{twojobs-mix-load40-gemm/orion/att/retinanet}{0.0}
\defR{twojobs-mix-load40-gemm/orion/att/bertlarge}{0.0}
\defR{twojobs-mix-load40-gemm/orion/thrabs/train_bert}{0.00}
\defR{twojobs-mix-load40-gemm/orion/thrabs/gemm}{0.00}
\defR{twojobs-mix-load40-gemm/orion/thr/train_bert}{0}
\defR{twojobs-mix-load40-gemm/orion/thr/gemm}{0}
\defR{pair-resnet50-load30/alone/attmin}{97.3}
\defR{pair-resnet50-load30/alone/attmean}{97.3}
\defR{pair-resnet50-load30/alone/att/resnet50}{97.3}
\defR{pair-resnet50-load30/alone/p99/resnet50}{18.2}
\defR{pair-resnet50-load30/alone/thrabs/train_bert}{1.62}
\defR{pair-resnet50-load50/alone/attmin}{90.7}
\defR{pair-resnet50-load50/alone/attmean}{90.7}
\defR{pair-resnet50-load50/alone/att/resnet50}{90.7}
\defR{pair-resnet50-load50/alone/p99/resnet50}{25.8}
\defR{pair-resnet50-load50/alone/thrabs/train_bert}{1.62}
\defR{pair-resnet50-load70/alone/attmin}{73.6}
\defR{pair-resnet50-load70/alone/attmean}{73.6}
\defR{pair-resnet50-load70/alone/att/resnet50}{73.6}
\defR{pair-resnet50-load70/alone/p99/resnet50}{45.7}
\defR{pair-resnet50-load70/alone/thrabs/train_bert}{1.62}
\defR{pair-retinanet-load30/alone/attmin}{100.0}
\defR{pair-retinanet-load30/alone/attmean}{100.0}
\defR{pair-retinanet-load30/alone/att/retinanet}{100.0}
\defR{pair-retinanet-load30/alone/p99/retinanet}{31.9}
\defR{pair-retinanet-load30/alone/thrabs/train_bert}{1.62}
\defR{pair-retinanet-load50/alone/attmin}{100.0}
\defR{pair-retinanet-load50/alone/attmean}{100.0}
\defR{pair-retinanet-load50/alone/att/retinanet}{100.0}
\defR{pair-retinanet-load50/alone/p99/retinanet}{46.5}
\defR{pair-retinanet-load50/alone/thrabs/train_bert}{1.62}
\defR{pair-retinanet-load70/alone/attmin}{99.8}
\defR{pair-retinanet-load70/alone/attmean}{99.8}
\defR{pair-retinanet-load70/alone/att/retinanet}{99.8}
\defR{pair-retinanet-load70/alone/p99/retinanet}{79.5}
\defR{pair-retinanet-load70/alone/thrabs/train_bert}{1.62}
\defR{pair-bertlarge-load30/alone/attmin}{100.0}
\defR{pair-bertlarge-load30/alone/attmean}{100.0}
\defR{pair-bertlarge-load30/alone/att/bertlarge}{100.0}
\defR{pair-bertlarge-load30/alone/p99/bertlarge}{58.5}
\defR{pair-bertlarge-load30/alone/thrabs/train_bert}{1.62}
\defR{pair-bertlarge-load50/alone/attmin}{99.9}
\defR{pair-bertlarge-load50/alone/attmean}{99.9}
\defR{pair-bertlarge-load50/alone/att/bertlarge}{99.9}
\defR{pair-bertlarge-load50/alone/p99/bertlarge}{92.0}
\defR{pair-bertlarge-load50/alone/thrabs/train_bert}{1.62}
\defR{pair-bertlarge-load70/alone/attmin}{96.6}
\defR{pair-bertlarge-load70/alone/attmean}{96.6}
\defR{pair-bertlarge-load70/alone/att/bertlarge}{96.6}
\defR{pair-bertlarge-load70/alone/p99/bertlarge}{184.0}
\defR{pair-bertlarge-load70/alone/thrabs/train_bert}{1.62}
\defR{pair-perception-30fps/alone/attmin}{99.6}
\defR{pair-perception-30fps/alone/attmean}{99.6}
\defR{pair-perception-30fps/alone/att/perception}{99.6}
\defR{pair-perception-30fps/alone/p99/perception}{13.7}
\defR{pair-perception-30fps/alone/thrabs/train_bert}{1.62}
\defR{mix-load40/alone/attmin}{99.4}
\defR{mix-load40/alone/attmean}{99.8}
\defR{mix-load40/alone/att/resnet50}{99.4}
\defR{mix-load40/alone/att/retinanet}{100.0}
\defR{mix-load40/alone/att/bertlarge}{100.0}
\defR{mix-load40/alone/p99/resnet50}{13.4}
\defR{mix-load40/alone/p99/retinanet}{34.8}
\defR{mix-load40/alone/p99/bertlarge}{72.3}
\defR{mix-load40/alone/thrabs/train_bert}{1.62}
\defR{mix-load70/alone/attmin}{97.9}
\defR{mix-load70/alone/attmean}{99.3}
\defR{mix-load70/alone/att/resnet50}{97.9}
\defR{mix-load70/alone/att/retinanet}{100.0}
\defR{mix-load70/alone/att/bertlarge}{100.0}
\defR{mix-load70/alone/p99/resnet50}{17.2}
\defR{mix-load70/alone/p99/retinanet}{35.1}
\defR{mix-load70/alone/p99/bertlarge}{65.8}
\defR{mix-load70/alone/thrabs/train_bert}{1.62}
\defR{step-load40/alone/attmin}{98.4}
\defR{step-load40/alone/attmean}{99.5}
\defR{step-load40/alone/att/resnet50}{98.4}
\defR{step-load40/alone/att/retinanet}{100.0}
\defR{step-load40/alone/att/bertlarge}{100.0}
\defR{step-load40/alone/p99/resnet50}{16.3}
\defR{step-load40/alone/p99/retinanet}{34.8}
\defR{step-load40/alone/p99/bertlarge}{72.3}
\defR{step-load40/alone/thrabs/train_bert}{1.62}
\defR{tight-x0.50/alone/attmin}{99.4}
\defR{tight-x0.50/alone/attmean}{99.8}
\defR{tight-x0.50/alone/att/resnet50}{99.4}
\defR{tight-x0.50/alone/att/retinanet}{100.0}
\defR{tight-x0.50/alone/att/bertlarge}{100.0}
\defR{tight-x0.50/alone/p99/resnet50}{13.4}
\defR{tight-x0.50/alone/p99/retinanet}{34.8}
\defR{tight-x0.50/alone/p99/bertlarge}{72.3}
\defR{tight-x0.50/alone/thrabs/train_bert}{1.62}
\defR{tight-x0.75/alone/attmin}{99.4}
\defR{tight-x0.75/alone/attmean}{99.8}
\defR{tight-x0.75/alone/att/resnet50}{99.4}
\defR{tight-x0.75/alone/att/retinanet}{100.0}
\defR{tight-x0.75/alone/att/bertlarge}{100.0}
\defR{tight-x0.75/alone/p99/resnet50}{13.4}
\defR{tight-x0.75/alone/p99/retinanet}{34.8}
\defR{tight-x0.75/alone/p99/bertlarge}{72.3}
\defR{tight-x0.75/alone/thrabs/train_bert}{1.62}
\defR{tight-x1.50/alone/attmin}{99.4}
\defR{tight-x1.50/alone/attmean}{99.8}
\defR{tight-x1.50/alone/att/resnet50}{99.4}
\defR{tight-x1.50/alone/att/retinanet}{100.0}
\defR{tight-x1.50/alone/att/bertlarge}{100.0}
\defR{tight-x1.50/alone/p99/resnet50}{13.4}
\defR{tight-x1.50/alone/p99/retinanet}{34.8}
\defR{tight-x1.50/alone/p99/bertlarge}{72.3}
\defR{tight-x1.50/alone/thrabs/train_bert}{1.62}
\defR{tight-x2.00/alone/attmin}{99.4}
\defR{tight-x2.00/alone/attmean}{99.8}
\defR{tight-x2.00/alone/att/resnet50}{99.4}
\defR{tight-x2.00/alone/att/retinanet}{100.0}
\defR{tight-x2.00/alone/att/bertlarge}{100.0}
\defR{tight-x2.00/alone/p99/resnet50}{13.4}
\defR{tight-x2.00/alone/p99/retinanet}{34.8}
\defR{tight-x2.00/alone/p99/bertlarge}{72.3}
\defR{tight-x2.00/alone/thrabs/train_bert}{1.62}
\defR{tight-x3.00/alone/attmin}{99.4}
\defR{tight-x3.00/alone/attmean}{99.8}
\defR{tight-x3.00/alone/att/resnet50}{99.4}
\defR{tight-x3.00/alone/att/retinanet}{100.0}
\defR{tight-x3.00/alone/att/bertlarge}{100.0}
\defR{tight-x3.00/alone/p99/resnet50}{13.4}
\defR{tight-x3.00/alone/p99/retinanet}{34.8}
\defR{tight-x3.00/alone/p99/bertlarge}{72.3}
\defR{tight-x3.00/alone/thrabs/train_bert}{1.62}
\defR{pair-resnet50-load30-trainresnet/alone/attmin}{97.3}
\defR{pair-resnet50-load30-trainresnet/alone/attmean}{97.3}
\defR{pair-resnet50-load30-trainresnet/alone/att/resnet50}{97.3}
\defR{pair-resnet50-load30-trainresnet/alone/p99/resnet50}{18.2}
\defR{pair-resnet50-load30-trainresnet/alone/thrabs/train_resnet}{13.29}
\defR{pair-resnet50-load30-gemm/alone/attmin}{97.3}
\defR{pair-resnet50-load30-gemm/alone/attmean}{97.3}
\defR{pair-resnet50-load30-gemm/alone/att/resnet50}{97.3}
\defR{pair-resnet50-load30-gemm/alone/p99/resnet50}{18.2}
\defR{pair-resnet50-load30-gemm/alone/thrabs/gemm}{15.81}
\defR{pair-bertlarge-load30-trainresnet/alone/attmin}{100.0}
\defR{pair-bertlarge-load30-trainresnet/alone/attmean}{100.0}
\defR{pair-bertlarge-load30-trainresnet/alone/att/bertlarge}{100.0}
\defR{pair-bertlarge-load30-trainresnet/alone/p99/bertlarge}{58.5}
\defR{pair-bertlarge-load30-trainresnet/alone/thrabs/train_resnet}{13.29}
\defR{pair-bertlarge-load30-gemm/alone/attmin}{100.0}
\defR{pair-bertlarge-load30-gemm/alone/attmean}{100.0}
\defR{pair-bertlarge-load30-gemm/alone/att/bertlarge}{100.0}
\defR{pair-bertlarge-load30-gemm/alone/p99/bertlarge}{58.5}
\defR{pair-bertlarge-load30-gemm/alone/thrabs/gemm}{15.81}
\defR{mix-load40-trainresnet/alone/attmin}{99.4}
\defR{mix-load40-trainresnet/alone/attmean}{99.8}
\defR{mix-load40-trainresnet/alone/att/resnet50}{99.4}
\defR{mix-load40-trainresnet/alone/att/retinanet}{100.0}
\defR{mix-load40-trainresnet/alone/att/bertlarge}{100.0}
\defR{mix-load40-trainresnet/alone/p99/resnet50}{13.4}
\defR{mix-load40-trainresnet/alone/p99/retinanet}{34.8}
\defR{mix-load40-trainresnet/alone/p99/bertlarge}{72.3}
\defR{mix-load40-trainresnet/alone/thrabs/train_resnet}{13.29}
\defR{mix-load40-gemm/alone/attmin}{99.4}
\defR{mix-load40-gemm/alone/attmean}{99.8}
\defR{mix-load40-gemm/alone/att/resnet50}{99.4}
\defR{mix-load40-gemm/alone/att/retinanet}{100.0}
\defR{mix-load40-gemm/alone/att/bertlarge}{100.0}
\defR{mix-load40-gemm/alone/p99/resnet50}{13.4}
\defR{mix-load40-gemm/alone/p99/retinanet}{34.8}
\defR{mix-load40-gemm/alone/p99/bertlarge}{72.3}
\defR{mix-load40-gemm/alone/thrabs/gemm}{15.81}
\defR{bursty-resnet50-load30/alone/attmin}{87.8}
\defR{bursty-resnet50-load30/alone/attmean}{87.8}
\defR{bursty-resnet50-load30/alone/att/resnet50}{87.8}
\defR{bursty-resnet50-load30/alone/p99/resnet50}{66.7}
\defR{bursty-resnet50-load30/alone/thrabs/train_bert}{1.62}
\defR{bursty-mix-load40/alone/attmin}{95.0}
\defR{bursty-mix-load40/alone/attmean}{98.3}
\defR{bursty-mix-load40/alone/att/resnet50}{95.0}
\defR{bursty-mix-load40/alone/att/retinanet}{100.0}
\defR{bursty-mix-load40/alone/att/bertlarge}{99.8}
\defR{bursty-mix-load40/alone/p99/resnet50}{21.8}
\defR{bursty-mix-load40/alone/p99/retinanet}{48.3}
\defR{bursty-mix-load40/alone/p99/bertlarge}{101.3}
\defR{bursty-mix-load40/alone/thrabs/train_bert}{1.62}
\defR{churn-load40/alone/attmin}{99.4}
\defR{churn-load40/alone/attmean}{99.8}
\defR{churn-load40/alone/att/resnet50}{99.4}
\defR{churn-load40/alone/att/bertlarge}{100.0}
\defR{churn-load40/alone/att/retinanet}{100.0}
\defR{churn-load40/alone/att/perception}{99.6}
\defR{churn-load40/alone/p99/resnet50}{13.4}
\defR{churn-load40/alone/p99/bertlarge}{72.3}
\defR{churn-load40/alone/p99/retinanet}{34.8}
\defR{churn-load40/alone/p99/perception}{13.7}
\defR{churn-load40/alone/thrabs/train_bert}{1.62}
\defR{scale-5/alone/attmin}{99.6}
\defR{scale-5/alone/attmean}{99.8}
\defR{scale-5/alone/att/resnet50}{99.6}
\defR{scale-5/alone/att/retinanet}{100.0}
\defR{scale-5/alone/att/bertlarge}{100.0}
\defR{scale-5/alone/att/perception}{99.6}
\defR{scale-5/alone/att/resnet50_b}{99.6}
\defR{scale-5/alone/p99/resnet50}{12.6}
\defR{scale-5/alone/p99/retinanet}{32.5}
\defR{scale-5/alone/p99/bertlarge}{70.5}
\defR{scale-5/alone/p99/perception}{13.7}
\defR{scale-5/alone/p99/resnet50_b}{12.6}
\defR{scale-5/alone/thrabs/train_bert}{1.62}
\defR{scale-10/alone/attmin}{99.6}
\defR{scale-10/alone/attmean}{100.0}
\defR{scale-10/alone/att/resnet50}{100.0}
\defR{scale-10/alone/att/retinanet}{100.0}
\defR{scale-10/alone/att/bertlarge}{100.0}
\defR{scale-10/alone/att/perception}{99.6}
\defR{scale-10/alone/att/resnet50_b}{100.0}
\defR{scale-10/alone/att/retinanet_b}{100.0}
\defR{scale-10/alone/att/bertlarge_b}{100.0}
\defR{scale-10/alone/att/resnet50_c}{100.0}
\defR{scale-10/alone/att/bertlarge_c}{100.0}
\defR{scale-10/alone/att/retinanet_c}{100.0}
\defR{scale-10/alone/p99/resnet50}{12.2}
\defR{scale-10/alone/p99/retinanet}{32.1}
\defR{scale-10/alone/p99/bertlarge}{71.0}
\defR{scale-10/alone/p99/perception}{13.7}
\defR{scale-10/alone/p99/resnet50_b}{12.2}
\defR{scale-10/alone/p99/retinanet_b}{32.1}
\defR{scale-10/alone/p99/bertlarge_b}{71.0}
\defR{scale-10/alone/p99/resnet50_c}{12.2}
\defR{scale-10/alone/p99/bertlarge_c}{71.0}
\defR{scale-10/alone/p99/retinanet_c}{32.1}
\defR{scale-10/alone/thrabs/train_bert}{1.62}
\defR{twobatch-mix-load40/alone/attmin}{99.4}
\defR{twobatch-mix-load40/alone/attmean}{99.8}
\defR{twobatch-mix-load40/alone/att/resnet50}{99.4}
\defR{twobatch-mix-load40/alone/att/retinanet}{100.0}
\defR{twobatch-mix-load40/alone/att/bertlarge}{100.0}
\defR{twobatch-mix-load40/alone/p99/resnet50}{13.4}
\defR{twobatch-mix-load40/alone/p99/retinanet}{34.8}
\defR{twobatch-mix-load40/alone/p99/bertlarge}{72.3}
\defR{twobatch-mix-load40/alone/thrabs/train_bert}{1.62}
\defR{twobatch-mix-load40/alone/thrabs/train_resnet}{13.29}
\defR{twojobs-mix-load40-gemm/alone/attmin}{99.4}
\defR{twojobs-mix-load40-gemm/alone/attmean}{99.8}
\defR{twojobs-mix-load40-gemm/alone/att/resnet50}{99.4}
\defR{twojobs-mix-load40-gemm/alone/att/retinanet}{100.0}
\defR{twojobs-mix-load40-gemm/alone/att/bertlarge}{100.0}
\defR{twojobs-mix-load40-gemm/alone/p99/resnet50}{13.4}
\defR{twojobs-mix-load40-gemm/alone/p99/retinanet}{34.8}
\defR{twojobs-mix-load40-gemm/alone/p99/bertlarge}{72.3}
\defR{twojobs-mix-load40-gemm/alone/thrabs/train_bert}{1.62}
\defR{twojobs-mix-load40-gemm/alone/thrabs/gemm}{15.81}
\defR{agg/all/berth-full/attmean}{95.8}
\defR{agg/all/berth-full/attmin}{94.1}
\defR{agg/all/berth-full/thr}{48}
\defR{agg/all/berth-full/close}{24}
\defR{agg/all/berth-full/ns}{0}
\defR{agg/all/berth-full/zero}{0}
\defR{agg/short/berth-full/attmean}{95.9}
\defR{agg/short/berth-full/attmin}{94.3}
\defR{agg/short/berth-full/thr}{48}
\defR{agg/short/berth-full/close}{21}
\defR{agg/short/berth-full/ns}{0}
\defR{agg/short/berth-full/zero}{0}
\defR{agg/gemm/berth-full/attmean}{95.2}
\defR{agg/gemm/berth-full/attmin}{93.0}
\defR{agg/gemm/berth-full/thr}{52}
\defR{agg/gemm/berth-full/close}{3}
\defR{agg/gemm/berth-full/ns}{0}
\defR{agg/gemm/berth-full/zero}{0}
\defR{agg/single/berth-full/attmean}{94.8}
\defR{agg/single/berth-full/attmin}{94.8}
\defR{agg/single/berth-full/thr}{54}
\defR{agg/single/berth-full/close}{14}
\defR{agg/single/berth-full/ns}{0}
\defR{agg/single/berth-full/zero}{0}
\defR{agg/multi/berth-full/attmean}{96.8}
\defR{agg/multi/berth-full/attmin}{93.5}
\defR{agg/multi/berth-full/thr}{44}
\defR{agg/multi/berth-full/close}{10}
\defR{agg/multi/berth-full/ns}{0}
\defR{agg/multi/berth-full/zero}{0}
\defR{agg/xs/berth-full/attmean}{95.5}
\defR{agg/xs/berth-full/attmin}{93.8}
\defR{agg/xs/berth-full/thr}{48}
\defR{agg/xs/berth-full/close}{21}
\defR{agg/xs/berth-full/ns}{0}
\defR{agg/xs/berth-full/zero}{0}
\defR{agg/all/berth-nsr/attmean}{95.8}
\defR{agg/all/berth-nsr/attmin}{94.1}
\defR{agg/all/berth-nsr/thr}{48}
\defR{agg/all/berth-nsr/close}{23}
\defR{agg/all/berth-nsr/ns}{0}
\defR{agg/all/berth-nsr/zero}{0}
\defR{agg/short/berth-nsr/attmean}{95.9}
\defR{agg/short/berth-nsr/attmin}{94.2}
\defR{agg/short/berth-nsr/thr}{48}
\defR{agg/short/berth-nsr/close}{20}
\defR{agg/short/berth-nsr/ns}{0}
\defR{agg/short/berth-nsr/zero}{0}
\defR{agg/gemm/berth-nsr/attmean}{95.4}
\defR{agg/gemm/berth-nsr/attmin}{93.3}
\defR{agg/gemm/berth-nsr/thr}{52}
\defR{agg/gemm/berth-nsr/close}{3}
\defR{agg/gemm/berth-nsr/ns}{0}
\defR{agg/gemm/berth-nsr/zero}{0}
\defR{agg/single/berth-nsr/attmean}{94.8}
\defR{agg/single/berth-nsr/attmin}{94.8}
\defR{agg/single/berth-nsr/thr}{53}
\defR{agg/single/berth-nsr/close}{14}
\defR{agg/single/berth-nsr/ns}{0}
\defR{agg/single/berth-nsr/zero}{0}
\defR{agg/multi/berth-nsr/attmean}{96.7}
\defR{agg/multi/berth-nsr/attmin}{93.4}
\defR{agg/multi/berth-nsr/thr}{43}
\defR{agg/multi/berth-nsr/close}{9}
\defR{agg/multi/berth-nsr/ns}{0}
\defR{agg/multi/berth-nsr/zero}{0}
\defR{agg/xs/berth-nsr/attmean}{95.5}
\defR{agg/xs/berth-nsr/attmin}{93.7}
\defR{agg/xs/berth-nsr/thr}{48}
\defR{agg/xs/berth-nsr/close}{20}
\defR{agg/xs/berth-nsr/ns}{0}
\defR{agg/xs/berth-nsr/zero}{0}
\defR{agg/all/berth-nba/attmean}{95.9}
\defR{agg/all/berth-nba/attmin}{94.3}
\defR{agg/all/berth-nba/thr}{49}
\defR{agg/all/berth-nba/close}{23}
\defR{agg/all/berth-nba/ns}{0}
\defR{agg/all/berth-nba/zero}{0}
\defR{agg/short/berth-nba/attmean}{96.0}
\defR{agg/short/berth-nba/attmin}{94.4}
\defR{agg/short/berth-nba/thr}{48}
\defR{agg/short/berth-nba/close}{20}
\defR{agg/short/berth-nba/ns}{0}
\defR{agg/short/berth-nba/zero}{0}
\defR{agg/gemm/berth-nba/attmean}{95.2}
\defR{agg/gemm/berth-nba/attmin}{93.4}
\defR{agg/gemm/berth-nba/thr}{52}
\defR{agg/gemm/berth-nba/close}{3}
\defR{agg/gemm/berth-nba/ns}{0}
\defR{agg/gemm/berth-nba/zero}{0}
\defR{agg/single/berth-nba/attmean}{94.9}
\defR{agg/single/berth-nba/attmin}{94.9}
\defR{agg/single/berth-nba/thr}{54}
\defR{agg/single/berth-nba/close}{14}
\defR{agg/single/berth-nba/ns}{0}
\defR{agg/single/berth-nba/zero}{0}
\defR{agg/multi/berth-nba/attmean}{96.8}
\defR{agg/multi/berth-nba/attmin}{93.7}
\defR{agg/multi/berth-nba/thr}{44}
\defR{agg/multi/berth-nba/close}{9}
\defR{agg/multi/berth-nba/ns}{0}
\defR{agg/multi/berth-nba/zero}{0}
\defR{agg/xs/berth-nba/attmean}{95.6}
\defR{agg/xs/berth-nba/attmin}{93.9}
\defR{agg/xs/berth-nba/thr}{48}
\defR{agg/xs/berth-nba/close}{20}
\defR{agg/xs/berth-nba/ns}{0}
\defR{agg/xs/berth-nba/zero}{0}
\defR{agg/all/berth-nsb/attmean}{95.9}
\defR{agg/all/berth-nsb/attmin}{94.2}
\defR{agg/all/berth-nsb/thr}{49}
\defR{agg/all/berth-nsb/close}{25}
\defR{agg/all/berth-nsb/ns}{0}
\defR{agg/all/berth-nsb/zero}{0}
\defR{agg/short/berth-nsb/attmean}{96.0}
\defR{agg/short/berth-nsb/attmin}{94.3}
\defR{agg/short/berth-nsb/thr}{48}
\defR{agg/short/berth-nsb/close}{22}
\defR{agg/short/berth-nsb/ns}{0}
\defR{agg/short/berth-nsb/zero}{0}
\defR{agg/gemm/berth-nsb/attmean}{95.2}
\defR{agg/gemm/berth-nsb/attmin}{93.2}
\defR{agg/gemm/berth-nsb/thr}{52}
\defR{agg/gemm/berth-nsb/close}{3}
\defR{agg/gemm/berth-nsb/ns}{0}
\defR{agg/gemm/berth-nsb/zero}{0}
\defR{agg/single/berth-nsb/attmean}{95.2}
\defR{agg/single/berth-nsb/attmin}{95.2}
\defR{agg/single/berth-nsb/thr}{54}
\defR{agg/single/berth-nsb/close}{15}
\defR{agg/single/berth-nsb/ns}{0}
\defR{agg/single/berth-nsb/zero}{0}
\defR{agg/multi/berth-nsb/attmean}{96.5}
\defR{agg/multi/berth-nsb/attmin}{93.2}
\defR{agg/multi/berth-nsb/thr}{44}
\defR{agg/multi/berth-nsb/close}{10}
\defR{agg/multi/berth-nsb/ns}{0}
\defR{agg/multi/berth-nsb/zero}{0}
\defR{agg/xs/berth-nsb/attmean}{95.7}
\defR{agg/xs/berth-nsb/attmin}{94.0}
\defR{agg/xs/berth-nsb/thr}{48}
\defR{agg/xs/berth-nsb/close}{22}
\defR{agg/xs/berth-nsb/ns}{0}
\defR{agg/xs/berth-nsb/zero}{0}
\defR{agg/all/berth-ng/attmean}{94.0}
\defR{agg/all/berth-ng/attmin}{91.9}
\defR{agg/all/berth-ng/thr}{49}
\defR{agg/all/berth-ng/close}{21}
\defR{agg/all/berth-ng/ns}{0}
\defR{agg/all/berth-ng/zero}{0}
\defR{agg/short/berth-ng/attmean}{93.9}
\defR{agg/short/berth-ng/attmin}{91.8}
\defR{agg/short/berth-ng/thr}{48}
\defR{agg/short/berth-ng/close}{18}
\defR{agg/short/berth-ng/ns}{0}
\defR{agg/short/berth-ng/zero}{0}
\defR{agg/gemm/berth-ng/attmean}{94.7}
\defR{agg/gemm/berth-ng/attmin}{92.5}
\defR{agg/gemm/berth-ng/thr}{52}
\defR{agg/gemm/berth-ng/close}{3}
\defR{agg/gemm/berth-ng/ns}{0}
\defR{agg/gemm/berth-ng/zero}{0}
\defR{agg/single/berth-ng/attmean}{92.1}
\defR{agg/single/berth-ng/attmin}{92.1}
\defR{agg/single/berth-ng/thr}{54}
\defR{agg/single/berth-ng/close}{12}
\defR{agg/single/berth-ng/ns}{0}
\defR{agg/single/berth-ng/zero}{0}
\defR{agg/multi/berth-ng/attmean}{95.7}
\defR{agg/multi/berth-ng/attmin}{91.7}
\defR{agg/multi/berth-ng/thr}{44}
\defR{agg/multi/berth-ng/close}{9}
\defR{agg/multi/berth-ng/ns}{0}
\defR{agg/multi/berth-ng/zero}{0}
\defR{agg/xs/berth-ng/attmean}{93.6}
\defR{agg/xs/berth-ng/attmin}{91.4}
\defR{agg/xs/berth-ng/thr}{48}
\defR{agg/xs/berth-ng/close}{18}
\defR{agg/xs/berth-ng/ns}{0}
\defR{agg/xs/berth-ng/zero}{0}
\defR{agg/all/berth-nsng/attmean}{94.1}
\defR{agg/all/berth-nsng/attmin}{92.0}
\defR{agg/all/berth-nsng/thr}{49}
\defR{agg/all/berth-nsng/close}{21}
\defR{agg/all/berth-nsng/ns}{0}
\defR{agg/all/berth-nsng/zero}{0}
\defR{agg/short/berth-nsng/attmean}{94.0}
\defR{agg/short/berth-nsng/attmin}{91.9}
\defR{agg/short/berth-nsng/thr}{48}
\defR{agg/short/berth-nsng/close}{18}
\defR{agg/short/berth-nsng/ns}{0}
\defR{agg/short/berth-nsng/zero}{0}
\defR{agg/gemm/berth-nsng/attmean}{94.8}
\defR{agg/gemm/berth-nsng/attmin}{92.6}
\defR{agg/gemm/berth-nsng/thr}{52}
\defR{agg/gemm/berth-nsng/close}{3}
\defR{agg/gemm/berth-nsng/ns}{0}
\defR{agg/gemm/berth-nsng/zero}{0}
\defR{agg/single/berth-nsng/attmean}{92.3}
\defR{agg/single/berth-nsng/attmin}{92.3}
\defR{agg/single/berth-nsng/thr}{54}
\defR{agg/single/berth-nsng/close}{12}
\defR{agg/single/berth-nsng/ns}{0}
\defR{agg/single/berth-nsng/zero}{0}
\defR{agg/multi/berth-nsng/attmean}{95.8}
\defR{agg/multi/berth-nsng/attmin}{91.7}
\defR{agg/multi/berth-nsng/thr}{44}
\defR{agg/multi/berth-nsng/close}{9}
\defR{agg/multi/berth-nsng/ns}{0}
\defR{agg/multi/berth-nsng/zero}{0}
\defR{agg/xs/berth-nsng/attmean}{93.7}
\defR{agg/xs/berth-nsng/attmin}{91.5}
\defR{agg/xs/berth-nsng/thr}{48}
\defR{agg/xs/berth-nsng/close}{18}
\defR{agg/xs/berth-nsng/ns}{0}
\defR{agg/xs/berth-nsng/zero}{0}
\defR{agg/all/berth-grouped/attmean}{90.1}
\defR{agg/all/berth-grouped/attmin}{82.9}
\defR{agg/all/berth-grouped/thr}{52}
\defR{agg/all/berth-grouped/close}{15}
\defR{agg/all/berth-grouped/ns}{0}
\defR{agg/all/berth-grouped/zero}{0}
\defR{agg/short/berth-grouped/attmean}{92.8}
\defR{agg/short/berth-grouped/attmin}{87.0}
\defR{agg/short/berth-grouped/thr}{52}
\defR{agg/short/berth-grouped/close}{15}
\defR{agg/short/berth-grouped/ns}{0}
\defR{agg/short/berth-grouped/zero}{0}
\defR{agg/gemm/berth-grouped/attmean}{72.4}
\defR{agg/gemm/berth-grouped/attmin}{54.7}
\defR{agg/gemm/berth-grouped/thr}{56}
\defR{agg/gemm/berth-grouped/close}{0}
\defR{agg/gemm/berth-grouped/ns}{0}
\defR{agg/gemm/berth-grouped/zero}{0}
\defR{agg/single/berth-grouped/attmean}{88.8}
\defR{agg/single/berth-grouped/attmin}{88.8}
\defR{agg/single/berth-grouped/thr}{56}
\defR{agg/single/berth-grouped/close}{10}
\defR{agg/single/berth-grouped/ns}{0}
\defR{agg/single/berth-grouped/zero}{0}
\defR{agg/multi/berth-grouped/attmean}{91.4}
\defR{agg/multi/berth-grouped/attmin}{77.3}
\defR{agg/multi/berth-grouped/thr}{49}
\defR{agg/multi/berth-grouped/close}{5}
\defR{agg/multi/berth-grouped/ns}{0}
\defR{agg/multi/berth-grouped/zero}{0}
\defR{agg/xs/berth-grouped/attmean}{89.5}
\defR{agg/xs/berth-grouped/attmin}{82.5}
\defR{agg/xs/berth-grouped/thr}{52}
\defR{agg/xs/berth-grouped/close}{12}
\defR{agg/xs/berth-grouped/ns}{0}
\defR{agg/xs/berth-grouped/zero}{0}
\defR{agg/all/berth/attmean}{89.7}
\defR{agg/all/berth/attmin}{84.1}
\defR{agg/all/berth/thr}{50}
\defR{agg/all/berth/close}{14}
\defR{agg/all/berth/ns}{0}
\defR{agg/all/berth/zero}{0}
\defR{agg/short/berth/attmean}{91.8}
\defR{agg/short/berth/attmin}{86.7}
\defR{agg/short/berth/thr}{49}
\defR{agg/short/berth/close}{14}
\defR{agg/short/berth/ns}{0}
\defR{agg/short/berth/zero}{0}
\defR{agg/gemm/berth/attmean}{76.0}
\defR{agg/gemm/berth/attmin}{66.6}
\defR{agg/gemm/berth/thr}{55}
\defR{agg/gemm/berth/close}{0}
\defR{agg/gemm/berth/ns}{0}
\defR{agg/gemm/berth/zero}{0}
\defR{agg/single/berth/attmean}{88.9}
\defR{agg/single/berth/attmin}{88.9}
\defR{agg/single/berth/thr}{56}
\defR{agg/single/berth/close}{10}
\defR{agg/single/berth/ns}{0}
\defR{agg/single/berth/zero}{0}
\defR{agg/multi/berth/attmean}{90.6}
\defR{agg/multi/berth/attmin}{79.7}
\defR{agg/multi/berth/thr}{45}
\defR{agg/multi/berth/close}{4}
\defR{agg/multi/berth/ns}{0}
\defR{agg/multi/berth/zero}{0}
\defR{agg/xs/berth/attmean}{89.0}
\defR{agg/xs/berth/attmin}{83.5}
\defR{agg/xs/berth/thr}{50}
\defR{agg/xs/berth/close}{11}
\defR{agg/xs/berth/ns}{0}
\defR{agg/xs/berth/zero}{0}
\defR{agg/all/timeslice/attmean}{48.9}
\defR{agg/all/timeslice/attmin}{38.4}
\defR{agg/all/timeslice/thr}{51}
\defR{agg/all/timeslice/close}{2}
\defR{agg/all/timeslice/ns}{0}
\defR{agg/all/timeslice/zero}{9}
\defR{agg/short/timeslice/attmean}{47.5}
\defR{agg/short/timeslice/attmin}{36.7}
\defR{agg/short/timeslice/thr}{50}
\defR{agg/short/timeslice/close}{2}
\defR{agg/short/timeslice/ns}{0}
\defR{agg/short/timeslice/zero}{9}
\defR{agg/gemm/timeslice/attmean}{58.4}
\defR{agg/gemm/timeslice/attmin}{50.2}
\defR{agg/gemm/timeslice/thr}{54}
\defR{agg/gemm/timeslice/close}{0}
\defR{agg/gemm/timeslice/ns}{0}
\defR{agg/gemm/timeslice/zero}{0}
\defR{agg/single/timeslice/attmean}{39.0}
\defR{agg/single/timeslice/attmin}{39.0}
\defR{agg/single/timeslice/thr}{58}
\defR{agg/single/timeslice/close}{1}
\defR{agg/single/timeslice/ns}{0}
\defR{agg/single/timeslice/zero}{5}
\defR{agg/multi/timeslice/attmean}{58.2}
\defR{agg/multi/timeslice/attmin}{37.9}
\defR{agg/multi/timeslice/thr}{44}
\defR{agg/multi/timeslice/close}{1}
\defR{agg/multi/timeslice/ns}{0}
\defR{agg/multi/timeslice/zero}{4}
\defR{agg/xs/timeslice/attmean}{47.5}
\defR{agg/xs/timeslice/attmin}{36.4}
\defR{agg/xs/timeslice/thr}{51}
\defR{agg/xs/timeslice/close}{2}
\defR{agg/xs/timeslice/ns}{0}
\defR{agg/xs/timeslice/zero}{9}
\defR{agg/all/mps/attmean}{3.8}
\defR{agg/all/mps/attmin}{2.3}
\defR{agg/all/mps/thr}{84}
\defR{agg/all/mps/close}{0}
\defR{agg/all/mps/ns}{0}
\defR{agg/all/mps/zero}{28}
\defR{agg/short/mps/attmean}{4.4}
\defR{agg/short/mps/attmin}{2.6}
\defR{agg/short/mps/thr}{82}
\defR{agg/short/mps/close}{0}
\defR{agg/short/mps/ns}{0}
\defR{agg/short/mps/zero}{24}
\defR{agg/gemm/mps/attmean}{0.0}
\defR{agg/gemm/mps/attmin}{0.0}
\defR{agg/gemm/mps/thr}{100}
\defR{agg/gemm/mps/close}{0}
\defR{agg/gemm/mps/ns}{0}
\defR{agg/gemm/mps/zero}{4}
\defR{agg/single/mps/attmean}{4.1}
\defR{agg/single/mps/attmin}{4.1}
\defR{agg/single/mps/thr}{91}
\defR{agg/single/mps/close}{0}
\defR{agg/single/mps/ns}{0}
\defR{agg/single/mps/zero}{13}
\defR{agg/multi/mps/attmean}{3.6}
\defR{agg/multi/mps/attmin}{0.6}
\defR{agg/multi/mps/thr}{77}
\defR{agg/multi/mps/close}{0}
\defR{agg/multi/mps/ns}{0}
\defR{agg/multi/mps/zero}{15}
\defR{agg/xs/mps/attmean}{0.1}
\defR{agg/xs/mps/attmin}{0.0}
\defR{agg/xs/mps/thr}{86}
\defR{agg/xs/mps/close}{0}
\defR{agg/xs/mps/ns}{0}
\defR{agg/xs/mps/zero}{27}
\defR{agg/all/mpsprio/attmean}{24.3}
\defR{agg/all/mpsprio/attmin}{11.8}
\defR{agg/all/mpsprio/thr}{72}
\defR{agg/all/mpsprio/close}{0}
\defR{agg/all/mpsprio/ns}{0}
\defR{agg/all/mpsprio/zero}{24}
\defR{agg/short/mpsprio/attmean}{27.9}
\defR{agg/short/mpsprio/attmin}{13.6}
\defR{agg/short/mpsprio/thr}{70}
\defR{agg/short/mpsprio/close}{0}
\defR{agg/short/mpsprio/ns}{0}
\defR{agg/short/mpsprio/zero}{20}
\defR{agg/gemm/mpsprio/attmean}{0.0}
\defR{agg/gemm/mpsprio/attmin}{0.0}
\defR{agg/gemm/mpsprio/thr}{83}
\defR{agg/gemm/mpsprio/close}{0}
\defR{agg/gemm/mpsprio/ns}{0}
\defR{agg/gemm/mpsprio/zero}{4}
\defR{agg/single/mpsprio/attmean}{10.0}
\defR{agg/single/mpsprio/attmin}{10.0}
\defR{agg/single/mpsprio/thr}{82}
\defR{agg/single/mpsprio/close}{0}
\defR{agg/single/mpsprio/ns}{0}
\defR{agg/single/mpsprio/zero}{12}
\defR{agg/multi/mpsprio/attmean}{37.7}
\defR{agg/multi/mpsprio/attmin}{13.5}
\defR{agg/multi/mpsprio/thr}{62}
\defR{agg/multi/mpsprio/close}{0}
\defR{agg/multi/mpsprio/ns}{0}
\defR{agg/multi/mpsprio/zero}{12}
\defR{agg/xs/mpsprio/attmean}{17.5}
\defR{agg/xs/mpsprio/attmin}{5.5}
\defR{agg/xs/mpsprio/thr}{72}
\defR{agg/xs/mpsprio/close}{0}
\defR{agg/xs/mpsprio/ns}{0}
\defR{agg/xs/mpsprio/zero}{23}
\defR{agg/all/mpslimits/attmean}{86.9}
\defR{agg/all/mpslimits/attmin}{79.7}
\defR{agg/all/mpslimits/thr}{16}
\defR{agg/all/mpslimits/close}{18}
\defR{agg/all/mpslimits/ns}{0}
\defR{agg/all/mpslimits/zero}{3}
\defR{agg/short/mpslimits/attmean}{85.4}
\defR{agg/short/mpslimits/attmin}{77.6}
\defR{agg/short/mpslimits/thr}{17}
\defR{agg/short/mpslimits/close}{16}
\defR{agg/short/mpslimits/ns}{0}
\defR{agg/short/mpslimits/zero}{3}
\defR{agg/gemm/mpslimits/attmean}{96.8}
\defR{agg/gemm/mpslimits/attmin}{93.8}
\defR{agg/gemm/mpslimits/thr}{12}
\defR{agg/gemm/mpslimits/close}{2}
\defR{agg/gemm/mpslimits/ns}{0}
\defR{agg/gemm/mpslimits/zero}{0}
\defR{agg/single/mpslimits/attmean}{95.8}
\defR{agg/single/mpslimits/attmin}{95.8}
\defR{agg/single/mpslimits/thr}{20}
\defR{agg/single/mpslimits/close}{15}
\defR{agg/single/mpslimits/ns}{0}
\defR{agg/single/mpslimits/zero}{0}
\defR{agg/multi/mpslimits/attmean}{78.6}
\defR{agg/multi/mpslimits/attmin}{64.6}
\defR{agg/multi/mpslimits/thr}{14}
\defR{agg/multi/mpslimits/close}{3}
\defR{agg/multi/mpslimits/ns}{0}
\defR{agg/multi/mpslimits/zero}{3}
\defR{agg/xs/mpslimits/attmean}{85.6}
\defR{agg/xs/mpslimits/attmin}{77.8}
\defR{agg/xs/mpslimits/thr}{16}
\defR{agg/xs/mpslimits/close}{16}
\defR{agg/xs/mpslimits/ns}{0}
\defR{agg/xs/mpslimits/zero}{3}
\defR{agg/all/mpsparts/attmean}{87.1}
\defR{agg/all/mpsparts/attmin}{79.1}
\defR{agg/all/mpsparts/thr}{17}
\defR{agg/all/mpsparts/close}{17}
\defR{agg/all/mpsparts/ns}{0}
\defR{agg/all/mpsparts/zero}{3}
\defR{agg/short/mpsparts/attmean}{85.9}
\defR{agg/short/mpsparts/attmin}{77.3}
\defR{agg/short/mpsparts/thr}{17}
\defR{agg/short/mpsparts/close}{15}
\defR{agg/short/mpsparts/ns}{0}
\defR{agg/short/mpsparts/zero}{3}
\defR{agg/gemm/mpsparts/attmean}{97.4}
\defR{agg/gemm/mpsparts/attmin}{95.5}
\defR{agg/gemm/mpsparts/thr}{15}
\defR{agg/gemm/mpsparts/close}{2}
\defR{agg/gemm/mpsparts/ns}{0}
\defR{agg/gemm/mpsparts/zero}{0}
\defR{agg/single/mpsparts/attmean}{95.6}
\defR{agg/single/mpsparts/attmin}{95.6}
\defR{agg/single/mpsparts/thr}{19}
\defR{agg/single/mpsparts/close}{14}
\defR{agg/single/mpsparts/ns}{0}
\defR{agg/single/mpsparts/zero}{0}
\defR{agg/multi/mpsparts/attmean}{78.5}
\defR{agg/multi/mpsparts/attmin}{62.5}
\defR{agg/multi/mpsparts/thr}{15}
\defR{agg/multi/mpsparts/close}{3}
\defR{agg/multi/mpsparts/ns}{0}
\defR{agg/multi/mpsparts/zero}{3}
\defR{agg/xs/mpsparts/attmean}{85.9}
\defR{agg/xs/mpsparts/attmin}{77.1}
\defR{agg/xs/mpsparts/thr}{16}
\defR{agg/xs/mpsparts/close}{15}
\defR{agg/xs/mpsparts/ns}{0}
\defR{agg/xs/mpsparts/zero}{3}
\defR{agg/all/lithos/attmean}{85.5}
\defR{agg/all/lithos/attmin}{78.4}
\defR{agg/all/lithos/thr}{18}
\defR{agg/all/lithos/close}{17}
\defR{agg/all/lithos/ns}{0}
\defR{agg/all/lithos/zero}{3}
\defR{agg/short/lithos/attmean}{83.9}
\defR{agg/short/lithos/attmin}{76.1}
\defR{agg/short/lithos/thr}{19}
\defR{agg/short/lithos/close}{15}
\defR{agg/short/lithos/ns}{0}
\defR{agg/short/lithos/zero}{3}
\defR{agg/gemm/lithos/attmean}{96.7}
\defR{agg/gemm/lithos/attmin}{93.6}
\defR{agg/gemm/lithos/thr}{11}
\defR{agg/gemm/lithos/close}{2}
\defR{agg/gemm/lithos/ns}{0}
\defR{agg/gemm/lithos/zero}{0}
\defR{agg/single/lithos/attmean}{94.7}
\defR{agg/single/lithos/attmin}{94.7}
\defR{agg/single/lithos/thr}{20}
\defR{agg/single/lithos/close}{14}
\defR{agg/single/lithos/ns}{0}
\defR{agg/single/lithos/zero}{0}
\defR{agg/multi/lithos/attmean}{76.9}
\defR{agg/multi/lithos/attmin}{63.1}
\defR{agg/multi/lithos/thr}{16}
\defR{agg/multi/lithos/close}{3}
\defR{agg/multi/lithos/ns}{0}
\defR{agg/multi/lithos/zero}{3}
\defR{agg/xs/lithos/attmean}{85.3}
\defR{agg/xs/lithos/attmin}{77.5}
\defR{agg/xs/lithos/thr}{15}
\defR{agg/xs/lithos/close}{16}
\defR{agg/xs/lithos/ns}{0}
\defR{agg/xs/lithos/zero}{3}
\defR{agg/all/dysm/attmean}{29.1}
\defR{agg/all/dysm/attmin}{14.9}
\defR{agg/all/dysm/thr}{76}
\defR{agg/all/dysm/close}{1}
\defR{agg/all/dysm/ns}{0}
\defR{agg/all/dysm/zero}{24}
\defR{agg/short/dysm/attmean}{29.1}
\defR{agg/short/dysm/attmin}{14.6}
\defR{agg/short/dysm/thr}{77}
\defR{agg/short/dysm/close}{1}
\defR{agg/short/dysm/ns}{0}
\defR{agg/short/dysm/zero}{21}
\defR{agg/gemm/dysm/attmean}{29.3}
\defR{agg/gemm/dysm/attmin}{17.2}
\defR{agg/gemm/dysm/thr}{73}
\defR{agg/gemm/dysm/close}{0}
\defR{agg/gemm/dysm/ns}{0}
\defR{agg/gemm/dysm/zero}{3}
\defR{agg/single/dysm/attmean}{27.3}
\defR{agg/single/dysm/attmin}{27.3}
\defR{agg/single/dysm/thr}{85}
\defR{agg/single/dysm/close}{1}
\defR{agg/single/dysm/ns}{0}
\defR{agg/single/dysm/zero}{9}
\defR{agg/multi/dysm/attmean}{30.8}
\defR{agg/multi/dysm/attmin}{3.4}
\defR{agg/multi/dysm/thr}{68}
\defR{agg/multi/dysm/close}{0}
\defR{agg/multi/dysm/ns}{0}
\defR{agg/multi/dysm/zero}{15}
\defR{agg/xs/dysm/attmean}{29.0}
\defR{agg/xs/dysm/attmin}{14.6}
\defR{agg/xs/dysm/thr}{77}
\defR{agg/xs/dysm/close}{1}
\defR{agg/xs/dysm/ns}{0}
\defR{agg/xs/dysm/zero}{21}
\defR{agg/all/xsched/attmean}{71.3}
\defR{agg/all/xsched/attmin}{64.4}
\defR{agg/all/xsched/thr}{42}
\defR{agg/all/xsched/close}{8}
\defR{agg/all/xsched/ns}{4}
\defR{agg/all/xsched/zero}{4}
\defR{agg/short/xsched/attmean}{73.3}
\defR{agg/short/xsched/attmin}{67.4}
\defR{agg/short/xsched/thr}{40}
\defR{agg/short/xsched/close}{8}
\defR{agg/short/xsched/ns}{4}
\defR{agg/short/xsched/zero}{4}
\defR{agg/gemm/xsched/attmean}{58.4}
\defR{agg/gemm/xsched/attmin}{44.6}
\defR{agg/gemm/xsched/thr}{53}
\defR{agg/gemm/xsched/close}{0}
\defR{agg/gemm/xsched/ns}{0}
\defR{agg/gemm/xsched/zero}{0}
\defR{agg/single/xsched/attmean}{66.8}
\defR{agg/single/xsched/attmin}{66.8}
\defR{agg/single/xsched/thr}{46}
\defR{agg/single/xsched/close}{5}
\defR{agg/single/xsched/ns}{2}
\defR{agg/single/xsched/zero}{2}
\defR{agg/multi/xsched/attmean}{75.6}
\defR{agg/multi/xsched/attmin}{62.2}
\defR{agg/multi/xsched/thr}{38}
\defR{agg/multi/xsched/close}{3}
\defR{agg/multi/xsched/ns}{2}
\defR{agg/multi/xsched/zero}{2}
\defR{agg/xs/xsched/attmean}{81.9}
\defR{agg/xs/xsched/attmin}{74.0}
\defR{agg/xs/xsched/thr}{48}
\defR{agg/xs/xsched/close}{8}
\defR{agg/xs/xsched/ns}{0}
\defR{agg/xs/xsched/zero}{0}
\defR{agg/all/multistream/attmean}{24.5}
\defR{agg/all/multistream/attmin}{11.5}
\defR{agg/all/multistream/thr}{72}
\defR{agg/all/multistream/close}{0}
\defR{agg/all/multistream/ns}{0}
\defR{agg/all/multistream/zero}{24}
\defR{agg/short/multistream/attmean}{28.1}
\defR{agg/short/multistream/attmin}{13.2}
\defR{agg/short/multistream/thr}{71}
\defR{agg/short/multistream/close}{0}
\defR{agg/short/multistream/ns}{0}
\defR{agg/short/multistream/zero}{20}
\defR{agg/gemm/multistream/attmean}{0.0}
\defR{agg/gemm/multistream/attmin}{0.0}
\defR{agg/gemm/multistream/thr}{84}
\defR{agg/gemm/multistream/close}{0}
\defR{agg/gemm/multistream/ns}{0}
\defR{agg/gemm/multistream/zero}{4}
\defR{agg/single/multistream/attmean}{10.3}
\defR{agg/single/multistream/attmin}{10.3}
\defR{agg/single/multistream/thr}{83}
\defR{agg/single/multistream/close}{0}
\defR{agg/single/multistream/ns}{0}
\defR{agg/single/multistream/zero}{12}
\defR{agg/multi/multistream/attmean}{37.8}
\defR{agg/multi/multistream/attmin}{12.7}
\defR{agg/multi/multistream/thr}{63}
\defR{agg/multi/multistream/close}{0}
\defR{agg/multi/multistream/ns}{0}
\defR{agg/multi/multistream/zero}{12}
\defR{agg/xs/multistream/attmean}{17.6}
\defR{agg/xs/multistream/attmin}{5.1}
\defR{agg/xs/multistream/thr}{73}
\defR{agg/xs/multistream/close}{0}
\defR{agg/xs/multistream/ns}{0}
\defR{agg/xs/multistream/zero}{23}
\defR{agg/all/tgs/attmean}{84.9}
\defR{agg/all/tgs/attmin}{81.5}
\defR{agg/all/tgs/thr}{9}
\defR{agg/all/tgs/close}{16}
\defR{agg/all/tgs/ns}{2}
\defR{agg/all/tgs/zero}{2}
\defR{agg/short/tgs/attmean}{90.1}
\defR{agg/short/tgs/attmin}{86.9}
\defR{agg/short/tgs/thr}{3}
\defR{agg/short/tgs/close}{16}
\defR{agg/short/tgs/ns}{1}
\defR{agg/short/tgs/zero}{1}
\defR{agg/gemm/tgs/attmean}{49.2}
\defR{agg/gemm/tgs/attmin}{45.2}
\defR{agg/gemm/tgs/thr}{47}
\defR{agg/gemm/tgs/close}{0}
\defR{agg/gemm/tgs/ns}{1}
\defR{agg/gemm/tgs/zero}{1}
\defR{agg/single/tgs/attmean}{87.8}
\defR{agg/single/tgs/attmin}{87.8}
\defR{agg/single/tgs/thr}{13}
\defR{agg/single/tgs/close}{10}
\defR{agg/single/tgs/ns}{0}
\defR{agg/single/tgs/zero}{0}
\defR{agg/multi/tgs/attmean}{82.2}
\defR{agg/multi/tgs/attmin}{75.6}
\defR{agg/multi/tgs/thr}{5}
\defR{agg/multi/tgs/close}{6}
\defR{agg/multi/tgs/ns}{2}
\defR{agg/multi/tgs/zero}{2}
\defR{agg/xs/tgs/attmean}{86.5}
\defR{agg/xs/tgs/attmin}{82.7}
\defR{agg/xs/tgs/thr}{10}
\defR{agg/xs/tgs/close}{13}
\defR{agg/xs/tgs/ns}{1}
\defR{agg/xs/tgs/zero}{1}
\defR{agg/all/mig/attmean}{88.4}
\defR{agg/all/mig/attmin}{80.3}
\defR{agg/all/mig/thr}{14}
\defR{agg/all/mig/close}{16}
\defR{agg/all/mig/ns}{0}
\defR{agg/all/mig/zero}{3}
\defR{agg/short/mig/attmean}{87.2}
\defR{agg/short/mig/attmin}{78.3}
\defR{agg/short/mig/thr}{13}
\defR{agg/short/mig/close}{14}
\defR{agg/short/mig/ns}{0}
\defR{agg/short/mig/zero}{3}
\defR{agg/gemm/mig/attmean}{96.2}
\defR{agg/gemm/mig/attmin}{93.6}
\defR{agg/gemm/mig/thr}{21}
\defR{agg/gemm/mig/close}{2}
\defR{agg/gemm/mig/ns}{0}
\defR{agg/gemm/mig/zero}{0}
\defR{agg/single/mig/attmean}{94.2}
\defR{agg/single/mig/attmin}{94.2}
\defR{agg/single/mig/thr}{29}
\defR{agg/single/mig/close}{13}
\defR{agg/single/mig/ns}{0}
\defR{agg/single/mig/zero}{0}
\defR{agg/multi/mig/attmean}{82.9}
\defR{agg/multi/mig/attmin}{67.2}
\defR{agg/multi/mig/thr}{0}
\defR{agg/multi/mig/close}{3}
\defR{agg/multi/mig/ns}{0}
\defR{agg/multi/mig/zero}{3}
\defR{agg/xs/mig/attmean}{87.2}
\defR{agg/xs/mig/attmin}{78.3}
\defR{agg/xs/mig/thr}{12}
\defR{agg/xs/mig/close}{14}
\defR{agg/xs/mig/ns}{0}
\defR{agg/xs/mig/zero}{3}
\defR{agg/all/tally/attmean}{24.5}
\defR{agg/all/tally/attmin}{10.2}
\defR{agg/all/tally/thr}{21}
\defR{agg/all/tally/close}{1}
\defR{agg/all/tally/ns}{0}
\defR{agg/all/tally/zero}{27}
\defR{agg/short/tally/attmean}{26.6}
\defR{agg/short/tally/attmin}{11.7}
\defR{agg/short/tally/thr}{15}
\defR{agg/short/tally/close}{1}
\defR{agg/short/tally/ns}{0}
\defR{agg/short/tally/zero}{23}
\defR{agg/gemm/tally/attmean}{9.9}
\defR{agg/gemm/tally/attmin}{0.0}
\defR{agg/gemm/tally/thr}{64}
\defR{agg/gemm/tally/close}{0}
\defR{agg/gemm/tally/ns}{0}
\defR{agg/gemm/tally/zero}{4}
\defR{agg/single/tally/attmean}{21.0}
\defR{agg/single/tally/attmin}{21.0}
\defR{agg/single/tally/thr}{18}
\defR{agg/single/tally/close}{1}
\defR{agg/single/tally/ns}{0}
\defR{agg/single/tally/zero}{11}
\defR{agg/multi/tally/attmean}{27.7}
\defR{agg/multi/tally/attmin}{0.0}
\defR{agg/multi/tally/thr}{24}
\defR{agg/multi/tally/close}{0}
\defR{agg/multi/tally/ns}{0}
\defR{agg/multi/tally/zero}{16}
\defR{agg/xs/tally/attmean}{22.4}
\defR{agg/xs/tally/attmin}{8.9}
\defR{agg/xs/tally/thr}{24}
\defR{agg/xs/tally/close}{1}
\defR{agg/xs/tally/ns}{0}
\defR{agg/xs/tally/zero}{24}
\defR{agg/all/orion/attmean}{52.8}
\defR{agg/all/orion/attmin}{39.7}
\defR{agg/all/orion/thr}{58}
\defR{agg/all/orion/close}{13}
\defR{agg/all/orion/ns}{3}
\defR{agg/all/orion/zero}{18}
\defR{agg/short/orion/attmean}{60.7}
\defR{agg/short/orion/attmin}{45.5}
\defR{agg/short/orion/thr}{57}
\defR{agg/short/orion/close}{13}
\defR{agg/short/orion/ns}{2}
\defR{agg/short/orion/zero}{14}
\defR{agg/gemm/orion/attmean}{0.1}
\defR{agg/gemm/orion/attmin}{0.1}
\defR{agg/gemm/orion/thr}{65}
\defR{agg/gemm/orion/close}{0}
\defR{agg/gemm/orion/ns}{1}
\defR{agg/gemm/orion/zero}{4}
\defR{agg/single/orion/attmean}{82.0}
\defR{agg/single/orion/attmin}{82.0}
\defR{agg/single/orion/thr}{62}
\defR{agg/single/orion/close}{13}
\defR{agg/single/orion/ns}{0}
\defR{agg/single/orion/zero}{2}
\defR{agg/multi/orion/attmean}{25.6}
\defR{agg/multi/orion/attmin}{0.0}
\defR{agg/multi/orion/thr}{55}
\defR{agg/multi/orion/close}{0}
\defR{agg/multi/orion/ns}{3}
\defR{agg/multi/orion/zero}{16}
\defR{agg/xs/orion/attmean}{51.2}
\defR{agg/xs/orion/attmin}{38.1}
\defR{agg/xs/orion/thr}{60}
\defR{agg/xs/orion/close}{11}
\defR{agg/xs/orion/ns}{2}
\defR{agg/xs/orion/zero}{16}
\defR{agg/all/alone/attmean}{97.9}
\defR{agg/all/alone/attmin}{97.5}
\defR{llm/llm-load40/berth-full/att}{100.0}
\defR{llm/llm-load40/berth-full/be}{38}
\defR{llm/llm-load40/berth-full/tpot}{16}
\defR{llm/llm-load40/berth-full/ttft}{76}
\defR{llm/llm-load40/berth-ng/att}{100.0}
\defR{llm/llm-load40/berth-ng/be}{38}
\defR{llm/llm-load40/berth-ng/tpot}{17}
\defR{llm/llm-load40/berth-ng/ttft}{79}
\defR{llm/llm-load40/berth-nsng/att}{100.0}
\defR{llm/llm-load40/berth-nsng/be}{5}
\defR{llm/llm-load40/berth-nsng/tpot}{10}
\defR{llm/llm-load40/berth-nsng/ttft}{62}
\defR{llm/llm-load40/timeslice/att}{2.8}
\defR{llm/llm-load40/timeslice/be}{49}
\defR{llm/llm-load40/timeslice/tpot}{35}
\defR{llm/llm-load40/timeslice/ttft}{199}
\defR{llm/llm-load40/mps/att}{0.0}
\defR{llm/llm-load40/mps/be}{84}
\defR{llm/llm-load40/mps/tpot}{118}
\defR{llm/llm-load40/mps/ttft}{41865}
\defR{llm/llm-load40/mpsprio/att}{2.8}
\defR{llm/llm-load40/mpsprio/be}{74}
\defR{llm/llm-load40/mpsprio/tpot}{40}
\defR{llm/llm-load40/mpsprio/ttft}{170}
\defR{llm/llm-load40/mpslimits/att}{99.8}
\defR{llm/llm-load40/mpslimits/be}{51}
\defR{llm/llm-load40/mpslimits/tpot}{19}
\defR{llm/llm-load40/mpslimits/ttft}{189}
\defR{llm/llm-load40/mpsparts/att}{99.8}
\defR{llm/llm-load40/mpsparts/be}{50}
\defR{llm/llm-load40/mpsparts/tpot}{16}
\defR{llm/llm-load40/mpsparts/ttft}{170}
\defR{llm/llm-load40/lithos/att}{99.5}
\defR{llm/llm-load40/lithos/be}{53}
\defR{llm/llm-load40/lithos/tpot}{20}
\defR{llm/llm-load40/lithos/ttft}{192}
\defR{llm/llm-load40/dysm/att}{82.6}
\defR{llm/llm-load40/dysm/be}{64}
\defR{llm/llm-load40/dysm/tpot}{30}
\defR{llm/llm-load40/dysm/ttft}{559}
\defR{llm/llm-load40/xsched/att}{99.8}
\defR{llm/llm-load40/xsched/be}{32}
\defR{llm/llm-load40/xsched/tpot}{23}
\defR{llm/llm-load40/xsched/ttft}{133}
\defR{llm/llm-load40/tgs/att}{100.0}
\defR{llm/llm-load40/tgs/be}{1}
\defR{llm/llm-load40/tgs/tpot}{12}
\defR{llm/llm-load40/tgs/ttft}{82}
\defR{llm/llm-load40/mig/att}{98.5}
\defR{llm/llm-load40/mig/be}{54}
\defR{llm/llm-load40/mig/tpot}{25}
\defR{llm/llm-load40/mig/ttft}{205}
\defR{llm/llm-load70/berth-full/att}{100.0}
\defR{llm/llm-load70/berth-full/be}{30}
\defR{llm/llm-load70/berth-full/tpot}{18}
\defR{llm/llm-load70/berth-full/ttft}{90}
\defR{llm/llm-load70/berth-ng/att}{100.0}
\defR{llm/llm-load70/berth-ng/be}{32}
\defR{llm/llm-load70/berth-ng/tpot}{18}
\defR{llm/llm-load70/berth-ng/ttft}{93}
\defR{llm/llm-load70/berth-nsng/att}{100.0}
\defR{llm/llm-load70/berth-nsng/be}{5}
\defR{llm/llm-load70/berth-nsng/tpot}{12}
\defR{llm/llm-load70/berth-nsng/ttft}{75}
\defR{llm/llm-load70/timeslice/att}{0.0}
\defR{llm/llm-load70/timeslice/be}{43}
\defR{llm/llm-load70/timeslice/tpot}{55}
\defR{llm/llm-load70/timeslice/ttft}{6008}
\defR{llm/llm-load70/mps/att}{0.0}
\defR{llm/llm-load70/mps/be}{84}
\defR{llm/llm-load70/mps/tpot}{110}
\defR{llm/llm-load70/mps/ttft}{102081}
\defR{llm/llm-load70/mpsprio/att}{0.0}
\defR{llm/llm-load70/mpsprio/be}{68}
\defR{llm/llm-load70/mpsprio/tpot}{66}
\defR{llm/llm-load70/mpsprio/ttft}{33416}
\defR{llm/llm-load70/mpslimits/att}{91.6}
\defR{llm/llm-load70/mpslimits/be}{50}
\defR{llm/llm-load70/mpslimits/tpot}{31}
\defR{llm/llm-load70/mpslimits/ttft}{260}
\defR{llm/llm-load70/mpsparts/att}{89.4}
\defR{llm/llm-load70/mpsparts/be}{49}
\defR{llm/llm-load70/mpsparts/tpot}{32}
\defR{llm/llm-load70/mpsparts/ttft}{261}
\defR{llm/llm-load70/lithos/att}{14.8}
\defR{llm/llm-load70/lithos/be}{52}
\defR{llm/llm-load70/lithos/tpot}{44}
\defR{llm/llm-load70/lithos/ttft}{6283}
\defR{llm/llm-load70/dysm/att}{81.4}
\defR{llm/llm-load70/dysm/be}{48}
\defR{llm/llm-load70/dysm/tpot}{33}
\defR{llm/llm-load70/dysm/ttft}{496}
\defR{llm/llm-load70/xsched/att}{93.6}
\defR{llm/llm-load70/xsched/be}{28}
\defR{llm/llm-load70/xsched/tpot}{28}
\defR{llm/llm-load70/xsched/ttft}{186}
\defR{llm/llm-load70/tgs/att}{100.0}
\defR{llm/llm-load70/tgs/be}{1}
\defR{llm/llm-load70/tgs/tpot}{16}
\defR{llm/llm-load70/tgs/ttft}{105}
\defR{llm/llm-load70/mig/att}{0.0}
\defR{llm/llm-load70/mig/be}{53}
\defR{llm/llm-load70/mig/tpot}{49}
\defR{llm/llm-load70/mig/ttft}{36715}
\defR{llm/llm-gemm/berth-full/att}{99.8}
\defR{llm/llm-gemm/berth-full/be}{46}
\defR{llm/llm-gemm/berth-full/tpot}{22}
\defR{llm/llm-gemm/berth-full/ttft}{126}
\defR{llm/llm-gemm/berth-ng/att}{99.5}
\defR{llm/llm-gemm/berth-ng/be}{46}
\defR{llm/llm-gemm/berth-ng/tpot}{24}
\defR{llm/llm-gemm/berth-ng/ttft}{125}
\defR{llm/llm-gemm/berth-nsng/att}{100.0}
\defR{llm/llm-gemm/berth-nsng/be}{17}
\defR{llm/llm-gemm/berth-nsng/tpot}{13}
\defR{llm/llm-gemm/berth-nsng/ttft}{88}
\defR{llm/llm-gemm/timeslice/att}{3.1}
\defR{llm/llm-gemm/timeslice/be}{49}
\defR{llm/llm-gemm/timeslice/tpot}{37}
\defR{llm/llm-gemm/timeslice/ttft}{201}
\defR{llm/llm-gemm/mpsprio/att}{0.0}
\defR{llm/llm-gemm/mpsprio/be}{94}
\defR{llm/llm-gemm/mpsprio/tpot}{431}
\defR{llm/llm-gemm/mpsprio/ttft}{142315}
\defR{llm/llm-gemm/mpslimits/att}{99.8}
\defR{llm/llm-gemm/mpslimits/be}{50}
\defR{llm/llm-gemm/mpslimits/tpot}{17}
\defR{llm/llm-gemm/mpslimits/ttft}{180}
\defR{llm/llm-gemm/mpsparts/att}{100.0}
\defR{llm/llm-gemm/mpsparts/be}{50}
\defR{llm/llm-gemm/mpsparts/tpot}{15}
\defR{llm/llm-gemm/mpsparts/ttft}{167}
\defR{llm/llm-gemm/lithos/att}{99.7}
\defR{llm/llm-gemm/lithos/be}{54}
\defR{llm/llm-gemm/lithos/tpot}{18}
\defR{llm/llm-gemm/lithos/ttft}{185}
\defR{llm/llm-gemm/dysm/att}{83.9}
\defR{llm/llm-gemm/dysm/be}{72}
\defR{llm/llm-gemm/dysm/tpot}{29}
\defR{llm/llm-gemm/dysm/ttft}{531}
\defR{llm/llm-gemm/xsched/att}{3.7}
\defR{llm/llm-gemm/xsched/be}{48}
\defR{llm/llm-gemm/xsched/tpot}{34}
\defR{llm/llm-gemm/xsched/ttft}{197}
\defR{llm/llm-gemm/tgs/att}{3.2}
\defR{llm/llm-gemm/tgs/be}{50}
\defR{llm/llm-gemm/tgs/tpot}{36}
\defR{llm/llm-gemm/tgs/ttft}{194}
\defR{llm/llm-gemm/mig/att}{98.4}
\defR{llm/llm-gemm/mig/be}{52}
\defR{llm/llm-gemm/mig/tpot}{26}
\defR{llm/llm-gemm/mig/ttft}{206}
\defR{llm/llm-scale-3/berth-full/att}{100.0}
\defR{llm/llm-scale-3/berth-full/be}{27}
\defR{llm/llm-scale-3/berth-full/tpot}{17}
\defR{llm/llm-scale-3/berth-full/ttft}{83}
\defR{llm/llm-scale-3/berth-ng/att}{100.0}
\defR{llm/llm-scale-3/berth-ng/be}{28}
\defR{llm/llm-scale-3/berth-ng/tpot}{18}
\defR{llm/llm-scale-3/berth-ng/ttft}{81}
\defR{llm/llm-scale-3/berth-nsng/att}{100.0}
\defR{llm/llm-scale-3/berth-nsng/be}{0}
\defR{llm/llm-scale-3/berth-nsng/tpot}{11}
\defR{llm/llm-scale-3/berth-nsng/ttft}{68}
\defR{llm/llm-scale-3/timeslice/att}{1.3}
\defR{llm/llm-scale-3/timeslice/be}{38}
\defR{llm/llm-scale-3/timeslice/tpot}{47}
\defR{llm/llm-scale-3/timeslice/ttft}{278}
\defR{llm/llm-scale-3/mps/att}{0.0}
\defR{llm/llm-scale-3/mps/be}{81}
\defR{llm/llm-scale-3/mps/tpot}{124}
\defR{llm/llm-scale-3/mps/ttft}{77495}
\defR{llm/llm-scale-3/mpsprio/att}{1.4}
\defR{llm/llm-scale-3/mpsprio/be}{65}
\defR{llm/llm-scale-3/mpsprio/tpot}{45}
\defR{llm/llm-scale-3/mpsprio/ttft}{185}
\defR{llm/llm-scale-3/mpslimits/att}{99.5}
\defR{llm/llm-scale-3/mpslimits/be}{38}
\defR{llm/llm-scale-3/mpslimits/tpot}{22}
\defR{llm/llm-scale-3/mpslimits/ttft}{197}
\defR{llm/llm-scale-3/mpsparts/att}{99.8}
\defR{llm/llm-scale-3/mpsparts/be}{38}
\defR{llm/llm-scale-3/mpsparts/tpot}{18}
\defR{llm/llm-scale-3/mpsparts/ttft}{178}
\defR{llm/llm-scale-3/lithos/att}{99.4}
\defR{llm/llm-scale-3/lithos/be}{39}
\defR{llm/llm-scale-3/lithos/tpot}{22}
\defR{llm/llm-scale-3/lithos/ttft}{202}
\defR{llm/llm-scale-3/dysm/att}{84.6}
\defR{llm/llm-scale-3/dysm/be}{41}
\defR{llm/llm-scale-3/dysm/tpot}{30}
\defR{llm/llm-scale-3/dysm/ttft}{482}
\defR{llm/llm-scale-3/xsched/att}{40.8}
\defR{llm/llm-scale-3/xsched/be}{23}
\defR{llm/llm-scale-3/xsched/tpot}{31}
\defR{llm/llm-scale-3/xsched/ttft}{187}
\defR{llm/llm-scale-3/tgs/att}{100.0}
\defR{llm/llm-scale-3/tgs/be}{1}
\defR{llm/llm-scale-3/tgs/tpot}{19}
\defR{llm/llm-scale-3/tgs/ttft}{116}
\defR{llm/llm-scale-3/mig/att}{98.0}
\defR{llm/llm-scale-3/mig/be}{42}
\defR{llm/llm-scale-3/mig/tpot}{27}
\defR{llm/llm-scale-3/mig/ttft}{212}
\defR{llm/llm-scale-5/berth-full/att}{100.0}
\defR{llm/llm-scale-5/berth-full/be}{17}
\defR{llm/llm-scale-5/berth-full/tpot}{15}
\defR{llm/llm-scale-5/berth-full/ttft}{74}
\defR{llm/llm-scale-5/berth-ng/att}{100.0}
\defR{llm/llm-scale-5/berth-ng/be}{18}
\defR{llm/llm-scale-5/berth-ng/tpot}{16}
\defR{llm/llm-scale-5/berth-ng/ttft}{78}
\defR{llm/llm-scale-5/berth-nsng/att}{100.0}
\defR{llm/llm-scale-5/berth-nsng/be}{0}
\defR{llm/llm-scale-5/berth-nsng/tpot}{12}
\defR{llm/llm-scale-5/berth-nsng/ttft}{66}
\defR{llm/llm-scale-5/timeslice/att}{0.8}
\defR{llm/llm-scale-5/timeslice/be}{30}
\defR{llm/llm-scale-5/timeslice/tpot}{53}
\defR{llm/llm-scale-5/timeslice/ttft}{306}
\defR{llm/llm-scale-5/mps/att}{0.0}
\defR{llm/llm-scale-5/mps/be}{76}
\defR{llm/llm-scale-5/mps/tpot}{140}
\defR{llm/llm-scale-5/mps/ttft}{40475}
\defR{llm/llm-scale-5/mpsprio/att}{2.0}
\defR{llm/llm-scale-5/mpsprio/be}{60}
\defR{llm/llm-scale-5/mpsprio/tpot}{39}
\defR{llm/llm-scale-5/mpsprio/ttft}{153}
\defR{llm/llm-scale-5/mpslimits/att}{100.0}
\defR{llm/llm-scale-5/mpslimits/be}{11}
\defR{llm/llm-scale-5/mpslimits/tpot}{15}
\defR{llm/llm-scale-5/mpslimits/ttft}{160}
\defR{llm/llm-scale-5/mpsparts/att}{100.0}
\defR{llm/llm-scale-5/mpsparts/be}{11}
\defR{llm/llm-scale-5/mpsparts/tpot}{15}
\defR{llm/llm-scale-5/mpsparts/ttft}{150}
\defR{llm/llm-scale-5/lithos/att}{99.7}
\defR{llm/llm-scale-5/lithos/be}{13}
\defR{llm/llm-scale-5/lithos/tpot}{15}
\defR{llm/llm-scale-5/lithos/ttft}{164}
\defR{llm/llm-scale-5/dysm/att}{43.2}
\defR{llm/llm-scale-5/dysm/be}{56}
\defR{llm/llm-scale-5/dysm/tpot}{46}
\defR{llm/llm-scale-5/dysm/ttft}{511}
\defR{llm/llm-scale-5/xsched/att}{1.9}
\defR{llm/llm-scale-5/xsched/be}{16}
\defR{llm/llm-scale-5/xsched/tpot}{38}
\defR{llm/llm-scale-5/xsched/ttft}{219}
\defR{llm/llm-scale-5/tgs/att}{8.7}
\defR{llm/llm-scale-5/tgs/be}{1}
\defR{llm/llm-scale-5/tgs/tpot}{31}
\defR{llm/llm-scale-5/tgs/ttft}{176}
\defR{llm/llm-scale-5/mig/att}{99.8}
\defR{llm/llm-scale-5/mig/be}{0}
\defR{llm/llm-scale-5/mig/tpot}{17}
\defR{llm/llm-scale-5/mig/ttft}{160}
\defR{llm/llm-load40-twojobs/berth-full/att}{100.0}
\defR{llm/llm-load40-twojobs/berth-full/be}{20}
\defR{llm/llm-load40-twojobs/berth-full/tpot}{18}
\defR{llm/llm-load40-twojobs/berth-full/ttft}{113}
\defR{llm/llm-load40-twojobs/berth-ng/att}{99.1}
\defR{llm/llm-load40-twojobs/berth-ng/be}{24}
\defR{llm/llm-load40-twojobs/berth-ng/tpot}{25}
\defR{llm/llm-load40-twojobs/berth-ng/ttft}{130}
\defR{llm/llm-load40-twojobs/berth-nsng/att}{100.0}
\defR{llm/llm-load40-twojobs/berth-nsng/be}{9}
\defR{llm/llm-load40-twojobs/berth-nsng/tpot}{14}
\defR{llm/llm-load40-twojobs/berth-nsng/ttft}{92}
\defR{llm/llm-load40-twojobs/timeslice/att}{1.6}
\defR{llm/llm-load40-twojobs/timeslice/be}{29}
\defR{llm/llm-load40-twojobs/timeslice/tpot}{61}
\defR{llm/llm-load40-twojobs/timeslice/ttft}{356}
\defR{llm/llm-load40-twojobs/mpsprio/att}{0.0}
\defR{llm/llm-load40-twojobs/mpsprio/be}{47}
\defR{llm/llm-load40-twojobs/mpsprio/tpot}{430}
\defR{llm/llm-load40-twojobs/mpsprio/ttft}{142863}
\defR{llm/llm-load40-twojobs/lithos/att}{99.7}
\defR{llm/llm-load40-twojobs/lithos/be}{27}
\defR{llm/llm-load40-twojobs/lithos/tpot}{19}
\defR{llm/llm-load40-twojobs/lithos/ttft}{186}
\defR{llm/llm-load40-twojobs/dysm/att}{84.3}
\defR{llm/llm-load40-twojobs/dysm/be}{36}
\defR{llm/llm-load40-twojobs/dysm/tpot}{29}
\defR{llm/llm-load40-twojobs/dysm/ttft}{503}
\defR{llm/llm-load40-twojobs/xsched/att}{2.4}
\defR{llm/llm-load40-twojobs/xsched/be}{27}
\defR{llm/llm-load40-twojobs/xsched/tpot}{45}
\defR{llm/llm-load40-twojobs/xsched/ttft}{265}
\defR{llm/llm-load40-twojobs/mig/att}{98.5}
\defR{llm/llm-load40-twojobs/mig/be}{27}
\defR{llm/llm-load40-twojobs/mig/tpot}{26}
\defR{llm/llm-load40-twojobs/mig/ttft}{207}
\defR{app/qshift/pair-resnet50-load30-gemm/shifts}{0}
\defR{app/qshift/pair-resnet50-load30-gemm/heldms}{0.0}
\defR{app/qshift/pair-resnet50-load30-gemm/met}{95.6}
\defR{app/qshift/pair-resnet50-load30-gemm/be}{71.9}
\defR{app/qshift/pair-resnet50-load30-gemm/metref}{95.6}
\defR{app/qshift/pair-resnet50-load30-gemm/beref}{71.8}
\defR{app/qshift/pair-resnet50-load30-trainresnet/shifts}{385,903}
\defR{app/qshift/pair-resnet50-load30-trainresnet/heldms}{28.6}
\defR{app/qshift/pair-resnet50-load30-trainresnet/met}{97.8}
\defR{app/qshift/pair-resnet50-load30-trainresnet/be}{62.8}
\defR{app/qshift/pair-resnet50-load30-trainresnet/metref}{97.9}
\defR{app/qshift/pair-resnet50-load30-trainresnet/beref}{66.4}
\defR{app/grace70/pair-resnet50-load70/g05}{65.4}
\defR{app/grace70/pair-resnet50-load70/g1}{67.0}
\defR{app/grace70/pair-resnet50-load70/g1p99}{53}
\defR{app/grace70/pair-bertlarge-load70/g05}{94.3}
\defR{app/grace70/pair-bertlarge-load70/g1}{95.6}
\defR{app/grace70/pair-bertlarge-load70/g1p99}{210}
\defR{app/sl100/grp/attmin}{40.2}
\defR{app/sl100/sep/attmin}{79.8}
\defR{app/sm/pair-resnet50-load30/lcshare}{24}
\defR{app/sm/pair-resnet50-load30/beshare}{70}
\defR{app/sm/pair-resnet50-load30/be}{69}
\defR{app/sm/mix-load40/lcshare}{38}
\defR{app/sm/mix-load40/beshare}{57}
\defR{app/sm/mix-load40/be}{57}
\defR{app/sm/llm-load40/lcshare}{56}
\defR{app/sm/llm-load40/beshare}{36}
\defR{app/sm/llm-load40/be}{36}
\defR{app/sm/llm-7b-load40/lcshare}{50}
\defR{app/sm/llm-7b-load40/beshare}{37}
\defR{app/sm/llm-7b-load40/be}{38}
\defR{app/englayout/gh/gemm}{3.0}
\defR{app/englayout/cap50/be}{44}
\defR{app/englayout/cap25/be}{23}
\defR{app/englayout/a/twojobs}{11.5}
\defR{app/englayout/splitC/met}{99.3}
\defR{app/hb/berth-full}{89.9}
\defR{app/hb/timeslice}{28.2}
\defR{app/hb/mpslimits}{47.4}
\defR{app/hb/mig}{64.7}
\defR{app/hb/tgs}{59.8}
\defR{app/hb/xsched}{54.2}

\newcommand{\para}[1]{\vspace{2pt}\noindent\textbf{#1}}
\newcommand{\srcnote}[1]{}   %
\usepackage{xcolor}
\usepackage{hyperref}
\hypersetup{colorlinks=true,linkcolor=black,citecolor=black,urlcolor=black,
  pdftitle={Berth: SLO-Preserving GPU Sharing Without New Preemption},
  pdfauthor={Mohammadali Khodabandehlou, Mahdi Alizadeh, Lei Gao, Murali Annavaram, Bhaskar Krishnamachari}}

\newcommand{\feedms}{$f$\xspace}
\newcommand{\cmark}{\ensuremath{\checkmark}}
\newcommand{\xmark}{\ensuremath{\times}}
\newcommand{\locShim}{1{,}355}   %
\newcommand{\locDaemon}{571}   %

\begin{document}

\date{}
\title{\Large \bf Berth: SLO-Preserving GPU Sharing Without New Preemption}

\author{
{\rm Mohammadali Khodabandehlou\qquad Mahdi Alizadeh\qquad Lei Gao}\\[1pt]
{\rm Murali Annavaram\qquad Bhaskar Krishnamachari}\\
University of Southern California
}

\maketitle

\begin{abstract}
Inference services keep GPUs reserved for their worst-case load, and most of that capacity sits idle. Filling it with
best-effort work such as training is the obvious remedy, and latency is why it rarely happens: on a shared GPU, a request
that must finish within milliseconds finds the best-effort job's kernels running or queued ahead of it. Spatial sharing
cannot evict them, temporal sharing gives the request only its turn, and hardware partitioning protects it by idling
what it does not use. Systems that keep both the SLO and the utilization have reached inside the stack (rewriting
kernels, patching the driver, instrumenting the framework), which the operator of a closed GPU stack cannot do.
This paper argues that reaching inside is unnecessary. The time-slicer built into commodity GPUs already preempts a
running kernel within a fraction of a millisecond and switches the moment a context runs out of work; what it lacks is
priority, and priority can be created from outside by controlling what the best-effort context has to run. Berth is a
user-space system built on this observation: a preloaded shim feeds the best-effort job to the GPU only when no request
needs it (feed control), a declared deadline reference extends the rule to continuous-batching LLM engines, and a
shortened best-effort time slice, set without privileges, bounds what a long kernel can cost. Berth changes no
application, framework, kernel or driver.
Across \nPoints{} scenario points with MLPerf SLOs and \nBaselines{} baselines, Berth meets SLOs within a few percentage points
of running alone, above every baseline, with \thrRatioRange$\times$ the best-effort throughput of the strongest
ones; beside LLM serving, it meets per-token targets for nearly every request while training keeps running. Sharing a
closed GPU under SLOs needs no new preemption mechanism, only control over what the existing one sees.
\end{abstract}

\section{Introduction}
\label{sec:intro}

\begin{figure}[t]
\centering
\includegraphics[width=\columnwidth]{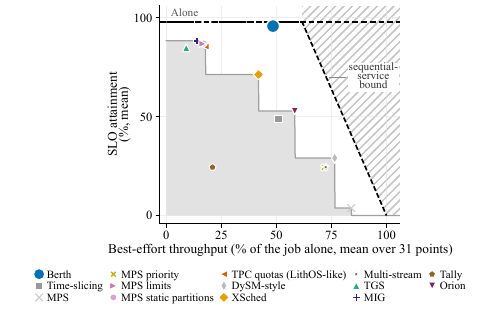}
\caption{Where Berth and every baseline land over our scenario suite: SLO attainment against the best-effort throughput
each keeps. Gray: the best any baseline reaches at that throughput or more; hatched: beyond the sequential-service bound
(\S\ref{sec:eval-overall}).}
\label{fig:intro}
\end{figure}

Inference services are provisioned for their peak load, so the GPUs that host them sit partly idle most of the time.
Production clusters fill such capacity with best-effort work, such as opportunistic training~\cite{tgs,antman} and
offline batch inference, but on the GPUs that serve latency-critical requests they rarely can without reaching inside
the stack~\cite{valve}, and the reason is latency. A request carries a service-level objective (SLO) of milliseconds, 15~ms for an MLPerf server-scenario
ResNet-50 query~\cite{mlperfrules} and a per-token target for every output token of an LLM service, and on a shared
GPU it finds the best-effort job's kernels already running or queued ahead of it. Whether the SLO holds then depends on
how the GPU interleaves the tenants' kernels.

\begin{figure}[t]
\centering
\includegraphics[width=\columnwidth]{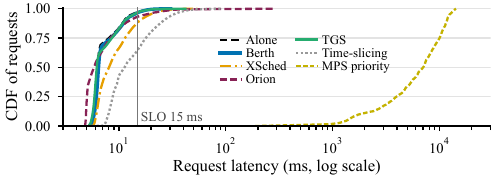}
\caption{Latency of ResNet-50 requests (batch 8, 15-ms SLO, Poisson arrivals at 30\% of its capacity alone) next to
BERT-large training under the stock mechanisms, three systems and Berth.}
\label{fig:cdf-pair}
\end{figure}

The sharing a GPU offers comes in three families, and each gives something up (Figure~\ref{fig:intro}). Spatial sharing runs both tenants'
kernels at once and cannot evict a running one: under NVIDIA's MPS, a request's thread blocks wait behind the training
job's, and even at a higher client priority none of the requests of Figure~\ref{fig:cdf-pair} meet their SLO
(\hlRnMpsprio\%). Temporal sharing gives each process its turn: under the default time-slicing, the request alternates
slices with the training job and runs at half speed, and \hlRnTs\% of requests meet the SLO, against \hlRnAlone\% when
the GPU serves the request alone. Hardware partitioning, such as MIG, gives the request an instance of its own, which
idles whenever the request does.

Research systems keep both the SLO and the utilization by reaching inside the stack: they run all tenants as
threads of one process~\cite{orion}, need them on non-default streams~\cite{xschedrel}, rewrite kernels and replace
closed-library ones with CUTLASS~\cite{tally}, split or recompile kernels~\cite{hummingbird,morphx}, or patch the driver~\cite{gpreempt,valve,gpuext}, all out of reach for an
operator who receives tenants as they are and cannot rewrite the closed kernels of cuBLAS or cuDNN. TGS~\cite{tgs}, which runs unmodified tenants, nearly stops the best-effort job instead: in
Figures~\ref{fig:cdf-pair} and~\ref{fig:bars-pairs} it meets the SLO for \hlRnTgs\% of requests while training runs at \hlRnTgsThr\% of its
speed.

This paper argues that reaching inside is unnecessary. Between processes, which run in separate contexts, the
time-slicer built into commodity GPUs is already a preemptive scheduler: it stops a running kernel within a fraction of
a millisecond once the running context has used its slice, and it switches the moment a context runs out of work
(\S\ref{sec:bg-timeslicer}). What it lacks is priority. It is fair, so a best-effort context that has work queued, as
a training framework that launches ahead of the GPU almost always does, receives every other slice and every idle
moment between a request's kernels. Priority can therefore be created from outside: if the best-effort context has
nothing to run while a request needs the GPU, the time-slicer gives the request the GPU and preempts the running
best-effort kernel.

Berth is a user-space system built on this observation. A shim preloaded into every tenant sees each kernel launch
inside the driver, including launches from closed libraries, and feeds the best-effort job to the GPU only when no
request needs it, which we call \emph{feed control}. While a latency-critical request is in flight, the shim holds the
best-effort process's launches (the \emph{request gate}); otherwise it keeps the queued best-effort work, predicted
from a profile recorded once, within one small bound per GPU (the \emph{queue bound}).

When a request needs the GPU depends on where its deadline starts. A request server's
deadline runs from each request's arrival, outside the shim's view, so Berth holds the best-effort job for the whole
request and a short grace period after it. A continuous-batching LLM engine decodes almost without pause, and its
per-token deadline runs from the start of each step, so Berth feeds the best-effort job while the step's remaining
slack, measured against the engine's observed step time, allows. Each tenant declares this \emph{deadline reference}
with its SLO, and a tenant with a looser SLO holds its launches while one with a tighter SLO has a request in flight
(the \emph{tiered feed}).

A best-effort kernel that
is already running when a request arrives can last tens of milliseconds; Berth shortens the best-effort context's time
slice through the unprivileged control call the driver itself uses to set it, so that such a kernel costs a request a
fraction of a millisecond per turn. Berth changes no application, framework, kernel or driver, and needs no
privileges.

Neither mechanism suffices alone. The slice alone bounds a running kernel, not the queue ahead of a request: with ten
tenants it leaves the worst tenant at \scSliceTen\% SLO attainment (Figure~\ref{fig:ablation}). The request gate alone lets a backlog build
between requests and the queue bound alone lets the job run during them; with tight SLOs or high load, either alone
lowers the worst tenant's SLO attainment by \fcLossLo--\fcLossHi{} points (Figure~\ref{fig:feed}). Feed control alone cannot
shorten a running kernel: next to the \emph{GEMM job}, back-to-back \kMaxGemm-ms matrix multiplications (GEMMs), ResNet-50
meets its SLO for \klFeedGrpGemm\% of requests under feed control alone and \scSliceRnGemm\% under the slice alone.

We compare Berth with \nBaselines{} baselines, including the released code of XSched, TGS, Orion and Tally, on \nPoints{}
scenario points with MLPerf SLOs and on \nLlmPoints{} points with vLLM services (\S\ref{sec:eval}); a tenant that a policy cannot run
counts as 0\% SLO attainment. Berth meets SLOs within a few percentage points of running alone (\aggBerthAtt\% against \aggAloneAtt\% on
average), above every baseline (Figure~\ref{fig:intro}). The baselines that come closest wall off or throttle the best-effort job; Berth keeps
\thrRatioRange$\times$ their best-effort throughput, about as much as plain time-slicing keeps while it misses
half of the requests. Beside vLLM services, Berth meets both targets for nearly every request while training
keeps running, where stock sharing meets almost none.

This paper contributes the observation that the stock time-slicer preempts running kernels and withholds only
priority (\S\ref{sec:background}); feed control and deadline references, which create that priority from outside with
a bound on the delay (\S\ref{sec:design}); a fixed short slice on the best-effort context alone, set by rewriting the
driver's own control call, that bounds what a running kernel costs under feed control (\S\ref{sec:impl}); and an
evaluation against \nBaselines{} baselines on \nPoints{} scenario and \nLlmPoints{} LLM points (\S\ref{sec:eval}).

\section{Background and Motivation}
\label{sec:background}

\subsection{How processes share a GPU}
\label{sec:bg-sharing}

On NVIDIA GPUs, each process has its own context, and the work of a context reaches the GPU through channels that
belong to one time-slice group (TSG)~\cite{bakita_demystify}. The GPU's runlist scheduler runs one TSG at a time in
work-conserving round-robin order: a TSG runs until its time slice expires, at which point the scheduler preempts its
in-progress work and switches to the next TSG with work, or until it runs out of work, which also triggers a
switch~\cite{bakita_partition}. Kernels of different processes therefore never run at the same time, but a long kernel
cannot keep another process off the GPU for longer than a slice. We call this temporal sharing the
\emph{time-slicer}. Four GPU-wide presets (\texttt{nvidia-smi compute-policy -{}-set-timeslice} 0--3)~\cite{nvsmi} set every
context's slice: in our probes the default and short presets both give about 2~ms, and medium and long make a probe
wait up to \tsMediumMax{} and \tsLongMax~ms (Table~\ref{tab:tsprobe}). The driver itself sets each context's slice
through an unprivileged control (\S\ref{sec:design-slice}). GPreempt and gpu\_ext set per-context slices in a modified
driver~\cite{gpreempt,gpuext}, and XSched's Level~3 preempts a context by zeroing its slice through the same control,
then restores it~\cite{xsched}; Berth gives the best-effort context alone a fixed short slice, set once by rewriting the
driver's own call rather than at each preemption, and composes it with feed control.

Under the Multi-Process Service (MPS), the kernels of all clients run concurrently on the streaming multiprocessors
(SMs) in one shared TSG, so the time-slicer never switches between clients~\cite{bakita_partition}; its client
priority (only a performance hint to the CUDA driver~\cite{mpsdoc}), thread caps and static SM partitions stop no running block. Multi-Instance GPU (MIG) splits a GPU into up to seven instances~\cite{migdoc}, each owned whether
used or not.

\subsection{What the time-slicer gives and withholds}
\label{sec:bg-timeslicer}

Prior work measured 2-ms slices on a consumer GPU~\cite{bakita_demystify}; we characterized our testbed's
data-center GPU (\S\ref{sec:eval-setup}) from a latency-critical tenant's view. A \emph{hog} process
runs back-to-back matrix multiplications of 8192$^2$ elements (one \tsHogLong-ms kernel each) or 2048$^2$ elements
(\tsHogShort~ms each, with a synchronization after each). A \emph{probe} process in another context, at random
intervals of 20--60~ms, launches a tiny kernel or a 4096$^2$ multiplication and records its completion time
(Table~\ref{tab:tsprobe}). The 4096$^2$ probe takes \tsRunUninterrupted~ms when it runs uninterrupted on the busy GPU (medium and long slices)
and \tsRunAlone~ms alone, after the GPU has idled to lower clocks.

\begin{table}[t]
\centering
\footnotesize
\begin{tabular}{@{}llrrrr@{}}
\toprule
Probe & Other process & Slice & p50 & p99 & max \\
\midrule
Tiny kernel & none & -- & 0.05 & 0.08 & 0.08 \\
Tiny kernel & 61-ms kernels & default & 0.27 & 1.62 & 2.26 \\
Tiny kernel & 61-ms kernels & short & 0.28 & 1.70 & 2.37 \\
Tiny kernel & 61-ms kernels & medium & 11.0 & 33.7 & 33.8 \\
Tiny kernel & 61-ms kernels & long & 22.9 & 42.6 & 42.8 \\
Tiny kernel & 1-ms kernels & default & 0.69 & 1.22 & 1.23 \\
4096$^2$ GEMM & none & -- & 13.3 & 13.4 & 13.4 \\
4096$^2$ GEMM & 61-ms kernels & default & 16.0 & 16.1 & 17.7 \\
4096$^2$ GEMM & 1-ms kernels & default & 13.0 & 13.6 & 13.6 \\
4096$^2$ GEMM & 1-ms kernels & medium & 8.5 & 9.1 & 9.2 \\
4096$^2$ GEMM & 1-ms kernels & long & 8.5 & 9.1 & 9.1 \\
\bottomrule
\end{tabular}

\caption{Completion time (ms) of a probe kernel while another process keeps the GPU busy, per time-slice setting.
``Tiny'': a 1,024-element addition; ``none'': the probe alone; \tsNmin--\tsNmax{} probes per row; the GEMM rows omit
the short setting, which behaves as the default.}
\label{tab:tsprobe}
\end{table}

\para{It preempts running kernels.} Against the hog of \tsHogLong-ms kernels, a tiny kernel completes in
\tsPreemptPfifty~ms at the median, against \tsPingAlone~$\mu$s alone: a switch costs about \tsSwitchCost~ms. The
scheduler preempts the hog as soon as the probe's context has work, unless the hog's slice is fresh, in which case the
probe waits out the rest of that slice: the worst case is \tsPreemptMax~ms, one slice plus the switch. The switches cost the hog \tsHogLoss\% of its throughput.

\para{It switches on idle.} Against the \tsHogShort-ms kernels with an idle gap after each, the tiny kernel waits only
for the hog kernel already running (median \tsIdlePfifty~ms, maximum \tsIdleMax~ms): the time-slicer switches as soon
as a context goes idle, without waiting for the slice to expire.

\para{Switching on idle cuts both ways.} A ResNet-50 request (batch 8) keeps the GPU busy for only \busyResNet\% of
its \svcResNet-ms service time; in the gaps between its kernels, a best-effort context with queued work takes the GPU
for up to a full slice. The time-slicer thus has the preemption a latency-critical tenant needs, for every kernel; what
defeats it is the other context's queue, which a training framework almost always keeps full. Table~\ref{tab:systems} compares
what existing systems require and provide.

\begin{table*}[t]
\centering
\scriptsize
\newcommand{\hd}[1]{\rotatebox{60}{\shortstack[l]{#1}}}
\setlength{\tabcolsep}{2pt}
\renewcommand{\arraystretch}{1.0}
\begin{tabular*}{\textwidth}{@{\extracolsep{\fill}}l*{9}{c}l@{\hspace{30pt}}}
 & \hd{App \& framework\\unmodified} & \hd{Closed kernels\\as shipped} & \hd{Stock driver} & \hd{No privileges} & \hd{Preempts\\running kernel} & \hd{Prioritizes\\the request} & \hd{Several LC\\tenants} & \hd{LLM engines\\(per-step)} & \hd{Best-effort\\keeps running} & \hd{In our\\comparison} \\
\midrule
Time-slicing & \cmark & \cmark & \cmark & \cmark & \cmark & \xmark & \cmark & \xmark & \cmark & \cmark \\
MPS variants & \cmark & \cmark & \cmark & \cmark & \xmark & \xmark$^{\mathrm{a}}$ & \cmark & \xmark & \cmark & \cmark \\
MIG~\cite{migdoc} & \cmark & \cmark & \cmark & \xmark & \xmark & \cmark & \cmark$^{\mathrm{b}}$ & \xmark & \cmark$^{\mathrm{b}}$ & \cmark \\
TGS~\cite{tgs} & \cmark & \cmark & \cmark & -- & \cmark$^{\mathrm{c}}$ & \cmark & -- & \xmark & \xmark$^{\mathrm{d}}$ & \cmark \\
XSched~\cite{xsched} & \cmark$^{\mathrm{e}}$ & \cmark & \cmark & \cmark$^{\mathrm{f}}$ & \cmark$^{\mathrm{f}}$ & \cmark & \cmark & -- & \cmark & \cmark \\
Orion~\cite{orion} & \xmark$^{\mathrm{g}}$ & \cmark & \cmark & -- & \xmark & \cmark & \xmark & \xmark & \cmark & \cmark \\
Tally~\cite{tally} & \xmark$^{\mathrm{h}}$ & \xmark & -- & -- & \cmark$^{\mathrm{i}}$ & \cmark & \xmark & -- & \cmark & \cmark \\
LithOS~\cite{lithos} & \cmark & \cmark & -- & -- & \cmark$^{\mathrm{i}}$ & \cmark & \cmark & \xmark$^{\mathrm{j}}$ & \cmark & re-impl.$^{\mathrm{k}}$ \\
Hummingbird~\cite{hummingbird} & -- & \xmark & \xmark$^{\mathrm{l}}$ & \xmark$^{\mathrm{l}}$ & \cmark$^{\mathrm{i}}$ & \cmark & -- & \cmark & \cmark & -- \\
MorphX~\cite{morphx} & \xmark & \xmark & -- & -- & \cmark$^{\mathrm{i}}$ & \cmark & -- & \xmark$^{\mathrm{j}}$ & \cmark & built \\
GPreempt~\cite{gpreempt} & \xmark & \cmark & \xmark & \xmark & \cmark & \cmark & \cmark & -- & \cmark & built \\
gpu\_ext~\cite{gpuext} & \cmark & \cmark$^{\mathrm{m}}$ & \xmark & \xmark & \cmark & \cmark & \cmark & \xmark$^{\mathrm{j}}$ & \cmark & -- \\
Green contexts~\cite{greenctx} & \xmark & -- & \cmark & -- & -- & -- & -- & -- & -- & not run \\
Valve~\cite{valve} & \xmark$^{\mathrm{n}}$ & \cmark & \xmark & \xmark & \cmark & \cmark & -- & \cmark & \cmark$^{\mathrm{o}}$ & -- \\
\textbf{Berth} & \cmark & \cmark & \cmark & \cmark & \cmark & \cmark & \cmark & \cmark & \cmark & \textbf{(this paper)} \\
\bottomrule
\end{tabular*}
\caption{What protecting a latency-critical (LC) tenant requires and provides, according to each system's paper
(vendor mechanisms: documentation and our measurements): \cmark{} as stated, \xmark{} the opposite, -- not stated; last
column: as released (\cmark), re-implemented, built, not run, or -- (not in our comparison) (\S\ref{sec:related}).
Superscripts mark partial support: $^{\mathrm{a}}$MPS
priority is only a performance hint to the CUDA driver; $^{\mathrm{b}}$at most seven instances, one needed for the best-effort job;
$^{\mathrm{c}}$by the stock time-slicer between its separate contexts~\cite{bakita_partition};
$^{\mathrm{d}}$at a launch rate limited by design; $^{\mathrm{e}}$tenants on non-default streams;
$^{\mathrm{f}}$Level~3 zeroes the context's slice, unprivileged in our runs; our configuration (\S\ref{sec:eval-setup}) is
Level~1; $^{\mathrm{g}}$tenants as threads of its
process; $^{\mathrm{h}}$its own PyTorch build; $^{\mathrm{i}}$at block or split-kernel boundaries; $^{\mathrm{j}}$LLMs
evaluated without per-step targets; $^{\mathrm{k}}$as TPC quotas (LithOS-like): quotas of texture processing clusters (TPCs) and stealing only (no atomizer,
right-sizing or frequency scaling); $^{\mathrm{l}}$its memory-offloading module runs on a patched unified-memory (UVM) driver; $^{\mathrm{m}}$device-side
policies need PTX, CUDA's virtual instruction set; $^{\mathrm{n}}$a patch to the offline engine; $^{\mathrm{o}}$after a cooldown once the online job is
idle.}
\label{tab:systems}
\end{table*}

\section{Design}
\label{sec:design}

\subsection{Setting and goals}

\begin{figure}[!b]
\centering
\includegraphics[width=\columnwidth]{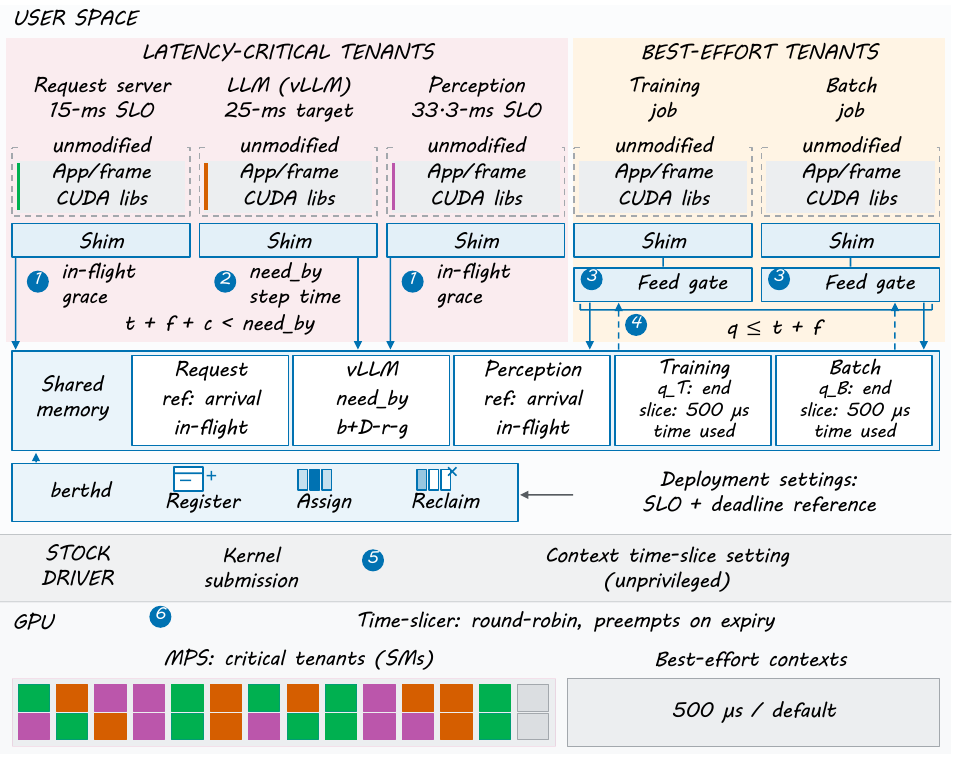}
\caption{Berth on a shared GPU: a shim per tenant and \texttt{berthd}'s shared-memory table.
Gray parts are untouched: applications, frameworks, CUDA libraries, the stock driver and the GPU. The mechanisms of
\S\ref{sec:design-feed}--\ref{sec:design-slice}, in the order a launch meets them: \ding{182}~the in-flight flag and
the grace period; \ding{183}~$\mathit{need\_by}$ from the declared step deadline and the observed step time;
\ding{184}~the feed gate; \ding{185}~one queue budget $f$ per GPU; \ding{186}~the 500-$\mu$s best-effort slice, set
through the driver's own control (the default while an engine is registered and no request is in flight). \ding{187}~The GPU's own time-slicer (\S\ref{sec:background}) does every
preemption.}
\label{fig:overview}
\end{figure}

\looseness=1 %
A GPU is shared by \emph{latency-critical tenants}, which serve requests or run engine steps under an SLO, and
\emph{best-effort tenants}, typically training jobs, each an unmodified process; one operator starts every tenant,
as Orion states~\cite{orion} and TGS and REEF imply~\cite{tgs,reef}. Berth's goal is that every request or
step completes about as fast as alone while best-effort tenants use the rest, without changing applications, frameworks, kernels or the
driver, and without privileges.

Berth keeps the time-slicer as the preemption mechanism and controls only what it sees, through a shim per tenant
and a daemon, \texttt{berthd} (Figure~\ref{fig:overview}). Its principle: never leave a tenant's work queued on the GPU ahead
of a request that has less slack than that tenant.

\subsection{Feed control}
\label{sec:design-feed}

\begin{figure}[t]
\centering
\includegraphics[width=\columnwidth,height=2.566in,keepaspectratio]{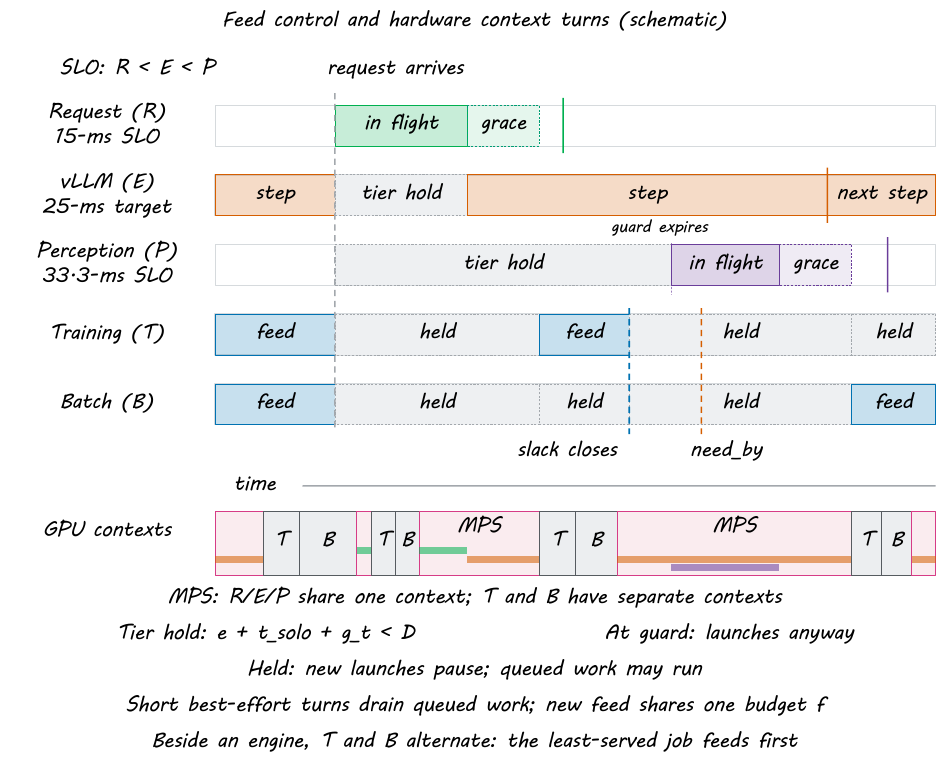}   %
\caption{A feed-control episode (schematic): latency-critical tenants R, E and P (SLOs 15~ms $<$ 25-ms step target
$<$ 33.3~ms) in the MPS context, best-effort jobs T and B in their own, and the GPU's context turns. \emph{Held}: no
new best-effort launch (queued work may run); \emph{tier hold}: a looser-SLO tenant waits while a tighter one is in
flight, until its guard expires; \emph{grace}: the flag stays up 0.5~ms after a request; \emph{slack closes}: feeding
stops when it would delay the step past $\mathit{need\_by}$.}
\label{fig:mechanism}
\end{figure}

A best-effort tenant's shim holds a launch while either condition holds:
\begin{enumerate}[label=(\alph*),leftmargin=*]
\item \emph{Request gate:} a latency-critical tenant needs the GPU.
\item \emph{Queue bound:} more than \feedms of best-effort work, in predicted GPU time (\feedMs~ms,
\S\ref{sec:design-calibration}), is queued.
\end{enumerate}
The shim tracks queued work as a FIFO queue: with $q$ the predicted completion of the work
queued, a launch at time $t$ is held while $q > t + f$; once released, its kernel of predicted duration $d$ starts at
$\max(t, q)$ and $q$ becomes $\max(t,q)+d$, so a kernel longer than $f$ is admitted whenever little is queued ahead. The queue is tracked per
launching thread. The budget is per GPU, not per
process (the \emph{shared budget}): $p$ best-effort processes, each with its own budget, could queue $pf$ ahead of a
request, so each publishes $q$ in its
shared-memory slot and holds a launch while $q$ plus the others' outstanding work exceeds $t+f$. With feed control
(Figure~\ref{fig:mechanism}), a request finds at most \feedms of queued best-effort work and the running kernel, after
which the best-effort context has nothing to run until the request no longer needs the GPU.

\subsection{Deadline references}
\label{sec:design-reference}

When a latency-critical tenant needs the GPU depends on where its deadline starts, which each tenant declares beside
its SLO as its \emph{deadline reference}: a request's arrival or the start of an iteration of its own loop.

\para{Arrival deadlines.} A request server's SLO runs from each request's arrival, and the shim sees a request only
from its first kernel launch, so its remaining slack cannot be known. Berth therefore holds the best-effort job for the whole request. Its shim sets the
tenant's \emph{in-flight} flag at the first kernel launched after a synchronization and clears it when the process
next synchronizes the device, which is how a framework waits for a request's result (\S\ref{sec:impl}); the gate holds while the flag is set and for
a \emph{grace period} of 0.5~ms after it clears. Under load the next request is already
queued and launches its first kernel within that period, so it finds no best-effort work newly queued ahead of it.

\para{Per-iteration deadlines.} A continuous-batching serving engine decodes whenever it holds requests, so it is in a
step almost all the time, and its per-token target runs from the start of each step. Holding the best-effort job for
every step would starve it, but a step usually needs less than its target, and Berth spends the slack. At each launch the engine's shim publishes
\[ \mathit{need\_by} = b + D - r - g, \]
where $b$ is the step's start, $D$ the per-step deadline, $g$ a guard of 20\% of $D$ (at least 2~ms), and $r$ the
step's remaining work: the engine's \emph{observed} step time (an exponentially weighted average, weight 0.2) times
the fraction of the step's profiled work not yet launched, or the whole observed step time if the engine has no
profile. The gate then follows a \emph{slack rule}: while an engine is in a step, a best-effort launch proceeds as long
as $t + f + c < \mathit{need\_by}$, where $c$ allows 0.3~ms for a switch, and the queue bound still applies. The
observed step time includes the slowdown that sharing causes, so the rule leaves more room when the GPU is idle and
less when it is busy. When both kinds of tenant share a GPU, a
request server's in-flight request holds the gate regardless of the engines' slack.

\para{Best-effort arbitration.} When the gate admits, a best-effort job that has received more than $f$ of predicted
GPU time beyond the least-served job waits; a job that has not launched for 50~ms (idle or exited) is left out of the
comparison. The arbitration acts only while an iteration-deadline tenant is registered: then the slack rule, not the
slice, decides which job runs, and one admitted launch gives a long-kernel job tens of milliseconds of GPU time, while a
short-kernel job must pass the gate thousands of times for as much.

\subsection{Tiered feed}
\label{sec:design-tiered}

Latency-critical tenants compete too: grouped under one MPS server (\S\ref{sec:design-layouts}), a ResNet-50
request (15-ms SLO) arriving during a BERT-large request (130-ms SLO) runs slower than alone, though BERT-large has more
than 100~ms of slack. The tiered feed applies the request gate among
latency-critical tenants: a tenant's shim holds its launches while a tenant with a \emph{tighter} SLO has a request in
progress. The hold lasts while $e + t_{\text{solo}} + g_t < D$, where $e$ is the time since the held tenant's current
request released its first kernel (or, while that kernel is held, since the hold began), $t_{\text{solo}}$ the held
tenant's service time alone, $D$ its SLO and $g_t$ a 2-ms guard; once the inequality fails, the tenant launches anyway.
Later holds in the same request keep measuring $e$ from that first release, so all holds of one request share one guard. $t_{\text{solo}}$ is the declared solo time, not an
observed one, so the hold a tenant allows does not grow with the slowdown that holding causes. A request that starts
within the grace period of the previous one begins its own $e$ at its first release; the grace period applies only to
the best-effort gate.

\subsection{Per-tenant time slice from user space}
\label{sec:design-slice}

\looseness=1 %
Feed control leaves at most $f$ of queued work, but a kernel already running can be long: a request that arrives during
one of the GEMM job's matrix multiplications (\kMaxGemm~ms in its profile) shares the GPU with it, slice by slice, until
the multiplication ends. The slice is a property of each time-slice group, and the CUDA driver sets the slice of each
context's compute group to 2{,}048~$\mu$s through the resource manager (RM)
control \texttt{NVA06C\_CTRL\_CMD\_SET\_TIMESLICE}, which the open kernel module's control table marks
\texttt{NON\_PRIVILEGED}~\cite{nvopenrm}. Berth's shim, in the best-effort process only, intercepts the driver's own call and
rewrites the value to 500~$\mu$s; latency-critical contexts keep the default. In a second run of the probe of
\S\ref{sec:bg-timeslicer}, next to its \ghHogKernel-ms GEMMs, this takes a tiny kernel's p99 from \ghPingPnnDef{} to
\ghPingPnnFive~ms. A 100-$\mu$s slice does worse than the default under feed control alone: queued best-effort work
takes more turns, and in the grouped layout every switch preempts all latency-critical kernels at once
(\S\ref{sec:eval-kernel}).

\emph{The slice rule:} while an iteration-deadline engine is registered, the best-effort context runs at the driver's
default slice, because the work the slack rule admits runs in turns inside the steps and a 500-$\mu$s slice would
split it into more turns; while an arrival-deadline tenant has a request in flight, the slice returns to 500~$\mu$s,
since that request cannot wait out a long turn.

\subsection{The delay a request can see}
\label{sec:design-bound}

Let $W$ be the best-effort work on the GPU when a request launches its first kernel. The queue bound admitted the most
recent best-effort launch only if at most $f$ was queued before it, so $W \le f + k_{\max}$, where $k_{\max}$ is the
job's longest kernel, and the request gate admits nothing new until the request ends. Our training jobs' kernels are short:
ResNet-50 training's longest kernel lasts \kMaxTrainResnet~ms; BERT-large training's median kernel, weighted by GPU
time, lasts \kTwTrainBert~ms and its longest \kMaxTrainBert~ms; the GEMM job runs one \kMaxGemm-ms kernel per step. The
best-effort context gets a turn when the request's own slice expires or its context goes idle while best-effort work
remains, so with $n$ such turns of at most a slice $s$ each (counting a turn in progress when the request arrives)
and a switch cost $c_s$ of about \tsSwitchCost~ms, the request finishes at most
\[ \min(f+k_{\max},\, n\,s) + n\,c_s \]
later than alone: beyond $n\,s$ the delay no longer grows with the kernel length. For an engine step, the slack rule
admits best-effort work only while $t + f + c < \mathit{need\_by}$, so the step is delayed by at most
$\min(f+k_{\max}, n\,s_d) + n\,c_s$ with the driver's default slice $s_d$ (2,048~$\mu$s), and meets its target while that delay fits in the guard and the observed step time predicts the
step.

\looseness=1 %
The bound rests on four assumptions: (1)~the profile predicts the kernels and $q$ tracks their completion: a kernel
unknown to the profile adds nothing to $q$, and a request that holds the GPU for longer than $W$ lets $q$ pass while that
work still waits, so the next request can meet up to $2(f+k_{\max})$ until a synchronization resets $q$; (2)~$f$ is
shared by the best-effort processes but tracked per launching thread within a process, so $m$ threads of one process put
up to $m(f+k_{\max})$ in flight; (3)~requests are separated by synchronizations, and one starting after the grace
period meets a fresh $W$; and (4)~only best-effort work competes with the request: work of other latency-critical
tenants is ordered by the tiered feed, whose guard gives no bound.

\subsection{Two layouts}
\label{sec:design-layouts}

Feed control needs the best-effort job in its own context. In the \emph{separate} layout every tenant is its own
context, and latency-critical tenants time-slice too; in the \emph{grouped} layout, which Berth uses,
they are clients of one MPS server, ordered by the tiered feed and sharing a fault domain~\cite{tgs}, and the
time-slicer acts between that context and the best-effort one.

\subsection{Calibration}
\label{sec:design-calibration}

Berth predicts a best-effort kernel's duration from its position in the iteration, counted per launching thread
from the last device synchronization: a new job runs alone for about 12~s with the shim recording each position's
median duration through CUPTI, CUDA's profiling interface; a position missing from the profile predicts zero. Engines
run without a profile. $f$ trades best-effort
throughput against $W$; we use \feedMs~ms.

\section{Implementation}
\label{sec:impl}

Berth is a shim (\locShim{} lines of C++), a daemon (\texttt{berthd}, \locDaemon{} lines) and a patch to
libsmctrl~\cite{libsmctrl}.

\para{Seeing every launch.} cuBLAS, cuDNN and other libraries launch through private paths that a wrapper of
\texttt{cudaLaunchKernel} misses. libsmctrl registers a callback in the CUDA driver's launch path; we extended it with
a hook the driver calls synchronously for every kernel launch. In a latency-critical tenant the hook raises the in-flight flag and publishes
$\mathit{need\_by}$ (after any tiered-feed hold); in a best-effort tenant it blocks the launching thread inside the
driver while a condition holds, spinning when the queue drains within 0.2~ms and sleeping in 20-$\mu$s steps otherwise.

\para{Request and step boundaries.} The shim interposes on \texttt{cudaDeviceSynchronize},
\texttt{cudaStreamSynchronize} and \texttt{cudaEventSynchronize} (not on blocking copies). Request servers synchronize the device after
each request (\texttt{torch.cuda.synchronize}); a device synchronization ends the request and resets $q$ and the kernel
positions of every thread of the process at their next launch. A serving engine that never synchronizes the
device, such as vLLM, which waits on events for its steps' output, ends a step at every event synchronization; stream
synchronizations end nothing. An engine's shim publishes its average step time in its slot. The shim also hooks graph launches
(\texttt{cudaGraphLaunch}, \texttt{cuGraphLaunch}) and treats a graph as one launch of its profiled duration in the
gate, the budget and $\mathit{need\_by}$. BERT-large training under \texttt{torch.compile}
launches two graphs per step (\cgKernFwd{} and \cgKernBwd{} kernels, \cgMsFwd{} and \cgMsBwd~ms). Next to it, the worst
tenant of the 30\%-load BERT-large pair and of the 40\% mix attains \cgPairOff\% and \cgMixOff\% when the trainer's
graphs are invisible to the gate and \cgPairOn\% and \cgMixOn\% when they are governed, against \cgPairPlain\% and
\cgMixPlain\% beside the plain trainer. The cost is the \cgMsBwd-ms backward graph arriving as one piece of queued work,
which the slice still cuts. On the GH200 (\S\ref{sec:eval-gpu2}) the tenants attain the same either way (the pair
100\%, the mix 98.5\% with the graphs invisible and 98.2\% with them governed), and vLLM's full-graph
decode still issues about 42 launches per step, so $\mathit{need\_by}$ stays current (100\% met either way).

\para{Setting the slice.} The best-effort shim defines \texttt{ioctl()} and, for the RM control that sets its compute
group's slice (\S\ref{sec:design-slice}), records the handles and overwrites the requested value; a watcher thread,
sampling every millisecond, restores the driver's value while an engine is registered and no arrival-deadline request is in flight.

\para{Registration and deployment.} The shim registers a named tenant with \texttt{berthd} at its first CUDA call;
\texttt{berthd} keeps 16 shared-memory slots per GPU and frees an exited process's slot within 500~$\mu$s. An operator
profiles a new best-effort job once (12~s here) and starts each tenant with \texttt{LD\_PRELOAD}, a name, a class and,
if latency-critical, its SLO, solo service time and deadline reference. No setting needs privileges.

\section{Evaluation}
\label{sec:eval}

\begin{figure*}[t]
\centering
\includegraphics[width=\textwidth]{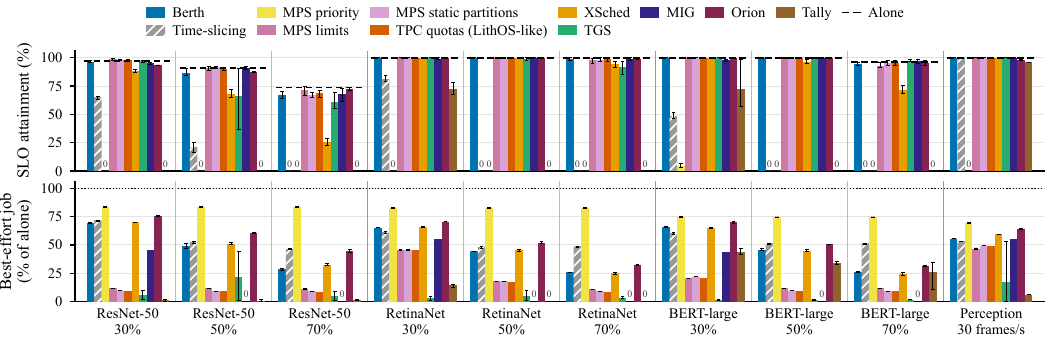}
\caption{Each latency-critical tenant next to BERT-large training at 30\%, 50\% and 70\% of its capacity alone
(perception: 30 frames/s). Top: SLO attainment (dashed: Alone); bottom: training, \% of alone. MPS, DySM-style and
multi-stream: Table~\ref{tab:summary}.}
\label{fig:bars-pairs}
\end{figure*}

\begin{figure*}[t]
\centering
\includegraphics[width=\textwidth]{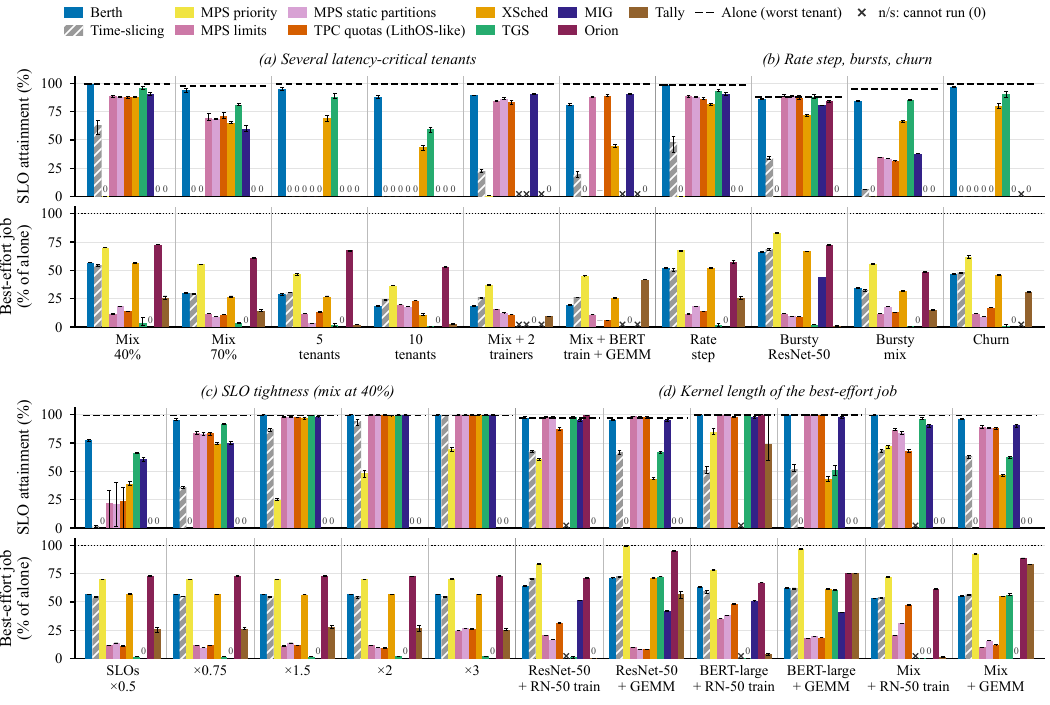}
\caption{The other \nOtherPoints{} points as in Figure~\ref{fig:bars-pairs}, worst tenant. (a)~Several tenants, the last
two with two best-effort jobs; (b)~rate step, bursts, churn; (c)~the 40\% mix with SLOs scaled; (d)~30\% pairs and the
mix next to ResNet-50 (RN-50) training or the GEMM job. --: not run (MPS static partitions beside BERT-large training
and the GEMM job).}
\label{fig:bars-multi}
\end{figure*}

\begin{table*}[t]
\centering
\small
\setlength{\tabcolsep}{1.0pt}
\begin{tabular*}{\textwidth}{@{\extracolsep{\fill}}lrrrrrrrrrrrrrrr@{}}
\toprule
 & \begin{tabular}[b]{@{}c@{}}Alone\end{tabular} & \begin{tabular}[b]{@{}c@{}}\textbf{Berth}\end{tabular} & \begin{tabular}[b]{@{}c@{}}Time-\\slicing\end{tabular} & \begin{tabular}[b]{@{}c@{}}MPS\end{tabular} & \begin{tabular}[b]{@{}c@{}}MPS\\priority\end{tabular} & \begin{tabular}[b]{@{}c@{}}MPS\\limits\end{tabular} & \begin{tabular}[b]{@{}c@{}}MPS static\\partitions\end{tabular} & \begin{tabular}[b]{@{}c@{}}TPC quotas\\(LithOS-like)\end{tabular} & \begin{tabular}[b]{@{}c@{}}DySM-\\style\end{tabular} & \begin{tabular}[b]{@{}c@{}}XSched\end{tabular} & \begin{tabular}[b]{@{}c@{}}Multi-\\stream\end{tabular} & \begin{tabular}[b]{@{}c@{}}TGS\end{tabular} & \begin{tabular}[b]{@{}c@{}}MIG\end{tabular} & \begin{tabular}[b]{@{}c@{}}Tally\end{tabular} & \begin{tabular}[b]{@{}c@{}}Orion\end{tabular} \\
\midrule
SLO attainment, mean (\%) & 97.9 & 95.8 & 48.9 & 3.8 & 24.3 & 86.9 & 87.1 & 85.5 & 29.1 & 71.3 & 24.5 & 84.9 & 88.4 & 24.5 & 52.8 \\
Worst tenant, mean (\%) & 97.5 & 94.1 & 38.4 & 2.3 & 11.8 & 79.7 & 79.1 & 78.4 & 14.9 & 64.4 & 11.5 & 81.5 & 80.3 & 10.2 & 39.7 \\
Points within 5 pp of Alone & -- & 24 & 2 & 0 & 0 & 18 & 17 & 17 & 1 & 8 & 0 & 16 & 16 & 1 & 13 \\
Points it cannot run & -- & 0 & 0 & 0 & 0 & 0 & 0 & 0 & 0 & 4 & 0 & 2 & 0 & 0 & 3 \\
Best-effort job (\% of alone) & 100 & 48 & 51 & 84 & 72 & 16 & 17 & 18 & 76 & 42 & 72 & 9 & 14 & 21 & 58 \\
\bottomrule
\end{tabular*}

\caption{Berth and the \nBaselines{} baselines over all \nPoints{} points (within 5~pp of Alone: every latency-critical
tenant within 5 percentage points of its attainment alone). Unserved tenants count as 0\tabNotRun}
\label{tab:summary}
\end{table*}

\subsection{Methodology}
\label{sec:eval-setup}

\para{Testbed.} One NVIDIA A100 PCIe 40\,GB on Chameleon Cloud with two \hostCpu{} CPUs,
\hostOs, driver \hostDriver, CUDA \hostCuda{} and PyTorch \hostTorch.

\para{Tenants.} Latency-critical tenants are unmodified PyTorch processes that serve open-loop requests first come,
first served; a request's latency runs from its arrival. ResNet-50 (batch 8), RetinaNet and BERT-large have Poisson
arrivals and MLPerf's server-scenario latency constraints of 15, 100 and 130~ms~\cite{mlperf,mlperfrules} as SLOs; a
perception loop (ResNet-50 on four frames) arrives at 30 per second with a 33.3-ms SLO. Alone, their requests take \svcResNet, \svcRetina,
\svcBert{} and \svcPercep~ms. Best-effort tenants run closed loop: BERT-large training (batch 8, sequence 512;
\thrAloneTrainBert{} steps/s alone), ResNet-50 training (batch 64; \thrAloneTrainResnet{} steps/s), and back-to-back
8192$^2$ fp32 matrix multiplications (\thrAloneGemm{} per second), the \emph{GEMM job}.

\para{Scenario points.} SLOs are fixed external values; load is the arrival rate as a fraction of the rate a tenant
sustains alone, summed over a point's tenants. The \nPoints{} points (Figures~\ref{fig:bars-pairs}
and~\ref{fig:bars-multi}) vary load, the tenant mix (\emph{the mix}: ResNet-50, RetinaNet and BERT-large together, each
at a third of the point's load), SLO tightness (0.5 to 3$\times$), the best-effort job, a rate step, bursts replayed from the Azure Functions 2021
trace~\cite{azuretrace} (peak minute \tracePeak$\times$ the mean), churn (\loadChurnPeak\% load
at its peak), five and ten tenants (\loadScaleFive\% and \loadScaleTen\% load), and two best-effort jobs beside the
40\% mix: two trainers, or BERT-large training and the GEMM job. Each
run scores the requests that arrive in a \windowS-s window after a \warmupS-s warm-up; requests still unserved 5~s after
the window ends count as missing their SLO. \emph{Alone} runs each tenant by itself, as the only client of an MPS server, at the point's rate.

\para{Berth and its ablations.} \emph{Berth} runs the grouped layout with $f=\feedMs$~ms shared by the best-effort
jobs, the arbitration, the deadline references with the slack rule and the 0.5-ms grace period, the tiered feed, and
the 500-$\mu$s best-effort slice with its rule. The suite's tenants declare arrival deadlines, the LLM engines of
\S\ref{sec:eval-llm} iteration deadlines, and one LLM point, the only one with a request server beside an engine, mixes
both. Each ablation removes one more mechanism, in this order: the slice rule, the
arbitration, the shared budget (only with two jobs), the grace period,
and the slack rule (every tenant then an arrival-deadline tenant); \emph{Feed only} is feed control with the arrival-deadline
gate (\S\ref{sec:design-reference}) and nothing else, in the grouped and the separate layout (\S\ref{sec:design-layouts}).
To separate the mechanisms (\S\ref{sec:eval-why}), \emph{Slice only} runs the 500-$\mu$s best-effort slice without feed
control or the tiered feed, \emph{Tiered + slice} adds the tiered feed to it, and \emph{Feed + tiered feed} adds the
tiered feed to Feed only (grouped).

\para{Baselines.} \emph{Time-slicing}: each tenant its own process. \emph{MPS}, \emph{MPS priority} (best-effort
clients at below-normal priority), \emph{MPS limits} (active-thread percentages) and \emph{MPS static partitions}
(4-SM chunks; driver 595.91.07, which they require); the last two give each latency-critical tenant the fewest TPCs (two SMs
each) on which an M/D/1 queue's p99 latency stays within 60\% of its SLO, leaving the best-effort job at least four.
\emph{TPC quotas (LithOS-like)}: our re-implementation of LithOS's quotas and stealing~\cite{lithos}, with the quotas of MPS
limits and idle quota lent to best-effort kernels predicted to finish within 0.5~ms; no atomizer, right-sizing or
frequency scaling. \emph{DySM-style}: partitions sized by the same model, resized on reported latency~\cite{dysm}. \emph{XSched}~\cite{xsched}: release v1.2.0, Level-1 queues, tenants on
non-default streams, best-effort in-flight threshold 2; its chip-independent Level~3 (time-slice groups) is worse on the A100
(worst tenant, \xsThreeAMean\% against \xsOneAMean\% over eight points) and better on the GH200 (\S\ref{sec:eval-gpu2}). \emph{Multi-stream}: all tenants
as threads of one process with stream priorities. \emph{TGS}~\cite{tgs}: its release, ported to CUDA~12, scheduling
untouched. \emph{MIG}: the smallest instance per tenant whose predicted SLO attainment reaches 95\%, the largest left
for the best-effort job. \emph{Orion}~\cite{orion} and \emph{Tally}~\cite{tally}: their releases, the tenant with the
tightest SLO as Orion's high-priority client.

\para{Metrics and protocol.} We report each latency-critical tenant's \emph{SLO attainment}, the share of its
requests that meet the SLO, and the best-effort job's throughput relative to the job alone. A point's \emph{worst
tenant} is its latency-critical tenant with the lowest SLO attainment. A tenant that a
policy does not serve counts as 0\%, as does every tenant of a point a policy cannot run (``n/s'': in their releases, XSched's
Level-1 queue fails on ResNet-50 training's launches, TGS supports one low-priority process, and Orion cannot model
churn or two training clients). Averages weight every point equally. Each cell, a point under one policy, is the mean of three runs (MPS static partitions: two, on
driver 595.91.07; one MIG cell: two), and error bars span the runs.

\subsection{Does Berth meet SLOs, and at what cost?}
\label{sec:eval-overall}

Averaged over all points and tenants (Table~\ref{tab:summary}), Berth attains \aggBerthAtt\% (Alone: \aggAloneAtt\%)
and keeps \aggBerthThr\% of the best-effort throughput. The four baselines that come closest in SLO attainment, MIG,
MPS static partitions, MPS limits and TPC quotas (LithOS-like), do so by keeping the best-effort job away from the GPU: Berth keeps
\thrRatioRange$\times$ their best-effort throughput, and every tenant stays within 5 percentage points of
its attainment alone on \aggBerthClose{} of the \nPoints{} points, against at most \aggMpslimitsClose{}.

\begin{figure}[t]
\centering
\includegraphics[width=\columnwidth]{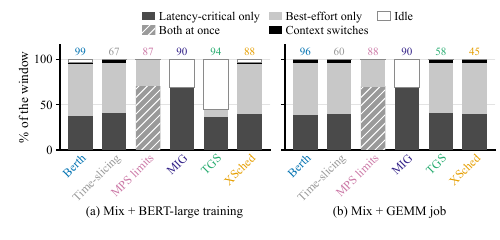}
\caption{Where the GPU's time goes on the 40\% mix (the GPU's compute engine, one traced run per cell, \% of the
window); above
each bar, that run's worst-tenant SLO attainment.}
\label{fig:split}
\end{figure}

A scheduler that serves requests one at a time with the whole GPU attains Alone's SLO attainment and leaves the
best-effort job $1-U$ of the GPU, where $U$ is the share the requests need alone (each tenant's arrival rate times its
request's GPU busy time, summed over the tenants). Averaged over the \nPoints{} points, $1-U$ is \feasCornerAll\%: the
sequential-service bound of Figure~\ref{fig:intro}. It is not a physical limit, since spatial sharing can overlap a
request's idle gaps with best-effort work. No baseline that keeps at least Berth's best-effort
throughput attains more than \frontierAll\% on average
(\frontierAllWho; Figure~\ref{fig:intro}). In Figure~\ref{fig:split}, Berth keeps the GPU busy
\splitBerthBusyLo--\splitBerthBusyHi\% of the window and splits it about as time-slicing does (\splitBerthLc\%
latency-critical, \splitBerthBe\% best-effort, switches \splitBerthCtxLo--\splitBerthCtxHi\%): it fits best-effort work
into the gaps between requests instead of idling the GPU. The job's turns yield
\splitBerthBeThrLo--\splitBerthBeThrHi\% of its throughput alone, slightly less than their share of the window: a
slice-sized turn runs colder than a continuous one, its caches refilled after every switch. MIG idles \splitMigIdle\% and
serves no best-effort job; TGS idles \splitTgsIdle\% beside training (best-effort \splitTgsBe\%) and cannot throttle
the GEMM job; MPS limits keeps best-effort kernels resident \splitLimitsResident\% of the time, yet the job runs at
\splitLimitsThrLo--\splitLimitsThrHi\% of its throughput alone. MPS and MPS priority keep more throughput, but a
request's blocks wait behind the training job's: the worst tenant attains under 1\% on \zeroMps{} points under MPS and
\zeroMpsprio{} under MPS priority. On the \nXsPoints{} points it
can run, XSched attains \R{agg/xs/xsched/attmean}\% at
\R{agg/xs/xsched/thr}\% of the best-effort job's throughput, Berth \R{agg/xs/berth-full/attmean}\% at
\R{agg/xs/berth-full/thr}\%; XSched decides per command, so best-effort work resumes between a request's launches.
Without an atomizer, TPC quotas (LithOS-like) lends idle TPCs to few training kernels and keeps about the best-effort
throughput MPS limits keeps (\aggLithosThr\% against \aggMpslimitsThr\%); Tally's PyTorch build lacks
cuDNN, so most tenants are overloaded even alone.
GPreempt~\cite{gpreempt}, built with its patched 550.120 driver and TVM commit, reproduces its artifact's
microbenchmark: VGG-19 at 100 requests/s beside continuous ResNet-152 is delayed 153~$\mu$s per request, against
631~$\mu$s with two plain contexts, at a 19\% best-effort cost. For ResNet-50 at the 30\% point's rate beside a dense GEMM job,
71.6\% of requests meet the SLO against 93.8\% alone: a per-context slice without feed control or GPreempt's early
execution hint (\S\ref{sec:related}) leaves a request behind the queued best-effort work.

\begin{figure}[t]
\centering
\includegraphics[width=\columnwidth]{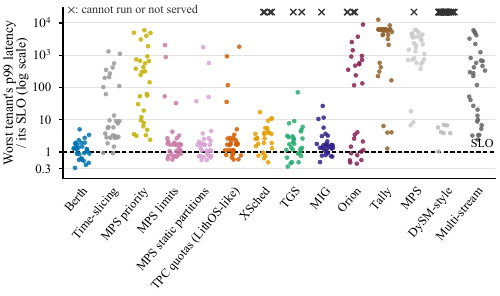}
\caption{The tail: the worst tenant's p99 latency over its SLO per point (log; dashed: the SLO).
$\times$: cannot run or not served.}
\label{fig:tail}
\end{figure}

Berth's tail exceeds the SLO: its worst tenant's p99 is \tailBerthMed$\times$ its SLO at the median point and at
most \tailBerthMax$\times$ (Figure~\ref{fig:tail}); MPS limits, MPS static partitions, TPC quotas (LithOS-like), MIG and TGS are close
at the median (\tailShareMedLo--\tailShareMedHi$\times$) but reach \tailShareMaxLo--\tailShareMaxHi$\times$.
XSched and DySM-style stay within \tailXsMax$\times$ and time-slicing's median is \tailTsMed$\times$; MPS, MPS priority,
multi-stream, Orion and Tally miss by orders of magnitude at the median.

\subsection{What does each mechanism contribute?}
\label{sec:eval-why}

\begin{figure}[t]
\centering
\includegraphics[width=\columnwidth]{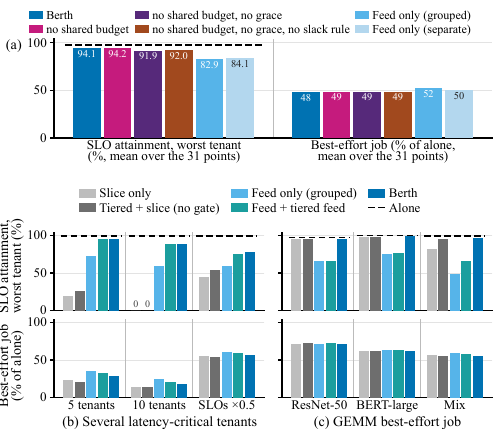}
\caption{What each mechanism contributes. (a)~Berth and five ablations over all \nPoints{} points (without the slice
rule and without the arbitration equal Berth here and are not shown). (b)~Five and ten tenants, the 40\% mix with SLOs halved; (c)~the 30\% ResNet-50 and BERT-large
pairs and the 40\% mix next to the GEMM job, mechanisms added one at a time (grouped; slice only on the ResNet-50
pair: separate). Bottom: best-effort job, \% of alone. Ablation names are cumulative: each also lacks the mechanisms
before it in \S\ref{sec:eval-setup}'s order.}
\label{fig:ablation}
\end{figure}

Each mechanism closes a gap the others leave (Figure~\ref{fig:ablation}(a)). Feed only (grouped) keeps the worst tenant at
\aggFeedGrpAttMin\% on average; the tiered feed and the slice raise it to \aggBerthNgAttMin\%, and the grace period to
\aggBerthNsbAttMin\%, while best-effort throughput goes from \aggFeedGrpThr\% to \aggBerthNsbThr\%. The slack rule changes
nothing on the suite, whose tenants declare arrival deadlines; it decides the LLM points (\S\ref{sec:eval-llm}). The
shared budget acts only with two best-effort jobs (\S\ref{sec:eval-robust}), the slice rule only beside an engine and a request server, and the arbitration only with
two jobs beside an engine (both \S\ref{sec:eval-llm}); over all points Berth attains \aggBerthAttMin\% at \aggBerthThr\%.

The tiered feed and the slice address different points (Figure~\ref{fig:ablation}(b, c)). With five and ten tenants
and with halved SLOs, the tiered feed raises the worst tenant from \FeedGrpScaleFiveMin\%, \FeedGrpScaleTenMin\% and
\R{tight-x0.50/berth-grouped/attmin}\% to \tierOnlyScaleFive\%, \tierOnlyScaleTen\% and \tierOnlyTightHalf\%; next to
the GEMM job the slice raises it (\S\ref{sec:eval-kernel}). For one tenant the slice suffices alone (\scSliceRn\% and
\scSliceRnGemm\% on the ResNet-50 pairs); for several it does not (\nScMulti{} multi-tenant points:
\scMultiSlice\% with the slice alone and \scMultiTierSlice\% with the slice and the tiered feed but no gate, against
\scMultiBerth\% for Berth).

\begin{figure}[t]
\centering
\includegraphics[width=\columnwidth]{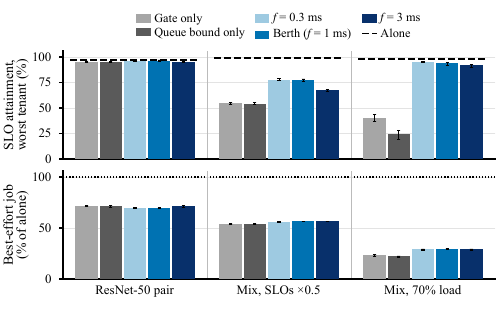}
\caption{Feed control's two conditions and $f$: worst tenant (top; dashed: Alone) and best-effort job (bottom), each
one change to Berth (on the ResNet-50 pair, to Berth without the grace period).}
\label{fig:feed}
\end{figure}

\para{The two conditions and $f$.} On the ResNet-50 pair either condition alone attains what both do
(\rnSubGateOnly\% and \rnSubQbOnly\%, Berth \R{pair-resnet50-load30/berth-full/attmin}\%), but where SLOs are tight or
load is high either alone lowers the worst tenant's SLO attainment by \fcLossLo--\fcLossHi{} points (Figure~\ref{fig:feed}). The gate protects a running request; the queue bound protects the next
one from a backlog built between requests. Attainment is flat for $f$ from 0.3
to 1~ms (with $f=0.3$~ms the worst tenant attains \fcFShortTight\% with SLOs halved and \fcFShortMix\% at 70\% load,
against Berth's \R{tight-x0.50/berth-full/attmin}\% and \R{mix-load70/berth-full/attmin}\%); $f=3$~ms costs
\fcLongCostTight{} points with SLOs halved and buys no throughput.

\subsection{What overhead does Berth add?}
\label{sec:eval-cost}

The launch hook lengthens a tenant's request or step, run alone, by \shimRnPct\%
(ResNet-50, \shimRnLaunches{} launches per request), \shimTrainRnPct\% (ResNet-50 training), or under 1\% (RetinaNet, BERT-large, the
perception loop, BERT-large training, the GEMM job). The gate sleeps in 20-$\mu$s steps, spinning only when the queue
drains within 0.2~ms: on the ResNet-50 pair Berth's processes use \cpuPairBerth\% of one core, against
\cpuPairRef\% under plain time-slicing, where the training job, never held, spins in the driver; \texttt{berthd} uses
\cpuBerthd\%, the slice watcher \cpuWatcher\%, and a latency-critical tenant's shim about \cpuLcShim{} point of one core.
Against Feed only (grouped), Berth gives up \costMedian\% of the job's throughput at
the median point, and most where many tenants hold the job back: \costScaleFive\% and \costScaleTen\% with five and ten
tenants and \costTwobatch\% with two trainers.

\subsection{How do load and tight SLOs affect Berth?}
\label{sec:eval-sweeps}

\begin{figure}[t]
\centering
\includegraphics[width=\columnwidth]{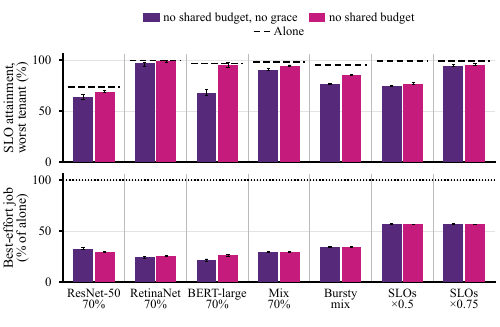}
\caption{The grace period at the highest loads and tightest SLOs: worst tenant (top; dashed: Alone) and best-effort
job (bottom), with and without the grace period. Ablation names are cumulative: each also lacks the mechanisms before
it in \S\ref{sec:eval-setup}'s order.}
\label{fig:grace}
\end{figure}

As SLOs tighten (Figure~\ref{fig:bars-multi}(c)), Berth stays within 5 points of Alone down to 0.75$\times$; at
0.5$\times$ it attains \R{tight-x0.50/berth-full/attmin}\% (TGS \R{tight-x0.50/tgs/attmin}\%).

Load is harder (Figure~\ref{fig:bars-pairs}): at high load, requests run back to back, and the in-flight flag drops
between them (\S\ref{sec:design-bound}'s bound assumes requests separated by synchronizations). The grace period keeps the gate closed across that gap
(Figure~\ref{fig:grace}): BERT-large at 70\% load attains \graceBertSeventy\% with it and \graceBertSeventyNo\%
without, and the 70\% mix \graceMixSeventy\% and \graceMixSeventyNo\%, at
about the same best-effort throughput.

\subsection{Does the slice cap a long kernel's cost?}
\label{sec:eval-kernel}

A long kernel defeats feed control alone, and the slice removes its cost (Figures~\ref{fig:bars-multi}(d)
and~\ref{fig:ablation}(c)). Next to the GEMM job, Feed only (grouped) serves ResNet-50 at \klFeedGrpGemm\%, no better
than time-slicing (\klTsGemm\%) or TGS (\klTgsGemm\%), and its p99 is \klFeedGrpGemmPX$\times$ Alone's, as
\S\ref{sec:design-bound}'s bound predicts for $k_{\max}=\kMaxGemm$~ms.

With the slice, Berth attains \klBerthGemm\% for ResNet-50 (p99
\klBerthGemmPX$\times$ Alone's), \klBerthBertGemm\% for BERT-large and \klBerthMixGemm\% for the mix, and the
GEMM job keeps \klBerthGemmThr\%, \klBerthBertGemmThr\% and \klBerthMixGemmThr\% of its throughput. Under Feed only (grouped), a 100-$\mu$s
best-effort slice in place of the driver's default drops the five-tenant point's two ResNet-50 tenants from
\matGrpRnLo--\matGrpRnHi\% to \hundredGrpRnLo--\hundredGrpRnHi\% and training from \matGrpThr{} to \hundredGrpThr{} steps/s.

\subsection{Does Berth handle many tenants and jobs?}
\label{sec:eval-robust}

Berth leads every baseline with many tenants, which compete too (\S\ref{sec:design-bound}'s bound covers only
best-effort work), and under churn; two best-effort jobs are its weakest case after halved SLOs. With five
and ten tenants, the worst tenant attains \BerthScaleFiveMin\% and \BerthScaleTenMin\% under Berth, TGS
\R{scale-5/tgs/attmin}\% and \R{scale-10/tgs/attmin}\%, XSched \R{scale-5/xsched/attmin}\% and \R{scale-10/xsched/attmin}\%. With two trainers beside the 40\% mix, the
shared budget raises the worst tenant from \R{twobatch-mix-load40/berth-nsb/attmin}\% to
\R{twobatch-mix-load40/berth-full/attmin}\% at
\R{twobatch-mix-load40/berth-nsb/thr}\% and \R{twobatch-mix-load40/berth-full/thr}\% of the jobs' throughput. With
training and the GEMM job, ResNet-50 attains \tjBerthWorst\% while training and the GEMM job keep \tjBerthTrain\% and
\tjBerthGemm\% of their throughput alone: two contexts take two turns per round, doubling \S\ref{sec:design-bound}'s $n$, and
the budget bounds the queue, not the turns. MPS
limits, TPC quotas (LithOS-like) and MIG attain \tjMpslimitsWorst\%, \tjLithosWorst\% and \tjMigWorst\% while keeping
\tjMpslimitsThr\%, \tjLithosThr\% and \tjMigThr\% of the jobs' throughput; time-slicing and XSched attain \tjTsWorst\%
and \tjXsWorst\%.

\begin{figure}[t]
\centering
\includegraphics[width=\columnwidth]{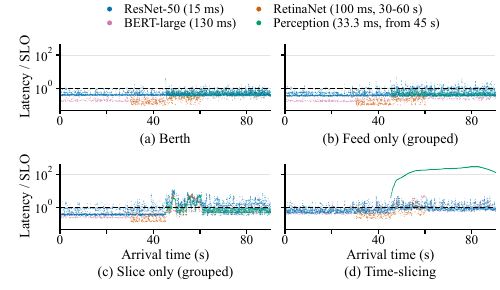}
\caption{Churn: per-request latency over the SLO (log scale; dashed: the SLO) under (a) Berth, (b) feed only
(grouped), (c) slice only (grouped), (d) time-slicing.}
\label{fig:churn}
\end{figure}

Under churn, Berth's worst tenant attains \R{churn-load40/berth-full/attmin}\% (Figure~\ref{fig:bars-multi}(b)); MPS
limits, MPS static partitions and TPC quotas (LithOS-like) leave the perception loop at 0\% once it joins, and MIG
leaves ResNet-50 at 0\% in its one-seventh instance. In Figure~\ref{fig:churn}, once the loop joins, ResNet-50
requests miss under Feed only (grouped), every tenant misses under Slice only while all four run (45--60~s), and under
time-slicing the loop's own latency never recovers.

\subsection{Can Berth co-serve LLM engines?}
\label{sec:eval-llm}

\begin{figure}[t]
\centering
\includegraphics[width=\columnwidth]{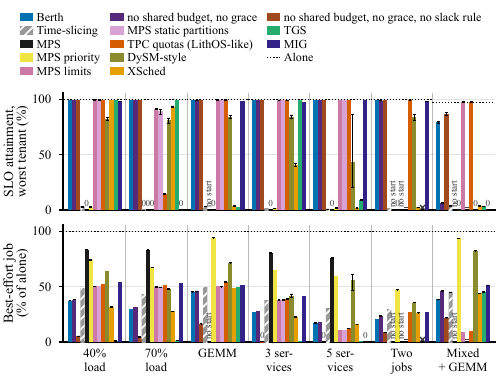}
\caption{LLM points: worst tenant's SLO attainment (top) and best-effort job (bottom; two jobs: mean); dotted:
Alone. Three, five services: 66\%, 67.5\% load. ``no start'': never started; $\times$: cannot
run\llmNotRunNote; ablations lack the slice rule. Multi-stream, Orion and Tally cannot host vLLM. Ablation names are cumulative, as in
Figure~\ref{fig:ablation}.}
\label{fig:llm}
\end{figure}

Berth co-serves vLLM engines with a training job, meeting both targets for nearly every request. We run
vLLM 0.11.0~\cite{vllm} services of Qwen2.5-1.5B and -0.5B~\cite{qwen25} with CUDA graphs and
torch.compile, with
arrivals and lengths from the Azure LLM inference trace 2023~\cite{splitwise}, for 180~s next to BERT-large training or
the GEMM job. Every request must meet a time to first token (TTFT) of 250~ms and a time per output token (TPOT) of
25~ms, tighter than MLPerf's interactive category for an 8B model~\cite{mlperfrules51}; the
services declare iteration deadlines with the TPOT target as $D$. Each point runs a 1.5B and a 0.5B service, except the
three- and five-service points, which add 0.5B services, and the mixed point, which runs one 1.5B service beside
ResNet-50. A service's load is its demanded output tokens per second over its solo capacity, summed over a point's
services.

\begin{figure}[t]
\centering
\includegraphics[width=\columnwidth]{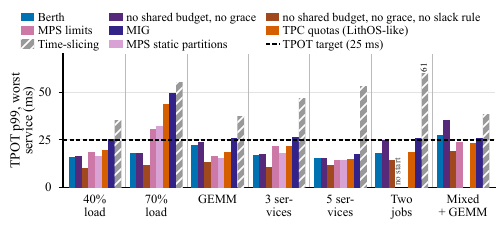}
\caption{TPOT p99 of the worst service per LLM point (``no start'': never started; a number: a bar beyond the axis).
Ablation names are cumulative, as in Figure~\ref{fig:ablation}.}
\label{fig:tpot}
\end{figure}

Under Berth the worst service attains \llmBerthWorstMin--\llmBerthWorstMax\% on every point without a request server (Figure~\ref{fig:llm}),
while training keeps \llmBerthBeLo--\llmBerthBeHi\% of its
throughput alone and the GEMM job \llmBerthBeGemm\%. The slack rule is what lets the best-effort job run: treating the
engines as arrival-deadline tenants, the last ablation also attains \llmBerthNsngWorstMin\% but leaves training
\llmBerthNsngBeLo--\llmBerthNsngBeHi\%, since an engine is in a step almost all the time. The work the slack
rule admits runs inside the steps' slack, as the TPOT tail shows (Figure~\ref{fig:tpot}): the worst service's
TPOT p99 is \llmBerthTpotLo--\llmBerthTpotHi~ms under Berth and \llmBerthNsngTpotLo--\llmBerthNsngTpotHi~ms when the gate
holds for every step, both within the target, and its TTFT p99 at most \llmBerthTtftHi~ms. The guard $g$ trades that tail against the job: from 10 to 30\% of $D$,
TPOT p99 falls from \sensTpotTenA{} to \sensTpotThirtyA~ms and the job's throughput from \sensBeTenA{} to
\sensBeThirtyA\% at 40\% load (\sensTpotTenB{} to \sensTpotThirtyB~ms, \sensBeTenB{} to \sensBeThirtyB\% at 70\%), every
target met; varying $c$ from 0.15 to 0.6~ms changes TPOT p99 by less than \sensCTpot~ms and the job's throughput by less
than 1 point.

The stock mechanisms cannot serve a decode loop: on the points without a request server, under time-slicing, MPS and MPS priority, the worst service's TPOT p99
reaches \llmStockTpotLo--\llmStockTpotHi~ms and it meets at most \llmStockWorstMax\% of its requests. \llmShareNames{} keep the worst service at \llmShareWorstLo--\llmShareWorstHi\% and \llmShareBeLo--\llmShareBeHi\% of
the best-effort job's throughput, with three exceptions. At 70\% load the 0.5B service is too slow on MIG's one-seventh
instance (\llmMigSeventy\%) and on its 8-TPC quota under TPC quotas (LithOS-like), \llmLithosSeventy\%; with five services
MIG has no instance left for training; with two jobs MPS limits does not start. At 40\% and 70\% load and with three
services the partitions keep more training than Berth, with five services less. TGS meets every target on the two- and three-service points beside training but leaves it at most
\llmTgsBeMax\%. With two jobs, the
arbitration decides the split: without it the GEMM job gets \tjlNbaGemm\% and training \tjlNbaTrain\% (\tjlNbaWorst\%
met); with it \tjlBerthGemm{} and \tjlBerthTrain\%, every request met, TPOT p99 \tjlBerthTpot{} instead of
\tjlNbaTpot~ms. MIG serves only training (\tjlMigTrain\%), TPC quotas (LithOS-like) only the GEMM job (\tjlLithosGemm\%);
time-slicing and XSched meet \tjlTsWorst\% and \tjlXsWorst\%.

On the mixed point, ResNet-50 (15-ms SLO) beside a 1.5B service and the GEMM job, the slice rule takes ResNet-50 from
\mxNsrRn{} to \mxBerthRn\% and the service from \mxNsrSvc{} to \mxBerthSvc\% met, at \mxBerthBe\% of the GEMM job's throughput,
\mxBeCost{} points less than without the slice rule; the service's per-step target binds (TPOT p99 \mxBerthTpot~ms against 25), as the job's turns
still take the step's slack. Without the declared reference, Berth meets \mxNsngRn\% and \mxNsngSvc\% but keeps only
\mxNsngBe\% of the job: the declared slack rule gives the GEMM job its room. MPS limits and TPC quotas (LithOS-like)
meet at least \mxSpatialLo\% for both tenants at \mxSpatialBeLo--\mxSpatialBeHi\% of the job; Berth trails them on attainment with about
\mxThrRatio$\times$ their best-effort throughput.

\subsection{Does Berth carry to another GPU?}
\label{sec:eval-gpu2}

\begin{figure}[t]
\centering
\includegraphics[width=\columnwidth]{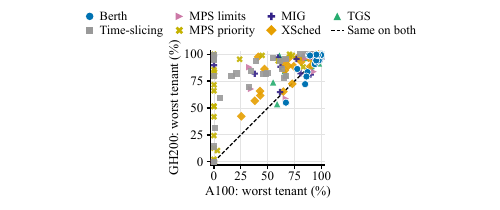}
\caption{Worst tenant per point, A100 (x) vs GH200 (y). Berth stays top right; the baselines rise with the GH200's shorter
kernels.}
\label{fig:gh200}
\end{figure}

On a GH200 (Grace CPU, H100 with 132 SMs) with the same 580.178.04 open driver, tenants,
points and deadlines, Berth's mechanisms carry over: the slice setting is honored (a tiny kernel's p99 under the GEMM hog falls from \ghPortPingDef{} to
\ghPortPingFive~ms, at a \ghPortHogCost\% cost to the hog), the tiered feed
holds launches on every multi-tenant point, and every launch uses all 132 SMs (after a Hopper fix in the mask library). Berth
attains \ghBerthWorst\% for the worst tenant and \ghBerthMean\% on average at \ghBerthThr\% of the best-effort job
(A100, the same \nGhPoints{} points, all but the one with BERT-large training and the GEMM job: \ghABerthWorst\%,
\ghABerthAtt\%, \ghABerthThr\%). Berth kept its A100 settings; MPS limits and MIG were re-sized for this host. Every baseline gains (Figure~\ref{fig:gh200}): the H100's
kernels are about half as long, so a 2-ms slice costs a request less. Berth still leads time-slicing, MPS limits and
MIG by \mbox{\ghLeadLo--\ghLeadHi} points of worst-tenant attainment, with \ghThrVsMig$\times$ MIG's and \ghThrVsLimits$\times$
MPS limits' best-effort throughput. On the points each supports, TGS comes within \ghTgsBehind{} points of Berth's worst tenant at a fifth of Berth's
best-effort throughput, and XSched trails by \ghXsBehind{} points while keeping \ghXsThr\% of the best-effort job against
Berth's \ghBerthOnXsThr\%. \mbox{ResNet-50}, launch-bound on the Grace CPU, misses its SLO even alone at 30\% load and
above; the many-tenant points are a looser test (ten tenants: \ghScaleTen\% on the GH200 against \ghScaleTenA\% on the
A100). On the five LLM points without a second job or a request server (vLLM 0.11's aarch64 wheel), Berth,
MPS limits and MIG meet every target, at \ghLlmBerthThr\%, \ghLlmMpslimitsThr\% and \ghLlmMigThr\% best-effort, and time-slicing attains \ghLlmTsWorst\%: on this GPU only time-slicing falls behind. With Qwen2.5-7B (at
40\% and 70\% load beside training, beside the GEMM job, and with a 1.5B service), Berth's worst service meets both
targets for \sevenBBerthMetLo--\sevenBBerthMetHi\% of requests at \sevenBBerthThrLo--\sevenBBerthThrHi\% best-effort
throughput; time-slicing reaches \sevenBTsMetLo--\sevenBTsMetHi\%, MPS limits
\sevenBMpslimitsMetLo--\sevenBMpslimitsMetHi\%, and MIG \sevenBMigMetLo--\sevenBMigMetHi\% with no instance left for the
best-effort job beside the 49-GB model. With two jobs beside the 7B, the arbitration lifts Berth's worst service from
\sevenTwoNoArbMet\% to \sevenTwoArbMet\% and training from \sevenTwoNoArbTrain\% to \sevenTwoArbTrain\%. XSched at Level~3
attains \ghXsThreeWorst\% at \ghXsThreeThr\% over eight points (Level~1: \ghXsOneWorst\% at \ghXsOneThr\%;
Berth \ghBerthEightWorst\% at \ghBerthEightThr\%).

\section{Discussion and Limitations}
\label{sec:discussion}

\para{Signals Berth does not see.} The shim sees launches and synchronizations, not request queues: a
request that starts after the grace period meets a fresh $W$, and under bursts a held tenant's requests back up.
Berth is weakest there and near saturation, and an exposed queue length would close both gaps.

\para{Profiles and graphs.} A new model, batch size or GPU calls for a new profile. A job whose kernels vary breaks
\S\ref{sec:design-bound}'s assumption that the profile predicts the kernels: BERT-large training with a sequence length drawn per step from 128 to 512
runs up to \vlUnder$\times$ longer than its profile predicts, yet next to it Berth's worst tenant stays within
\vlPairDiff{} points (the 30\% BERT-large pair) and \vlMixDiff{} points (the 40\% mix) of its attainment beside the
fixed-shape job (\vlPair\% against \vlPairFixed\% and \vlMix\% against \vlMixFixed\%): the slice bounds what a misprediction does to a request, so the queue bound degrades and
the SLO does not. A graph is gated as one
launch (\S\ref{sec:impl}); graphs launched from device code are not seen.

\para{What Berth does not govern.} The grouped layout puts the latency-critical tenants in one MPS fault domain, which
the separate layout avoids at the cost of slicing between them, and orders them by the tiered feed's guard, a
heuristic without a bound. Berth does not manage memory capacity (every footprint fits here), copy-engine and PCIe traffic, or L2 and memory bandwidth, of which
latency-critical turns use \smLcDramLlm\% of DRAM read bandwidth on the 40\%-load LLM point (\smLcDramSeven\% at 7B).

\para{Portability.} TPC masks are less portable: the 64-bit
interface reaches 54 of the GH200's 66 TPCs, so TPC quotas (LithOS-like) need 128-bit masks there;
Berth sets no mask.

\para{Trust model.} One operator starts every tenant. The slice control is
unprivileged, so a tenant running its own code could raise its slice: Berth isolates tenants from interference,
not from a hostile tenant. Containment (a filter on that RM control, or \texttt{berthd} checking each
tenant's slice) is future work.

\para{Future drivers.} The control stayed unprivileged on every driver we probed: GPreempt's patched 550, the 560
proprietary and the 580 open module~\cite{gpreempt}, on Ampere and Hopper GPUs; Tegra GPUs remain unchecked. A
driver that privileged it would cost Berth the slice and its long-kernel results (\S\ref{sec:eval-kernel}); feed
control needs nothing from the driver.

\section{Related Work}
\label{sec:related}

\para{Colocation.} CPU colocation relies on interference prediction~\cite{bubbleup}, tail-latency
feedback~\cite{heracles,parties} and microsecond-scale core reallocation and policy switching~\cite{shenango,caladan,akts}; cluster managers fill
spare capacity~\cite{borg}, which tail latency makes delicate~\cite{tailatscale}.

\para{Controlling what reaches the GPU.} OS work schedules GPU submission by priority~\cite{timegraph}, manages GPUs as
resources~\cite{gdev,ptask} or enforces fair shares~\cite{disengaged}. Baymax reorders accelerator tasks by
predicted duration and Prophet predicts co-location interference~\cite{baymax,prophet}; AntMan throttles the
opportunistic job in the framework~\cite{antman}, and Salus and PipeSwitch time-share one GPU between
jobs~\cite{salus,pipeswitch}. TGS, closest to Berth in deployment, adapts the
opportunistic job's launch rate to the production job's throughput~\cite{tgs}; XSched turns streams into preemptible
queues~\cite{xsched}. Orion launches a best-effort kernel only when the high-priority client is idle, or the kernel is
small and complementary in resource profile, bounding outstanding work by predicted duration~\cite{orion}; feed
control applies that intuition across processes and leaves the preemption to the hardware. REEF resets idempotent
kernels on AMD GPUs~\cite{reef} and Picker validates idempotency at each launch~\cite{picker}; Paella schedules kernels
in the serving system~\cite{paella}.

\para{Preemption.} GPU preemption has been proposed in hardware~\cite{tanasic,chimera} and through kernel
transformations~\cite{flep,effisha}. GPreempt's reservation needs an early execution hint, a
blocking kernel launched on the critical stream before each request (100~$\mu$s ahead in its release), which its
scheduled loads allow~\cite{gpreempt}; open-loop arrivals give none, so for our tenants it could only launch at arrival. Its limits (one
process, two contexts, TVM graphs replayed as CUDA graphs, a patched driver; no PyTorch, vLLM, training or second
process) are those of its release, not of its idea. Berth's Slice only configuration (\S\ref{sec:eval-why}) isolates the
mechanism both share, an asymmetric per-context slice (500~$\mu$s; GPreempt's release: 1~s and 1~$\mu$s). gpu\_ext exposes slices and priorities to eBPF policies in the
module~\cite{gpuext}, Valve disables the offline job's channels while an online request is active~\cite{valve}, and GCAPS adds
context-level priorities on embedded GPUs~\cite{gcaps}. XSched's Level~3 preempts a context by zeroing its slice
through the control Berth uses~\cite{xsched}; Berth instead fixes a short best-effort slice once and composes it with
feed control. Tally and Hummingbird rewrite or split best-effort kernels from
PTX~\cite{tally,hummingbird}, LithOS's atomizer splits them without it~\cite{lithos}, and MorphX recompiles them into
persistent workers~\cite{morphx}; closed-library kernels have no PTX to rewrite or recompile, while the slice applies to
every kernel.

\para{Spatial sharing and scheduling above the GPU.} MPS limits, static partitions, MIG~\cite{migdoc,migchar} and
TPC masks~\cite{libsmctrl,bakita_partition} divide the SMs; CUDA's green contexts (12.4 and later) partition SMs among contexts of one process, and an
application must create its streams in them through the driver API~\cite{greenctx}. Systems size or resize such shares for
inference~\cite{gslice,gpulet,usher,sgdrc,lithos,vitamine}, real-time tasks~\cite{dysm}, LLM
serving~\cite{muxserve,bullet,nexusllm} and production clusters~\cite{muxflow}; a
partition idles when its owner does (Figure~\ref{fig:split}).
Serving systems~\cite{clipper,clockwork,nexus,infaas,shepherd,alpaserve}, iteration-level LLM
engines~\cite{orca,vllm,sarathi}, phase-disaggregated serving~\cite{distserve,splitwise} and cluster
schedulers~\cite{gandiva,tiresias,gavel,pollux} work above Berth, which builds on measurements
of NVIDIA's scheduling~\cite{amert,pitfalls,bakita_demystify}.

\section{Conclusion}
\label{sec:conclusion}

The stock time-slicer already preempts kernels within a fraction of a millisecond; it fails latency-critical tenants
only because it shares the GPU with whatever best-effort work is queued. Berth's feed control keeps that queue short,
deadline references extend feed control to serving engines, and a short slice bounds what a long kernel costs. On the \nPoints{} scenario points Berth attains
higher SLO attainment than all \nBaselines{} baselines, with \thrRatioRange$\times$ the best-effort throughput
of the four closest.

\para{AI use.} Claude (Anthropic), used through Claude Code, contributed substantially to this work under the authors'
direction: it wrote much of the shim, daemon and evaluation-suite code, the scripts that run the experiments and
generate the figures, and drafts of the text, which the authors revised. The system's design decisions, the choice of
baselines and scenarios, and the interpretation of the results are the authors'. All numbers were produced by the
code described in \S\ref{sec:impl} on the authors' hardware; the authors verified them and accept full responsibility for the paper's
contents.

{\footnotesize
\bibliographystyle{plain}
\bibliography{refs}}

\end{document}